\documentclass[3p]{elsarticle}
\usepackage{amsmath}

\usepackage{array}
\usepackage{graphicx}

\usepackage{subfigure}
\usepackage{subfigmat}

\usepackage{xcolor}

\usepackage{setspace}

\usepackage{tikz}
\usetikzlibrary{shapes,arrows}

\usepackage{etoolbox}

\usepackage{float}
\begin{document}

\begin{frontmatter}

\title{Combustion Dynamics of Ten-injector Rocket Engine
Using Flamelet Progress Variable}

\author[rvt]{Lei~Zhan}
\author[focal]{Tuan~M.~Nguyen}
\author[rvt]{Juntao~Xiong}
\author[rvt]{Feng~Liu}
\author[rvt]{William~A.~Sirignano}

\address[rvt]{Department of Mechanical and Aerospace Engineering, University of California, Irvine, California 92697, USA}
\address[focal]{Lawrence Livermore National Laboratory, Livermore, CA 94550, USA}

\begin{abstract}

The combustion instability is investigated computationally for a ten-injector rocket engine using the compressible flamelet progress variable (FPV) model and detached eddy simulation (DES). A C++ code is developed based on OpenFOAM 4.1 to apply the combustion model. Flamelet tables are generated for methane/oxygen combustion at the background pressure of $200$ bar using a 12-species chemical mechanism. The flames at this high pressure level have similar structures as those at much lower pressures. A power law is determined to rescale the reaction rate for the progress variable to address the pressure effect. The combustion is also simulated by the one-step-kinetics (OSK) model for comparison with the FPV model. 
Premixed and diffusion flames are identified locally for both the FPV and OSK models. Study of combustion instability shows that a combined first-longitudinal and first-tangential mode of $3200$ Hz is dominant for the FPV model while the OSK model favors a pure first-tangential mode of $2600$ Hz. The coupling among pressure oscillation, unsteady transverse flow and helicity fluctuation is discussed. A preliminary study of the resonance in the injectors, which is driven by the acoustic oscillation in the combustion chamber, is also presented.   

\end{abstract}

\end{frontmatter}

\section{Introduction}

Combustion instability is an acoustical phenomenon in which an unstable pressure oscillation is excited and sustained by combustion. In most high-power propulsion systems for rockets and airplanes, the high-energy release rate reinforces high-amplitude acoustical oscillations. Such oscillations can affect thrust in an undesirable way and sometimes lead to engine destruction. 

For more than a half century, it has been known that the relation between the characteristic time
(rate) for combustion in liquid-propellant rocket engines and the resonant oscillation period
(frequency) is critical in determining the engine stability. Early pioneering work by Crocco and Cheng 
\cite{crocco_1,crocco_2,crocco_3} established the principles using the sensitive time-lag concept in their analyses.
In recent years, combustion instability has been examined with more detailed description of the
combustion process. However, in many large-eddy simulations of the rocket combustion
dynamics, such as those adopting the Laminar Closure Model (LCM) for the reaction source terms, the combustion process was modelled only on the resolved scale although we know the
sub-grid combustion behavior in turbulent flows is dominant in determining burning rate and
energy-release rate. In some later works, the sub-grid effect on combustion was considered but modelling of significant phenomena like local extinction and reignition was missing. Most recently, Nguyen and Sirignano \cite{tuancnf, tuanaiaaj}, Nguyen et al. \cite{tuanjpp} and Shadram et al. \cite{zeinabaiaaj, zeinabcnf} made the effort to apply a cost-effective sub-grid combustion model to simulations of the rocket combustion dynamics where local extinction and reignition can be addressed; however, those works were limited to a description of a
particular experimental one-injector rocket motor. Furthermore, no works have made
comparisons between the predictions for resolved-scale combustion modelling (based on LCM) versus sub-grid
combustion modelling. Here, therefore, we aim to examine a multi-injector rocket engine and
compare the different results for resolved-scale modelling and sub-scale modelling of the
combustion.

In recent years, there is an increasing need for three-dimensional high-fidelity numerical study of combustion instability in multi-injector rocket engines. Compared to the single-injector configuration \cite{tuancnf, tuanjpp}, the combustion instability mechanism is more complex in a multi-injector combustion chamber because of the possible interactions among injectors and interactions with the complicated feed system and upstream manifold. Systematic experimental investigations of combustion instability for such configurations are quite expensive and technically difficult. Compared to experimental measurement, numerical simulations based on high-fidelity turbulence modeling and detailed chemical reaction mechanisms are able to provide more details about the flow physics as well as the combustion dynamics and hence gain a deeper insight into the combustion instability mechanisms. Among various instability modes, the transverse acoustic modes with large amplitude were observed in experiments under certain conditions \cite{dlrbkdexp_a,dlrbkdexp_b,dlrbkdexp_c}. Urbano et al. \cite{multicnf,pci_multi_inj} carried out combustion simulations using an in-house code for a complete small-scale rocket engine with 42 coaxial injectors and a nozzle outlet. Large eddy simulation (LES) was employed and a four-species $H_{2}$/$O_{2}$ reacting mechanism was considered. Combustion instability was studied with emphasis on the interaction between transverse and radial modes. Under the experimental condition which leads to self-sustained oscillations, computational results showed that tangential instability can be triggered when the system is perturbed by superimposing a large-amplitude pressure disturbance with a first transverse modal distribution. Hwang et al. \cite{energies_multi_inj} employed their OpenFOAM-based code to study the combustion instability for the same multi-injector rocket engine. The same turbulence model was used while the combustion model was modified to consider a more detailed chemical mechanism with 8 species and 12 reaction steps. The same type of pressure perturbation was superimposed under the experimentally unstable condition and a pressure disturbance greater than a threshold value was needed to trigger the limit-cycle tangential oscillations. Guo et al. \cite{fier_multi_inj} investigated the transverse combustion instability of a $91$-injector rocket engine using the standard $k-\epsilon$ two-equation turbulence model and the eddy-dissipation combustion model. A single-step global chemical reaction mechanism was adopted to model the chemical reaction process of the Dodecane/oxygen combustion. Self-excited high-frequency combustion instability of first-order tangential (1T) mode was observed. Two different kinds of 1T patterns, standing wave mode and traveling wave mode, were captured for two different oxygen-to-fuel ratios.

Xiong et al. \cite{xiongaiaaj2020, Juntao_CST, xiongaiaaj2022} presented
computational studies of nonlinear rocket-engine combustion instability with 10-, 19-, 30-, and 82-coaxial methane-oxygen injector ports using
a three-dimensional, unsteady, $k-\omega$ shear-stress transport delayed-detached-eddy simulation method with
a one-step chemical kinetic model. The triggered tangential and longitudinal instability modes are obtained
for the 19-injector geometry with combustion chamber diameter of 43 cm (where first-tangential and first-
longitudinal frequencies equate) by pulsing the injector mass flux. It is shown that an oscillating
combustion-chamber flow can be triggered to a new mode with a larger disturbance amplitude. For the 30-
injector combustion engine, the spontaneous longitudinal-mode and tangential-mode instabilities are
observed. The strengths of the two instability modes alternate during the simulation time. Energy appears
to transfer from one mode to the other and back again such that domination alternates between the two
modes. As the mixture becomes less fuel rich and moves towards the stoichiometric value, the oscillation
amplitude decreases. For the 82-injector combustion engine simulations of the Rocketdyne experiment
\cite{Jensen1989}, four different one-step chemical kinetic rates are simulated on the
resolved scale to determine the effect of burning rate. When the normalized kinetic rate is one, implying
the nominal Westbrook-Dryer rate, only the longitudinal-mode instability is observed. As the kinetic rate
decreases, less of the combustion occurs upstream near the longitudinal-mode pressure antinode. The
longitudinal-mode instability becomes weak and the first-tangential-mode instability emerges, which fits
better with experiment. Thus, an implication is that use of kinetics at the resolved scale produces a burning
rate which is too fast.

Due to the finite-rate chemistry assumption, most high-fidelity calculations of combustion instability, even ones with advanced turbulent combustion models \cite{srinivasan,Garby}, employed quite simple chemical
mechanisms that involve few species and reaction steps. Only a limited number of studies,
such as those conducted by Sardeshmukh et al. \cite{sardeshmukh} for 2D axisymmetric and Harvazinski et al. \cite{harvazinski_aiaa_c}
for fully 3D domain, employed the detailed GRI-Mech 1.2 mechanism to study longitudinal
combustion instability in a single element rocket engine. Sardeshmukh et al. \cite{sardeshmukh} observed an
improvement in pressure oscillation amplitude compared to previous axisymmetric calculations
performed by Harvazinski et al. \cite{harvazinski_phd_diss} and Garby et al. \cite{Garby}. However, these studies were computationally expensive because the authors employed LCM, which
directly transport and integrate the chemical source terms for all 31 species. Motivated by the
findings of Sardeshmukh et al. \cite{sardeshmukh}, Nguyen et al. \cite{tuanjpp} developed a new 2D axisymmetric compressible solver using Delayed-Detached-Eddy-Simulation (DDES) turbulence model together with the
compressible version \cite{pecnikaiaaj2012, saghafiancnf} of the Flamelet Progress Variable (FPV) combustion model \cite{piercemoinjfm}. A detailed
27-species mechanism \cite{petersnotes} was employed. Consistent with findings in \cite{sardeshmukh}, the new solver predicted
pressure fluctuation amplitude that is in good agreement with experimental findings. In contrast
to previous LCM simulations, computational cost associated with the new solver is much lower
due to the FPV model. Subsequently, Nguyen and Sirignano \cite{tuancnf}, by examining in detail the
flamelet solutions employed by the solver, found that local flame extinction and re-ignition was a
major mechanism in driving the combustion instability. In the most unstable case, depending on
the time in an oscillation cycle, the flame can either be attached to the backstep or lifted away
from it. This complex flame dynamics cannot be observed, despite also using a complex chemical
mechanism, by Pants et al. \cite{pant}, who used the Steady Laminar Flamelet (SLF) combustion model \cite{nppecs, nptc}.
By parameterizing its flamelet solutions using the scalar dissipation rate instead of the progress
variable used by the FPV, the SLF model could only access the stable burning branch. Therefore,
the flame in these calculations was attached to the backstep where local extinction and re-ignition
does not occur, leading to an under-prediction of the oscillation amplitude. Finally, Nguyen and
Sirignano \cite{tuanaiaaj} compared spontaneous and triggered instability of the single
element rocket engine. By applying an iso-thermal boundary condition, the spontaneous instability occurring in the chamber was suppressed. However, triggered instability could occur in
the otherwise stable domain if sufficiently large perturbation, in the form of inlet mass flow rate
disturbance, was introduced into the chamber.

In this paper, we extend our previous works by applying the flamelet model to the combustion instability
analysis for a realistic rocket engine geometry. Specifically, we perform a three-dimensional high-fidelity turbulent combustion
simulation of a 10-injector rocket engine and investigate the spontaneous combustion instability
using the efficient compressible FPV combustion model and detailed chemical reaction mechanism.
In the following sections, we will first describe the combustion models, followed by an introduction
of the simulation tools that we developed recently. Based on the simulation results, we will
discuss flamelet solutions at the high background pressures, partially premixed turbulent
combustion, and combustion instability analysis for both the FPV and the LCM models. Finally,
we briefly analyze the resonance phenomenon inside the injectors.

\section{The Flamelet Progress Variable Combustion Model}

\subsection{The Governing Equations}
\label{sec:governequ}
For a mixture of multi-species, the Favre-averaged Navier-Stokes equations are written in the following conservative form as 

\begin{equation}
\frac{\partial \bar{\rho}}{\partial t} + \frac{\partial \bar{\rho} \widetilde v_{j} }{\partial x_{j}}  = 0
\label{continuityequation}
\end{equation}

\begin{equation}
\frac{\partial \bar{\rho} \widetilde v_{i}}{\partial t} + \frac{\partial \bar{\rho} \widetilde v_{i} \widetilde v_{j}}{\partial x_{j}} = -\frac{\partial{\bar{p}}}{\partial x_{i}} + \frac{\partial(\tau_{i,j}+\tau^{R}_{i,j})}{\partial x_{j}}
\label{momentumequation}
\end{equation}

\begin{equation}
\frac{\partial \bar{\rho} \widetilde{h_{a}}}{\partial t} + \frac{\partial \bar{\rho} \widetilde v_{j} \widetilde{h_{a}}}{\partial x_{j}} + \frac{\partial \bar{\rho} \widetilde{K}}{\partial t} + \frac{\partial \bar{\rho} \widetilde v_{j} \widetilde{K}}{\partial x_{j}} - \frac{\partial \bar{p}}{\partial t} = \frac{\partial [\widetilde v_{i}(\tau_{i,j}+\tau^{R}_{i,j})]}{\partial x_{j}} + \frac{\partial}{\partial x_{j}}\left[\left(\frac{\lambda}{c_{p}}+\frac{\mu_{t}}{Pr_{t}}\right)\frac{\partial \widetilde{h_{a}}}{\partial x_{j}}\right]
\label{haequation}
\end{equation}
where $\bar{\rho}$ is the mean density, $\widetilde v_{j}$ is the mean velocity vector and $\bar{p}$ is the mean pressure. In the energy equation, $\widetilde{h_{a}}$ is the mean total (or absolute) enthalpy, which is the summation of the mean sensible enthalpy $\widetilde{h_{s}}$ and the mean enthalpy of formation $\widetilde{h_{c}}$, while $\widetilde{K} = \frac{1}{2}(\sum^{n}_{j=1}\widetilde v_{j}\widetilde v_{j})$ is the mean kinetic energy. $\lambda$ and $C_{p}$ are the heat conductivity coefficient and the specific heat at constant pressure. $Pr_{t}$ is turbulent Prandtl number. $\tau_{i,j}$ and $\tau^{R}_{i,j}$ are the molecular and turbulent viscous stress tensors and expressed as 
$\tau_{i,j} = \mu(\frac{\partial \widetilde v_{i}}{\partial x_{j}}+\frac{\partial \widetilde v_{j}}{\partial x_{i}}-\frac{2}{3}\frac{\partial \widetilde v_{k}}{\partial x_{k}}\delta_{i,j})$
and 
$\tau^{R}_{i,j} = \mu_{t}(\frac{\partial \widetilde v_{i}}{\partial x_{j}}+\frac{\partial \widetilde v_{j}}{\partial x_{i}}-\frac{2}{3}\frac{\partial \widetilde v_{k}}{\partial x_{k}}\delta_{i,j})$
with $\mu$ and $\mu_{t}$ being the molecular and turbulent viscosity. The detached-eddy simulation (DES) \cite{sstdes1} based on the $k-\omega$ shear-stress transport (SST) \cite{Mentersstkw} is adopted as the turbulence closure model.

\subsection{The Turbulent Combustion Model}
\label{sec:turcombusmodel}

For the computations of turbulent combustion in this paper, only the mean quantities are transported. Hence, the laminar flamelet solutions need to be convoluted with assumed probability density functions of the independent scalars, including mean mixture fraction $\widetilde{Z}$ and the mean variance of mixture fraction $\widetilde{Z^{''2}}$. The convolution generates flamelet libraries of mean quantities which can be accessed efficiently during the computation of the flow equations. In the FPV model, the flamelet tables are also parameterized by a mean reaction progress variable $\widetilde{C}$. In this paper, the progress variable is defined as the total mass fraction of four major species, which are $CO_{2}$, $H_{2}O$, $H_{2}$ and $CO$. Hence, a mean thermal and chemical quantity $\widetilde{\psi}_{i}$, such as composition or temperature, is given by

\begin{equation}
\widetilde{\psi}_{i}(\widetilde{Z},\widetilde{Z^{''2}},\widetilde{C}) = \int^{1}_{0}\int^{1}_{0}\psi_{i}(Z,C)\tilde{P}(Z,Z^{''2},C)dZdC
\label{convolution}
\end{equation}
where $\tilde{P}$ is the PDF function for $\widetilde{\psi}_{i}$. The PDF for the progress variable is frequently assumed to be the Dirac $\delta$ function of $C$; thus, \eqref{convolution} is reduced to 
\begin{equation}
\widetilde{\psi}_{i}(\widetilde{Z},\widetilde{Z^{''2}},\widetilde{C}) = \int^{1}_{0}\psi_{i}(Z,\widetilde{C})\tilde{P}(Z,Z^{''2})dZ
\label{reduceconvolution}
\end{equation}
Finally, the $\beta$ PDF is assumed for $\tilde{P}$, which gives
\begin{equation}
\tilde{P}(Z,Z^{''2}) = \frac{Z^{\alpha-1}(1-Z)^{\beta-1}}{\Gamma(\alpha)\Gamma(\beta)}\Gamma(\alpha+\beta)
\label{betapdf}
\end{equation}
where $\Gamma$ is the gamma function. $\alpha$ and $\beta$ are shape functions which are defined as $\alpha = Z\gamma$ and $\beta = (1-Z)\gamma$ with $\gamma = [Z(1-Z)/Z^{''2}]-1 \ge 0$. 
  
For compressible combustion, the tables should be generated at different background pressures and pressure should be included as an input to access the tables. Pecnik et al. \cite{pecnikaiaaj2012} showed that the composition of the reacting mixture is not sensitive to the  background pressure and only the reaction rate changes dramatically. This fact justifies the method of using tables under one background pressure (usually the averaged pressure) and rescaling the reaction rate for other pressures. In this paper, we adopt this approach to simplify preparation of the tables and the table lookup process.

At each time step in the CFD computation, the transport equations for scalars $\widetilde{Z}$, $\widetilde{Z^{''2}}$ and $\widetilde{C}$ are solved. When the Lewis number is equal to one, these equations are given as  

\begin{equation}
\frac{\partial \bar{\rho} \widetilde{Z}}{\partial t} + \frac{\partial \bar{\rho} \widetilde v_{j} \widetilde{Z}}{\partial x_{j}} = \frac{\partial}{\partial x_{j}}\left[\left(\frac{\mu}{Sc}+\frac{\mu_{t}}{Sc_{t}}\right)\frac{\partial \widetilde{Z}}{\partial x_{j}}\right]
\label{zequation}
\end{equation}

\begin{equation}
\frac{\partial \bar{\rho} \widetilde{Z^{''2}}}{\partial t} + \frac{\partial \bar{\rho} \widetilde v_{j} \widetilde{Z^{''2}}}{\partial x_{j}} = \frac{\partial}{\partial x_{j}}\left[\left(\frac{\mu}{Sc}+\frac{\mu_{t}}{Sc_{t}}\right)\frac{\partial \widetilde{Z^{''2}}}{\partial x_{j}}\right]+
2\frac{\mu_{t}}{Sc_{t}}\frac{\partial \widetilde{Z}}{\partial x_{j}}\frac{\partial \widetilde{Z}}{\partial x_{j}}-\bar{\rho}\widetilde{\chi}
\label{varzequation}
\end{equation}

\begin{equation}
\frac{\partial \bar{\rho} \widetilde{C}}{\partial t} + \frac{\partial \bar{\rho} \widetilde v_{j} \widetilde{C}}{\partial x_{j}} = \frac{\partial}{\partial x_{j}}\left[\left(\frac{\mu}{Sc}+\frac{\mu_{t}}{Sc_{t}}\right)\frac{\partial \widetilde{C}}{\partial x_{j}}\right]+\widetilde{\omega_{C}}
\label{cequation}
\end{equation}
where $\widetilde \chi$ is the Favre-averaged scalar dissipation rate. By comparing the integral scalar time scale and the turbulent flow time scale, $\widetilde \chi$ is frequently expressed as $\widetilde \chi = C_{\chi}\frac{\tilde \epsilon}{\tilde k}\widetilde{Z^{''2}}$ \cite{nppecs,nptc} with $\tilde k$ and $\tilde \epsilon$ being the mean turbulent kinetic energy and mean turbulent dissipation. A constant value of $2.0$ is often used for $C_{\chi}$. The laminar and turbulent Schmidt number $Sc$ and $Sc_{t}$ are $1.0$ and $0.7$, respectively. The source term $\widetilde{\omega_{C}}$ in equation \eqref{cequation} is the reaction rate for the progress variable. It is the summation of the reaction rates for the individual species involved in the definition of the progress variable and is prepared in the flamelet tables.  

The FPV approach is by nature a flamelet model. It also assumes that the chemical time scales are shorter than the turbulent time scales so that the flame can be viewed as a collection of laminar flamelets. For this reason, the FPV approach is regarded as assuming infinitely fast reactions. This assumption is valid in many turbulent combustion problems and the FPV approach has even been successfully applied to highly turbulent flames like Sandia Flames D and E as shown in the work by Ihme et al. \cite{ihme_1,ihme_2}. In combustion instability problems, the time scales of dominant acoustics are associated with longitudinal and transverse flows in the combustion chamber, and hence are even larger than the turbulent time scales. This fact further justifies the use of the FPV approach in prediction of combustion instability.

It should be noted that ignition delay is an important factor for combustion instability. Sardeshmukh et al. \cite{sardeshmukh} pointed out that ignition delay is closely related to mixing, local extinction and reignition, which are key to the instability mechanism. The timing of these events is directly coupled with the chamber acoustics in the unstable cases. The FPV model is quasi-steady; however, by inclusion of both the unstable and stable burning branches, it has been shown \cite{piercemoinjfm, tuancnf} to reflect very well the extinction, re-ignition, and associated flame-holding that occurs in a dynamic situation, such as turbulent combustion with or without large-amplitude acoustic oscillations. The use of the FPV model has been justified via comparison with a well known Purdue experiment \cite{tuancnf, tuanaiaaj, tuanjpp, zeinabaiaaj, zeinabcnf}. Despite some shortcomings that have been well presented \cite{tuancnf}, the FPV approach is currently the best option available for connection with flow-field simulation as a sub-grid combustion model.

\subsection{The laminar flame calculator}
\label{sec:flamemasterffcmy12}

Here, the laminar flamelet solutions are generated using FlameMaster, an open source C++ program package for zero-dimensional combustion and one-dimensional laminar flame calculations \cite{flamemaster}. The code can be used for steady and unsteady computations of premixed as well as non-premixed flames. For the methane/oxygen combustion that we consider, we use the module of the code that solves the steady flamelet equations in the mixture fraction space by assuming counterflow configuration. To run the code, a chemical mechanism and corresponding thermo and transport data are needed. The background pressure, composition and temperature on both the fuel and oxidizer sides are required. For each given stoichiometric scalar dissipation rate ($\chi_{st}$), the code generates a flame solution in the form of distributions of temperature, composition, reaction rates and heat release rate in the mixture fraction space. Hence, the laminar flamelet solutions are already parameterized by the scalar dissipation rate and the mixture fraction.

Since chamber pressure in the current study is around $200$ bar, a chemical mechanism suitable 
for high-pressure combustion is desired. We adopt a skeletal model of 12 species, which is 
generated by selecting relevant species and reaction pathways from version y of a 119-species 
Foundation Fuel Chemistry Model (FFCM-y) \cite{stanfordffcmy,ffcmcnf}.
The FFCM models were recently developed with uncertainty minimization against up-to-date 
experimental data for the combustion of small hydrocarbon fuels. The models were tested for 
fuels, such as methane, over the pressure range of $10$-$100$ atm. The 12-species skeletal model is called FFCMy-12 mechanism in this paper and the included species are $H_{2}$,  
$H$, $O_{2}$, $O$, $OH$, $HO_{2}$, $H_{2}O$, $CH_{3}$, $CH_{4}$, $CO$, $CO_{2}$, and 
$CH_{2}O$. Reaction rate constants in the model are optimized for the same wide range of pressure, using the FFCM-y computation results as the optimization targets. 
Note that ignition delay is among the optimization targets and it is identified an key factor for combustion instability. Sardeshmukh et al. \cite{sardeshmukh} pointed out the global one-step and two-step kinetics have too few tunable parameters to calibrate for targets like ignition delay over a wide range of operating conditions. The FFCMy-12 mechanism is much more detailed and has sufficiently many tunable parameters which support wide calibration range. Meanwhile, the mechanism is still compact in size compared to even more detailed kinetics. For these reasons, the FFCMy-12 mechanism is an cost-effective choice for high-pressure combustion simulation and combustion instability analysis in this paper.

\subsection{Numerical Solver}
\label{sec:numsol}

To numerically simulate combustion using the FPV approach, a C++ code is developed based on OpenFOAM 4.1. There is a similar open-source code called flameletFoam \cite{mullerflameletfoam}. However, it only provides the original version of the flamelet model. The progress variable approach is not available. Moreover, tabulated temperature is directly used without any correction, which makes the code applicable only to incompressible combustion. Last but not least, since the code is based on the outdated 2.4 version of OpenFOAM, selection of turbulence model is limited. For example, the hybrid RANS/LES based on the SST $k-\omega$ model is not available. In the newly developed code, transport equations \eqref{zequation}, \eqref{varzequation} and \eqref{cequation} are solved for $\widetilde{Z}$, $\widetilde{Z^{''2}}$ and $\widetilde{C}$. These quantities are then used to retrieve composition, $\widetilde{\omega_{C}}$ and $h_{c}$ from the tables. Since $\widetilde{\omega_{C}}$ is tabulated only for background pressure of $200$ bar, it is rescaled using a power law for other values of pressure. We will discuss it in the next section. In this paper, the energy equation is solved for $h_{a}$. After that, sensible enthalpy $h_{s}$ is obtained by subtracting $h_{c}$ from $h_{a}$. Temperature is then calculated from sensible enthalpy $h_{s}$ using the Joint Army Navy NASA Air Force (JANAF) polynomials and tables of thermodynamics \cite{nasa1993}. 

For turbulence modeling, the DES \cite{sstdes1} based on the SST $k-\omega$ model \cite{Mentersstkw}, a standard choice embedded in OpenFOAM 4.1, is selected for both the FPV approach and the one-step-kinetics method. The DES modification in the SST model is implemented by replacing the dissipation term $\beta^{\ast}\rho k \omega$ in the $k-$equation with $\beta^{\ast}\rho k \omega F_{DES}$. $k$, $\omega$ and $\beta^{\ast}$ are the turbulent kinetic energy, specific rate of dissipation and a constant of the model, respectively. The DES limiter $F_{DES}$ is given by:
\begin{equation}
F_{DES} = max \left(\frac{L_{t}}{C_{DES}\Delta}(1-F_{SST}),1 \right)
\label{Fdes}
\end{equation}
where $L_{t}=\frac{\sqrt{k}}{\beta^{\ast}\omega}$ and $L_{grid} = C_{DES}\Delta$ are the turbulent and grid length scales. $\Delta$ is the maximum local grid dimension and the model constant $C_{DES}$ is calibrated to be $0.61$. The coefficient $F_{SST}$ is selected to be the $F_{2}$ blending function \cite{Mentersstkw} of the SST model (by default in OpenFOAM 4.1) in order to reduce the undesired influence of the DES limiter on boundary layers that usually associated with the model of Strelets \cite{sstdes2} (when $F_{SST} = 0$). 

As temperature for each individual species in the flamelet is typically higher than its critical point of gas, all species are in the gaseous phase. The ideal-gas equation of state (EOS) is employed. 
Pressure and temperature of the burning mixture are both high in the present case. At these conditions, the perfect-gas
law describes the real gases well. Hence, the EOS for real gas is replaced by the ideal-gas EOS for simplicity without significant loss of accuracy in the current computations. 
The reacting-mixtures model is adopted to calculate the properties of the mixture. Since compressibility is considered in the combustion simulation, a density-based thermodynamics package is used. The differencing schemes are second-order accurate in both space and time. Implicit backward differencing is selected for time discretization. For spatial discretization, Gaussian integration is chosen. Second-order derivatives are approximated using linear interpolation from cell centers to cell faces.


For comparison, we also implement the one-step-kinetics (OSK) method, which applies the LCM combustion model and only considers the one-step global chemical reaction, by a canonical OpenFOAM code called rhoReactingFoam. The turbulence model, thermodynamics model, and numerical schemes are described above. Xiong et al. used this approach in simulations concerning rocket engines with 10, 19, 30, and 82 co-axial injectors. The code validity and accuracy were verified in those simulations, especially through the favorable comparison with the 82-injector Rocketdyne experiment based on a properly reduced global reaction rate in the OSK model. The fact that the global reaction in the OSK model has to be slowed identifies the need to use a more detailed chemical mechanism, which can be efficiently integrated in combustion simulations by the FPV approach.

Although the other configurations with fewer injectors have no matching experiments in the literature, there still exist the key features of the combustion dynamics in a multi-injector configuration. Meanwhile, the computational cost of the numerical analysis is reduced due to less geometry complexity. Last but not least, the flow within the injector ports can be simulated with a better resolution
compared to that for configurations with many injectors \cite{xiongaiaaj2020,Juntao_CST,xiongaiaaj2022}. To best leverage these benefits, a ten-injector rocket engine is examined in this paper.

\section{Results and Discussions}

\subsection{The Flamelet Solution}
\label{sec:flamelet}

\begin{figure}
    \begin{subfigmatrix}{2}
        \subfigure[S curve]
{\includegraphics{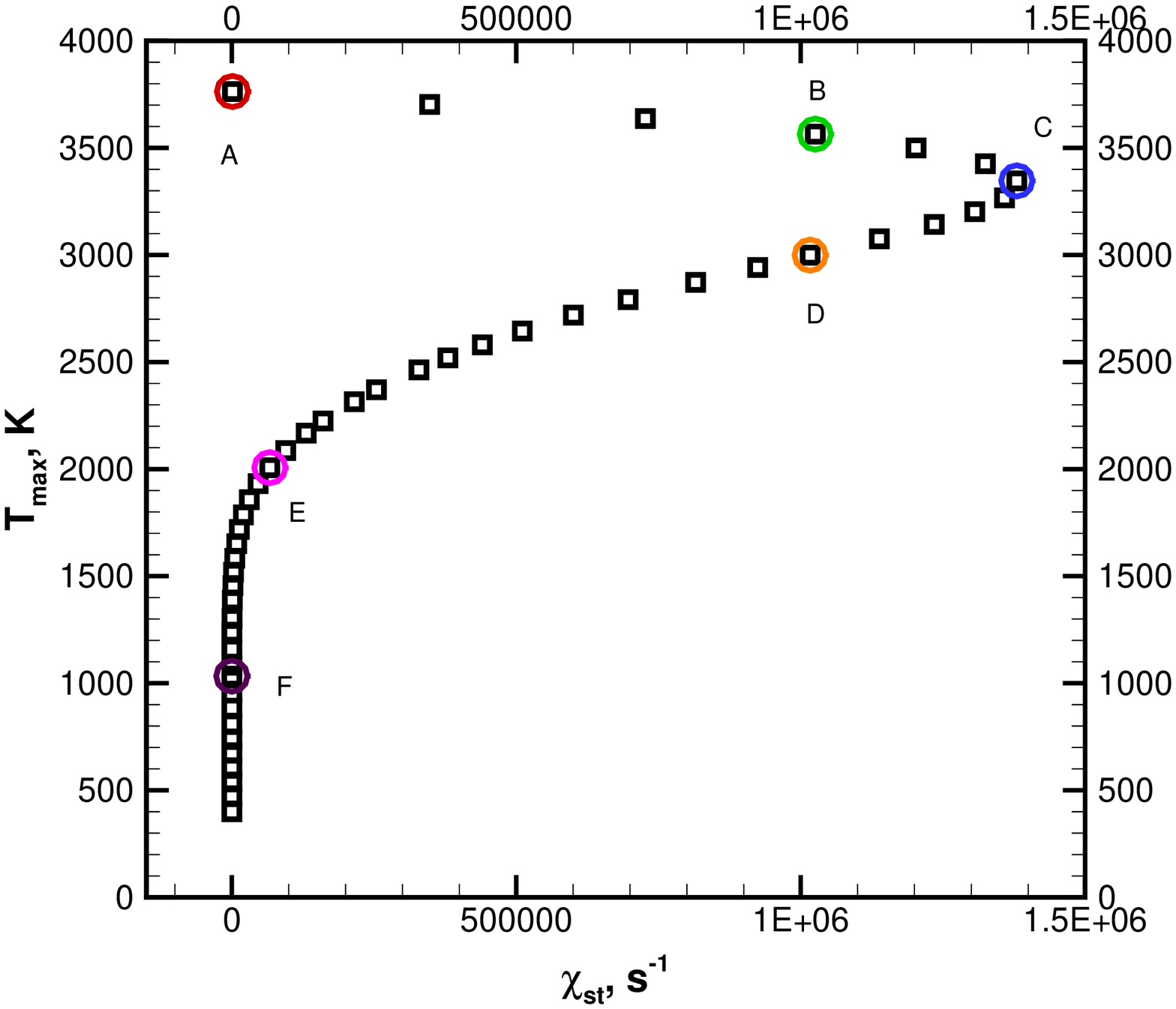}
        \label{fig:scurve}}
        \subfigure[$T_{max}$ vs. $C$]
{\includegraphics{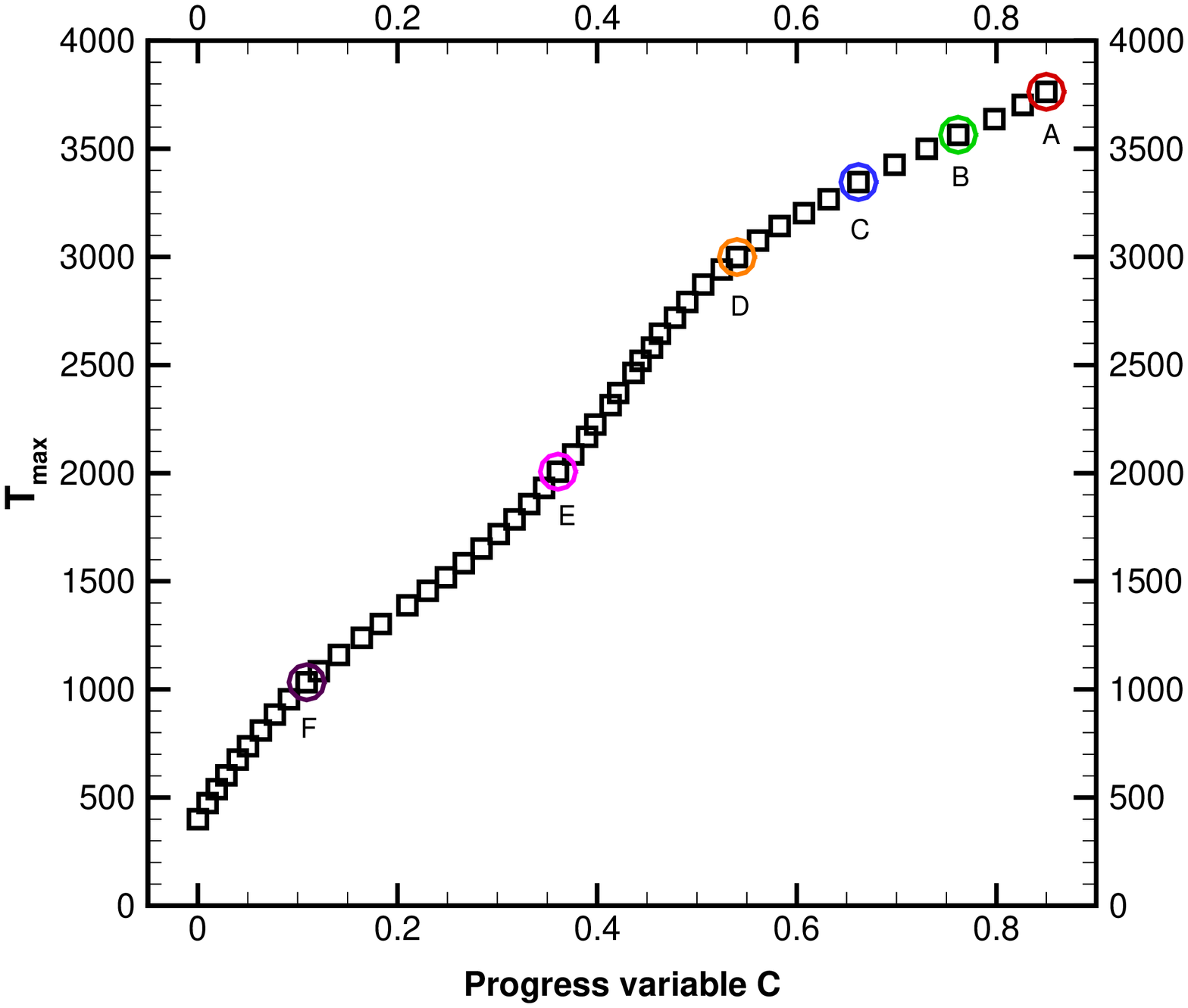}
        \label{fig:tmaxc}}
    \end{subfigmatrix}
    \caption{Steady flamelet solutions for methane/oxygen combustion with $T_{f} = T_{o} = 400 K$}
    \label{fig:tmaxflamelet}
\end{figure}

To apply the FPV model to the turbulent combustion of the ten-injector chamber, steady-state laminar flamelet solutions 
are obtained first by feeding the FFCMy-12 mechanism into the FlameMaster code. The fuel and oxidizer are pure methane and oxygen, respectively; operating pressure is $200$ bar; and the temperature on both fuel and oxidizer sides is $400 K$. Figure~\ref{fig:scurve} shows the maximum temperature in a flamelet solution as a function of $\chi_{st}$, which is frequently called the S curve. It is multi-valued as mentioned above and only the upper branch can be used in the SLF model.
In the progress variable approach, then the curve becomes single-valued and monotonic as shown in Figure~\ref{fig:tmaxc}. Hence, both the upper and lower branches can be explored in the flow computation.

\begin{figure}
    \begin{subfigmatrix}{2}
        \subfigure[$T(Z)$ at various progress variable]
{\includegraphics{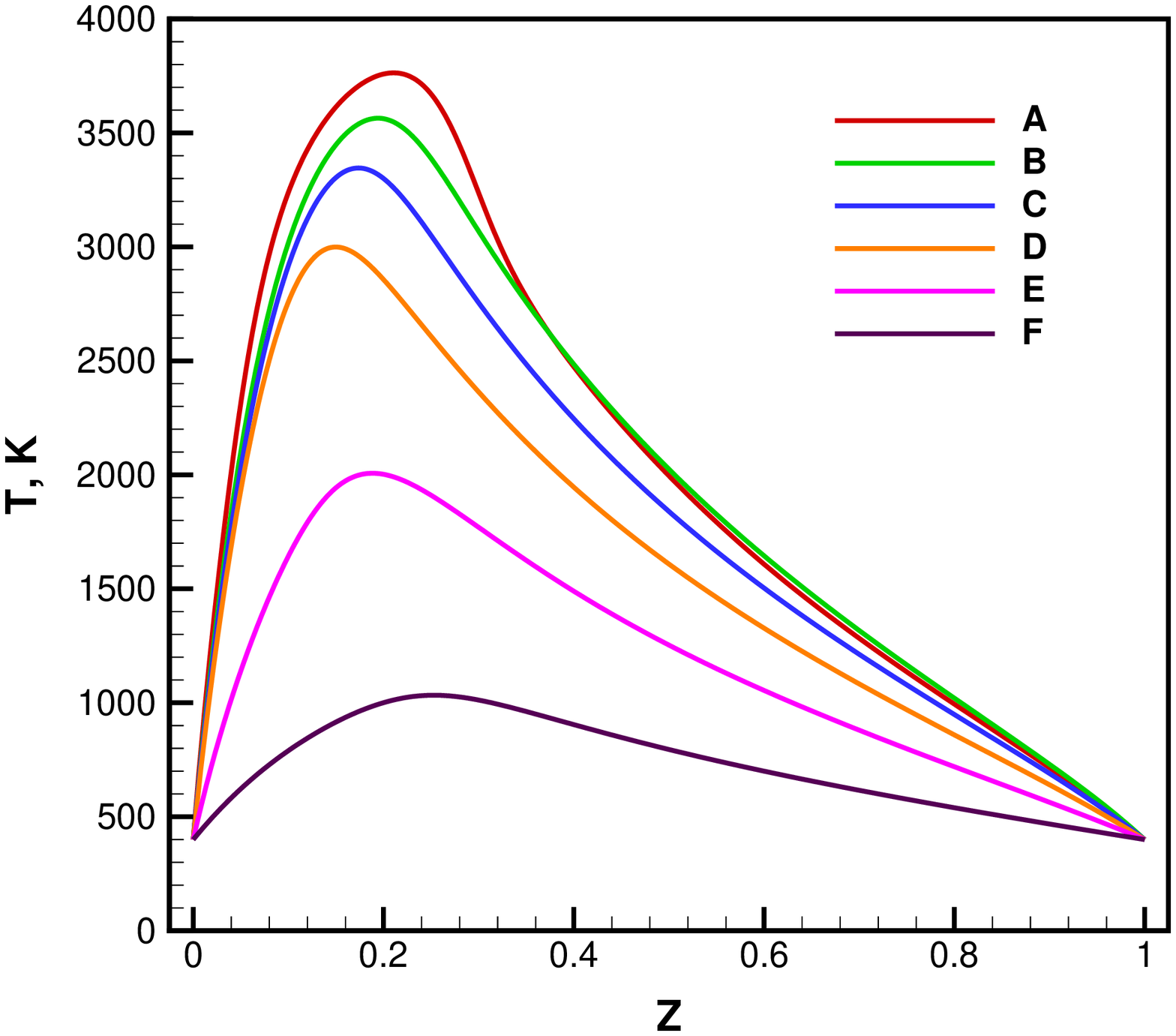}
        \label{fig:t(z)_c}}
        \subfigure[$HRR(Z)$ at and near quenching limit]
{\includegraphics{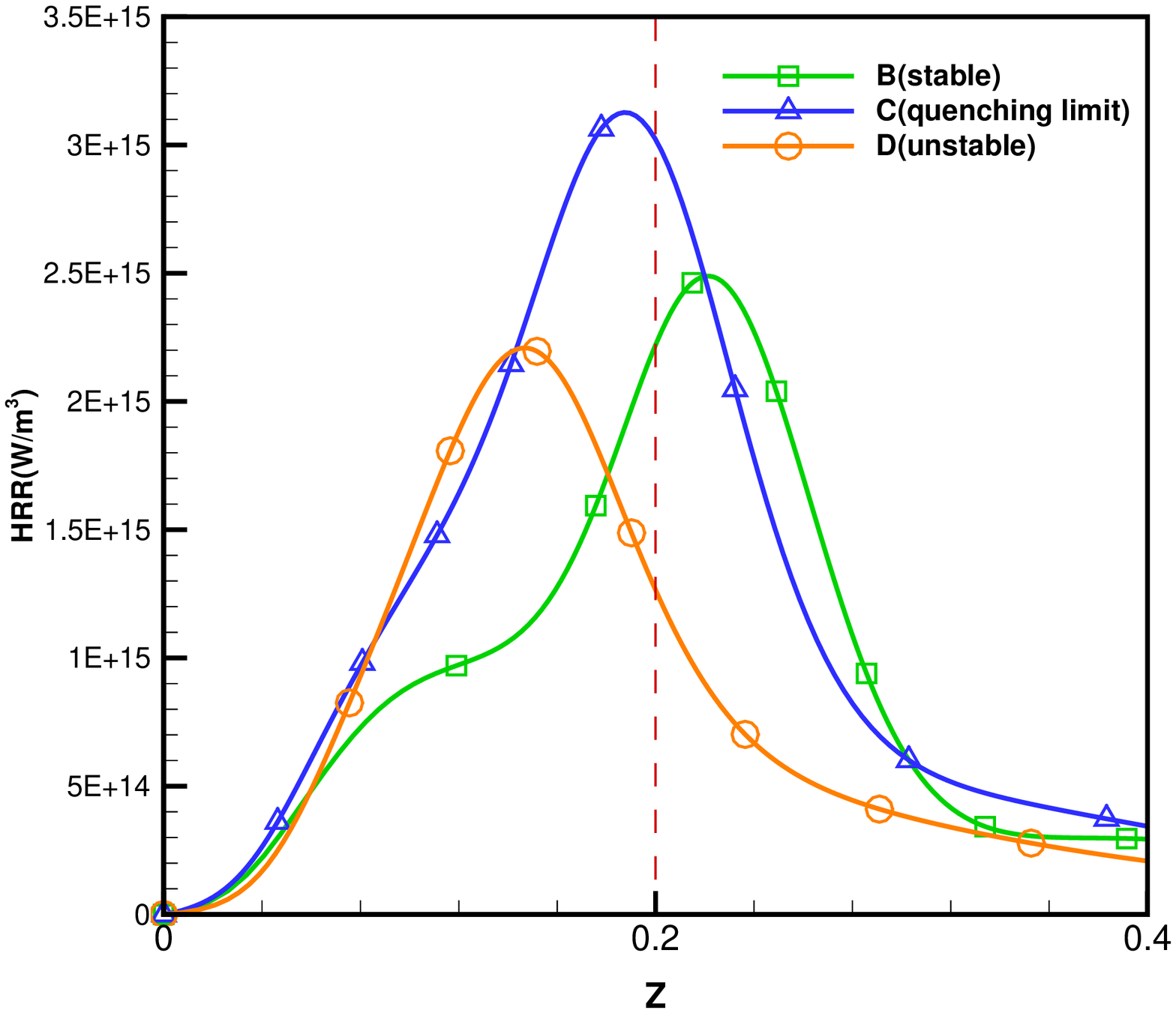}
        \label{fig:hrr(z)_quenching}}
    \end{subfigmatrix}
    \caption{Representative flamelet solutions over the complete S-curve.}
    \label{fig:solution_diff_C}
\end{figure}

Selected particular flamelet solutions along the S-shaped curve are shown in Figure~\ref{fig:t(z)_c}.
The mixture fraction is zero on the oxidizer side and one on the fuel side, indicating 
the flow is solely composed of the oxidizer or the fuel. The top two curves (A and B)  
are on the upper branch of the S-shaped curve and represent the stable burning flamelet solutions. 
The third curve (C) is the flamelet solution at the quenching limit.
The next three curves (D, E, and F) below it are on the lower branch of the S-shaped curve and represent unstable 
burning flamelet solutions. These
solutions are classified as unstable due to their sensitivity to small
perturbations and easily moving either toward the stable upper branch or
toward a stable quenched solution \cite{multi_regime_pitsh}.

\begin{figure}
    \begin{subfigmatrix}{2}
        \subfigure[Unstable]
{\includegraphics{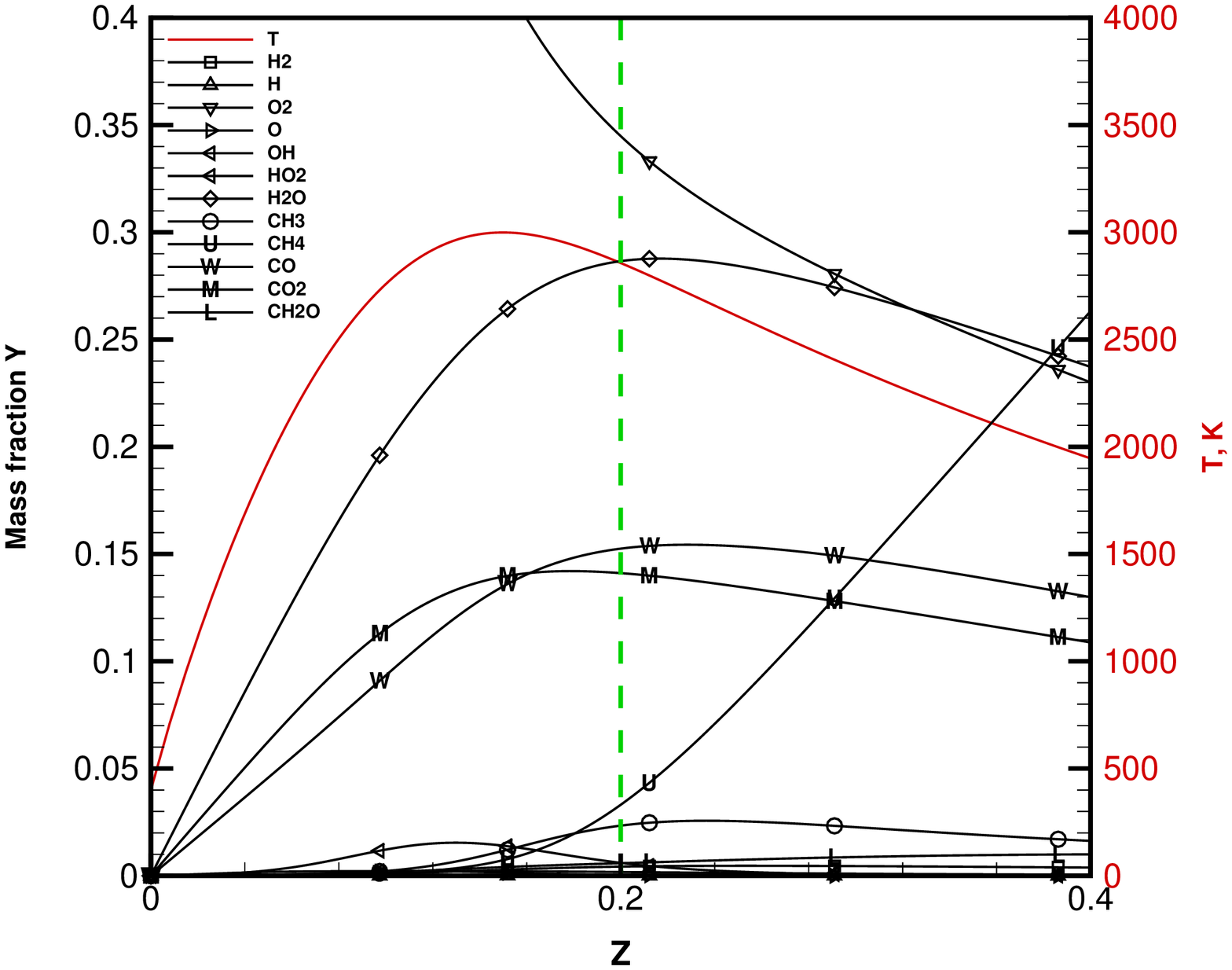}
        \label{fig:quenching_unstable}}
        \subfigure[stable]
{\includegraphics{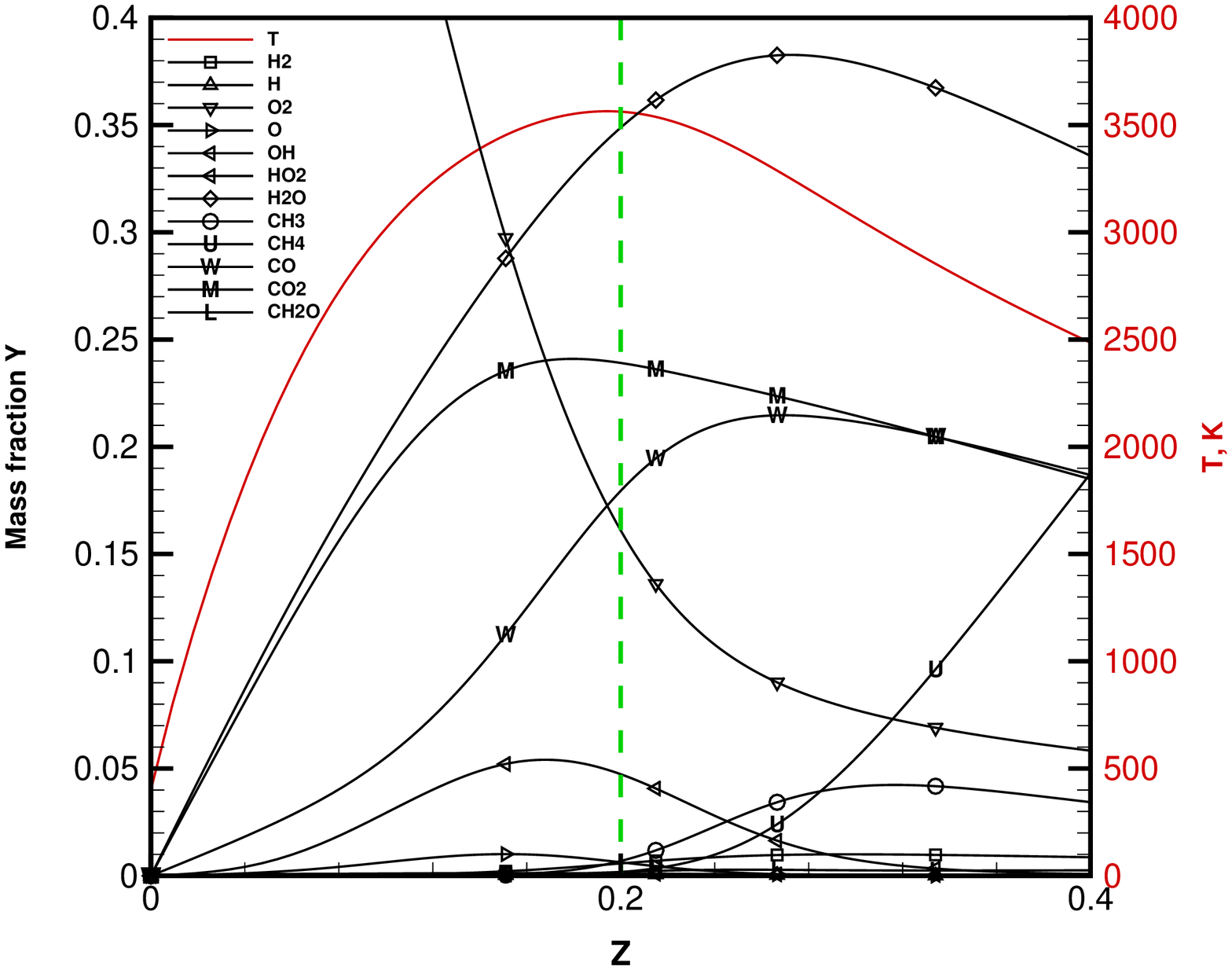}
        \label{fig:quenching_stable}}
    \end{subfigmatrix}
    \caption{Flamelet solutions near the quenching limit}
    \label{fig:near_quenching}
\end{figure}

Figure~\ref{fig:hrr(z)_quenching} shows the heat release rate (HRR) of the stable burning solution B, the quenching limit solution C and the unstable burning solution D. Note that solutions B and D are obtained using approximately the same stoichiometric dissipation rate. The vertical dash line marks the stoichiometric mixture
fraction $Z_{st} = 0.2$. A region immediately on the left is the approximated oxidation layer
\cite{flame_structure}. 
In the unstable burning limit, heat is intensively released 
on the fuel-lean side within the oxidation layer. 
This region of intense HRR is mainly associated with the reactions
involving $H_{2}O$ and $CO_{2}$. In the stable burning limit, however, the strong HRR region moves across the stoichiometric line to the slightly fuel-rich side.
Reactions in this branch are much
more stable and the reduction of $CH_{4}$ accounts for a larger
fraction of the total HRR. Figure~\ref{fig:near_quenching} shows mass fractions of individual species and temperature profiles as functions of the mixture fraction on the left for the unstable
burning flamelet solution B and on the right for the 
stable burning flamelet solution D. In both cases, due to extremely high scalar dissipation rates 
in the reaction zone (where characteristic diffusion time is quite small),
there is remarkable reactant leakage through the zone, as
shown in Figure~\ref{fig:near_quenching}. In the stable burning case, oxygen is 
consumed faster across the oxidation layer. Hence, oxygen
leakage to the fuel-rich side is less. Despite the much higher background pressure in this paper, 
The flame structures described above
are still qualitatively similar to those reported by previous work for methane/air
diffusion flame \cite{tuancnf, flame_structure}.

\begin{figure}
    \begin{subfigmatrix}{2}
        \subfigure[$\omega_{C}$ vs. $Z$]
{\includegraphics{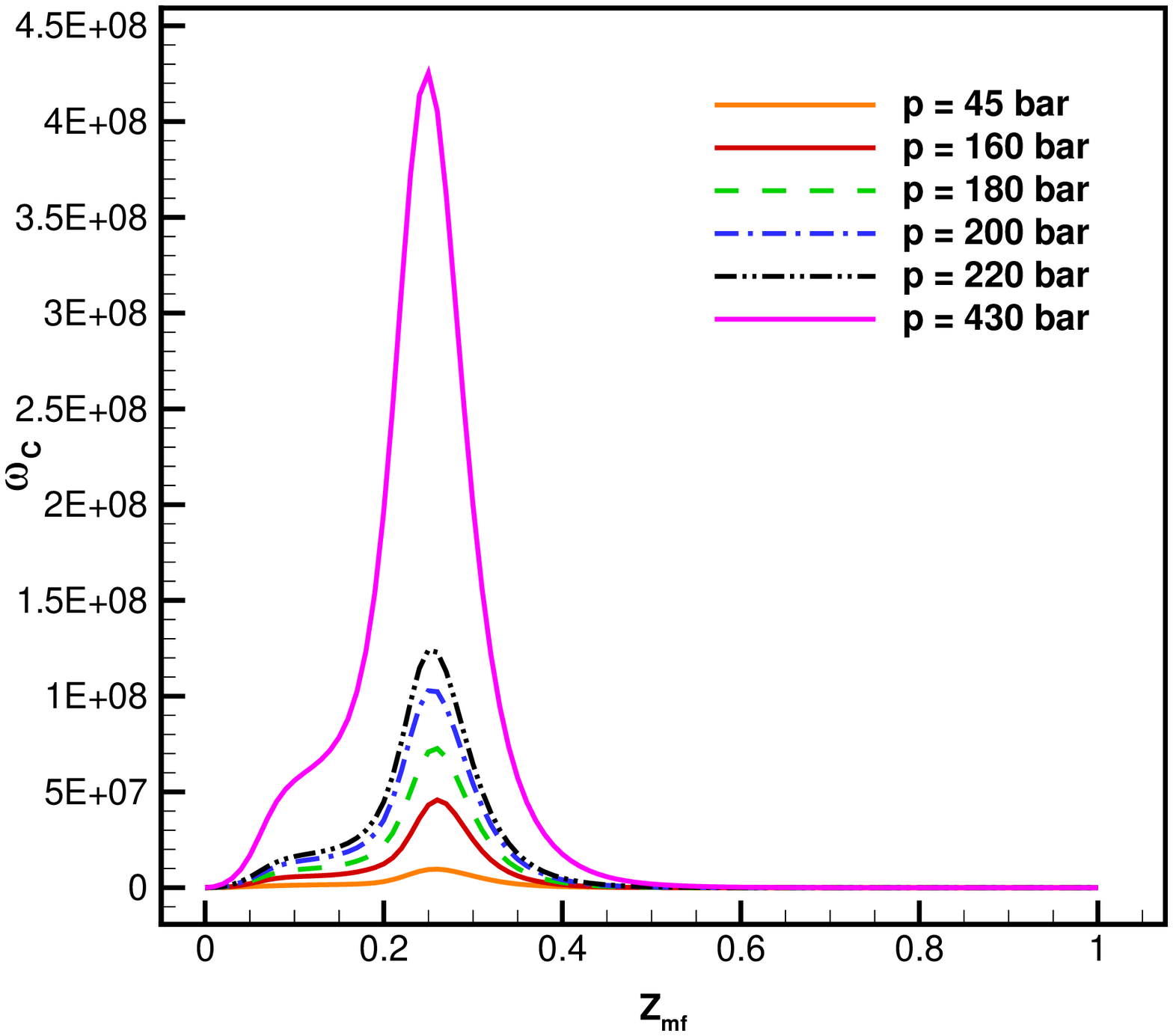}
        \label{fig:omc_Z}}
        \subfigure[$\omega_{C,max}$ vs. p]
{\includegraphics{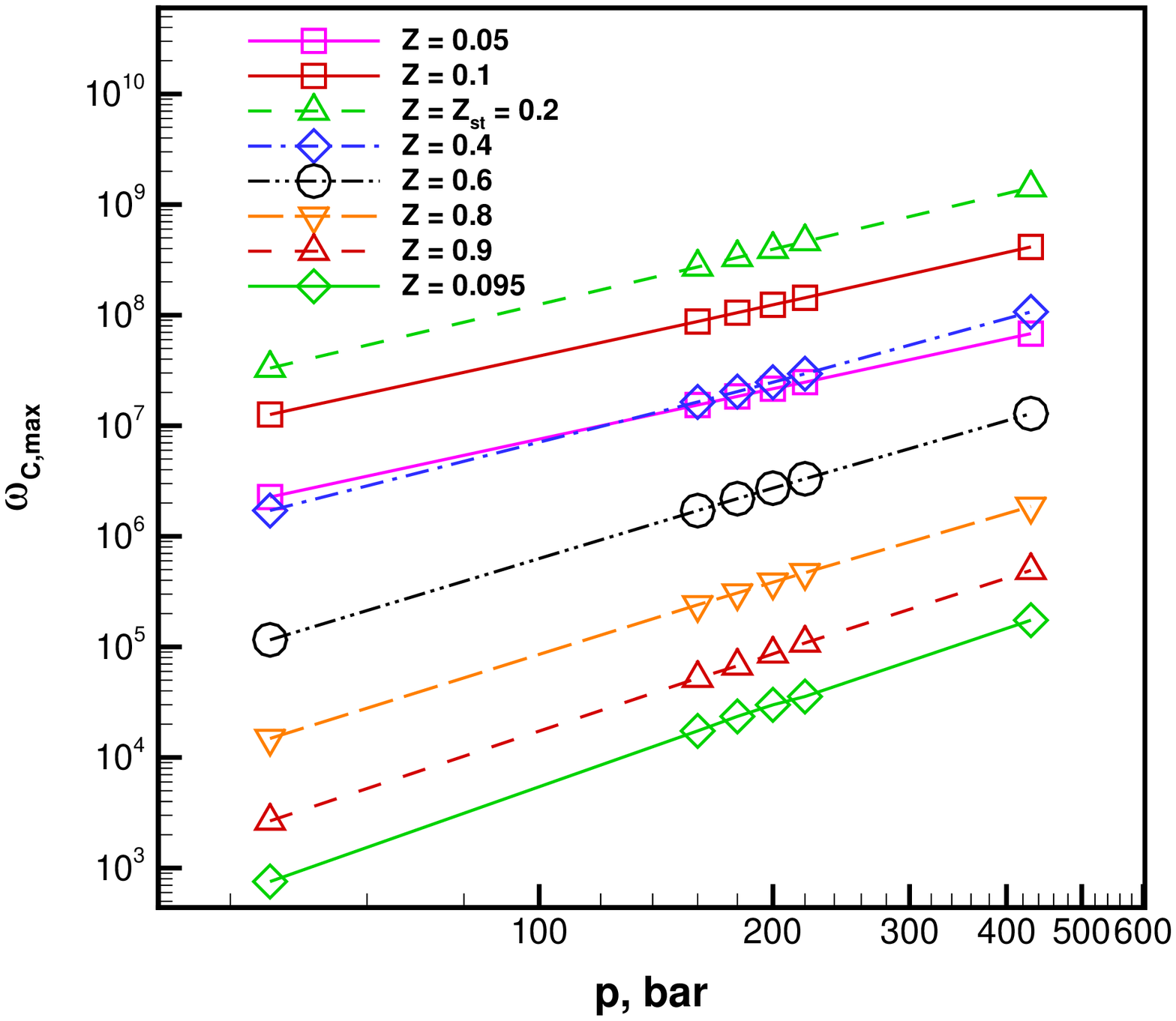}
        \label{fig:maxomc_p}}
    \end{subfigmatrix}
    \caption{Pressure effect on the reaction rate for the progress variable $\omega_{C}$}
    \label{fig:pressureomc}
\end{figure}

In this paper, the flamelet solutions for the background pressure of $200$ bar are actually used only in the table-look-up process during flow computations. Our observation confirms that composition changes little with pressure, whereas the reaction rate for the progress variable ($\omega_{C}$) varies dramatically. Figure~\ref{fig:omc_Z} shows $\omega_{C}$ as a function of $Z$ for different pressures with $C$ fixed at the maximum value $0.85$, which corresponds to flamelet solutions at chemical equilibrium.

Plotted in a log-log graph as shown in Figure~\ref{fig:maxomc_p}, $\omega_{C,max}$ changes linearly with pressure, indicating that $\omega_{C}$ is approximately a power function of pressure. $\widetilde{\omega_{C}}$ at a given pressure $\bar{p}$ can be obtained by  rescaling $\widetilde{\omega_{C_{0}}}$ at pressure of $200$ bar ($\bar{p_{0}}$) as 

\begin{equation}
\widetilde{\omega_{C}} = \widetilde{\omega_{C_{0}}}\frac{\bar{p}^{\eta}}{\bar{p_{0}}^{\eta}}
\label{omegacvsp}
\end{equation}

Note that the power $\eta$ changes with $Z$ and its variance $Z^{''}$. For a given combination of $Z$ and $Z^{''}$, $\eta$ can be calculated using \eqref{omegacvsp}. The dependence of $\eta$ on $Z$ and $Z^{''}$ can then be determined rigorously through regression analysis. Among frequently used regression functions, a quadratic function of $Z$ and a linear function of $Z^{''}$ seem to be the most cost-effective. The final approximation of $\eta$ is given in equation \eqref{rescalingalpha}.

\begin{equation}
\eta = (3.0882Z^{''}+0.0101)Z^{2}+(-3.7851Z^{''}+0.9567)Z+(0.6638Z^{''}+1.4589)
\label{rescalingalpha}
\end{equation}

Compared to the direct interpolation of the reaction rate based on pressure, the use of rescaling law \eqref{omegacvsp} and \eqref{rescalingalpha} avoids parameterization of the flamelet tables with respect to pressure and reduces the effort of table look-up process in combustion simulation. A similar approach was reported by Pecnik et al. \cite{pecnikaiaaj2012} for hydrogen combustion. Our work differs from it in two ways: 1) The rescaling law is determined for methane combustion under much higher background pressure; 2) the power's dependency on both $Z$ and $Z^{''}$ is considered.

\subsection{Combustion simulation of a ten-injector liquid rocket engine}
\label{sec:com_simu_10inj_rocket_engine}

\begin{figure}
    \begin{subfigmatrix}{2}
        \subfigure[Isometric view]
{\includegraphics{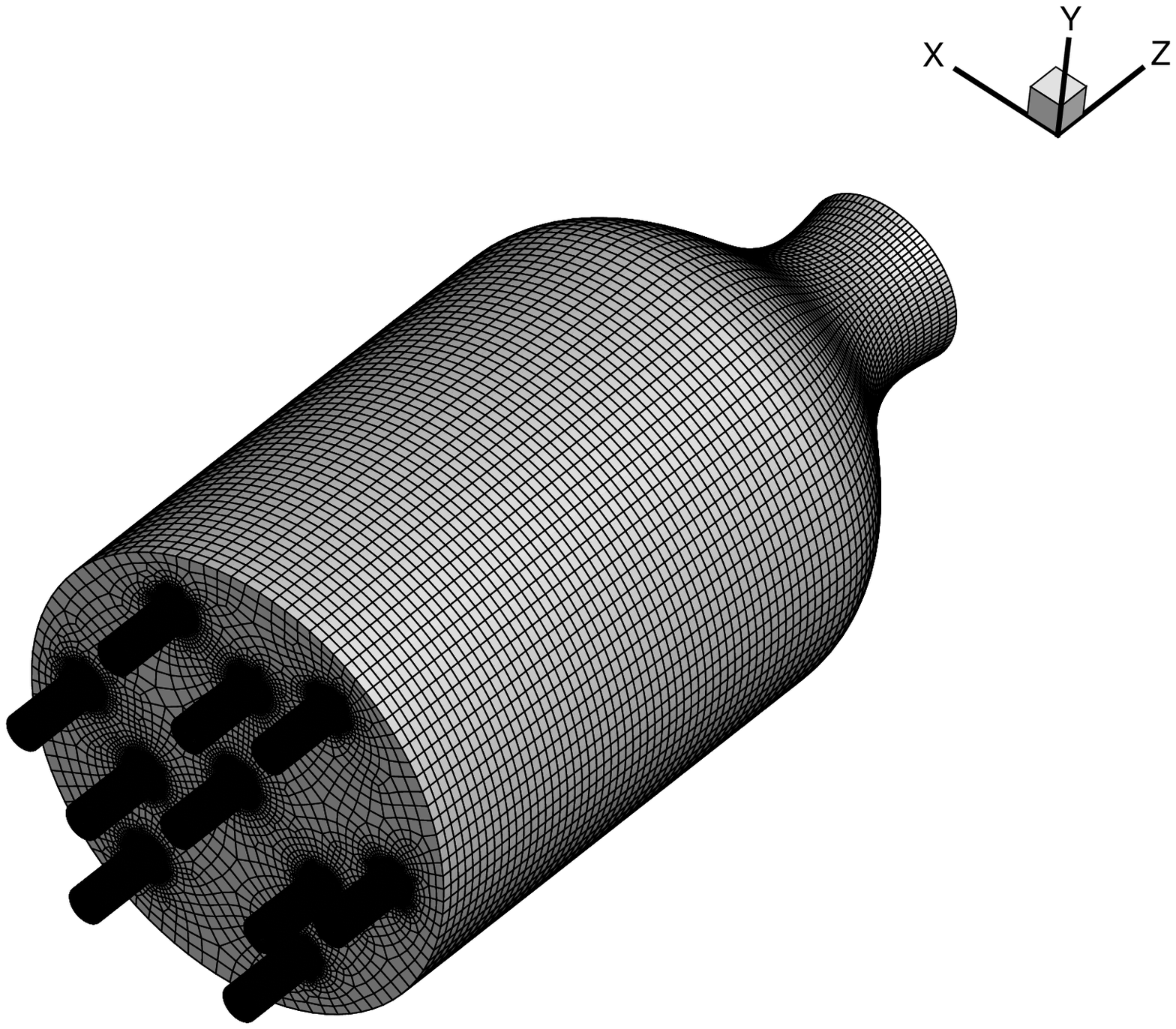}
        \label{fig:3dmesh}}
        \subfigure[Front view]
{\includegraphics{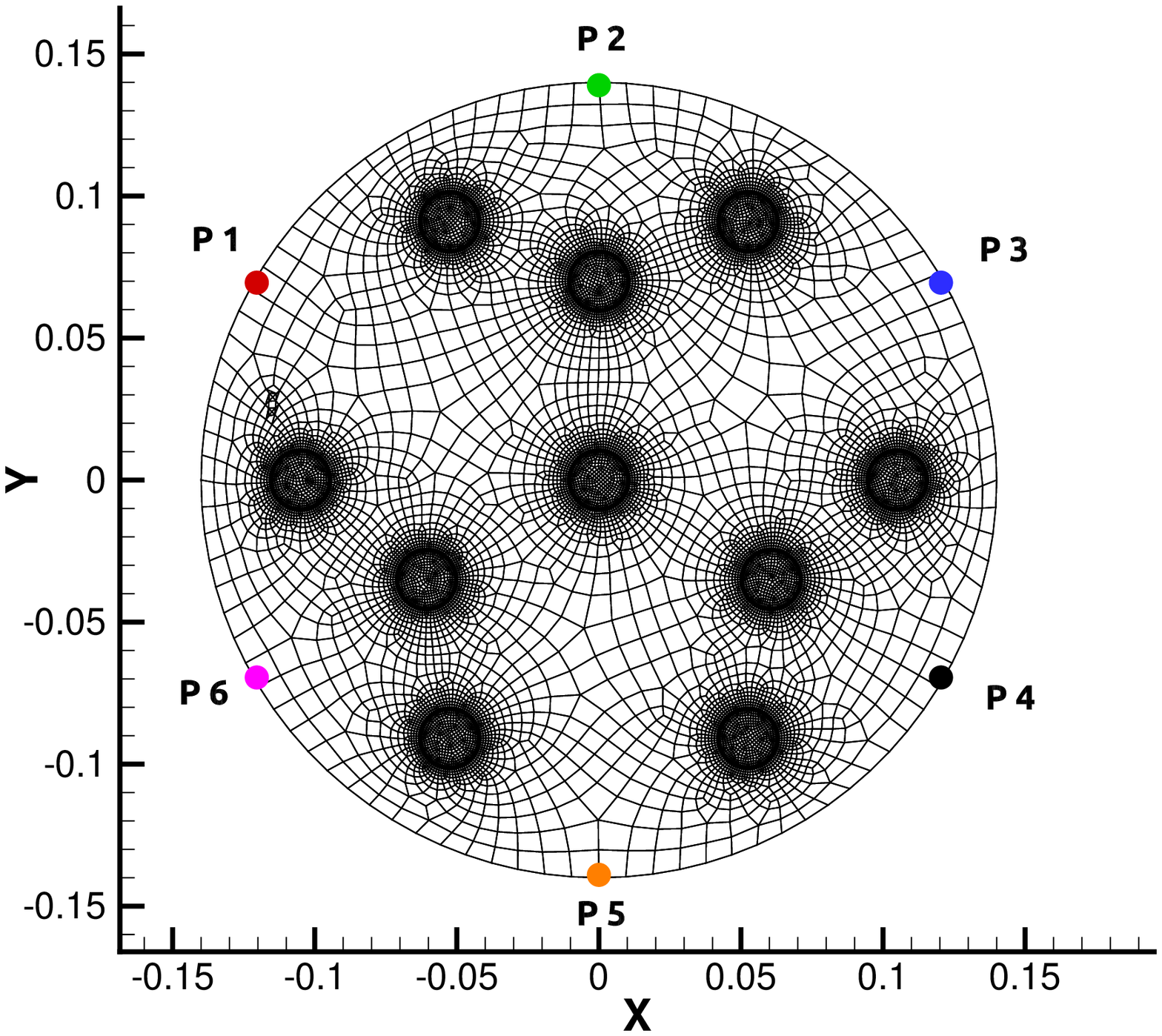}
        \label{fig:2dmesh}}
    \end{subfigmatrix}
    \caption{Unstructured mesh for the ten-injector rocket engine}
    \label{fig:mesh}
\end{figure}

In this paper, turbulent combustion is numerically simulated in a liquid rocket engine with an convergent-divergent nozzle and ten coaxial fuel/oxidizer injectors\cite{xiongaiaaj2020}. The three-dimensional unstructured mesh with $1.608502 \times 10^{6}$ grid cells is shown in Figure~\ref{fig:3dmesh}. The combustion chamber has a diameter of $28$ cm and a length of $53$ cm. The distance between the injector plate and the choked nozzle is $33$ cm. One of the ten coaxial injector ports mounted into the injector plate is located at the center. Three injector ports are equally distributed on the ring of radius $7$ cm. The remaining six ports are evenly positioned on an outer ring of radius $10.5$ cm. In the longitudinal direction, $55$, $40$ and $10$ grid cells are distributed in the combustion chamber, nozzle and injector,respectively. In the radial direction of the combustion chamber, depending on whether a non-central injector falls in, there are about $40$ or $80$ grid cells. The oxidizer, which is pure oxygen, is injected through the center of each coaxial port. The fuel, which is pure methane, is injected through the outside annulus. The mass fluxes for the fuel and oxidizer are $15$ kg/s and $60$ kg/s, respectively, leading to a propellant injection at the stoichiometric ratio. The diameter of the fuel injector is $1.96$ cm and that of the oxidizer is $2.2$ cm. For the given mass flux, this port geometry makes the mean injection velocity ratio between the fuel and the oxidizer $2.28$. Temperature at both the fuel and oxidizer injector inlets is $400$ K. The injection of propellants at $400$ K can match practice where propellants flow through the combustion-chamber walls, serving as a coolant before injection in a manner that brings the system closer to adiabatic operation. For the low-temperature flows inside injectors, ideal-gas EOS is not an accurate approximation. However, for the major portion of the volume of the chamber and nozzle, real-gas effects can be considered negligible. For this reason, ideal-gas EOS is applied everywhere in the rocket engine to reduce computational cost.

On the walls, the no-slip boundary condition is applied and the normal pressure gradient is zero. Furthermore, the walls are assumed adiabatic and free of surface chemistry. At the injector inlets, the mean mass fluxes for methane and oxygen are fixed, yielding the steady-state mass flow rate for sonic throat of the nozzle. The wave-transmissive boundary condition is used at the outlet boundary due to the supersonic nozzle flow. This boundary condition prevents waves from propagating backward into the system and hence guarantees that neither shock nor backflow occurs in the combustion.

The initial pressure in the combustion chamber is $200$ bar. The simulations start from an initially quiescent flow at $3229$ K and an chamber pressure at 200 bar. The higher chamber temperature shortens the ignition process and reduces the computational time as no explicit ignition model is included in the current simulation. 

\begin{figure}
    \begin{subfigmatrix}{3}
        \subfigure[$Y_{CH_{4}}$]
{\includegraphics{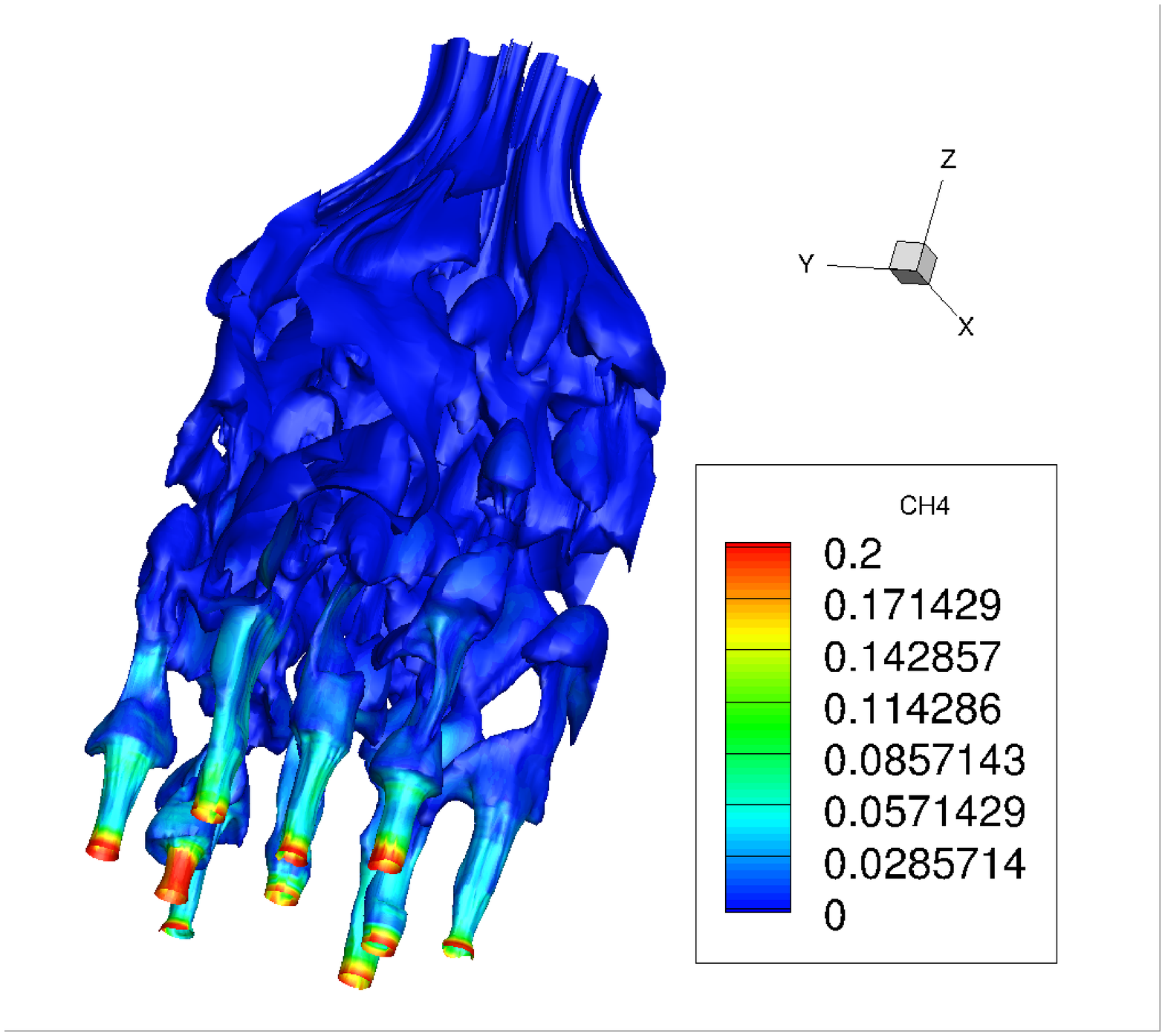}
        \label{fig:iso_zst_ch4}}
        \subfigure[$Y_{O_{2}}$]
{\includegraphics{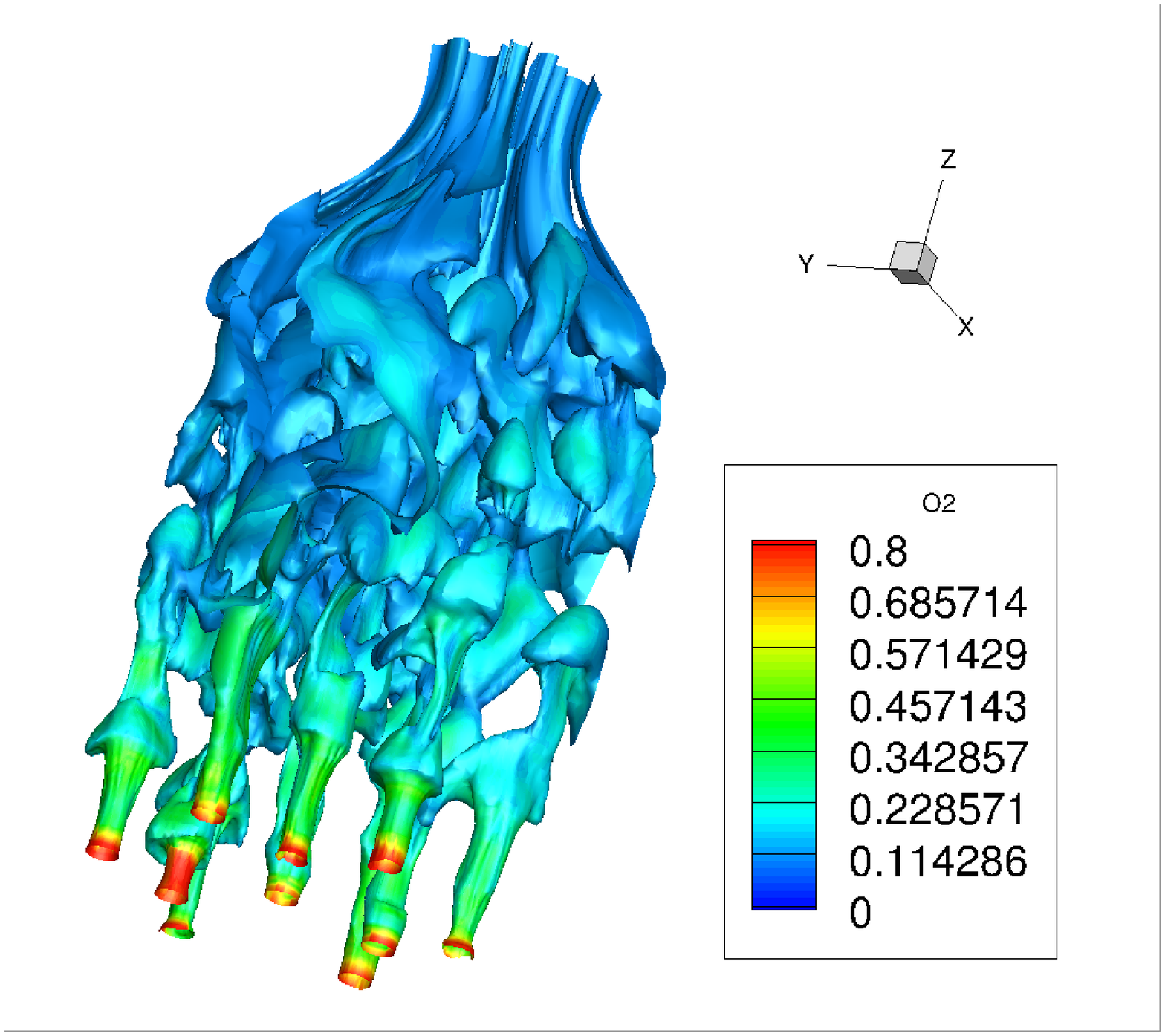}
        \label{fig:iso_zst_o2}}
        \subfigure[$Y_{CO_{2}}$]
{\includegraphics{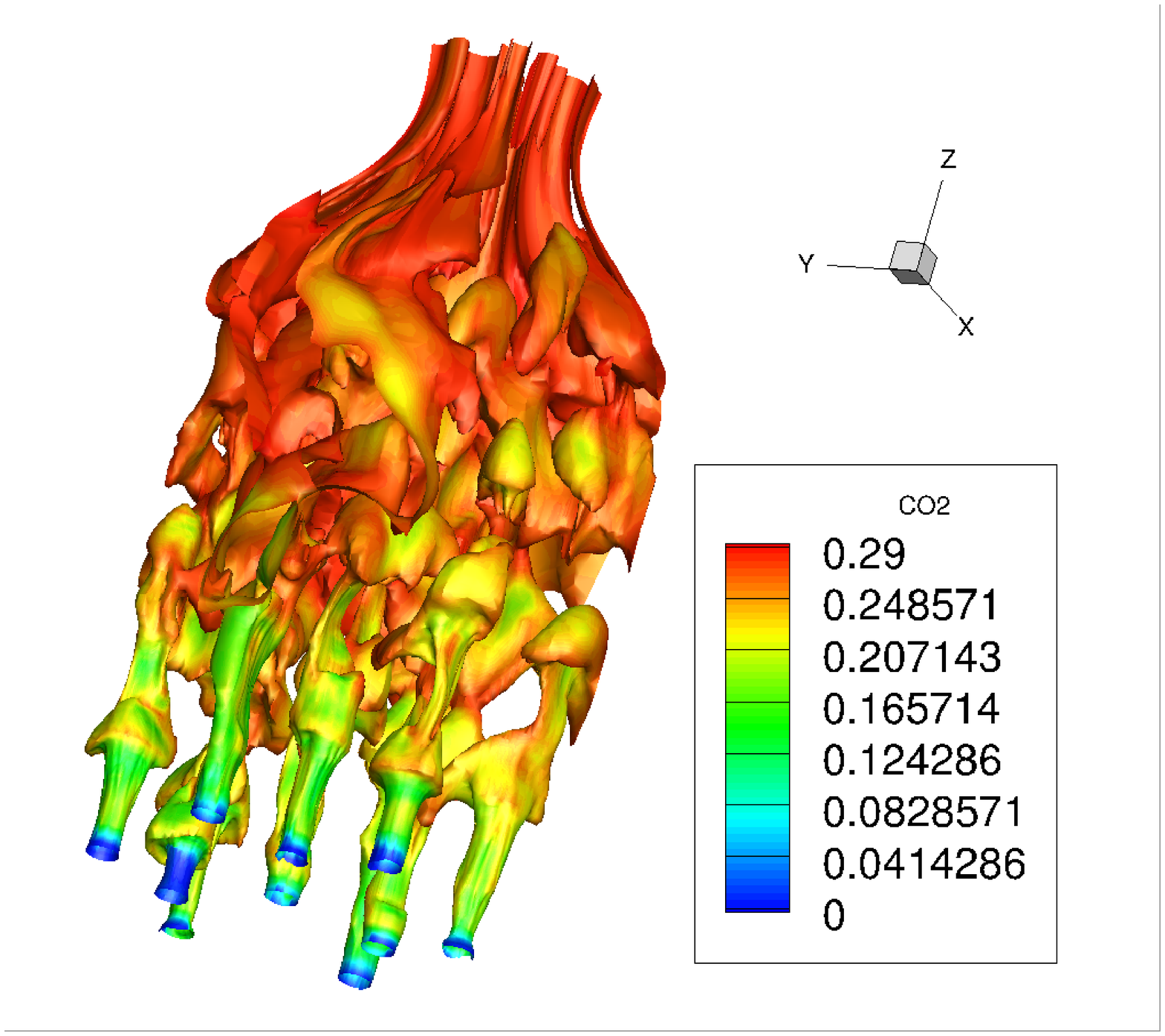}
        \label{fig:iso_zst_co2}}
        \subfigure[$Y_{H_{2}O}$]
{\includegraphics{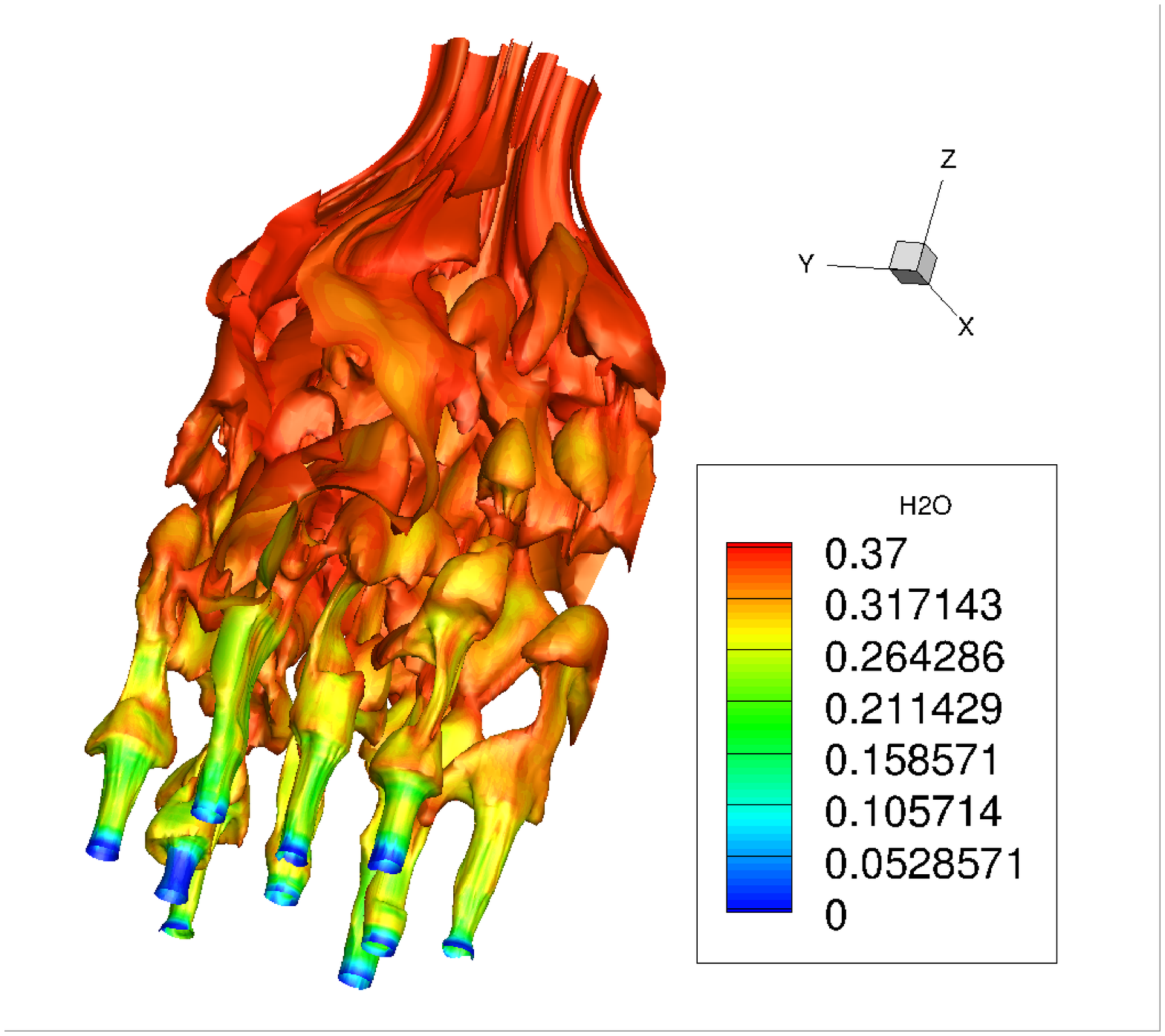}
        \label{fig:iso_zst_h2o}}
        \subfigure[$Y_{CO}$]
{\includegraphics{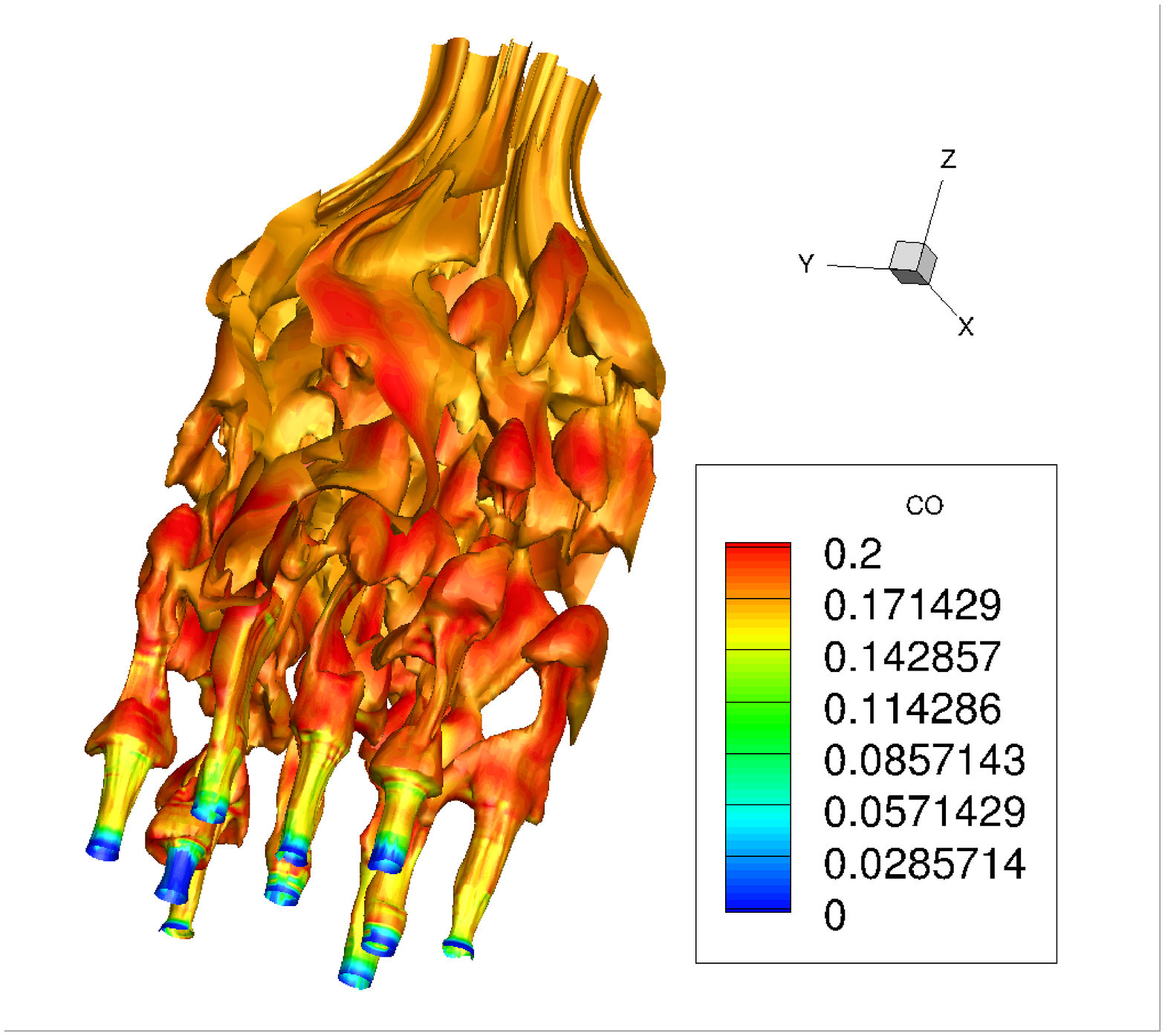}
        \label{fig:iso_zst_co}}
        \subfigure[$Y_{H_{2}}$]
{\includegraphics{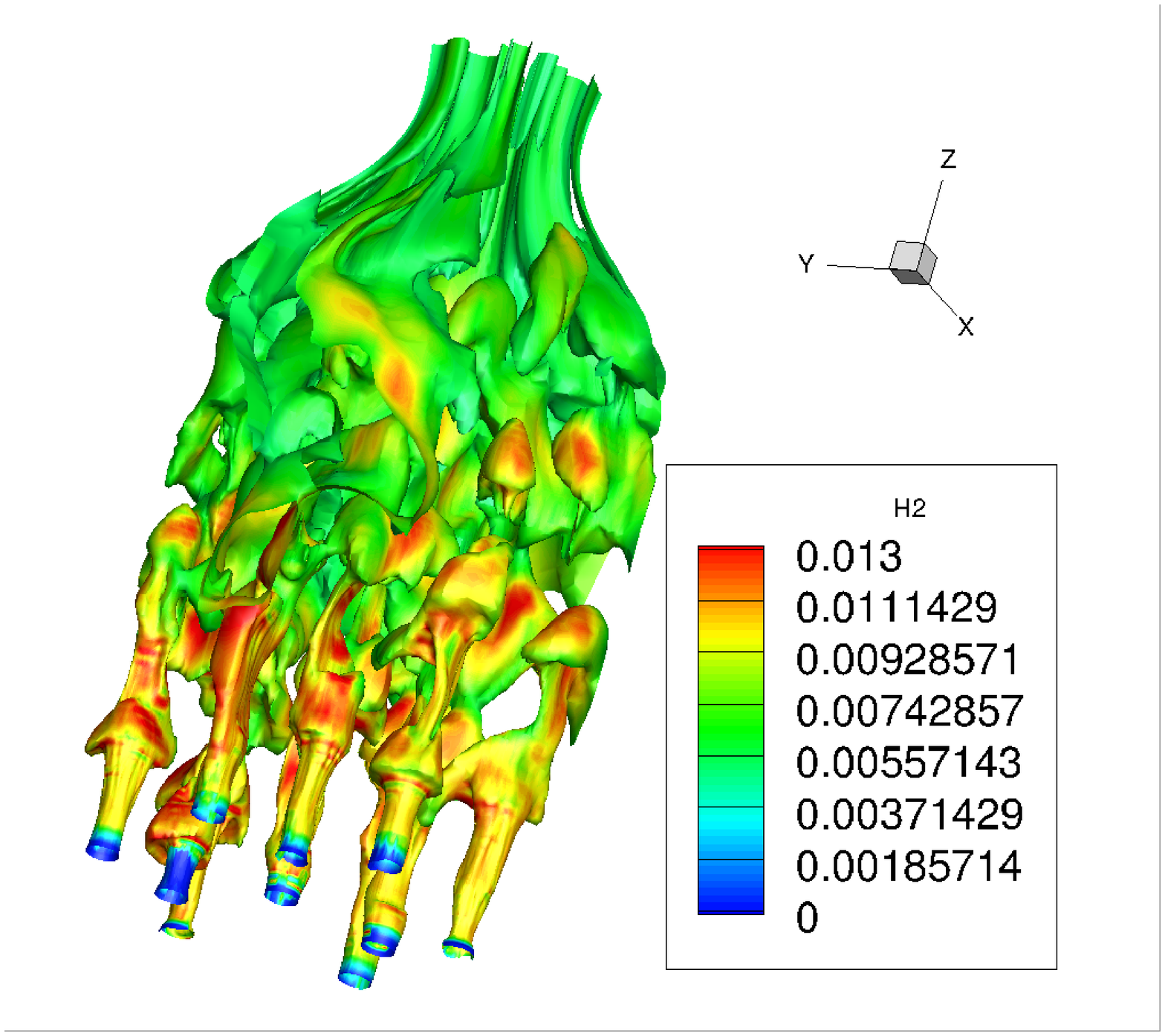}
        \label{fig:iso_zst_h2}}
        \subfigure[$T, K$]
{\includegraphics{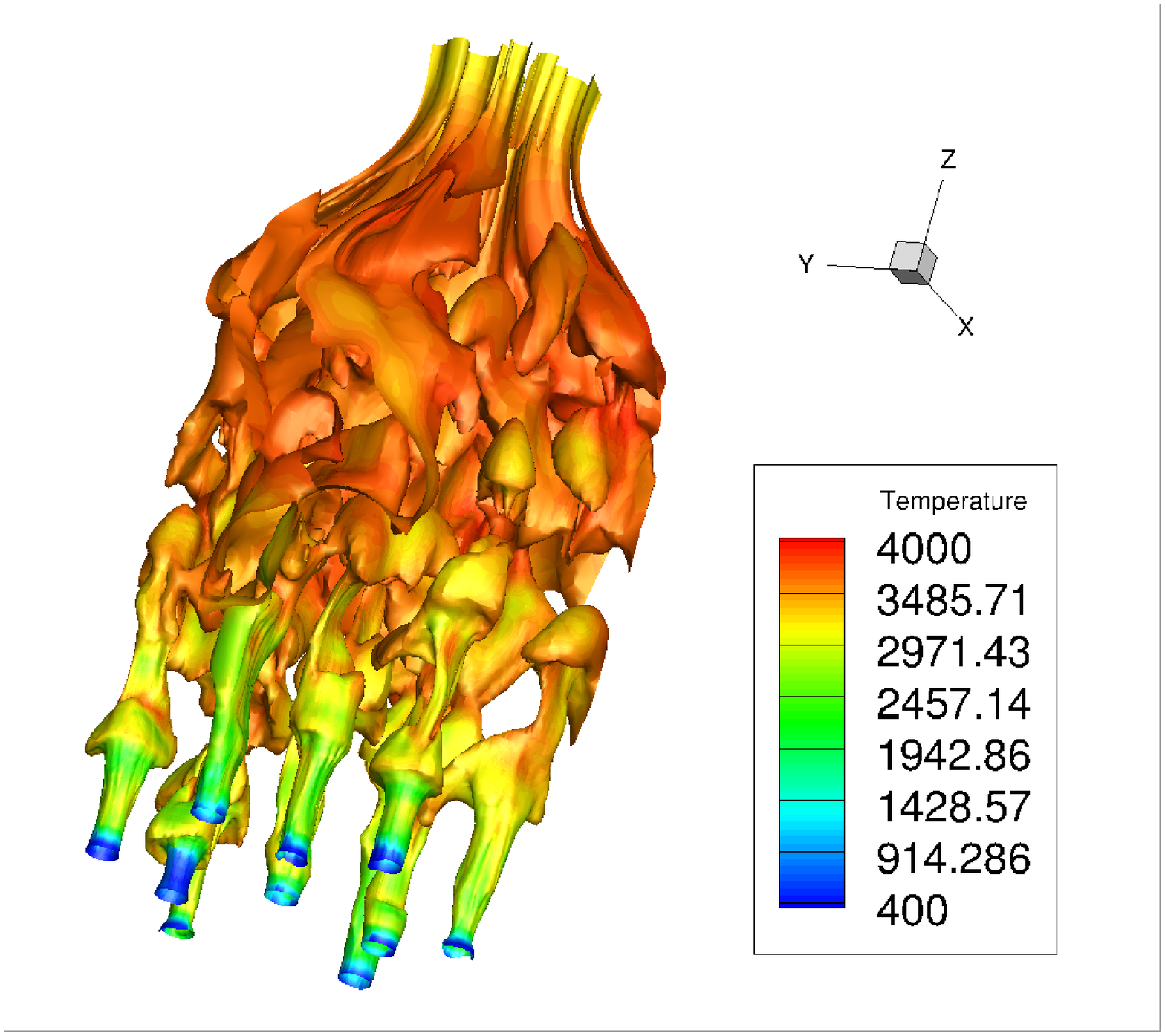}
        \label{fig:iso_zst_T}}
    \end{subfigmatrix}
    \caption{Instantaneous isosurface of $Z_{st} = 0.2$ colored by mass fraction of major reaction species and temperature for the
FPV model at t = 50 ms}
    \label{fig:iso_zst}
\end{figure}

\begin{figure}
    \begin{subfigmatrix}{3}
        \subfigure[$\bar{Y}_{CH4}$ for FPV]
{\includegraphics{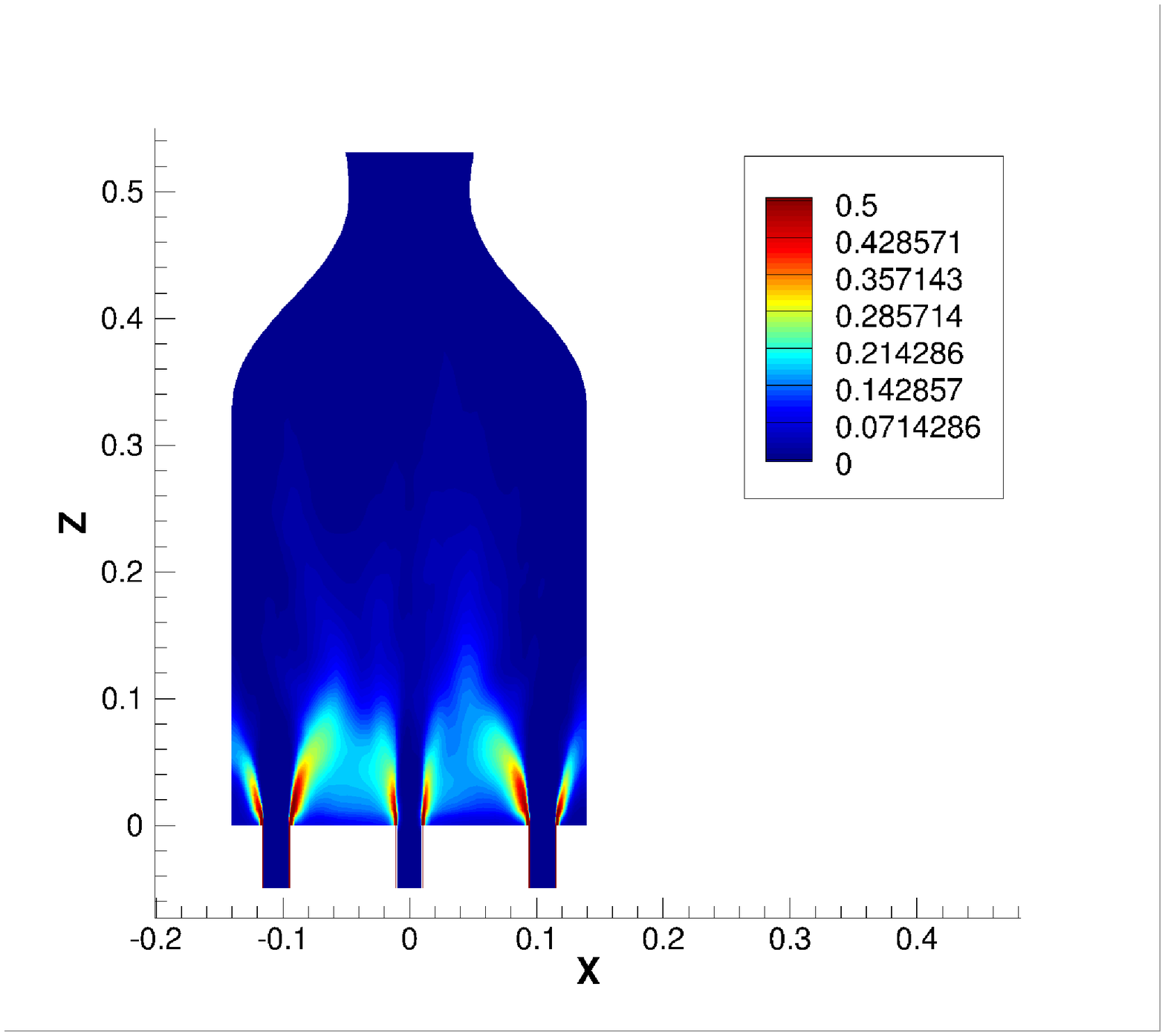}
        \label{fig:tave_ch4_fpv}}
        \subfigure[$\bar{T}$, K for FPV]
{\includegraphics{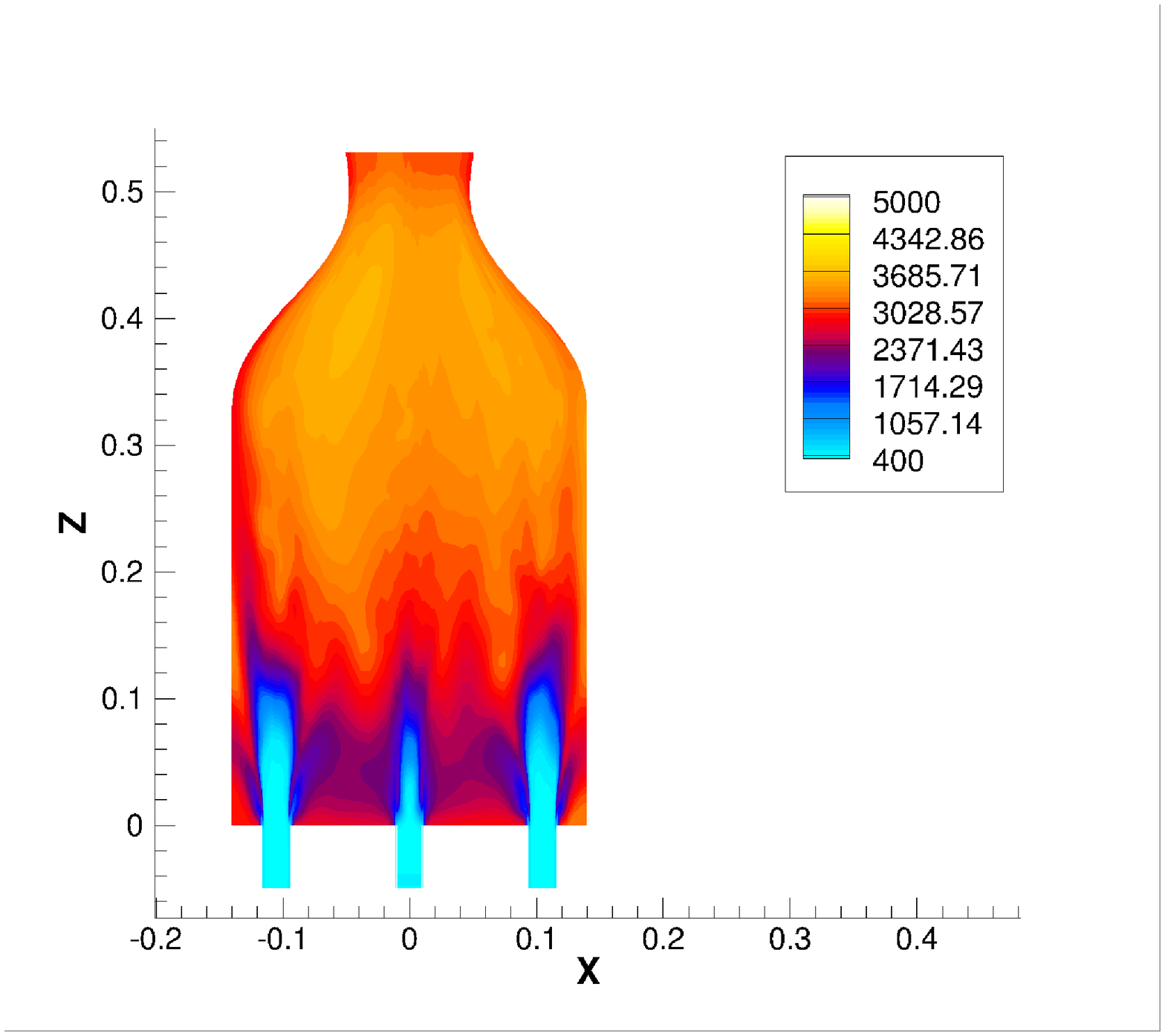}
        \label{fig:tave_T_fpv}}
        \subfigure[$\overline{Ma}$ for FPV]
{\includegraphics{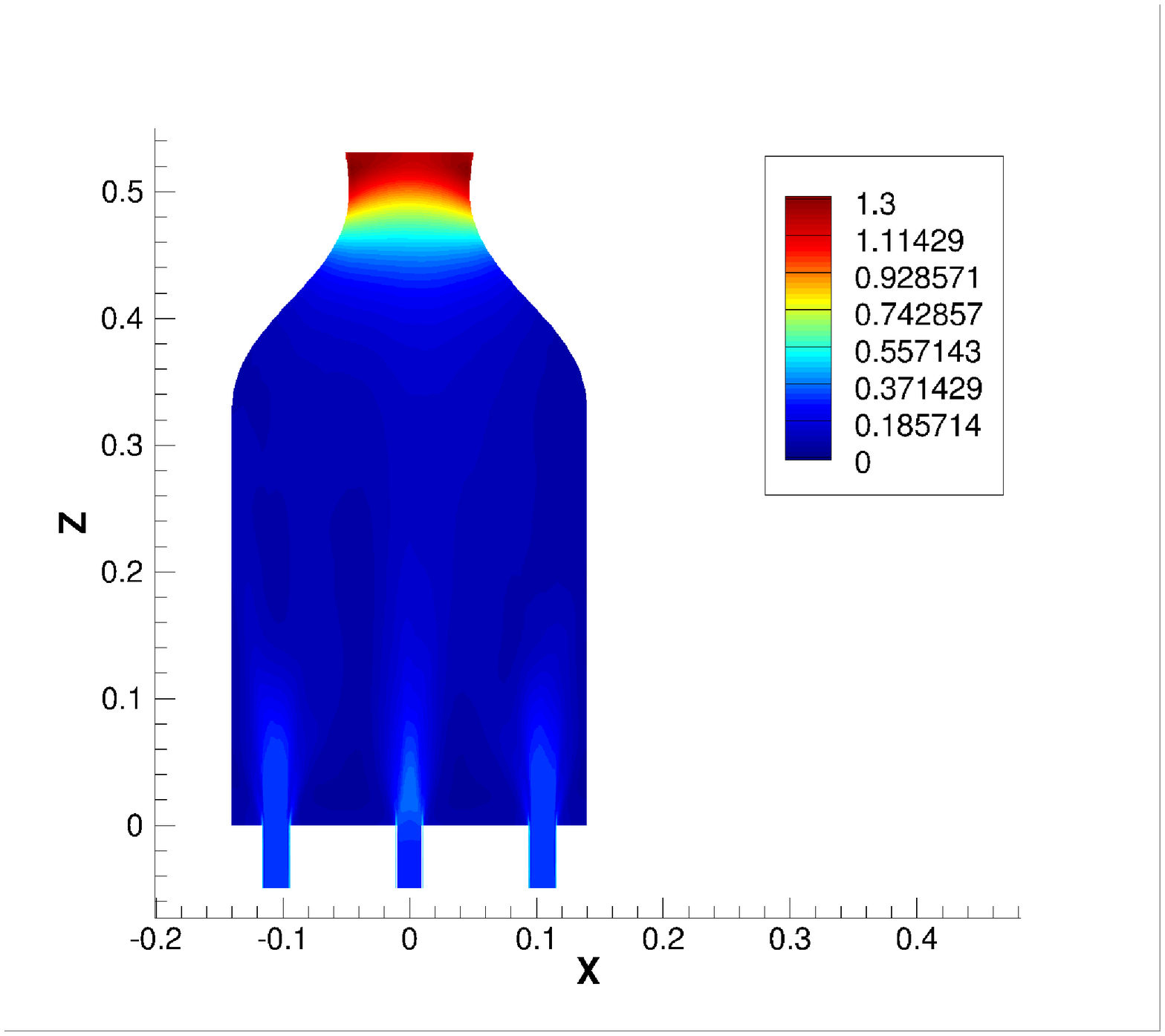}
        \label{fig:tave_mach_fpv}}
        \subfigure[$\bar{Y}_{CH4}$ for OSK]
{\includegraphics{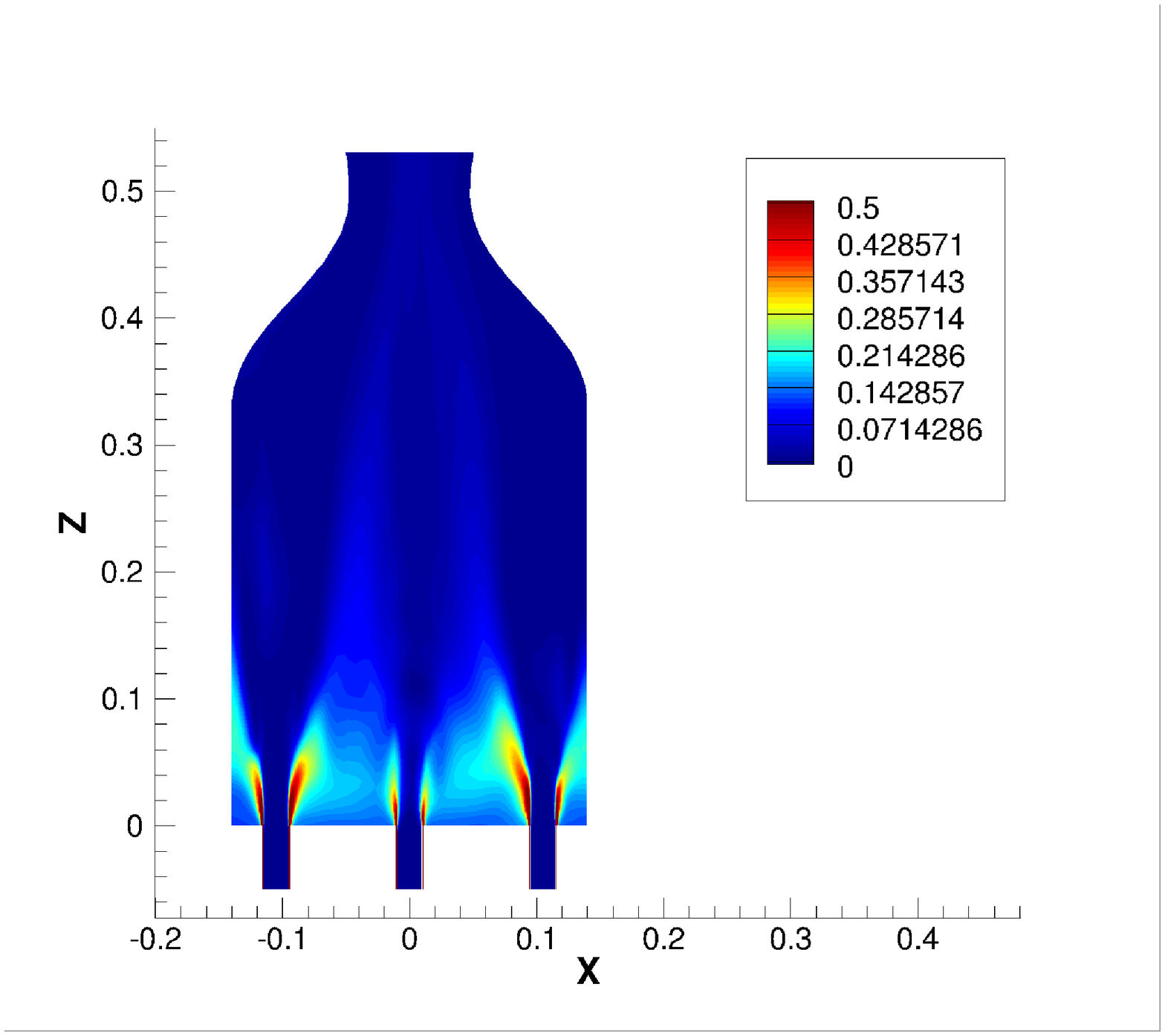}
        \label{fig:tave_ch4_osk}}
        \subfigure[$\bar{T}$, K for OSK]
{\includegraphics{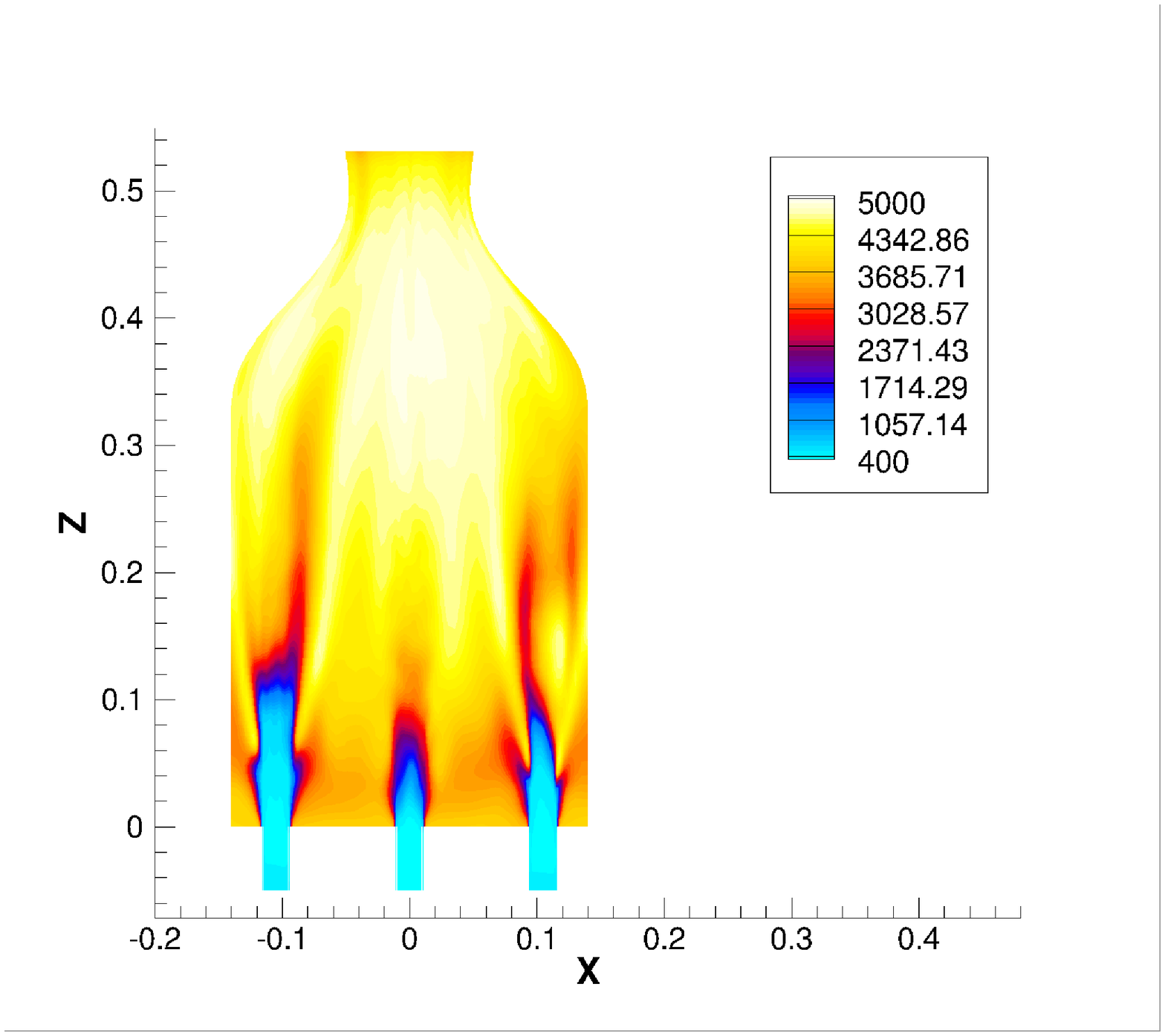}
        \label{fig:tave_T_osk}}
        \subfigure[$\overline{Ma}$ for OSK]
{\includegraphics{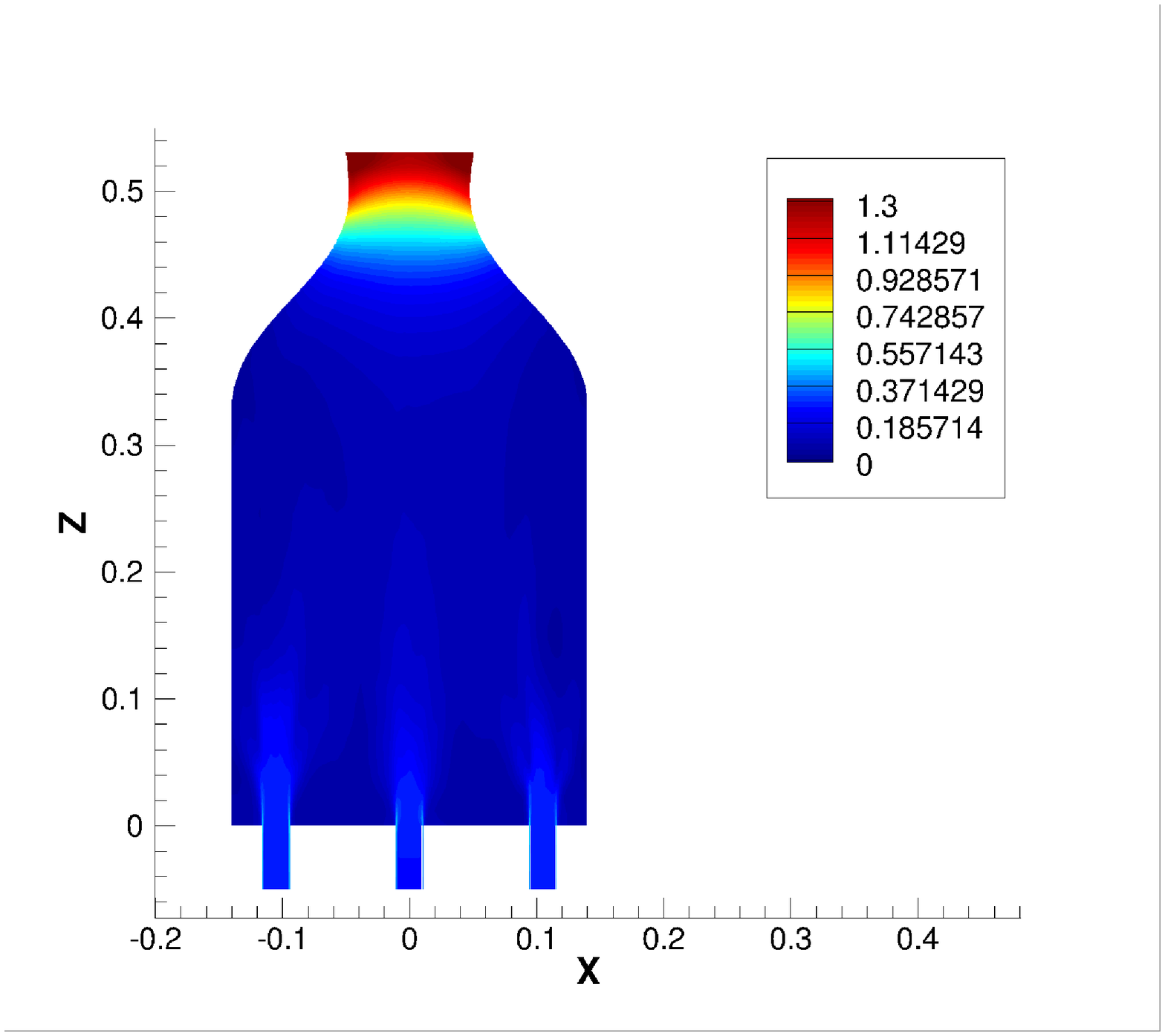}
        \label{fig:tave_mach_osk}}
    \end{subfigmatrix}
    \caption{Time-averaged contours on the $y = 0$ cm meridian plane}
    \label{fig:tave_contours}
\end{figure}

Figure~\ref{fig:iso_zst} shows the instantaneous isosurface of the stoichiometric mixture fraction $Z_{st} = 0.2$ colored by the mass fraction of the major species in the reactions and temperature for the FPV model. Both $CH_{4}$ and $O_{2}$ are consumed fast downstream of the injector plate, producing most of the $CO$ and $H_{2}$ in the upstream half of the combustion chamber. Remarkable reduction of $CO$ and $H_{2}$ occurs in the downstream half of the combustion chamber as further oxidation of the two intermediate species greatly outweighs their replenishment from the burning fuel, which is hardly seen in the same region. However, the continued oxidation of $CO$ and $H_{2}$ in the downstream half of the combustion chamber and the exhaust nozzle leads to the largest local accumulation of the final products $CO_{2}$ and $H_{2}O$. The temperature reaches a maximum as the mixture is about to enter the nozzle due to the continued heat release of the lasting oxidation process. However, the hot mixture cools in the entire nozzle, which is accompanied by the monotonic pressure drop and continued flow acceleration to supersonic speed. The instantaneous isosurface represents the dynamic structures in the flow associated with the propellent jets, indicating quite dynamic flow turbulence and chemical process. One can observe the propellent jet structures close to the injector, and a rapid transition towards a remarkably disorganized flow. This highly turbulent flow interacts with the reactive layer and wrinkles the flame front. As turbulence enhances mixing, the reaction layer is thickened as shown in Figures~\ref{fig:tave_ch4_fpv} and ~\ref{fig:tave_ch4_osk}.

\begin{figure}
    \begin{subfigmatrix}{2}
        \subfigure[$T$ vs. $Z$ at $z = 1$ cm]
{\includegraphics[width=0.3\textwidth]{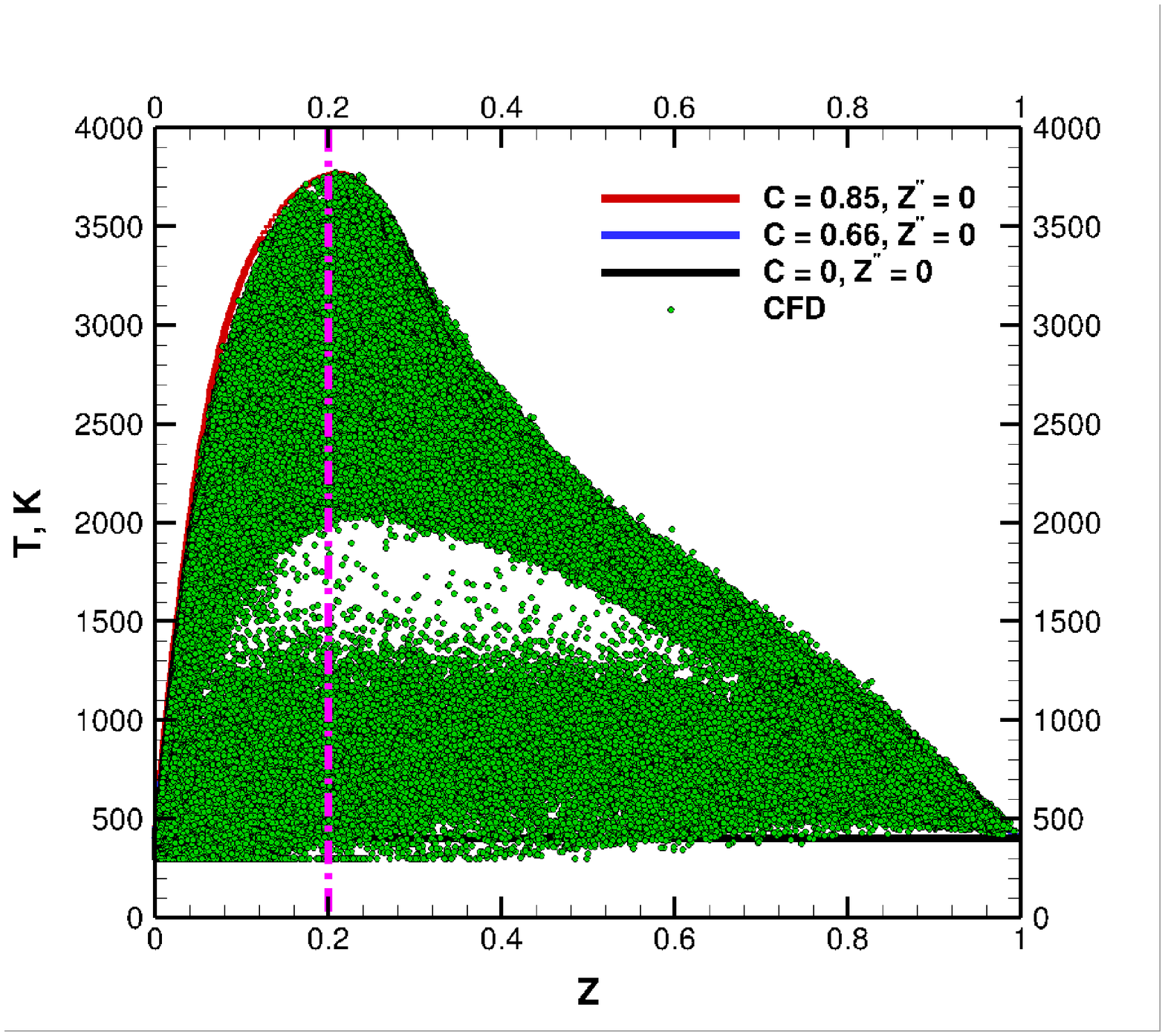}
        \label{fig:T_vs_Z_z_1cm}}
        \subfigure[$C$ vs. $Z$ at $z = 1$ cm]
{\includegraphics[width=0.3\textwidth]{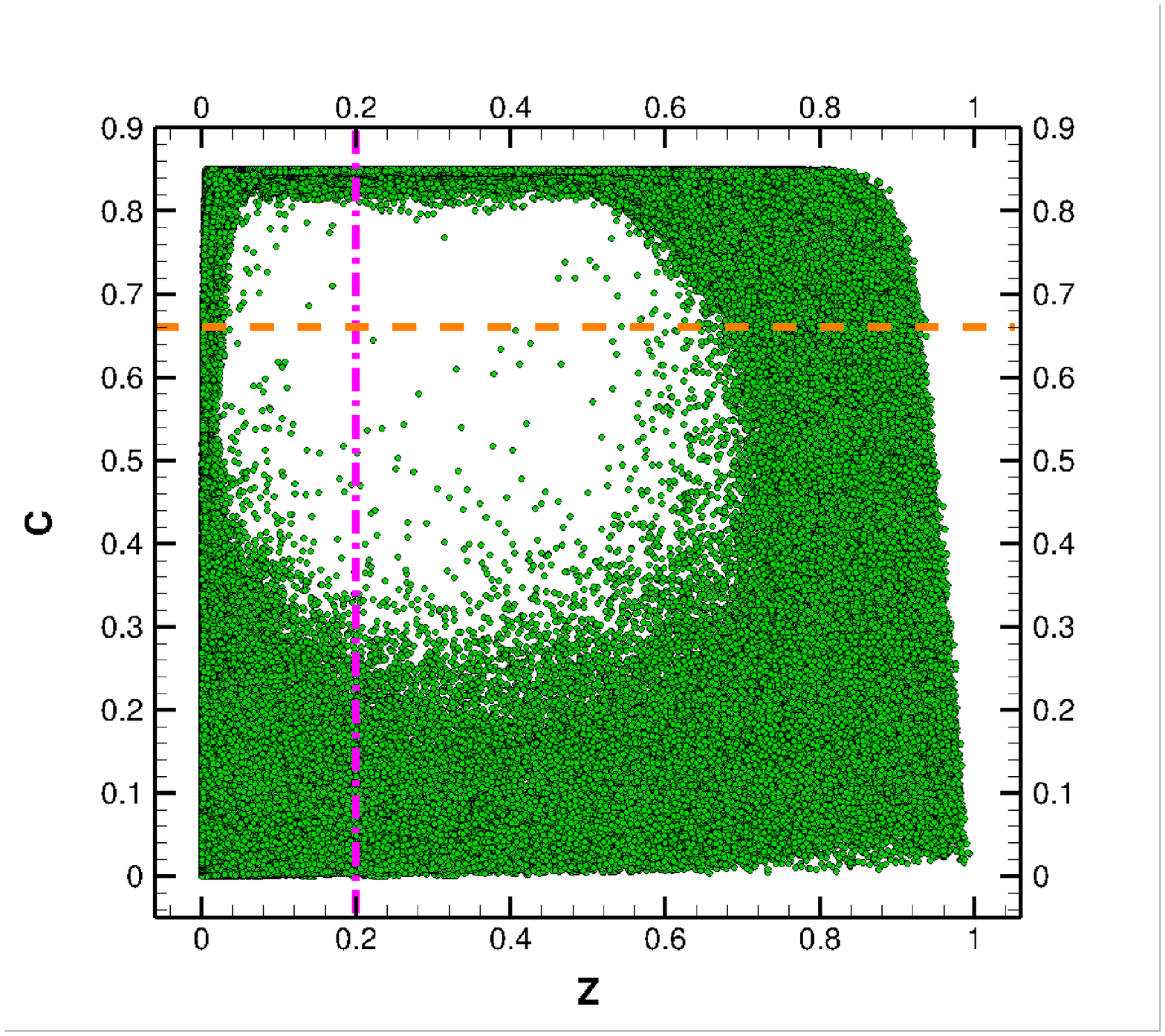}
        \label{fig:C_vs_Z_z_1cm}}
        \subfigure[$T$ vs. $Z$ at $z = 4$ cm]
{\includegraphics[width=0.3\textwidth]{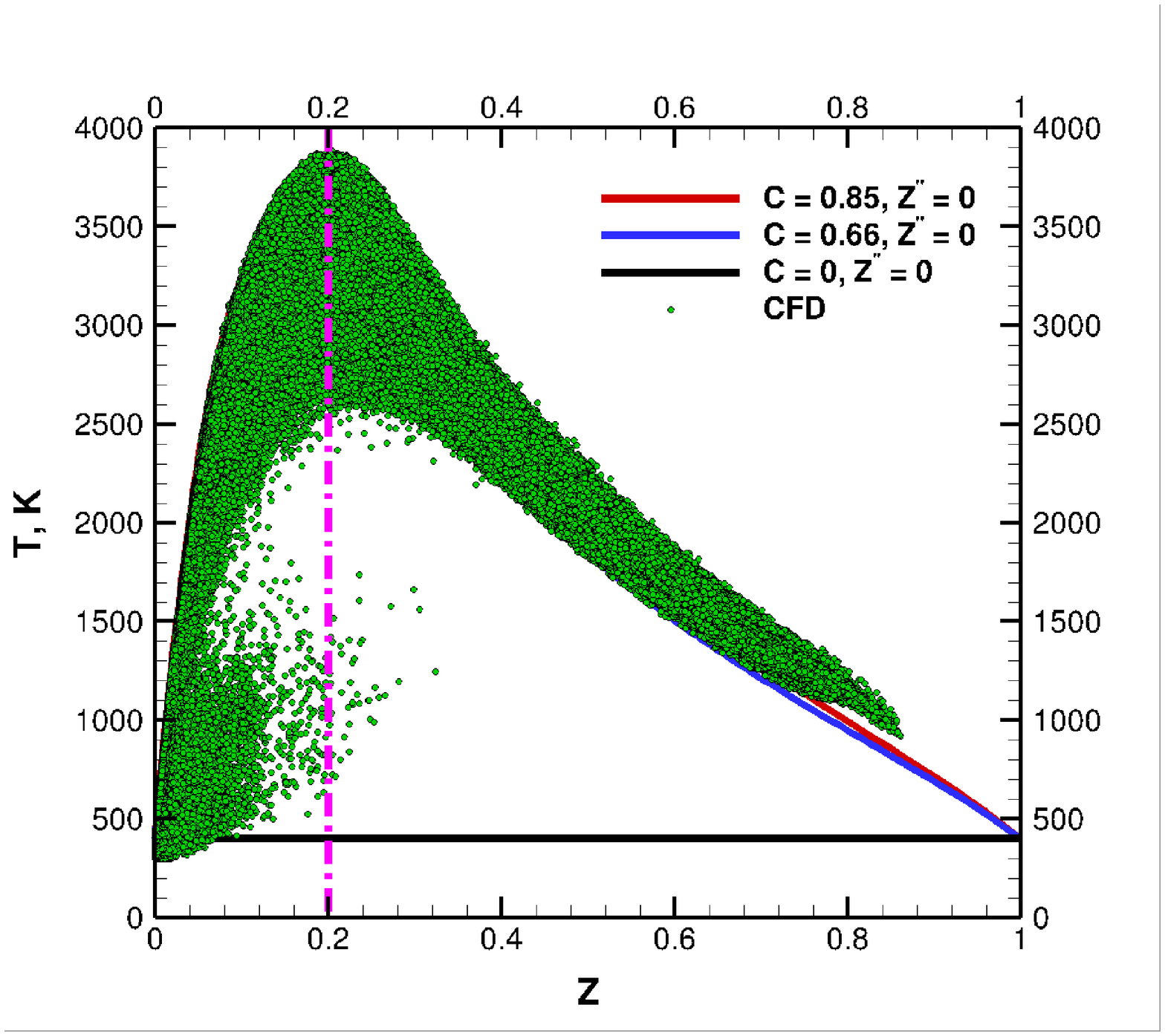}
        \label{fig:T_vs_Z_z_4cm}}
        \subfigure[$C$ vs. $Z$ at $z = 4$ cm]
{\includegraphics[width=0.3\textwidth]{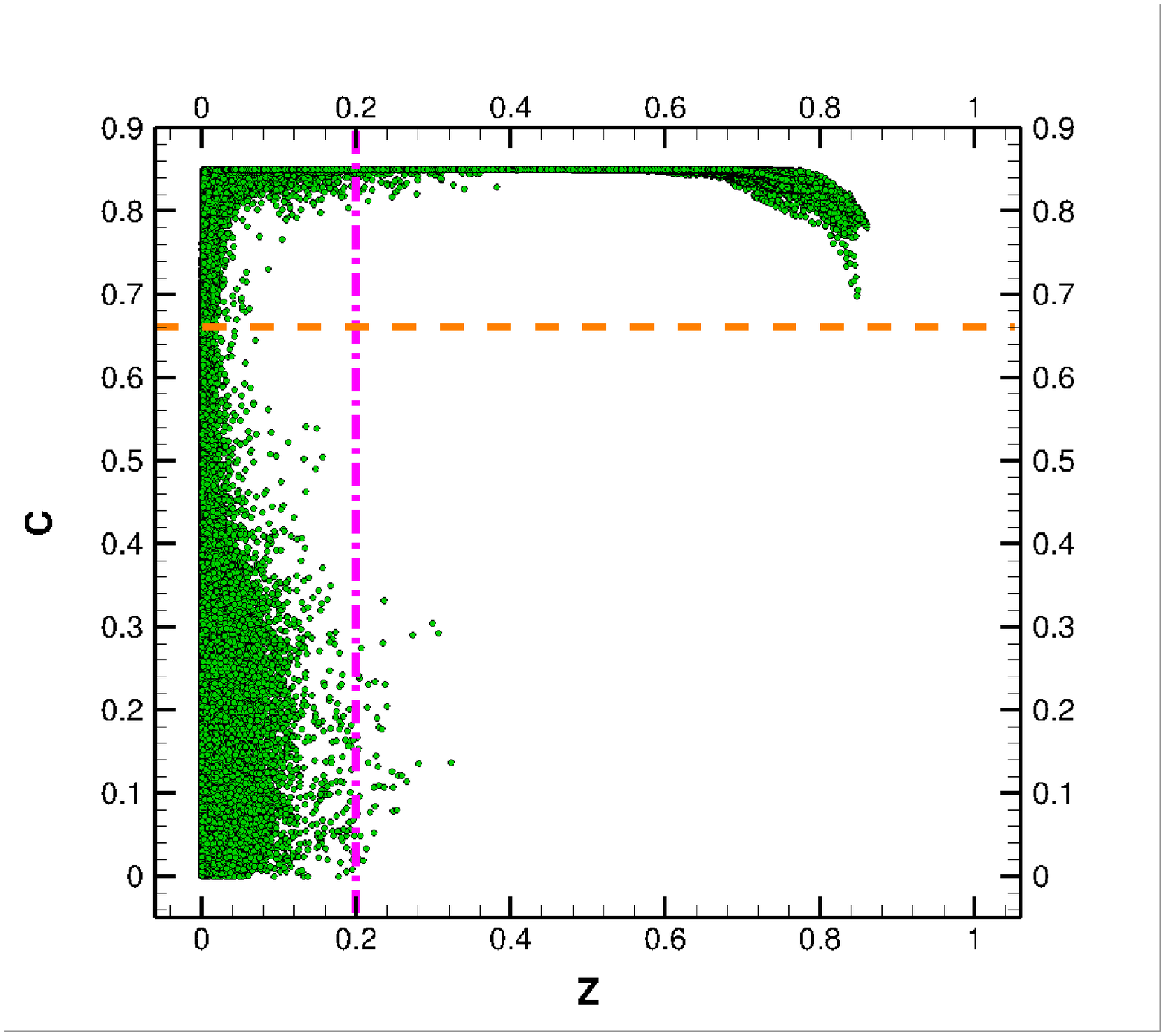}
        \label{fig:C_vs_Z_z_4cm}}
        \subfigure[$T$ vs. $Z$ at $z = 15$ cm]
{\includegraphics[width=0.3\textwidth]{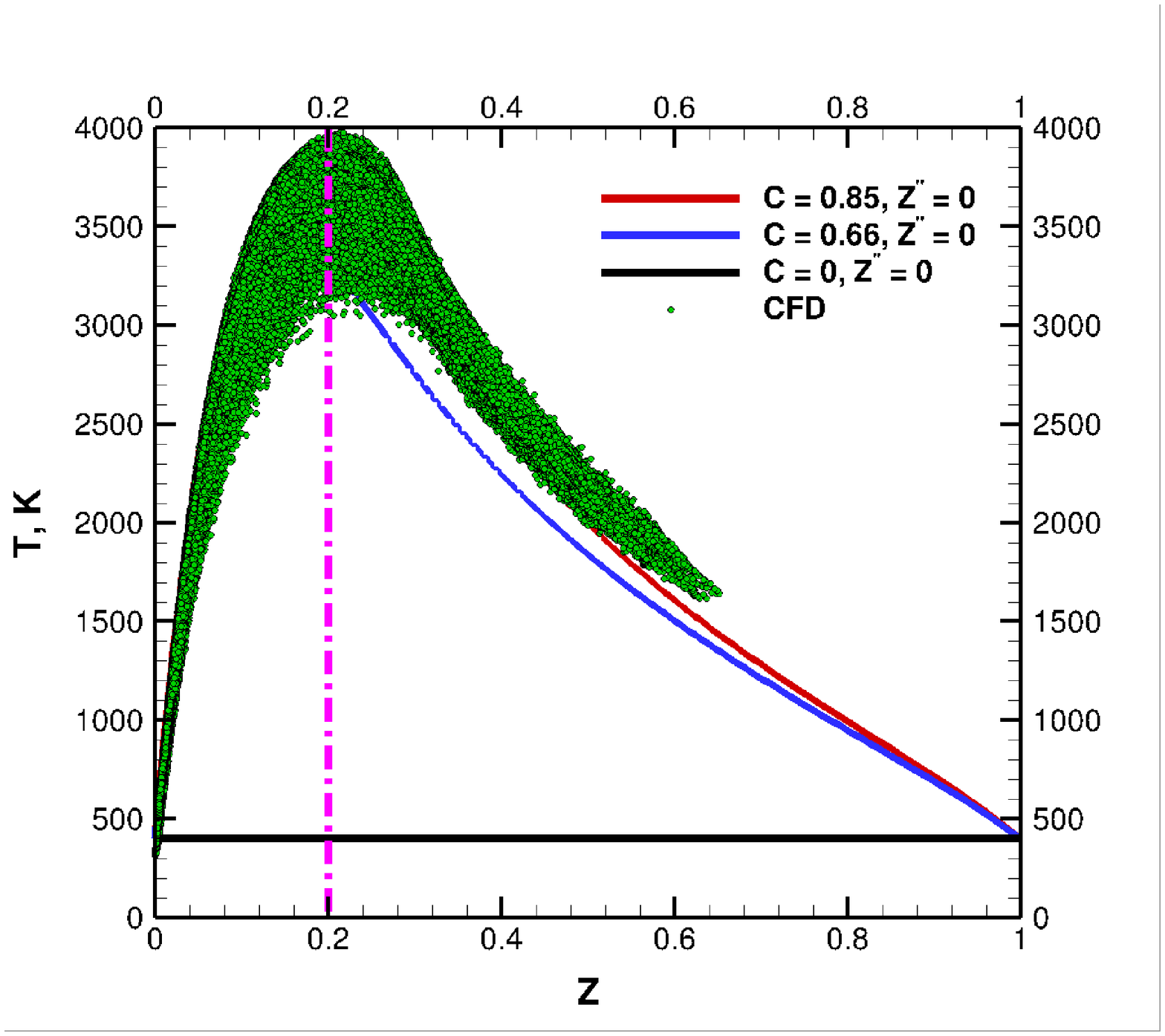}
        \label{fig:T_vs_Z_z_15cm}}
        \subfigure[$C$ vs. $Z$ at $z = 15$ cm]
{\includegraphics[width=0.3\textwidth]{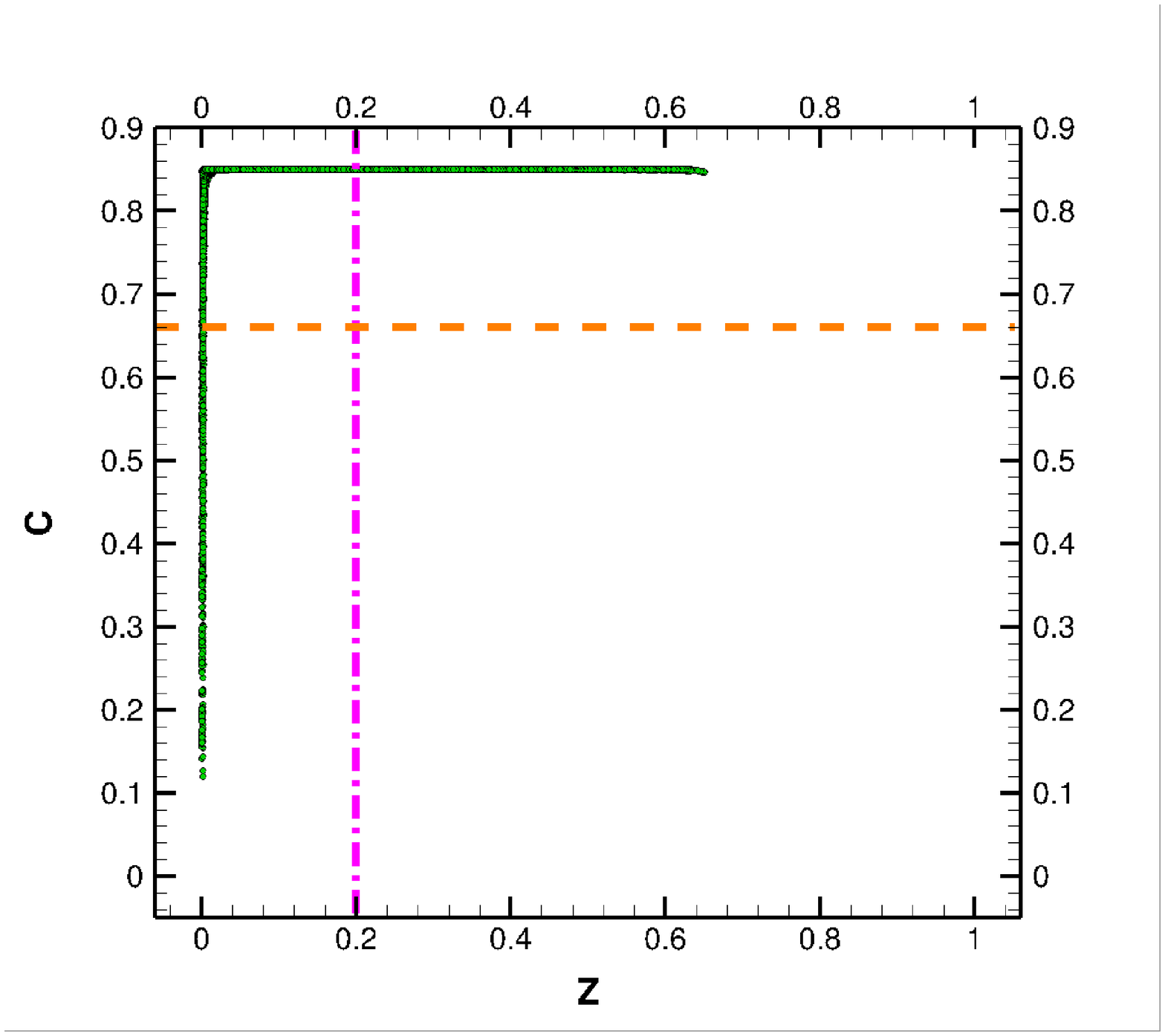}
        \label{fig:C_vs_Z_z_15cm}}
        \subfigure[$T$ vs. $Z$ at $z = 33$ cm]
{\includegraphics[width=0.3\textwidth]{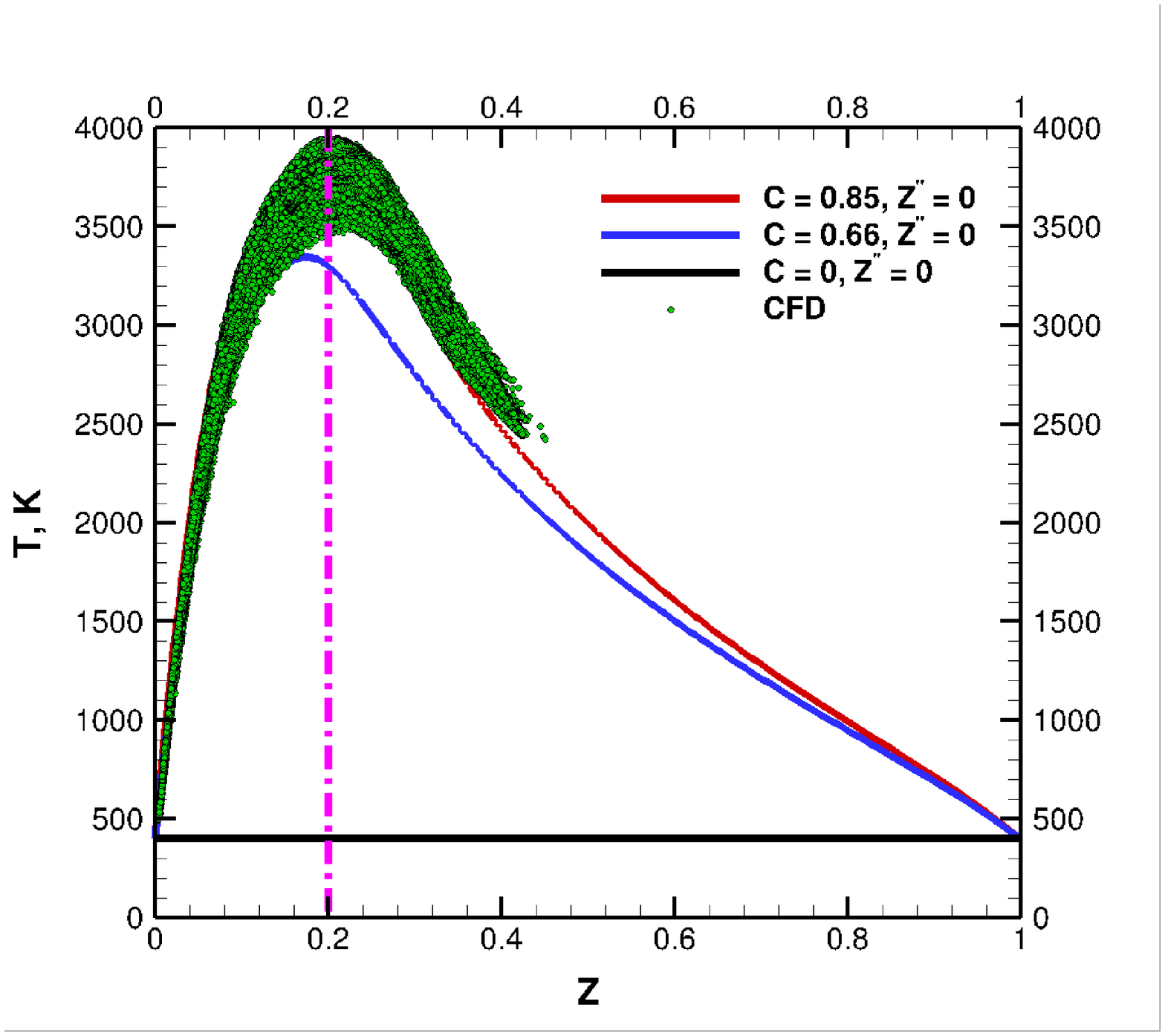}
        \label{fig:T_vs_Z_z_33cm}}
        \subfigure[$C$ vs. $Z$ at $z = 33$ cm]
{\includegraphics[width=0.3\textwidth]{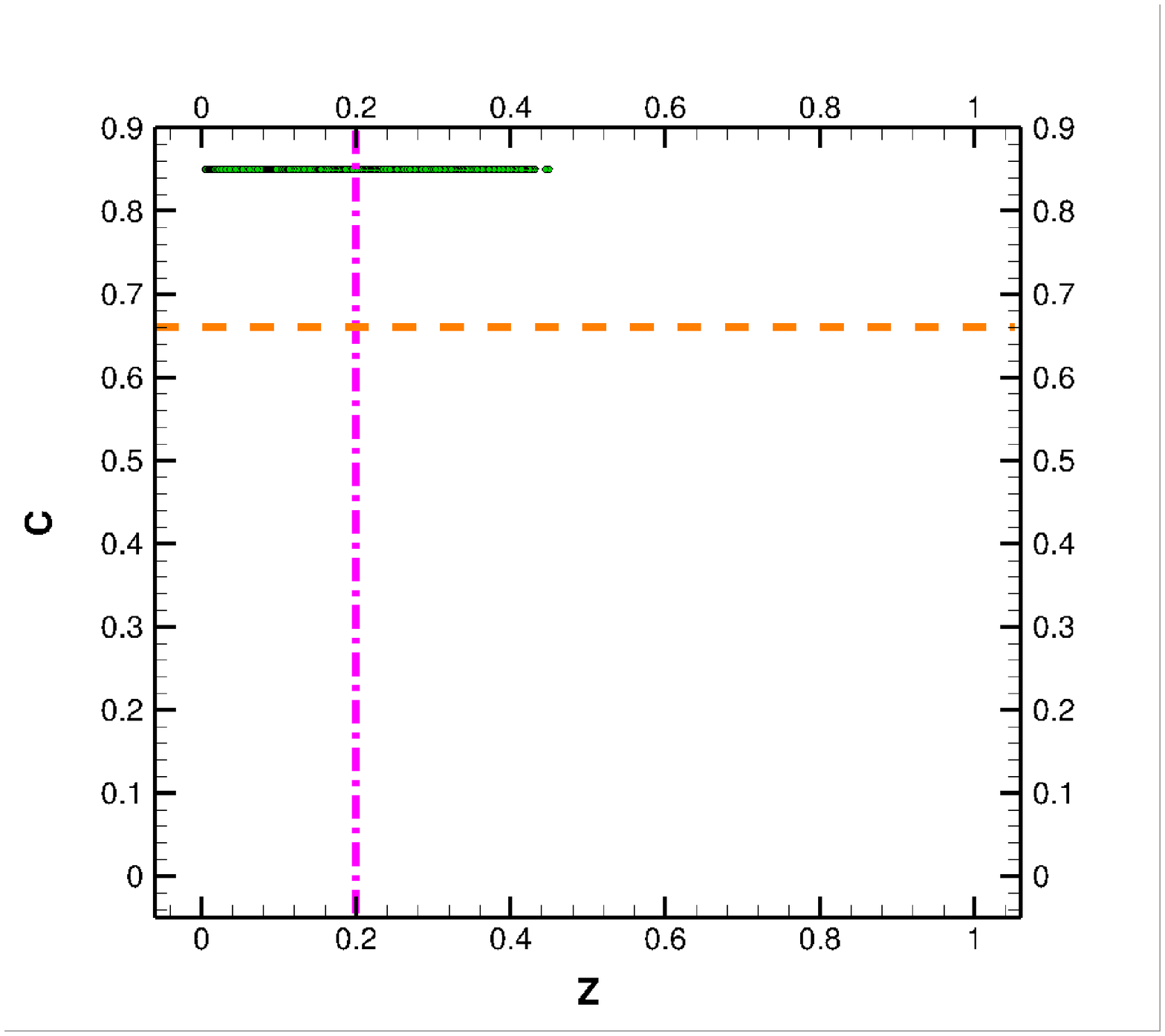}
        \label{fig:C_vs_Z_z_33cm}}
    \end{subfigmatrix}
    \caption{Scatter plot of the flamelet solutions used in CFD
on cross section planes}
    \label{fig:scatter_plot}
\end{figure}

Figure~\ref{fig:scatter_plot} shows scatter plots of the actual flamelet solution points that have been explored by the flow computations. Each of the green
points represents a flamelet solution. As references, the laminar flamelet solutions
for equilibrium burning, quenching limit and pure mixing are also plotted for background pressure
of 200 bar. 
Some scatter points may not appear in the region enveloped by these reference solutions.
A major reason is that the computed temperature may differ from that in the flamelet solutions due to
compressibility. 
The pink dash-dot line separates the fuel-lean ($Z \le 0.2$) and fuel-rich ($Z \ge 0.2$) burning regions. The
quenching limit for the current case is $C = 0.66$. The orange dash line
divides the stable ($C \ge 0.66$) and unstable ($C \le 0.66$) burning branches.
Along the mean flow direction/axial direction, local fuel-rich burning becomes more
and more stable, the locally unstable burning tends to be fuel-lean. Near the exhaust nozzle, both fuel-rich and fuel-lean burnings are stable.

\patchcmd{\subfigmatrix}{\hfill}{\hspace{0.8cm}}{}{}

\patchcmd{\subfigmatrix}{\hspace{0.8cm}}{\hfill}{}{}

\begin{figure}
    \begin{subfigmatrix}{3}
        \subfigure[Time history of the total $\dot{m}$ at the nozzle outlet]
{\includegraphics{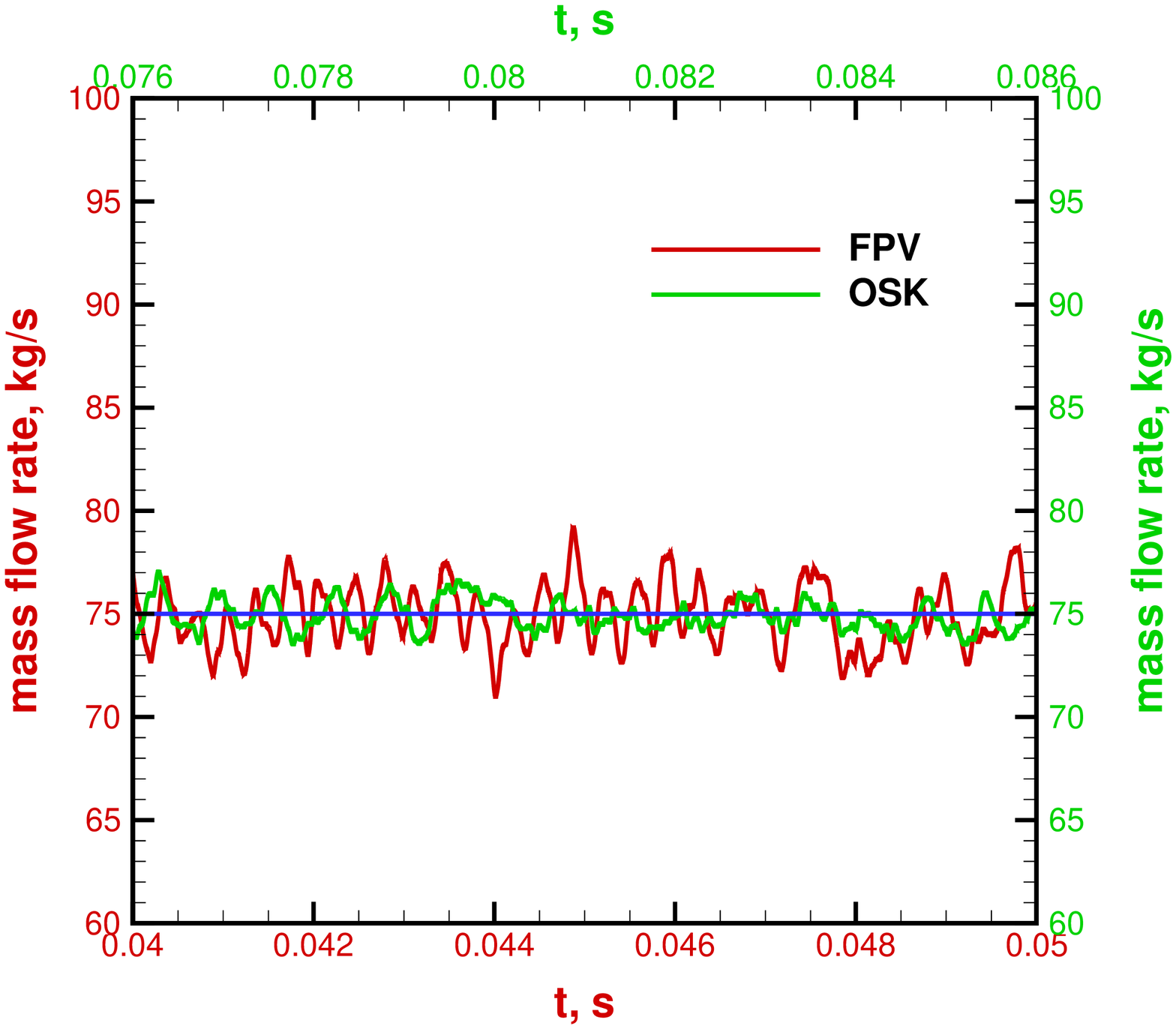}
        \label{fig:mdot_outlet}}
        \subfigure[Time-averaged $\dot{m}$ of individual species vs. axial position for the FPV model]
{\includegraphics{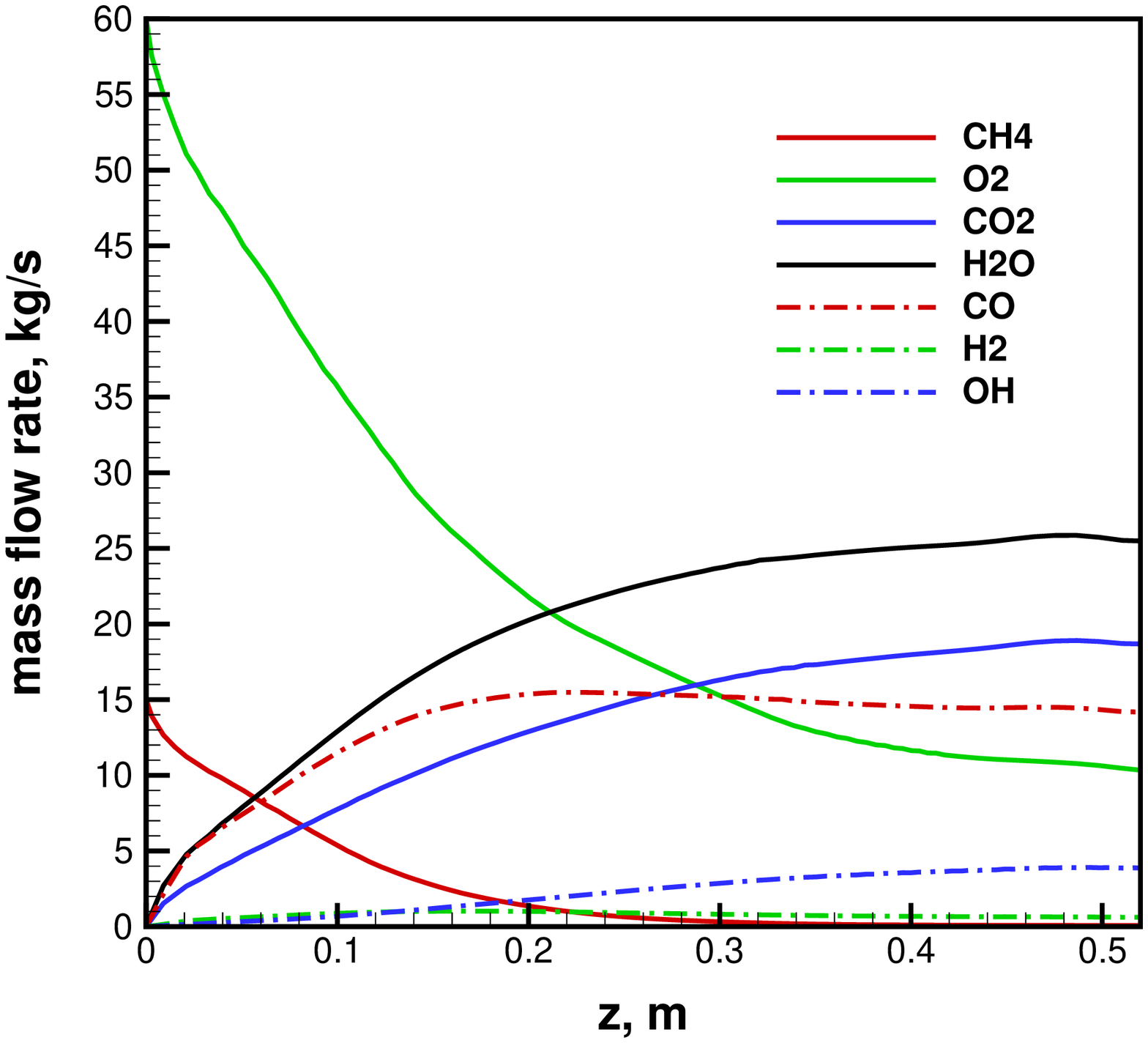}
        \label{fig:mdot_z_fpv}}
        \subfigure[Time-averaged $\dot{m}$ of individual species vs. axial position for the OSK model]
{\includegraphics{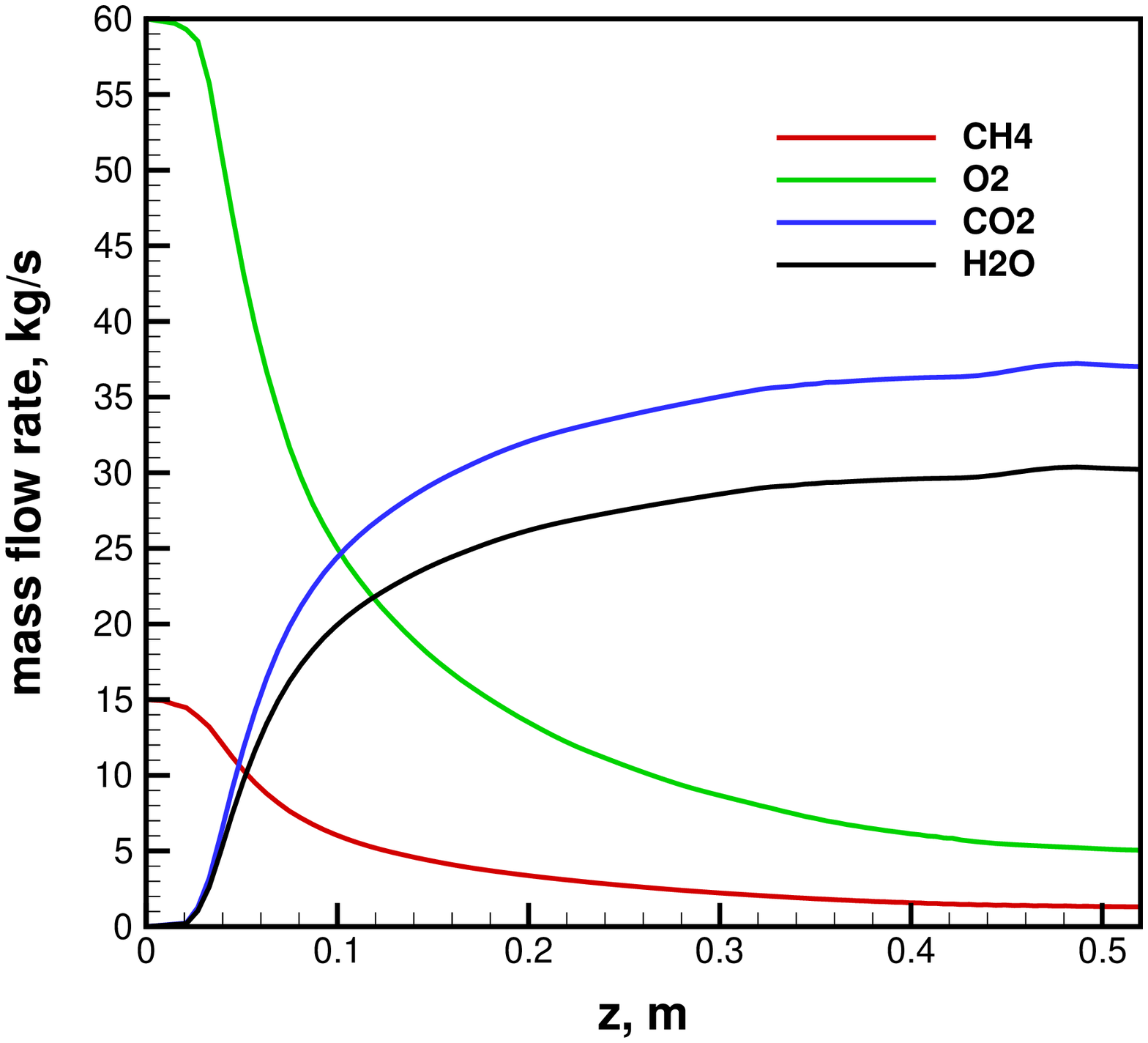}
        \label{fig:mdot_z_osk}}
    \end{subfigmatrix}
    \caption{Cross-sectional mass flow rate $\dot{m}$}
    \label{fig:mdot}
\end{figure}

For comparison, we also implement the OSK model \cite{xiongaiaaj2020} in our computations. Since only four species equations are solved in this combustion model, the overall computational cost is similar to that of the FPV model.
As shown in Figure~\ref{fig:mdot_outlet}, the total mass flow rate at the nozzle outlet remains fluctuating around 
the value at the injector inlets (blue line) with notable amplitude in the last ten milliseconds. 
The limit-cycle behavior indicates that combustion instability occurs.
Figures~\ref{fig:mdot_z_fpv} and ~\ref{fig:mdot_z_osk} show that the OSK model predicts an evident mixing region immediately downstream of the injector plate. Within this region, the global reaction is weak. After sufficient mixing is achieved, the fast global reaction takes place and the major production of $CO_{2}$ and $H_{2}O$ is completed within the upstream half of the combustion chamber. For the FPV model, fast dissociation of the fuel and further reactions are found just downstream of the injector plate as the chemical time scale is assumed to be much less than the turbulent time scales in the combustion model. Throughout the upstream half of the combustion chamber, lasting dissociation and oxidation of the fuel lead to continued accumulation of the major species including $CO$, $CO_{2}$, $H_{2}$, $H_{2}O$ and $OH$. However, the mass flow rates of $CO$ and $H_{2}$ start to decrease in the middle of the combustion chamber and they continue reducing further downstream. One of the two reasons is the full consumption of the fuel beyond the cross-sectional plane at the chamber midpoint. The oxidation of the carbon element into $CO$ and $H_{2}$ generation from the dissociation of the fuel almost completely stopped. The other reason is that, at the same time, further oxidation of $CO$ into $CO_{2}$ continues, as does the oxidation of $H_{2}$ into $OH$ and $H_{2}O$. For the reacting flow in the entire rocket engine, the global production rates of $CO_{2}$ and $H_{2}O$ are lower in the FPV model compared to the OSK model as the two species are found with larger mass flow rates in the exhaust nozzle for the latter model. Another major difference is that significant oxidation and hence heat release occur in the entire rocket engine including the downstream half of the combustion chamber for the FPV model. For the OSK model, the evident reaction is only found in the upstream half of the combustion chamber. This major difference suggests distinct acoustic instability behavior between the two models as different locations of significant heat release are predicted by the two models.

Figure~\ref{fig:tave_contours} shows the time-averaged contours over the last ten milliseconds. The exhaust nozzle is choked with both combustion models.
There is some unburnt fuel in the downstream part of the combustion chamber and the exhaust nozzle for the OSK model.
Temperature is overpredicted in the combustion chamber by the OSK model due to the one-step reaction assumption, which is unrealistic because dissociation is not allowed. The temperature
prediction using the FPV model is lower and more reasonable because much more detailed chemical kinetics is used.

To identify levels of local premixed flames and local diffusion flames, the following simple flame index can be defined:

\begin{equation}
Flame Index = \omega_{T}\frac{\nabla Y_{CH_{4}} \cdot \nabla Y_{O_{2}}}{\left | \nabla Y_{CH_{4}} \right | \cdot \left | \nabla Y_{O_{2}}\right |}
\label{FI}
\end{equation}

\begin{figure}
    \begin{subfigmatrix}{3}
        \subfigure[$y = 0$ cm plane for FPV]
{\includegraphics{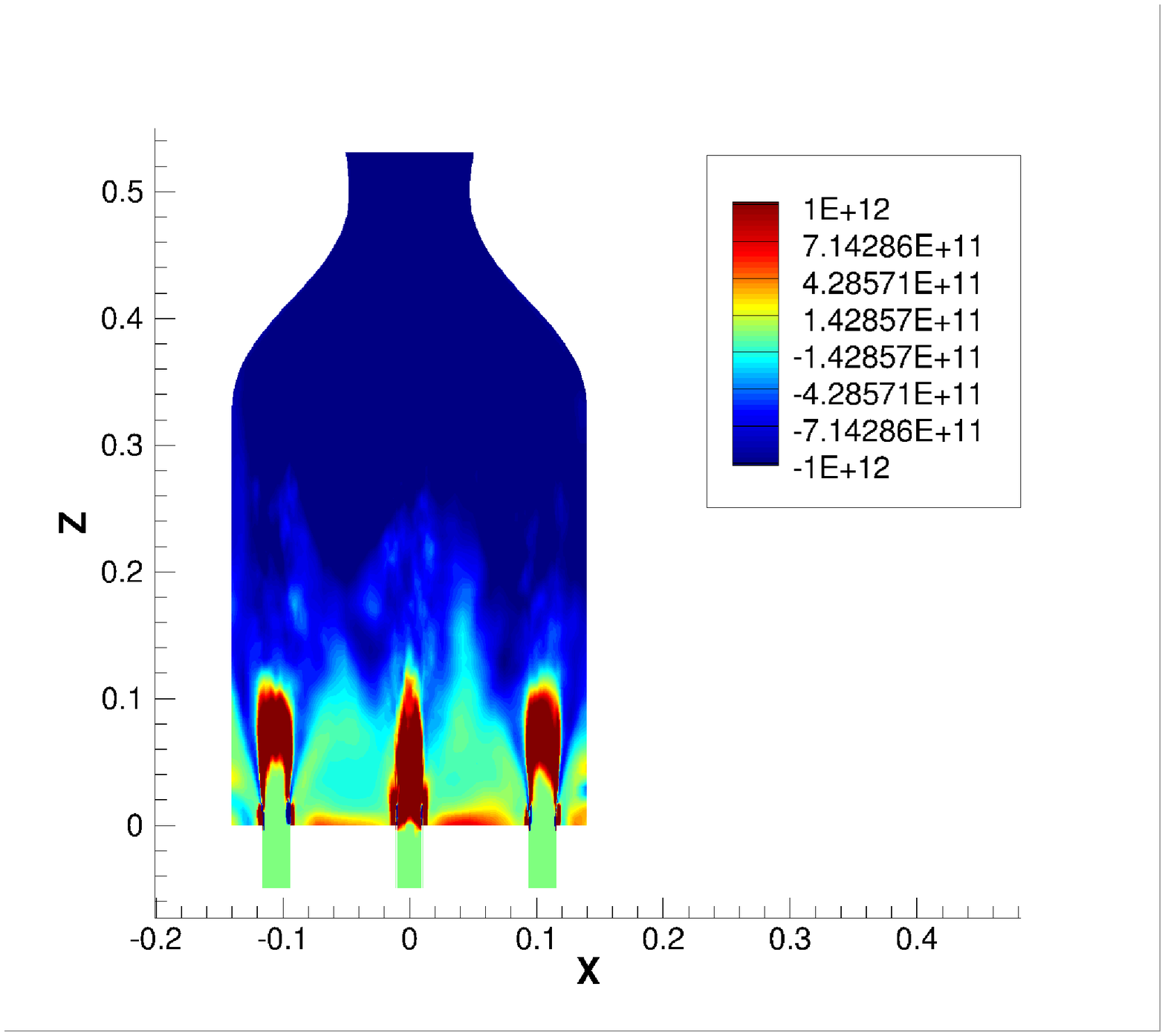}
        \label{fig:tave_FI_y_0cm_fpv}}
        \subfigure[$z = 4$ cm plane for FPV]
{\includegraphics{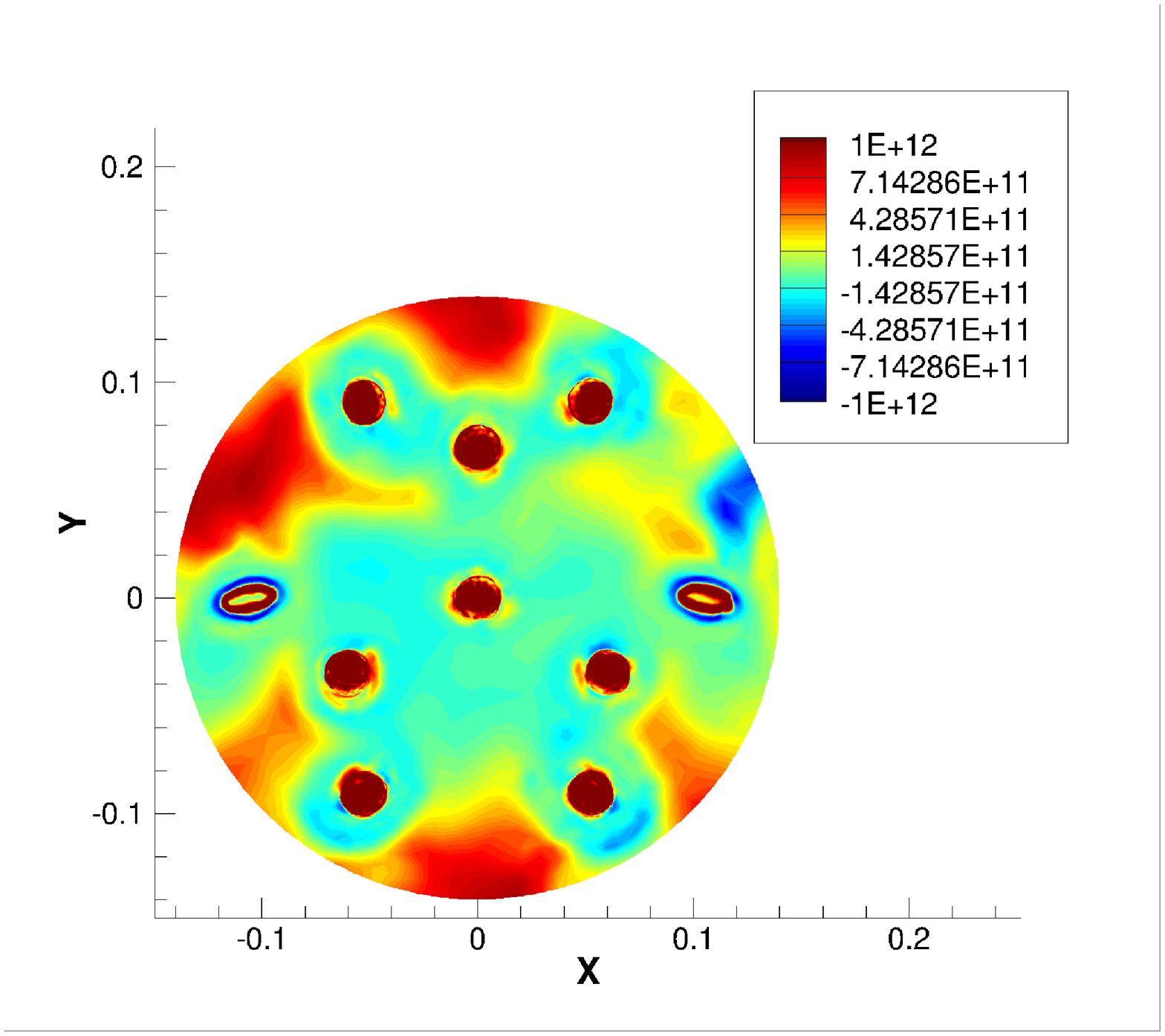}
        \label{fig:tave_FI_z_4cm_fpv}}
        \subfigure[$z = 33$ cm plane for FPV]
{\includegraphics{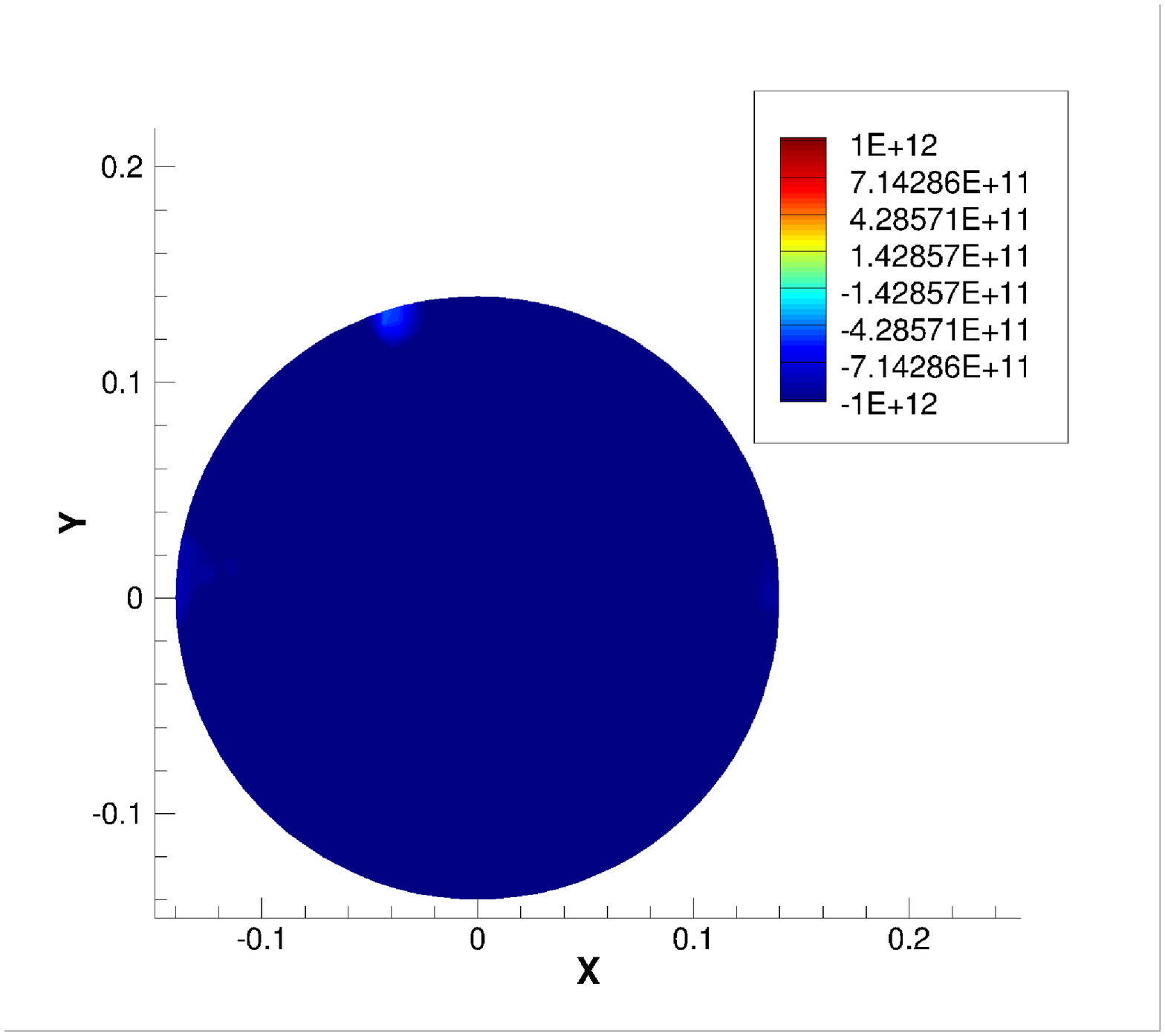}
        \label{fig:tave_FI_z_33cm_fpv}}
        \subfigure[$y = 0$ cm plane for OSK]
{\includegraphics{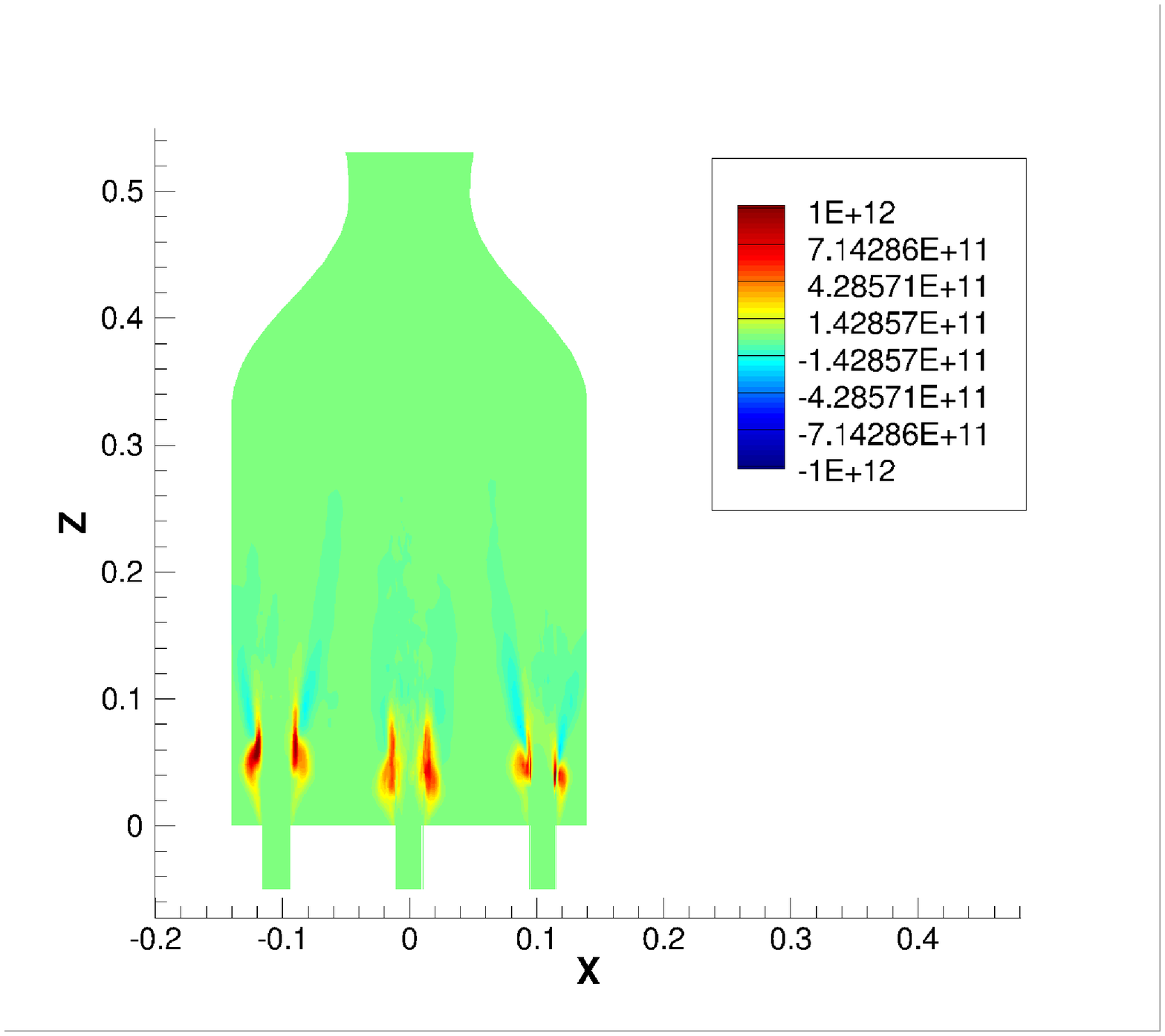}
        \label{fig:tave_FI_y_0cm_osk}}
        \subfigure[$z = 4$ cm plane for OSK]
{\includegraphics{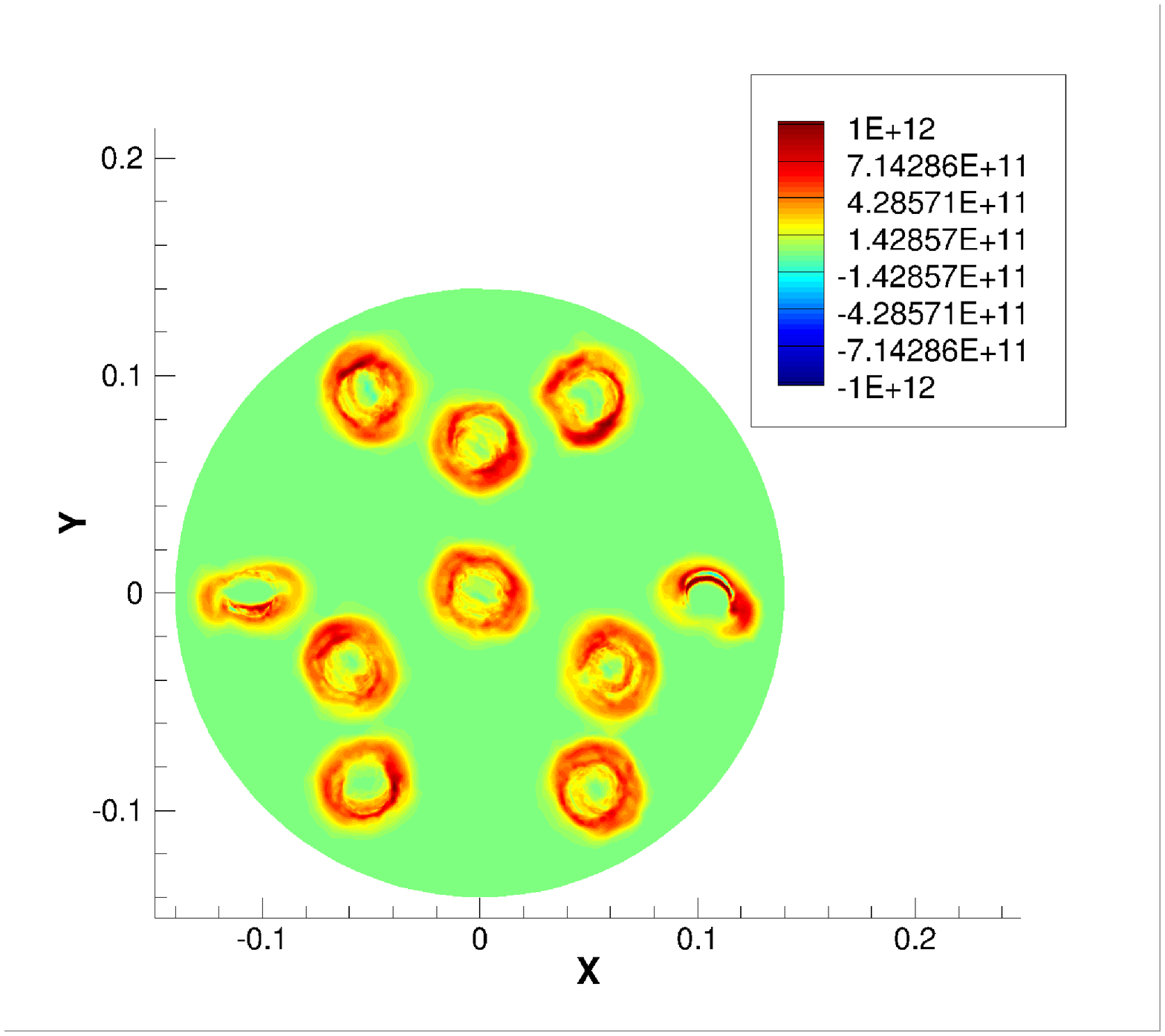}
        \label{fig:tave_FI_z_4cm_osk}}
        \subfigure[$z = 33$ cm plane for OSK]
{\includegraphics{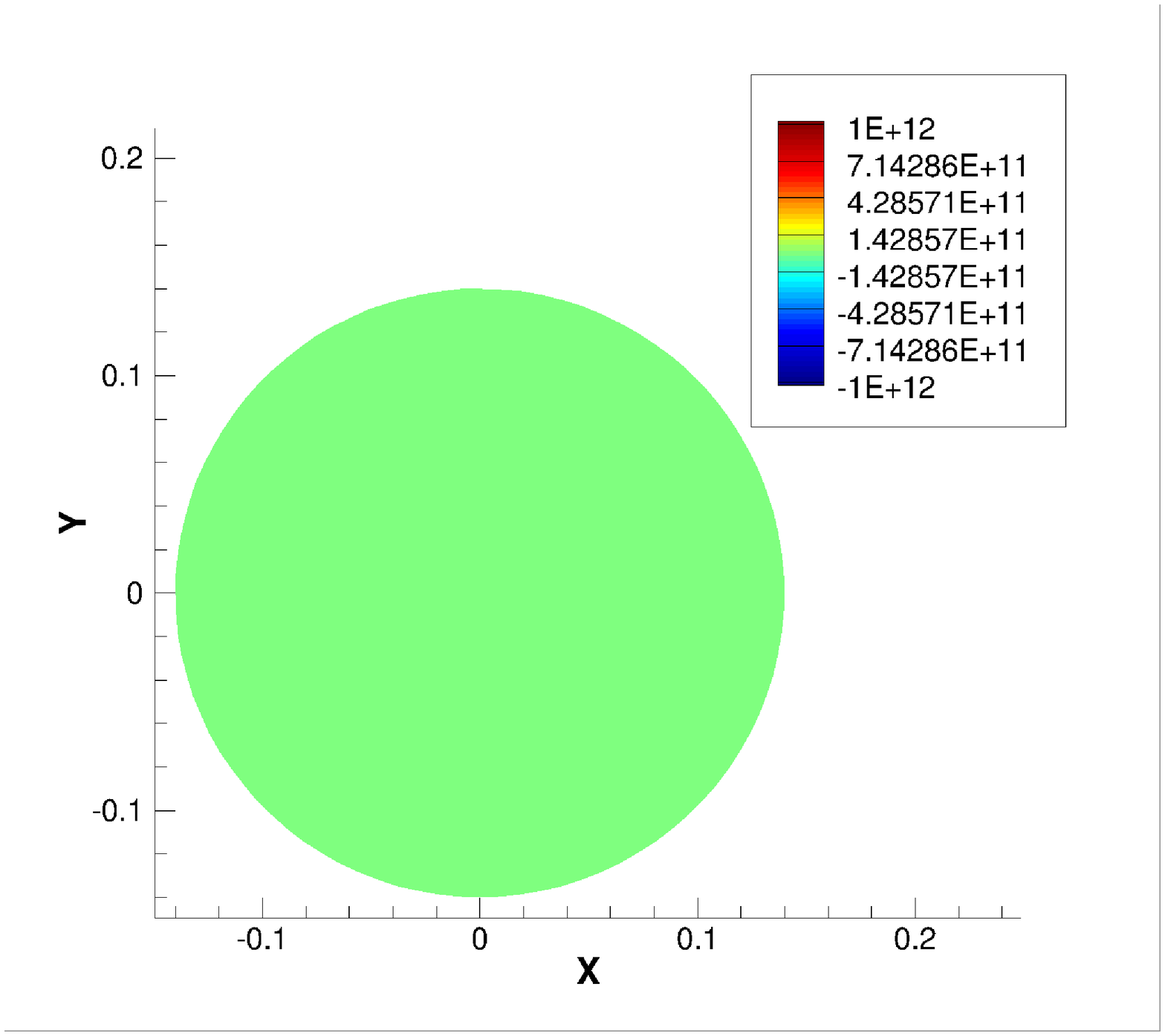}
        \label{fig:tave_FI_z_33cm_osk}}
    \end{subfigmatrix}
    \caption{Time-averaged flame index}
    \label{fig:tave_FI}
\end{figure}

Aligned gradients of the fuel mass fraction and the oxidizer mass fraction produce positive values of the flame index, which indicates premixed flames. When the two gradients are generally in the opposite direction, the flame index becomes negative, which suggests non-premixed or diffusion flames. As shown in Figure~\ref{fig:tave_FI}, although the fuel and oxidizer just begin to mix after they enter the combustion chamber, the current computation using the FPV model shows partially premixed combustion. 
Note that there is a contradiction with the use of the FPV model which is constructed on the assumption of a diffusion flame but yet yields results for premixed flames in some cases. 
The simulation by the OSK model also shows the partially premixed flames. In the surrounding region of the propellant jets, the premixed flames dominate for both combustion models. 
However, Figures~\ref{fig:tave_FI_y_0cm_fpv} and Figures~\ref{fig:tave_FI_y_0cm_osk} show that the OSK model predicts strong premixed flames away from the injector lips while the FPV model favors strong premixed flames that are attached to the injector lips. This is consistent with the observations from Figures~\ref{fig:mdot_z_fpv} and ~\ref{fig:mdot_z_osk} that significant one-step global reaction in the OSK model requires sufficient mixing before taking place and the FPV model assumes infinitely fast reactions. 
The contours on the $z = 4$ cm plane reveal a key feature of some near-wall flames that is often associated with nearby pressure nodes in combustion instability. In Figures~\ref{fig:tave_FI_z_4cm_fpv}, the premixed flames surrounding the left-most and the right-most injectors are noticeably stretched almost in the same direction, indicating transverse velocity fluctuation in the direction orthogonal to the stretch. Such flattened flames are called the ``N-flame" by Urbano et al. \cite{multicnf}. In addition, the stretched premixed flames are further enclosed by flattened diffusion flames. At the locations corresponding to the rest injectors, only the premixed flames are significant and their round shapes are maintained. On the same cross-sectional plane, similar premixed flame patterns are observed in Figures~\ref{fig:tave_FI_z_4cm_osk} for the OSK model except that the ``N-flame" is less stretched, indicating a slight azimuthal shift of the pressure nodes. The stretched premixed flames are not found surrounded by the flattened diffusion flames on this plane but it occurs on further downstream planes, which can be verified in Figures~\ref{fig:tave_FI_y_0cm_osk}. In the downstream half of the combustion chamber, the diffusion flames prevail for both combustion models. However, for the FPV model, continuous consumption of $CO$ and $H_{2}$ as shown in Figures~\ref{fig:mdot_z_fpv}, and hence associated heat release results in much stronger diffusion flames in the region. $CH_{4}$ and $O_{2}$ in the downstream half of the combustion chamber indeed diffuse in the opposite directions for the OSK model but the heat release there is negligible as no significant global reaction occurs in the region.

\begin{figure}
    \begin{subfigmatrix}{3}
        \subfigure[$\bar{C}$]
{\includegraphics{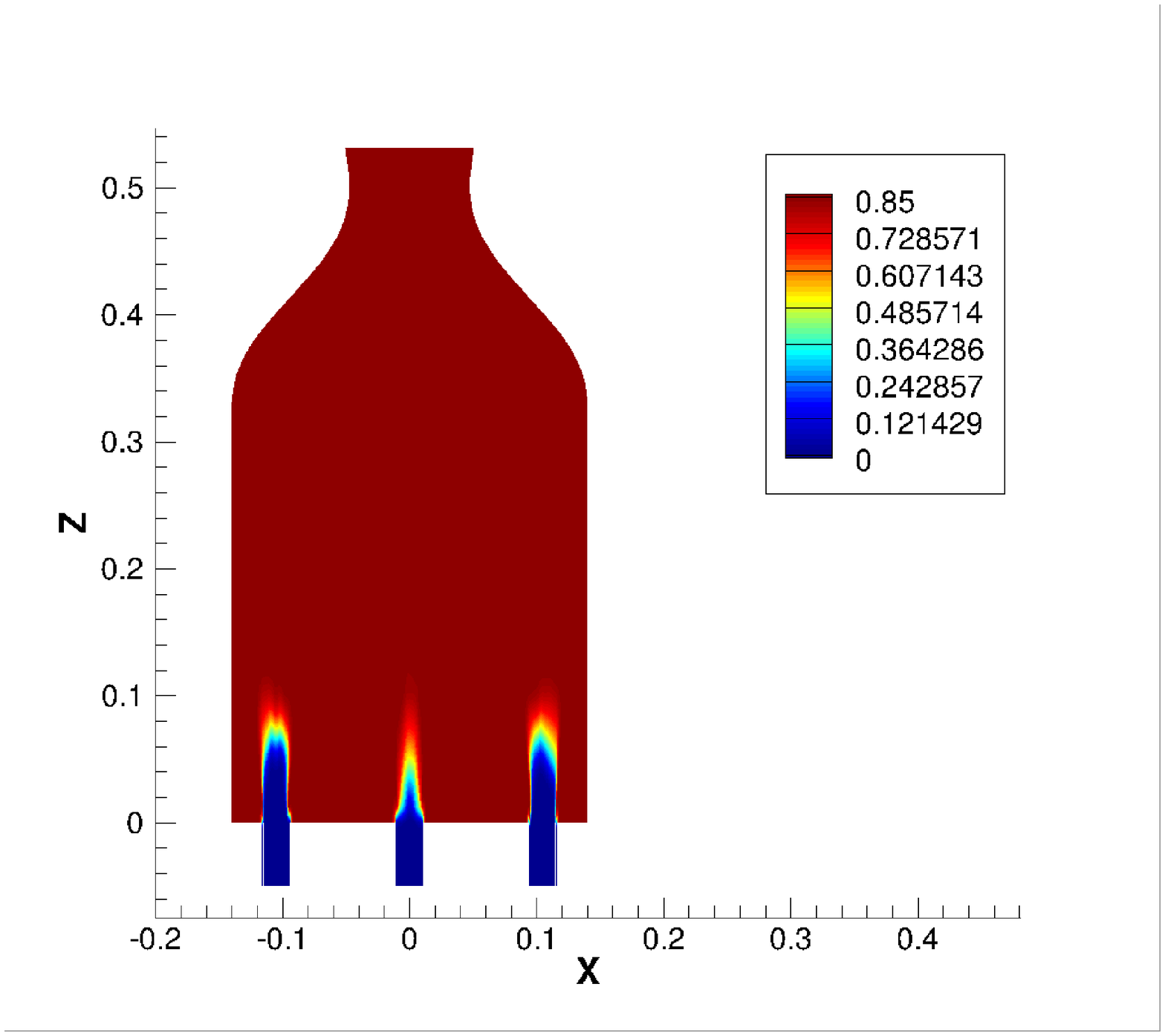}
        \label{fig:tave_progc_fpv}}
        \subfigure[$\bar{\omega_{C}}$, $kg/m^{3}/s$]
{\includegraphics{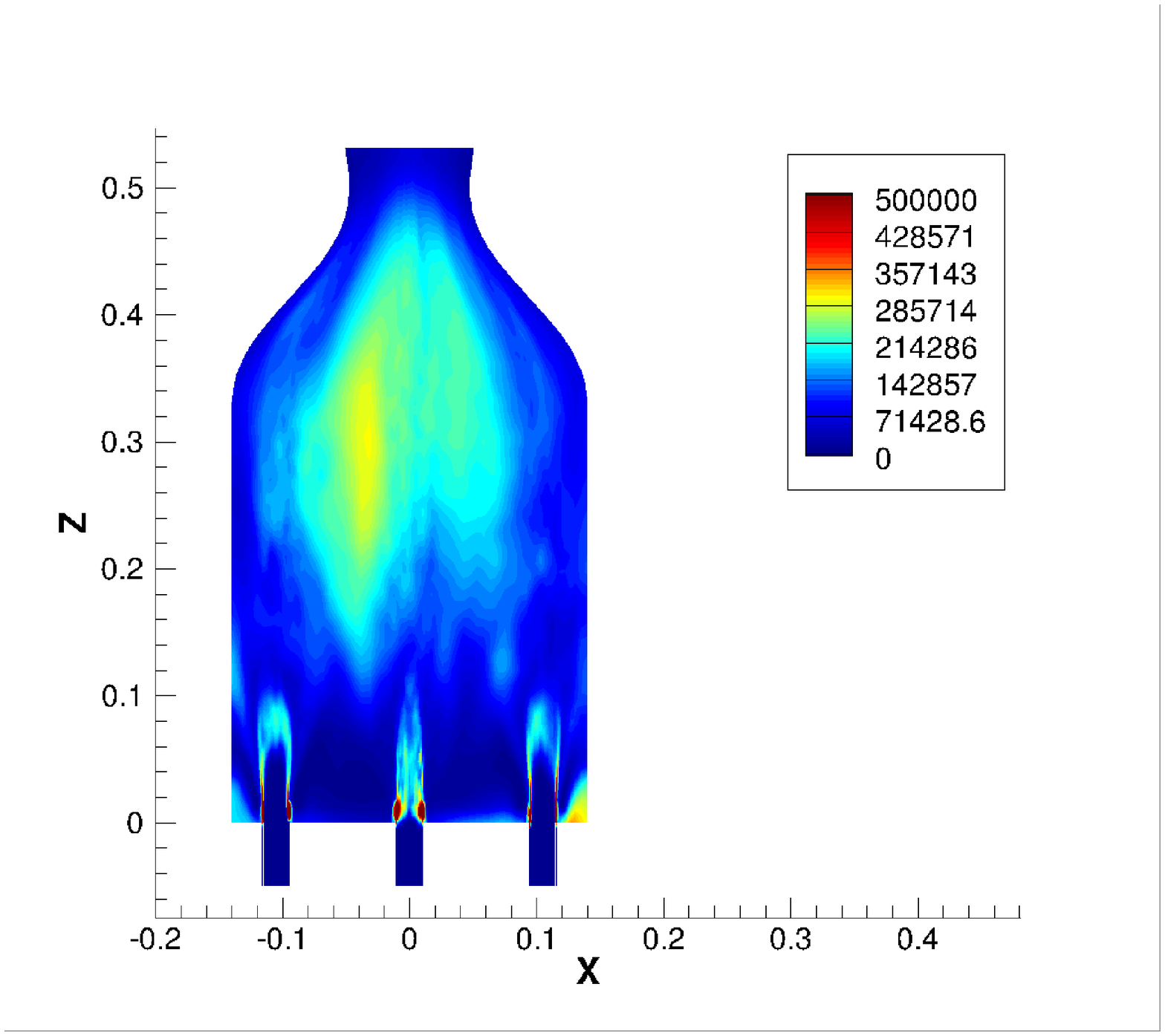}
        \label{fig:tave_omc_fpv}}
        \subfigure[$\bar{Z^{''}}$]
{\includegraphics{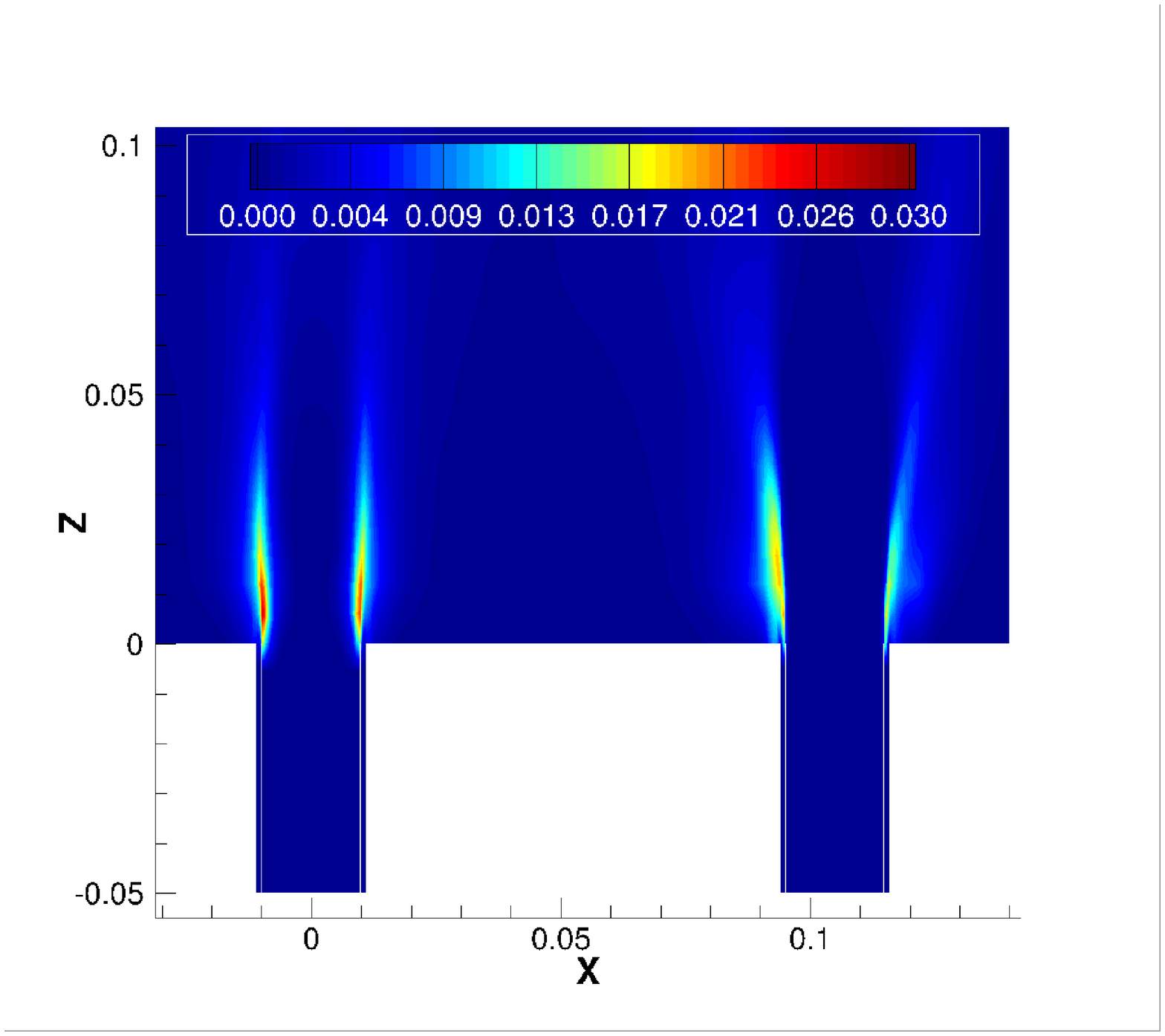}
        \label{fig:tave_varz_fpv}}
    \end{subfigmatrix}
    \caption{Time-averaged contours on the $y = 0$ cm meridian plane for FPV}
    \label{fig:tave_contours_fpv}
\end{figure}

To provide more insight into the flame anchoring, contours of the time-averaged progress variable and its reaction rate are shown in Figures~\ref{fig:tave_progc_fpv} and ~\ref{fig:tave_omc_fpv}. The premixed flame is anchored in the fuel/oxidizer shear layers. Despite the low total number of grid cells, the current mesh clusters grid cells in the jet regions and seems to be able to roughly predict effects of shear layer dynamics and unsteady mixing. Nguyen and Sirignano \cite{tuancnf} pointed out that turbulent mixing and turbulence/flame interaction of the mean mixture fraction $Z$
are modeled by solving equation of its variance $Z^{''}$. Figure~\ref{fig:tave_varz_fpv} shows the contour of time-averaged $Z^{''}$ in the exit vicinity of the central and an outer-annulus injectors. The present simulation finds that relatively large and fluctuating $Z^{''}$ appear in the fuel/oxidizer shear layers, reflecting active unsteady mixing and shear-layer/flame interaction in the region. In principle, low grid resolution in the flame-anchoring region will effectively mischaracterize the flame structure and hence the combustion instability. However, Nguyen and Sirignano \cite{tuanaiaaj} pointed out that relatively coarse mesh of roughly $60,000$ cells in axisymmetric simulations can still reasonably capture the correct unstable behavior of the Purdue CVRC experiments. Xiong et al. \cite{xiongaiaaj2020} performed the grid independence study for a 19-injector configuration and found further mesh refinement has little impact on pressure oscillation frequencies. The resolution of the present mesh is about identical to that for the 19-injector configuration. It should be sufficient for the combustion instability study in this paper. Figure~\ref{fig:tave_omc_fpv} also shows reactions in the downstream half of the chamber as well as in the nozzle for the FPV model and confirms the existence of the diffusion flames in those regions.

\subsection{Combustion instability analysis of the ten-injector liquid rocket engine}

\label{sec:com_inst_ana_10inj_rocket_engine}

\begin{figure}
    \begin{subfigmatrix}{3}
        \subfigure[Probes at $z = 1$ cm for FPV]
{\includegraphics{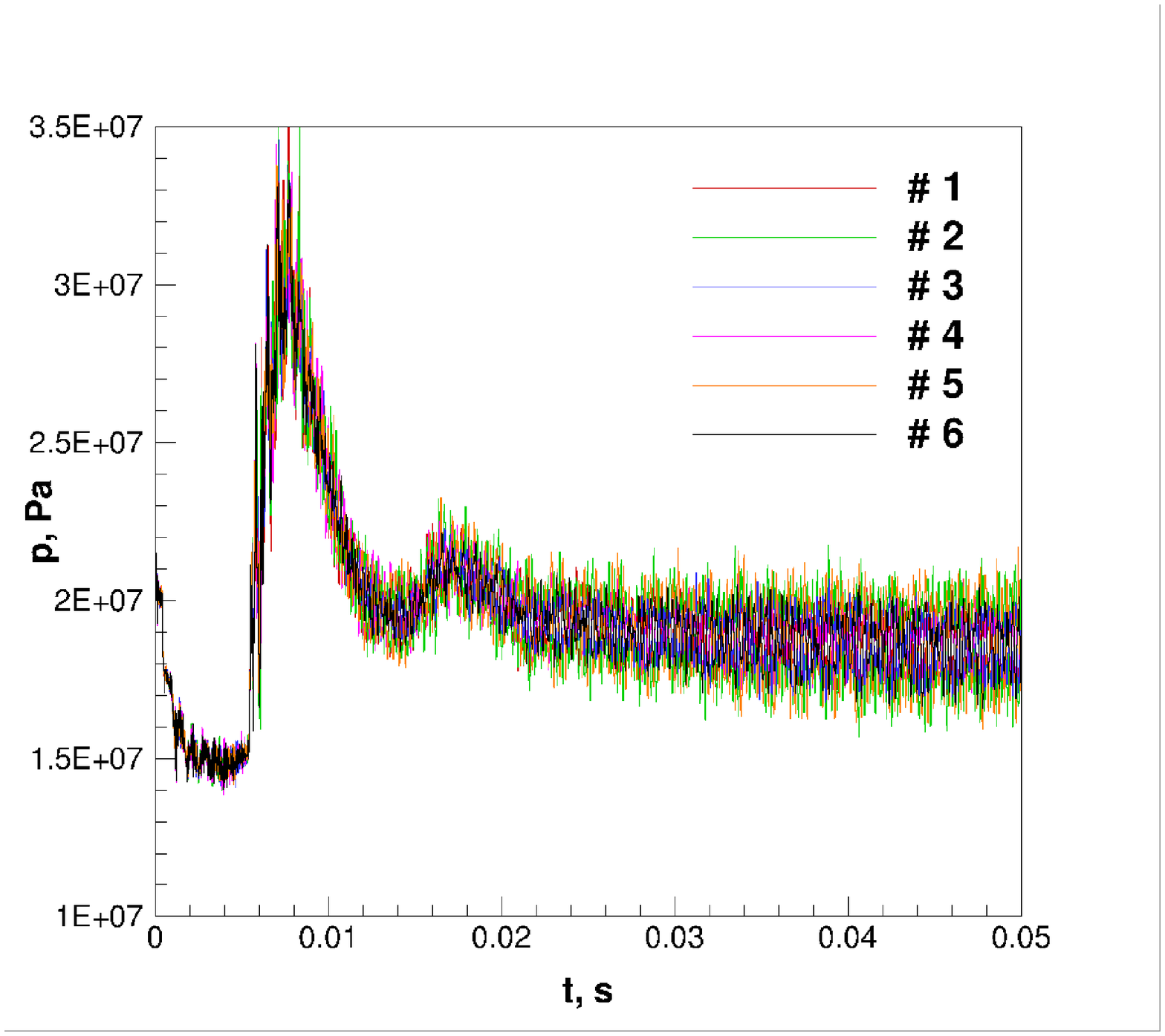}
        \label{fig:p_his_z_1cm_fpv}}
        \subfigure[Probes at $z = 18$ cm for FPV]
{\includegraphics{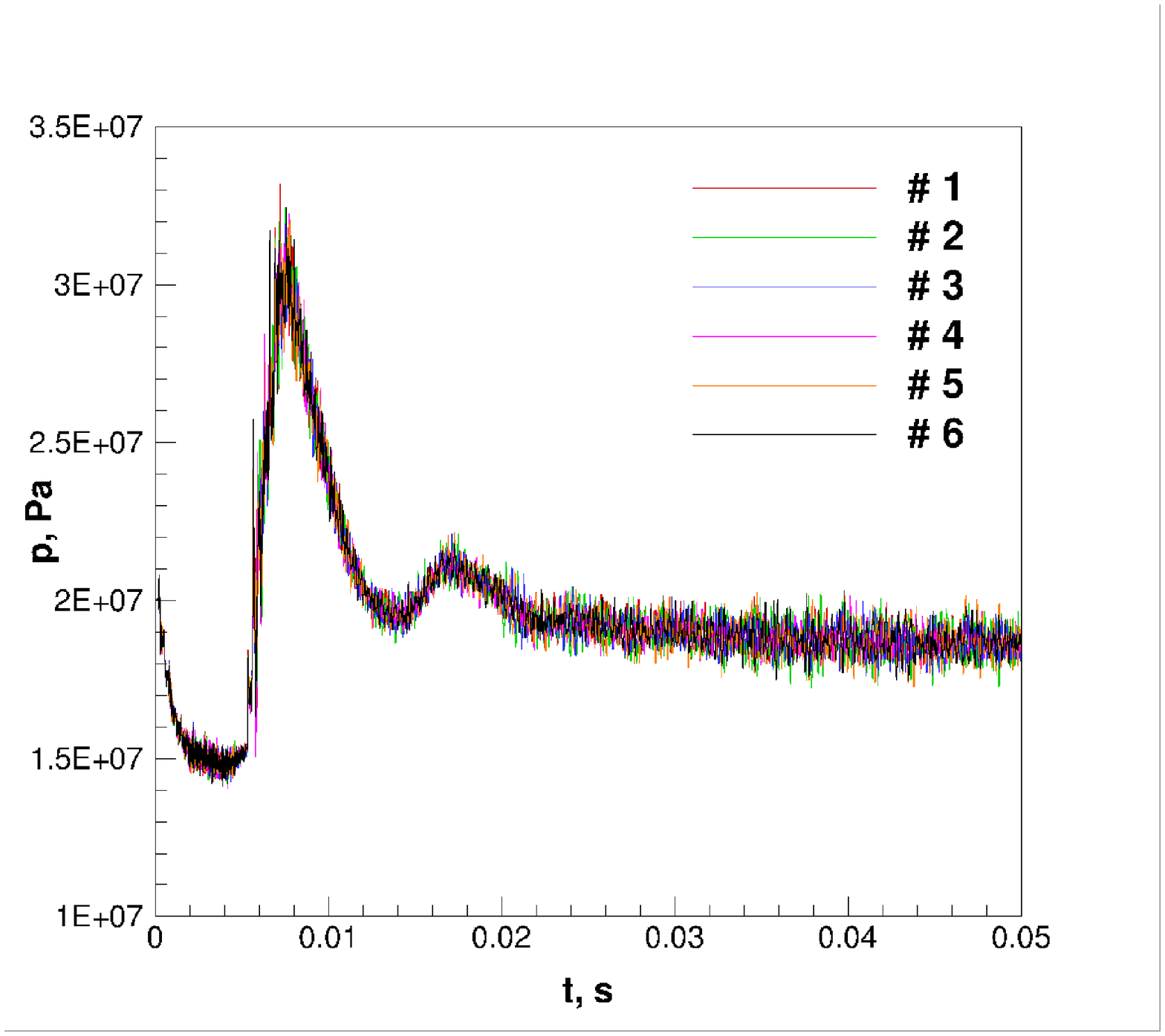}
        \label{fig:p_his_z_18cm_fpv}}
        \subfigure[Probes at $z = 33$ cm for FPV]
{\includegraphics{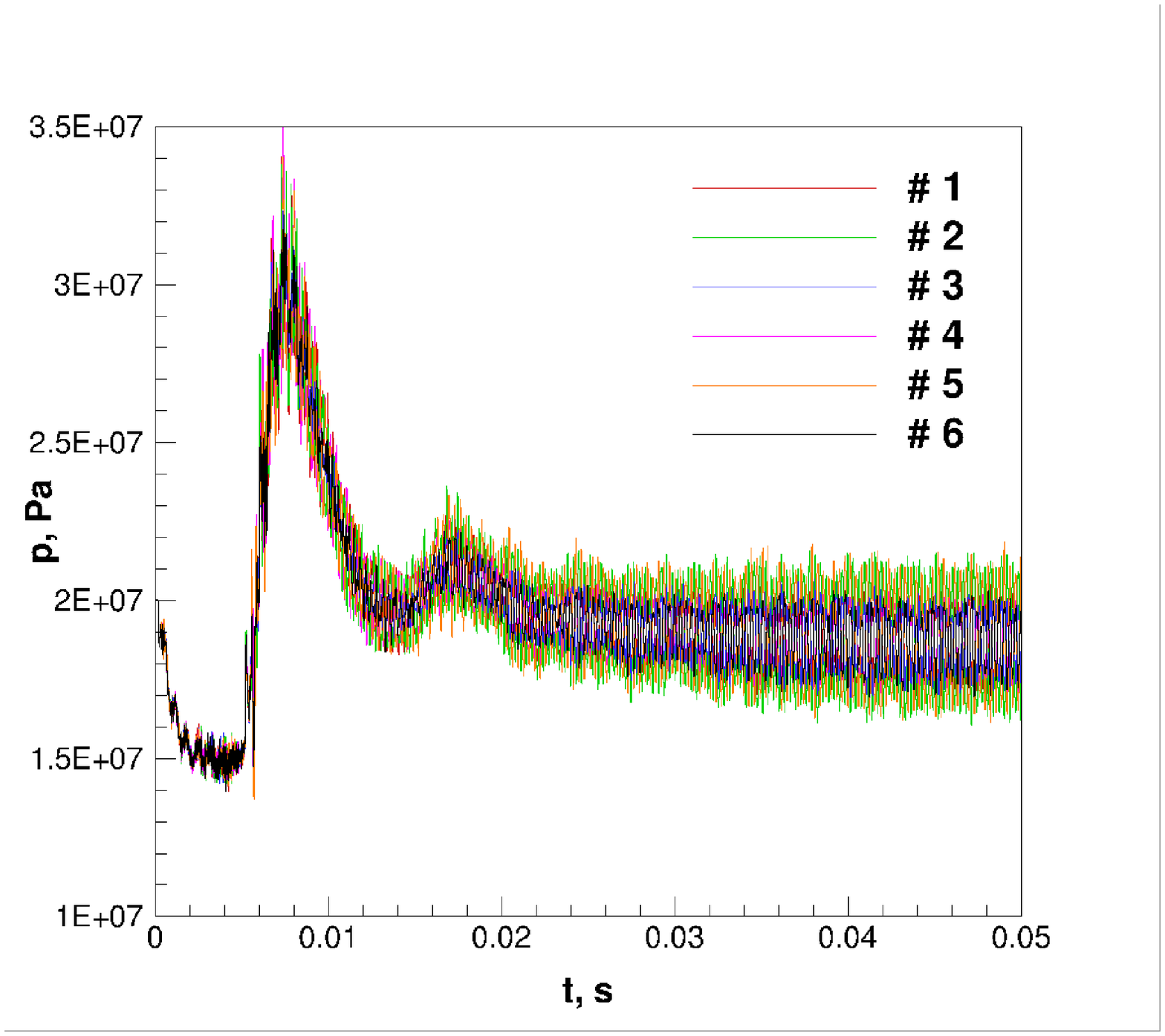}
        \label{fig:p_his_z_33cm_fpv}}
        \subfigure[Probes at $z = 1$ cm for OSK]
{\includegraphics{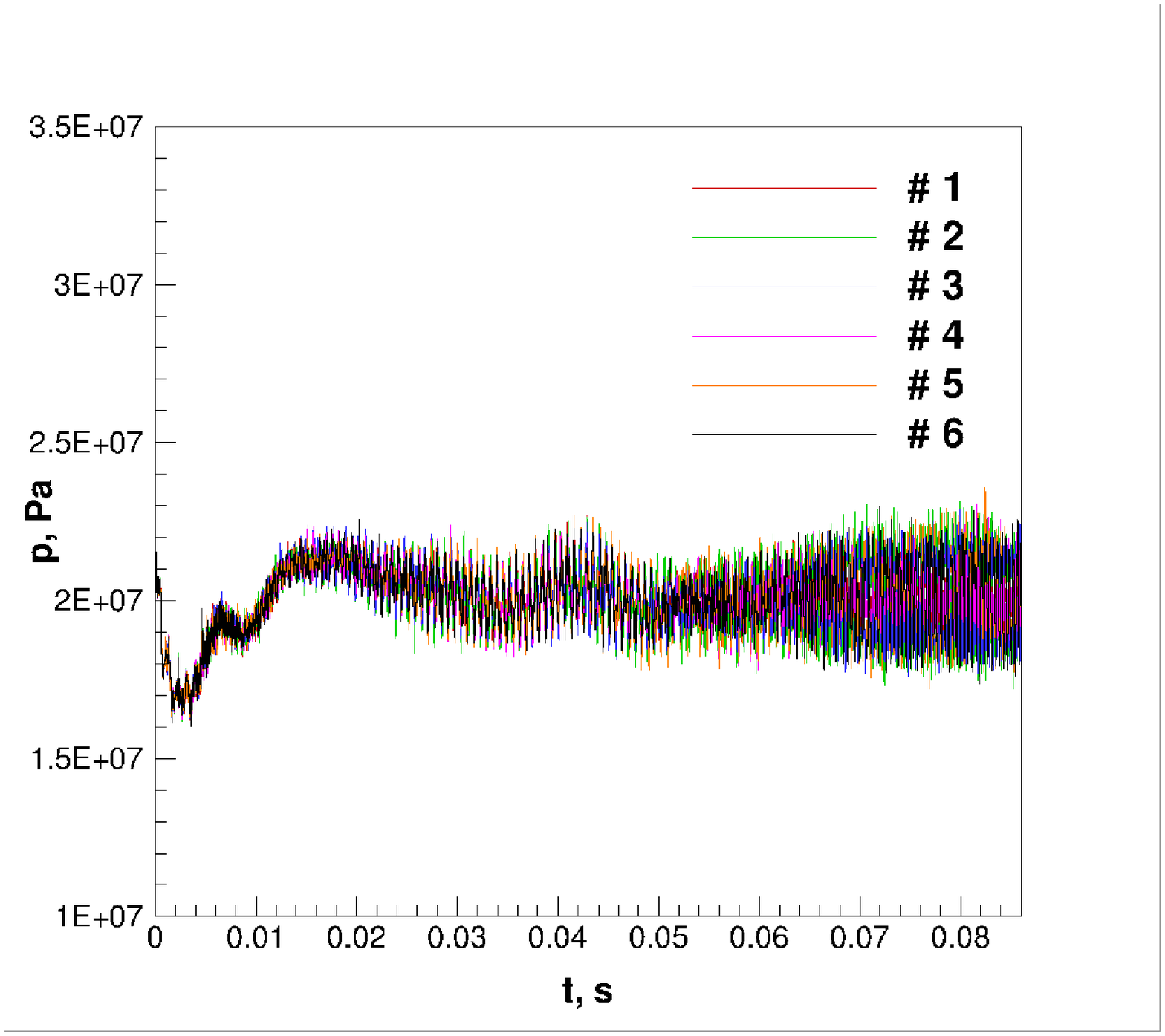}
        \label{fig:p_his_z_1cm_osk}}
        \subfigure[Probes at $z = 18$ cm for OSK]
{\includegraphics{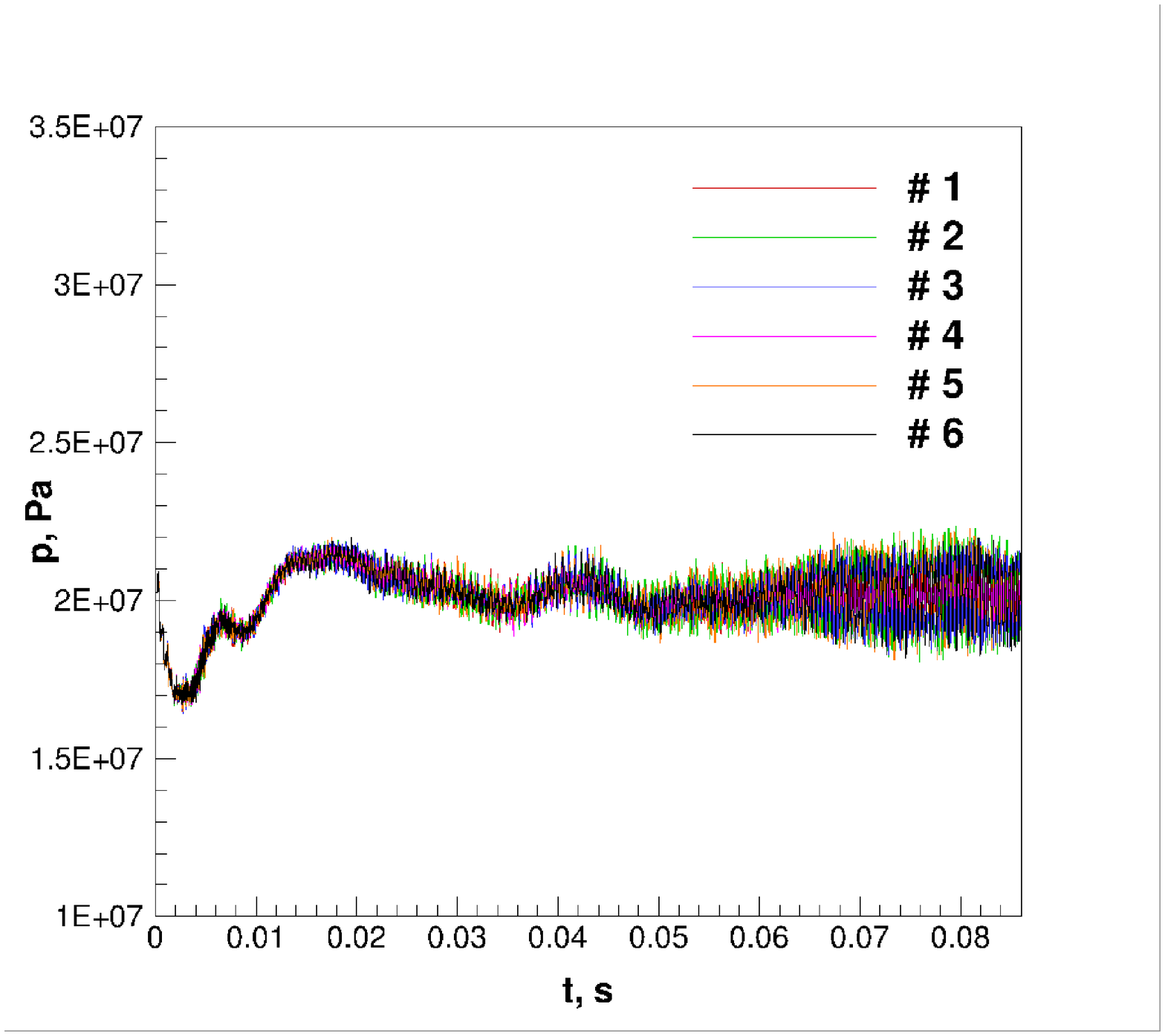}
        \label{fig:p_his_z_18cm_osk}}
        \subfigure[Probes at $z = 33$ cm for OSK]
{\includegraphics{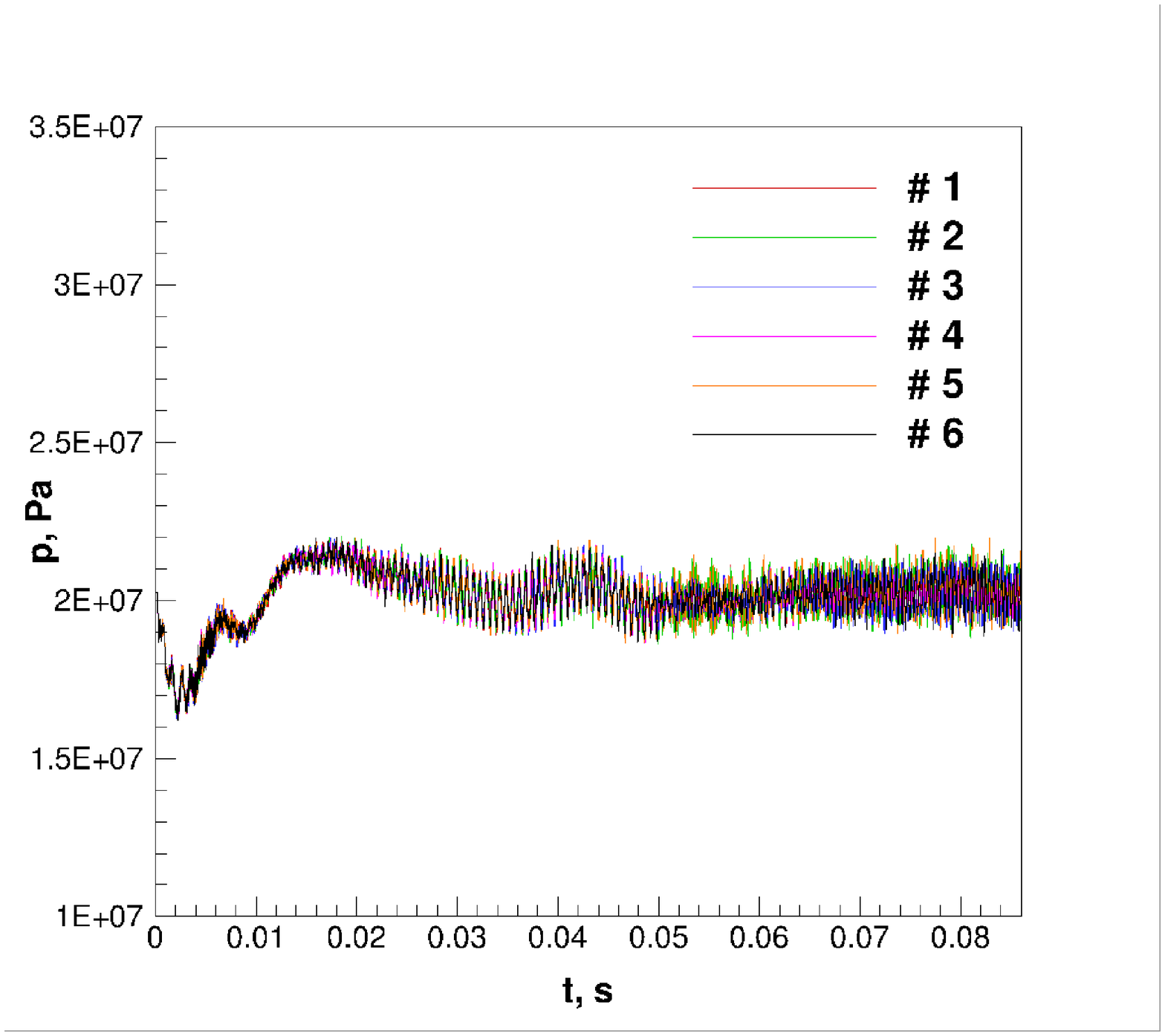}
        \label{fig:p_his_z_33cm_osk}}
    \end{subfigmatrix}
    \caption{Time history of pressure at the near-wall probes}
    \label{fig:p_his}
\end{figure}

\begin{figure}
    \begin{subfigmatrix}{3}
        \subfigure[Probes at $z = 1$ cm for FPV]
{\includegraphics{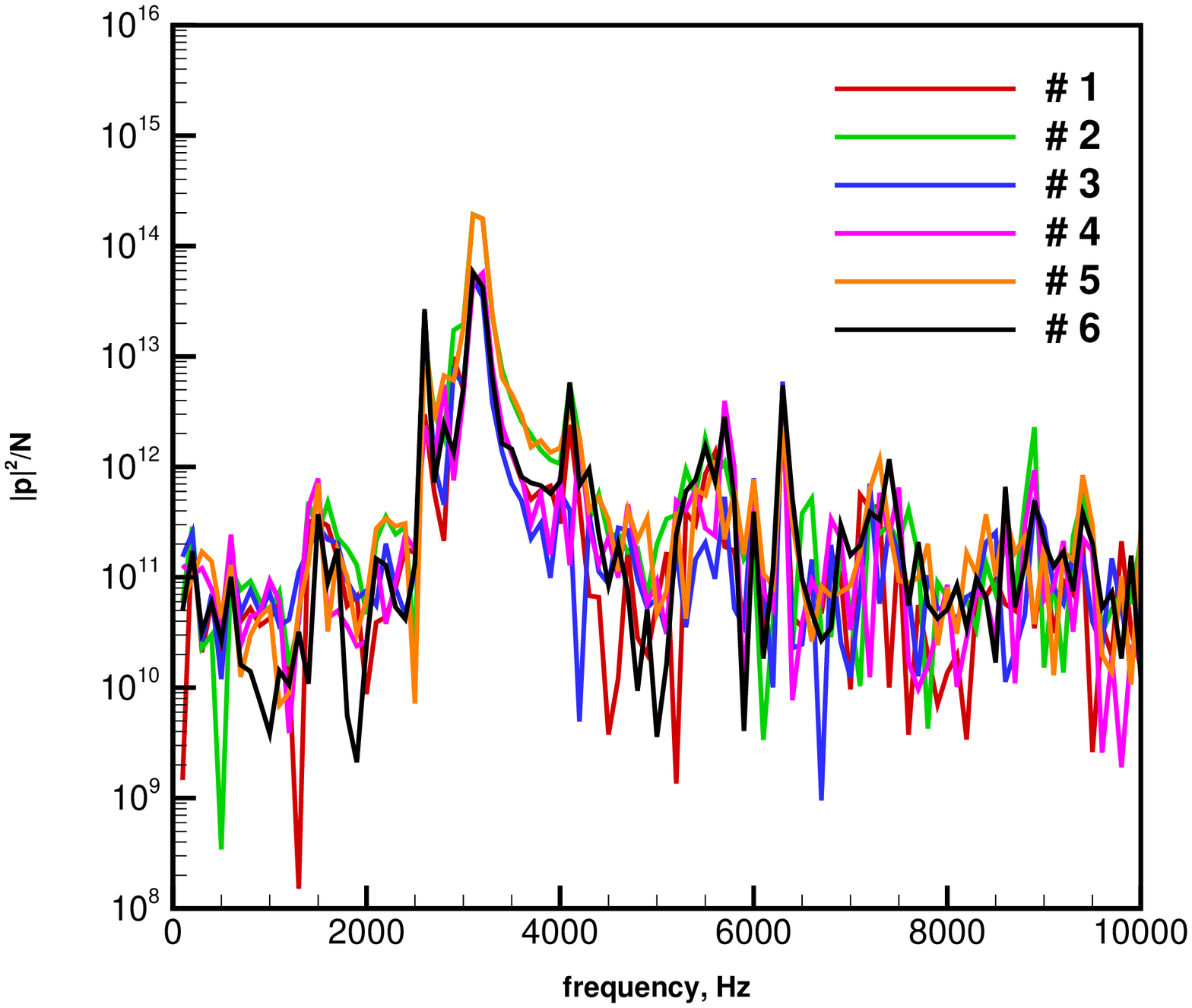}
        \label{fig:p_spec_z_1cm_fpv}}
        \subfigure[Probes at $z = 18$ cm for FPV]
{\includegraphics{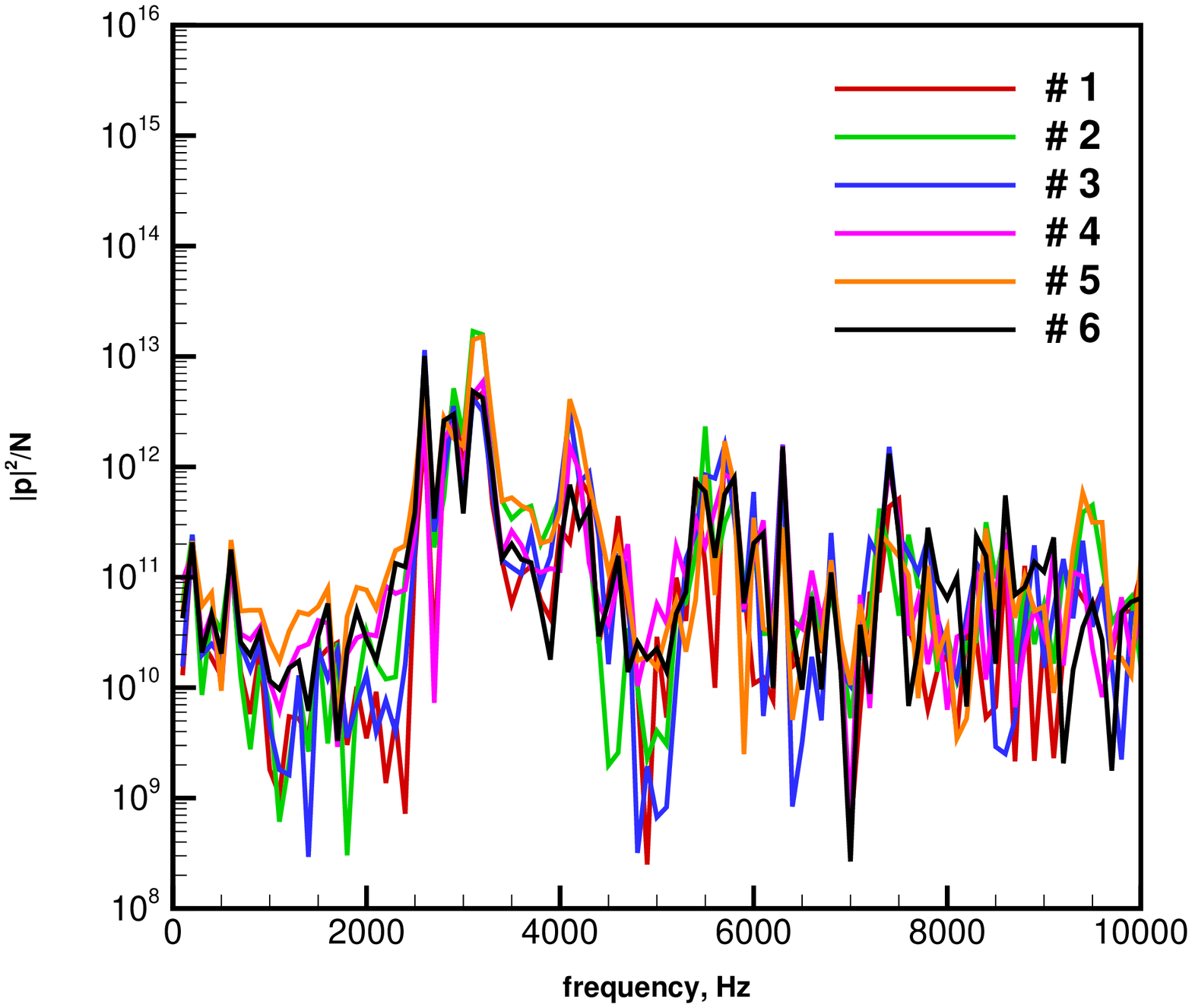}
        \label{fig:p_spec_z_18cm_fpv}}
        \subfigure[Probes at $z = 33$ cm for FPV]
{\includegraphics{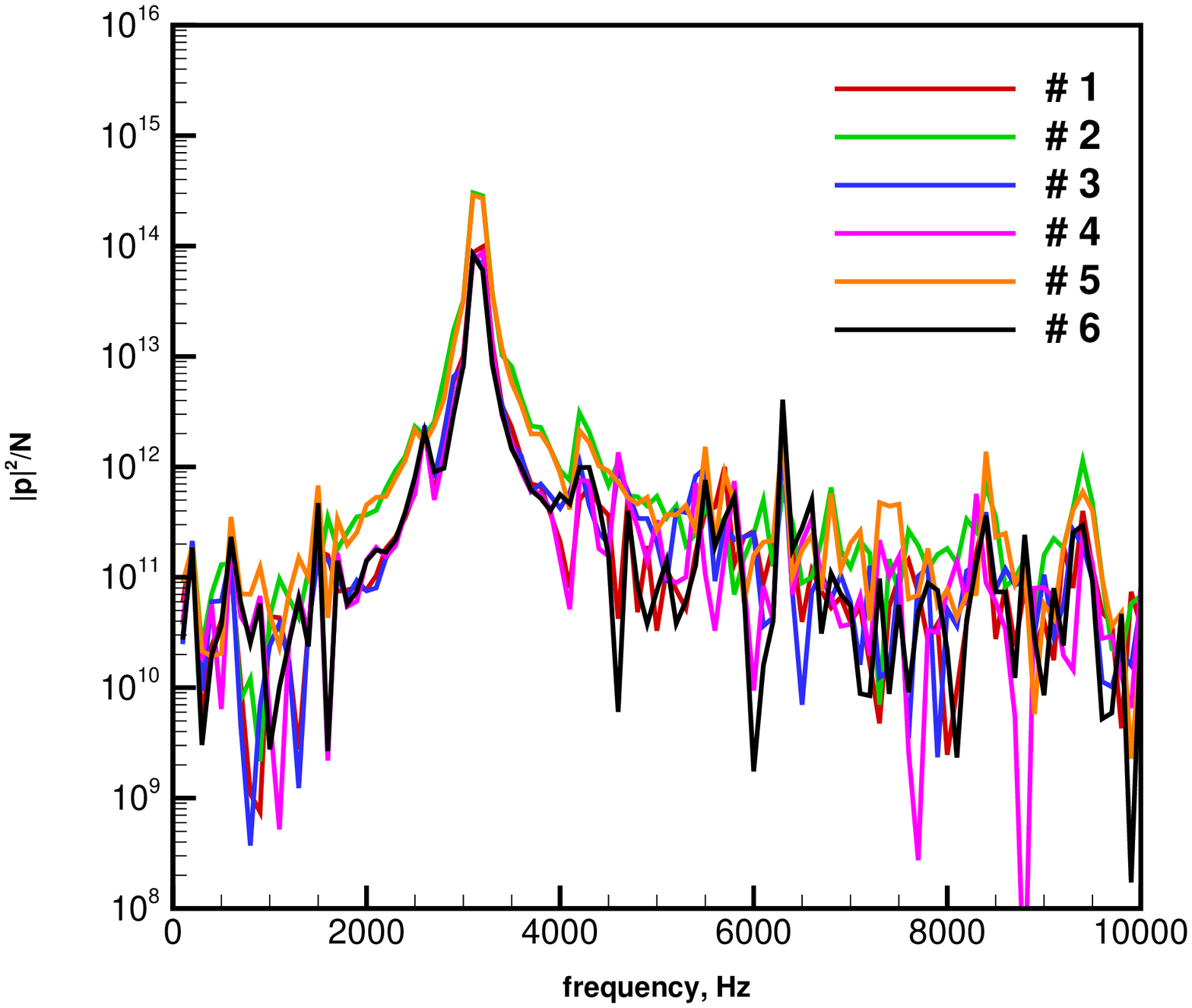}
        \label{fig:p_spec_z_33cm_fpv}}
        \subfigure[Probes at $z = 1$ cm for OSK]
{\includegraphics{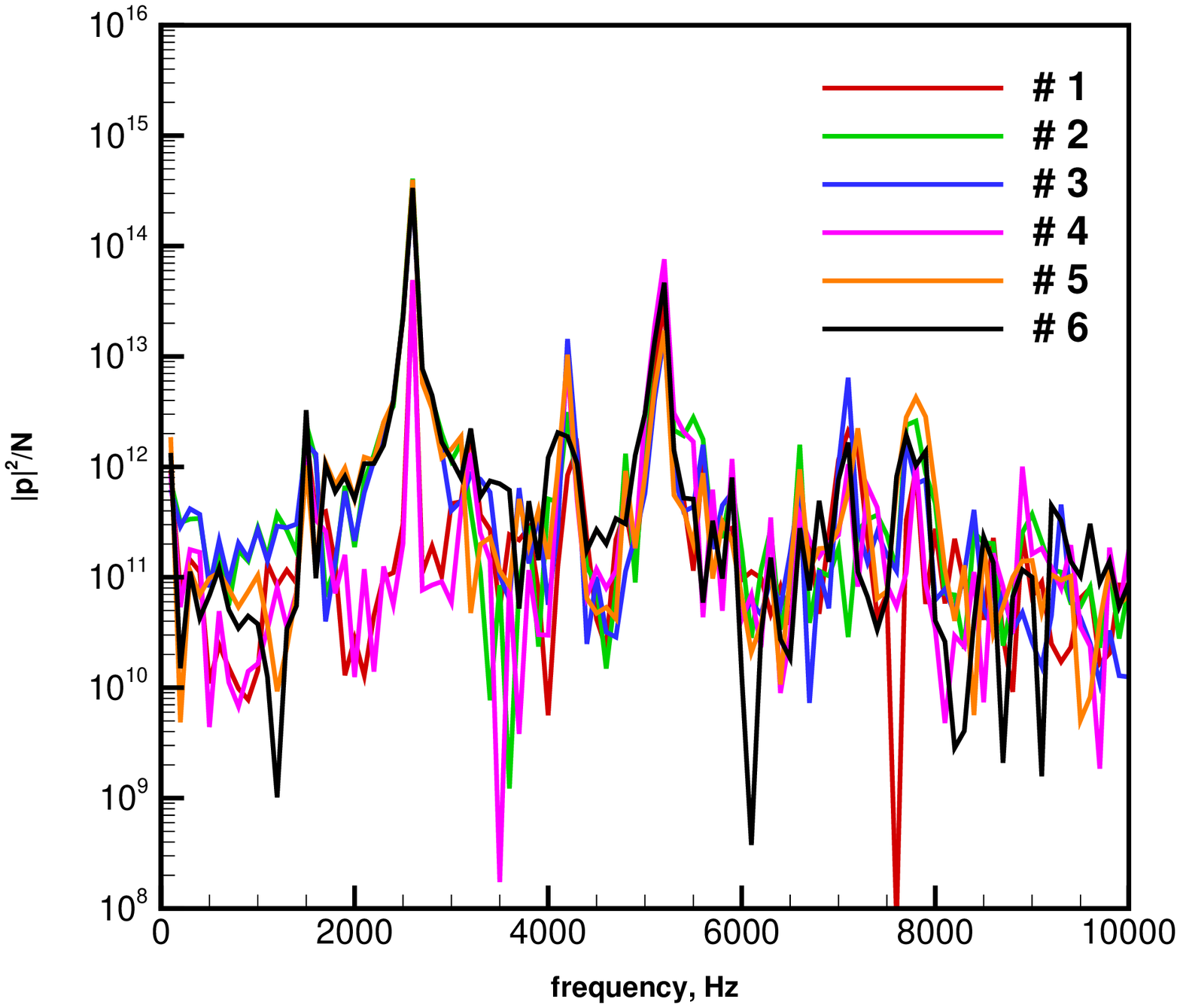}
        \label{fig:p_spec_z_1cm_osk}}
        \subfigure[Probes at $z = 18$ cm for OSK]
{\includegraphics{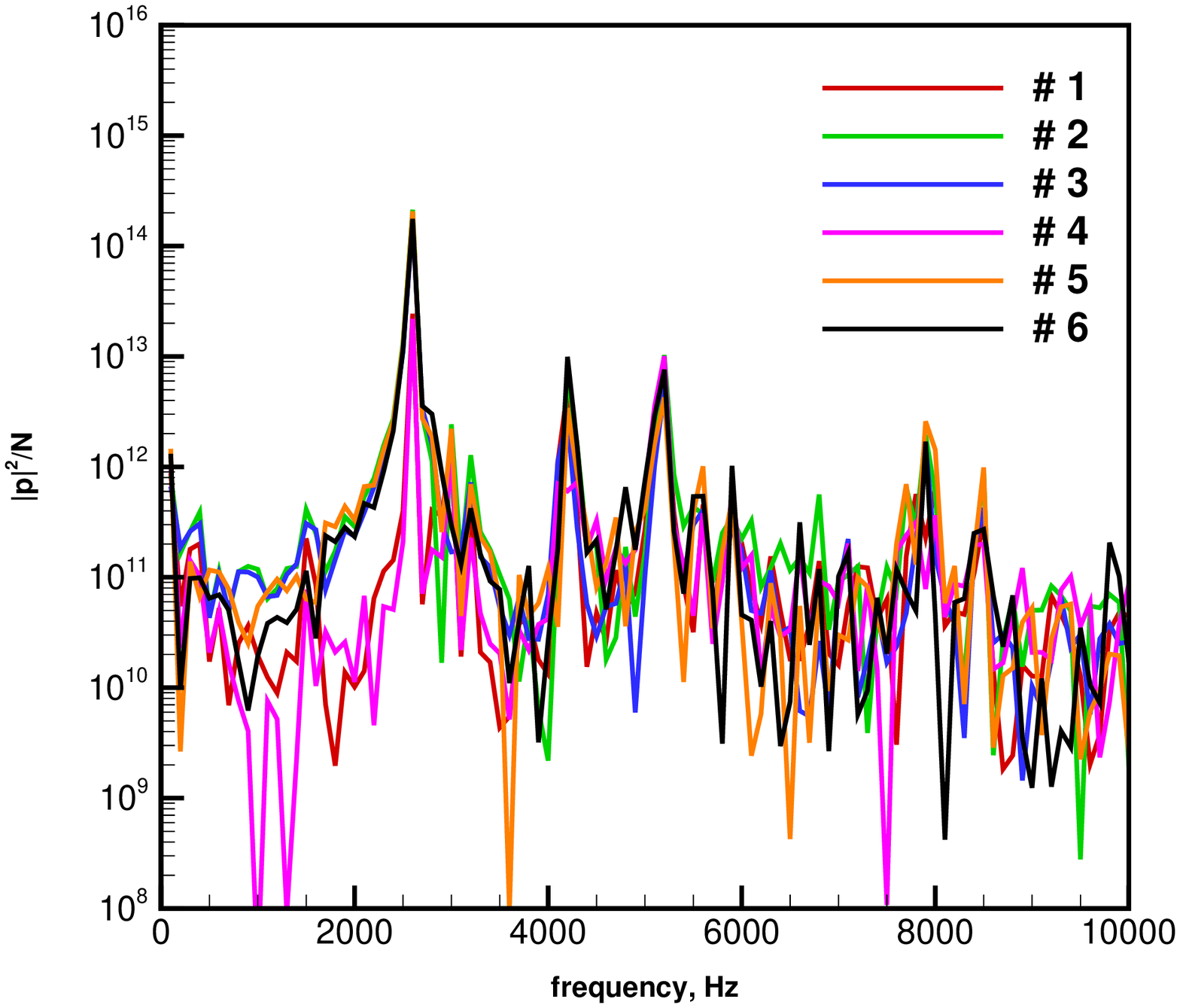}
        \label{fig:p_spec_z_18cm_osk}}
        \subfigure[Probes at $z = 33$ cm for OSK]
{\includegraphics{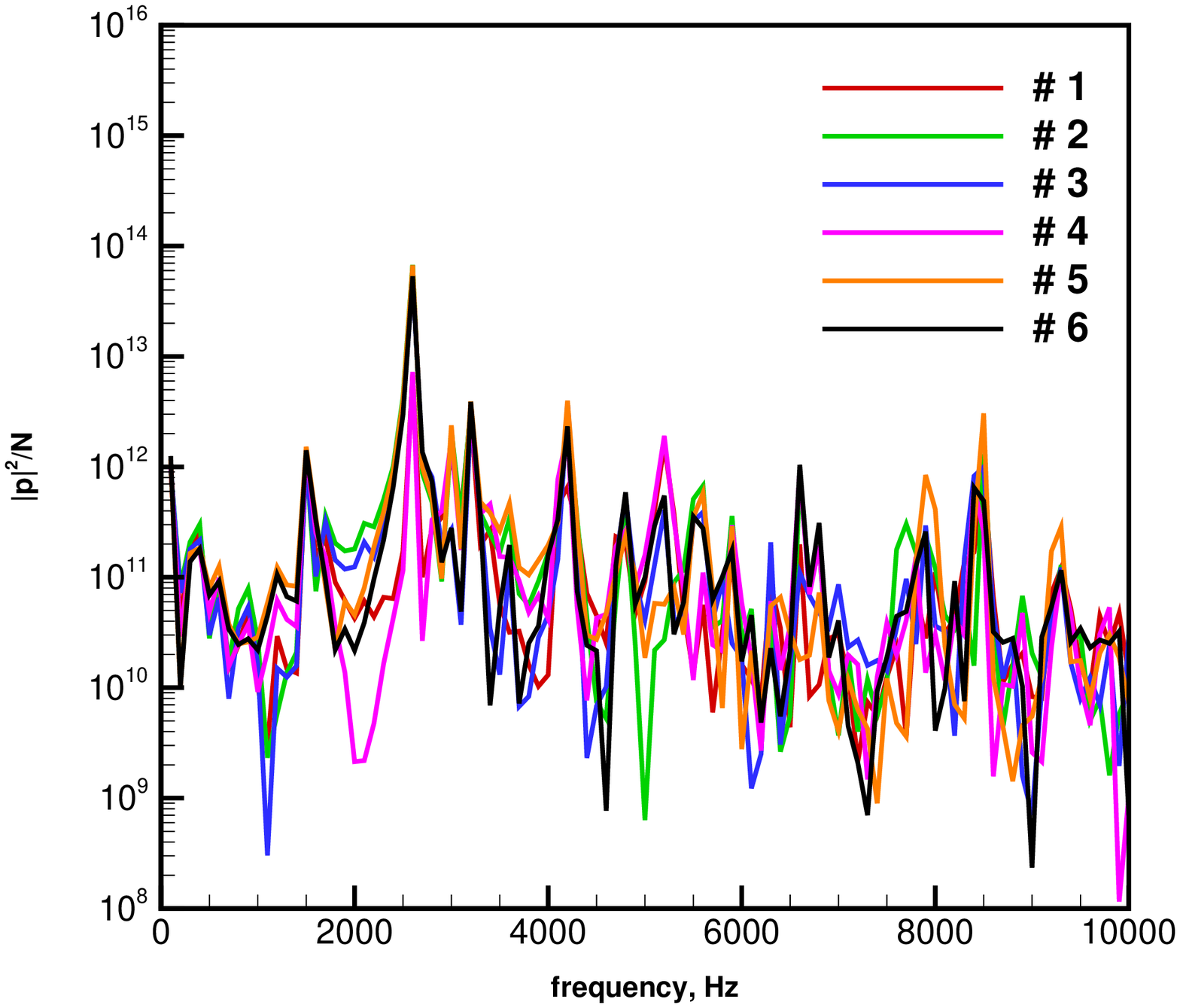}
        \label{fig:p_spec_z_33cm_osk}}
    \end{subfigmatrix}
    \caption{Spectrum of pressure at the near-wall probes}
    \label{fig:p_spec}
\end{figure}

As shown in Figure~\ref{fig:mdot_outlet}, the oscillatory temporal variation of the total mass flow rate at the nozzle outlet indicates combustion instability.
To monitor the possible tangential behavior of the combustion instability, we record the time histories of pressure on six probes placed $1$ mm away from the chamber wall and equally distributed in the circumferential direction as shown in Figure~\ref{fig:2dmesh}. To facilitate the analysis of possible longitudinal instability, we put three groups of such probes at three different axial locations downstream from the injector plate, which are $z = 0.01$ m, $z = 0.18$ m and $z = 0.33$ m. Figure~\ref{fig:p_his} shows that the pressure at all near-wall probes reaches limit-cycle oscillation in the last ten milliseconds for both the FPV and OSK models. The Fourier spectrum analysis of the pressure time histories, which is shown in Figure~\ref{fig:p_spec}, identifies that the mode of $3200$ Hz is dominant for the FPV model while the OSK model favors the mode of $2600$ Hz. Note that we just focus on the highest peak in the pressure spectrum. The rest peaks are at least ten times lower and hence has much less impact on the acoustic instability. The frequency of the dominant acoustic mode can be also estimated theoretically as summarized in the appendix. For the FPV model, the mean specific heat ratio $\bar \gamma$, the mean gas constant $\bar R$, the mean temperature $\bar T$ in the chamber are $1.20$, $426.5$ J/kg/K, and $2990.8$ K. The mean speed of sound $\bar c$ in the chamber is $1229.77$ m/s, leading to an estimated frequency of $3179.1$ Hz for the first-order combined tangential and longitudinal mode. Although the OSK model predicts a much higher $\bar T$ of $4167.6$ K, it yields lower $\bar \gamma$ and $\bar R$, which are $1.15$, $323.8$ J/kg/K. As a result, $\bar c$ for the OSK model is $1235.38$ m/s, which is just a little higher than that for the FPV model and yields an estimated frequency of $2585.4$ Hz for the first-order pure tangential mode. The computational results of the dominant frequency matches the theoretical estimation very well for both the FPV model and the OSK model. This indicates the resonant acoustic modes predicted by both models are physically allowed.

\begin{figure}
    \begin{subfigmatrix}{2}
        \subfigure[$3200$ Hz mode at $t = T/4$ for FPV]
{\includegraphics[width=0.3\textwidth]{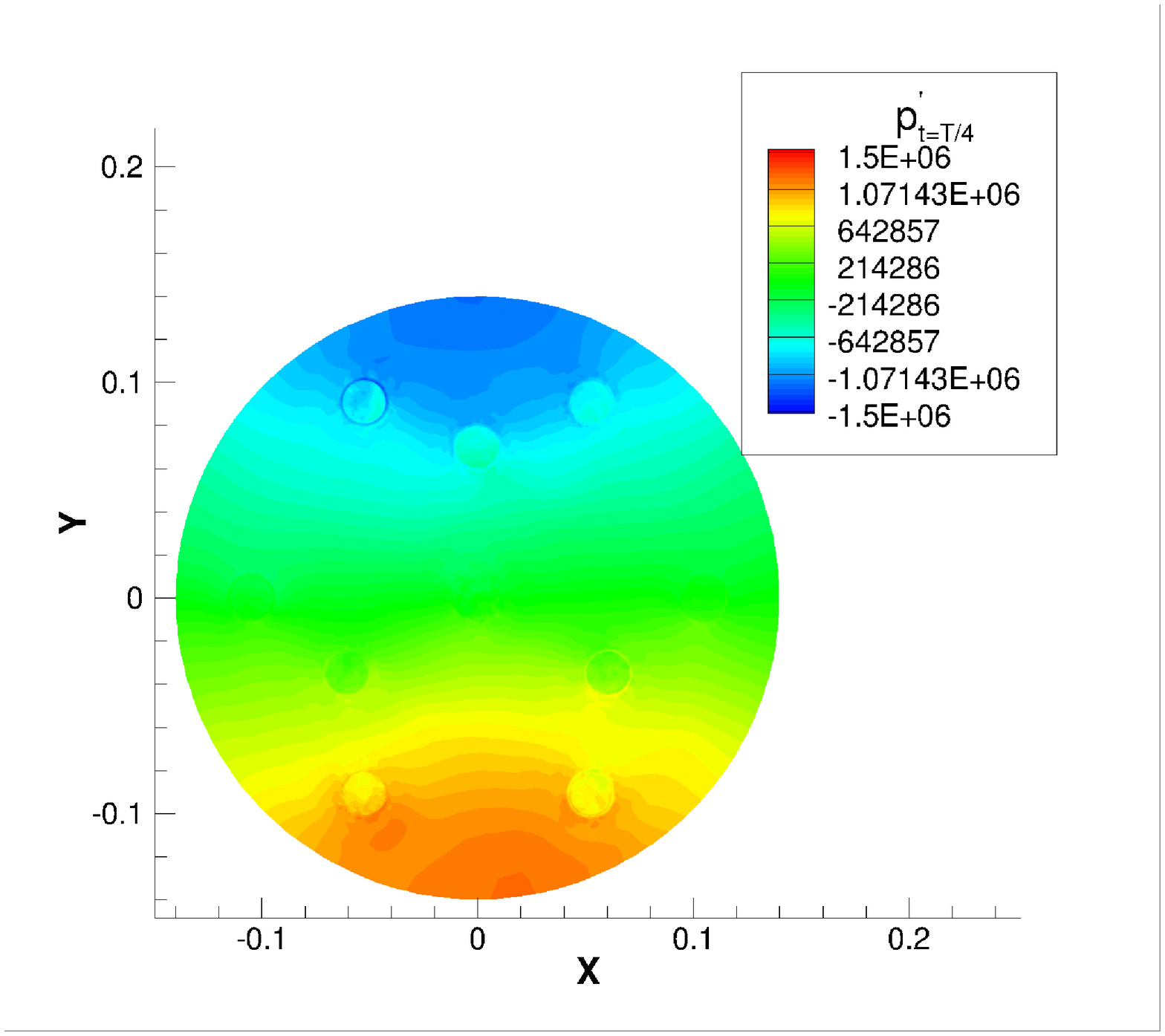}
        \label{fig:fpv_p_1qT_3200Hz_z_0cm}}
        \subfigure[$2600$ Hz mode at $t = T/4$ for OSK]
{\includegraphics[width=0.3\textwidth]{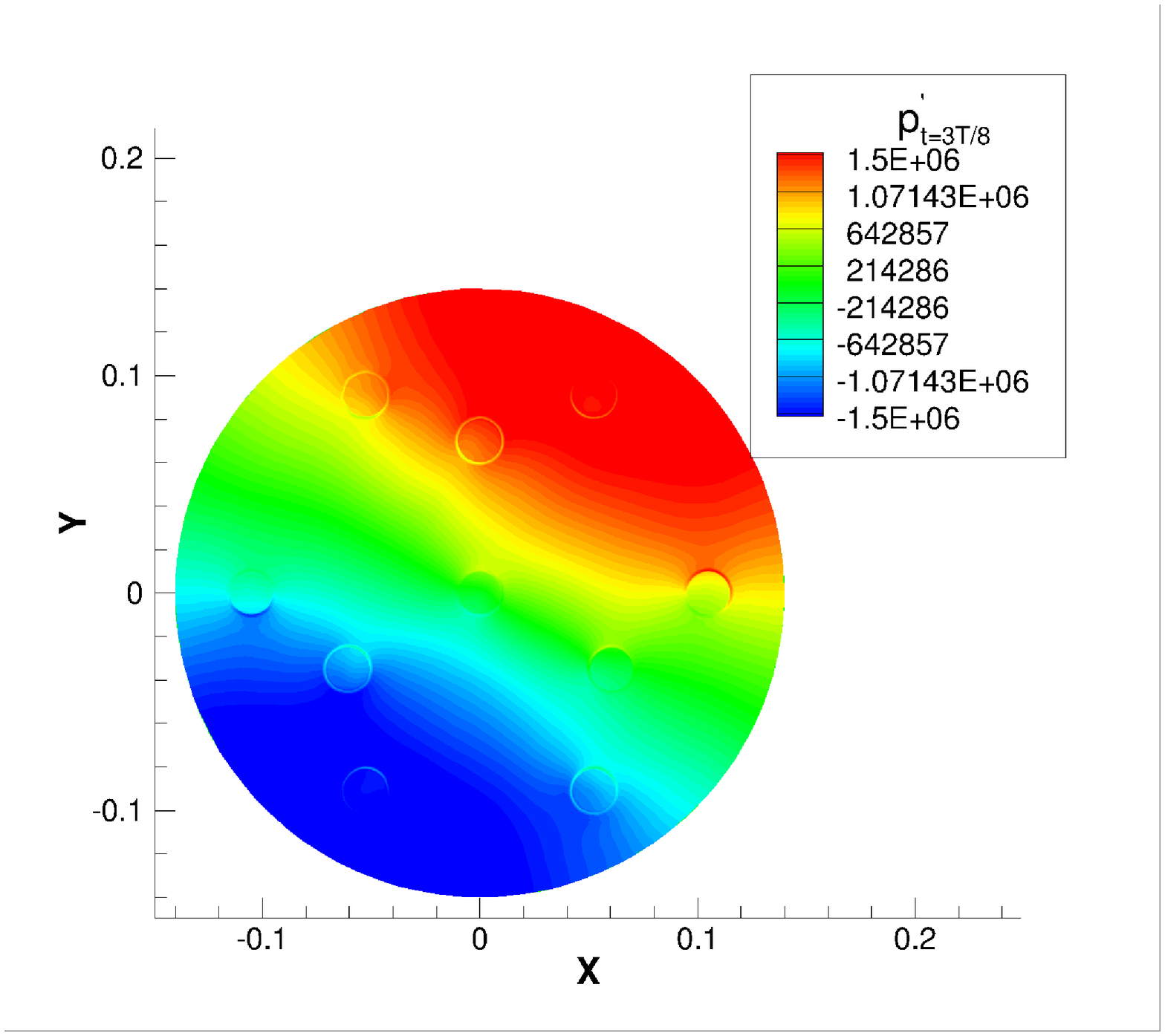}
        \label{fig:osk_p_1qT_2600Hz_z_0cm}}
        \subfigure[$3200$ Hz mode at $t = T/2$ for FPV]
{\includegraphics[width=0.3\textwidth]{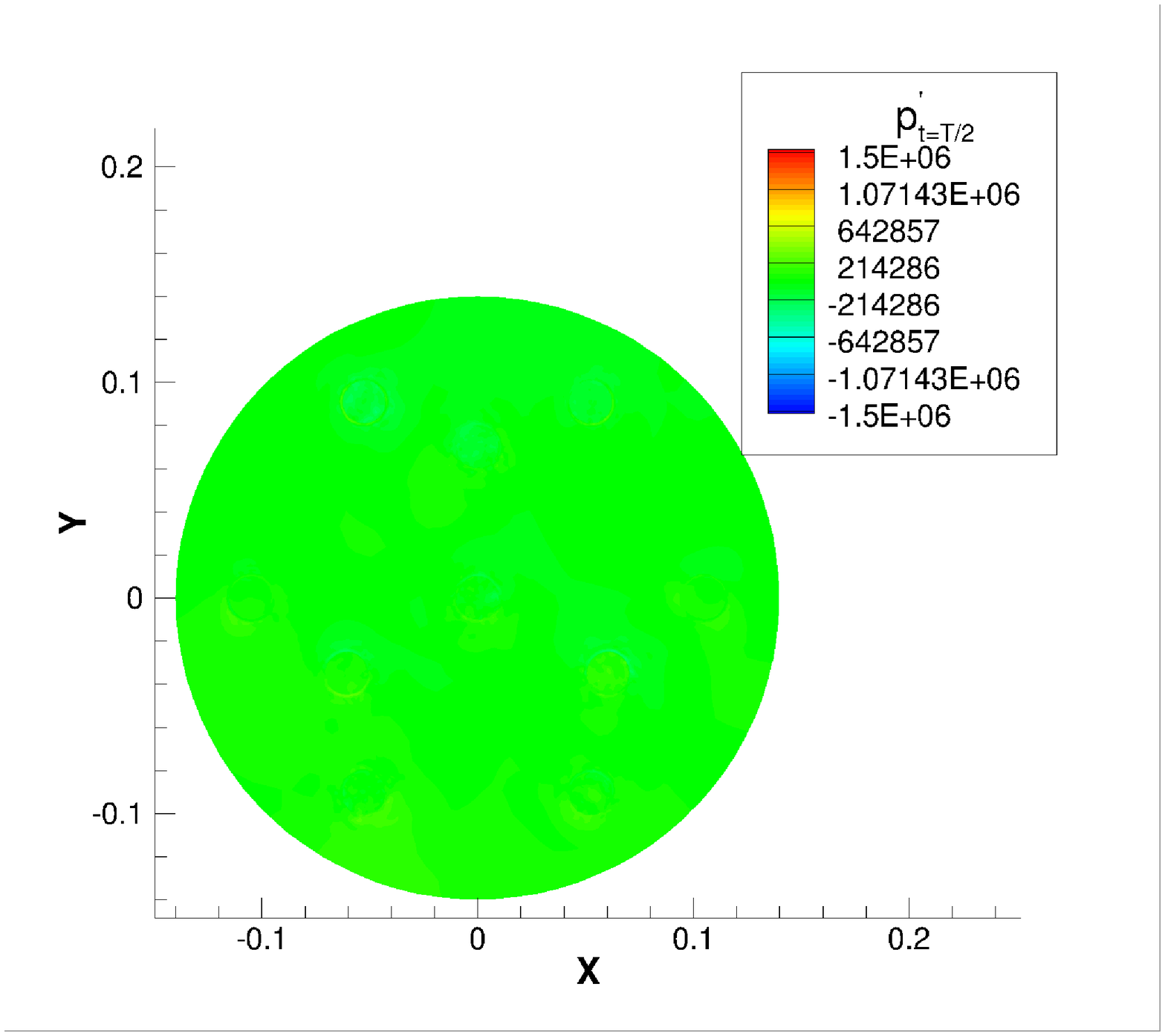}
        \label{fig:fpv_p_2qT_3200Hz_z_0cm}}
        \subfigure[$2600$ Hz mode at $t = T/2$ for OSK]
{\includegraphics[width=0.3\textwidth]{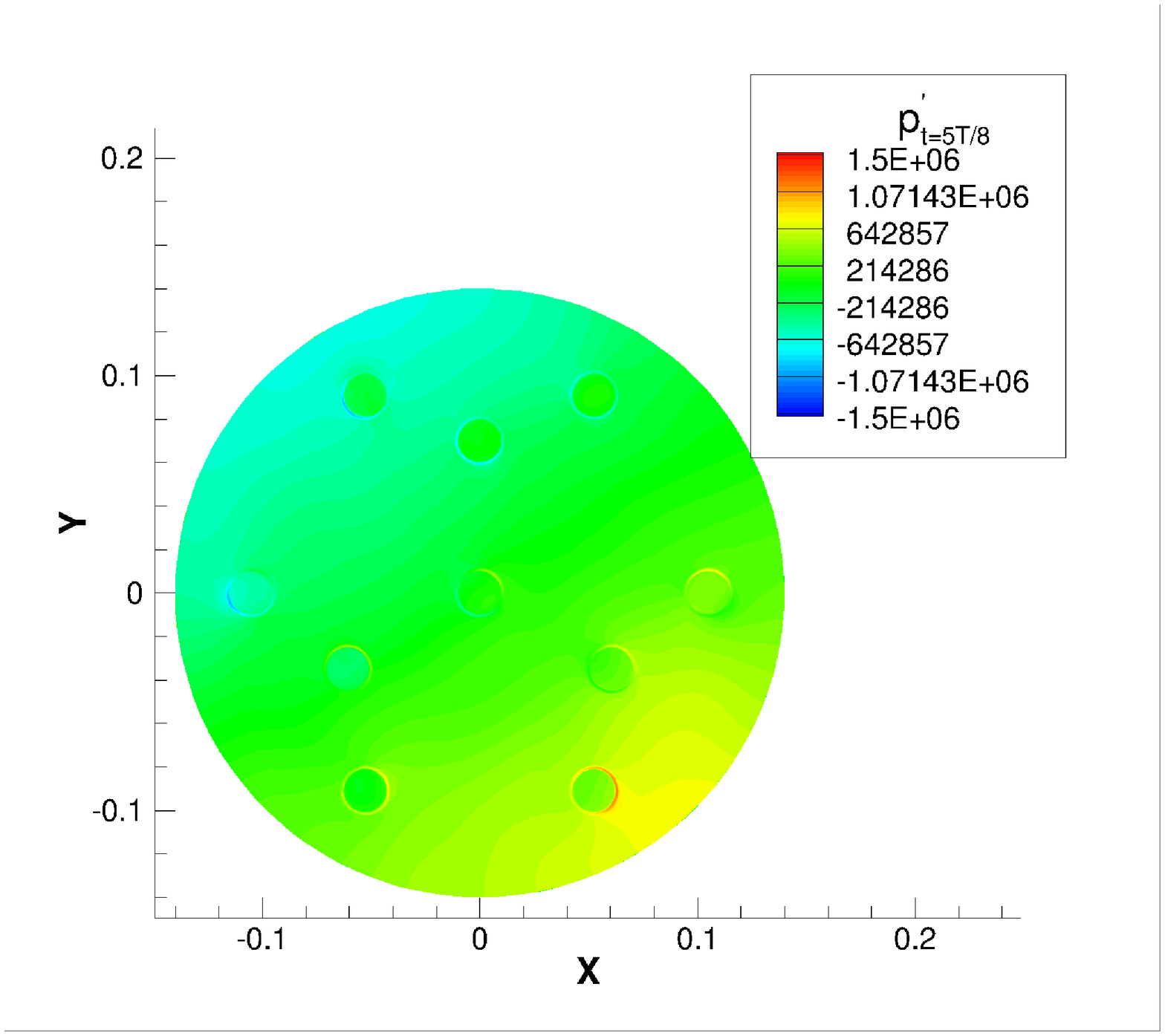}
        \label{fig:osk_p_2qT_2600Hz_z_0cm}}
        \subfigure[$3200$ Hz mode at $t = 3T/4$ for FPV]
{\includegraphics[width=0.3\textwidth]{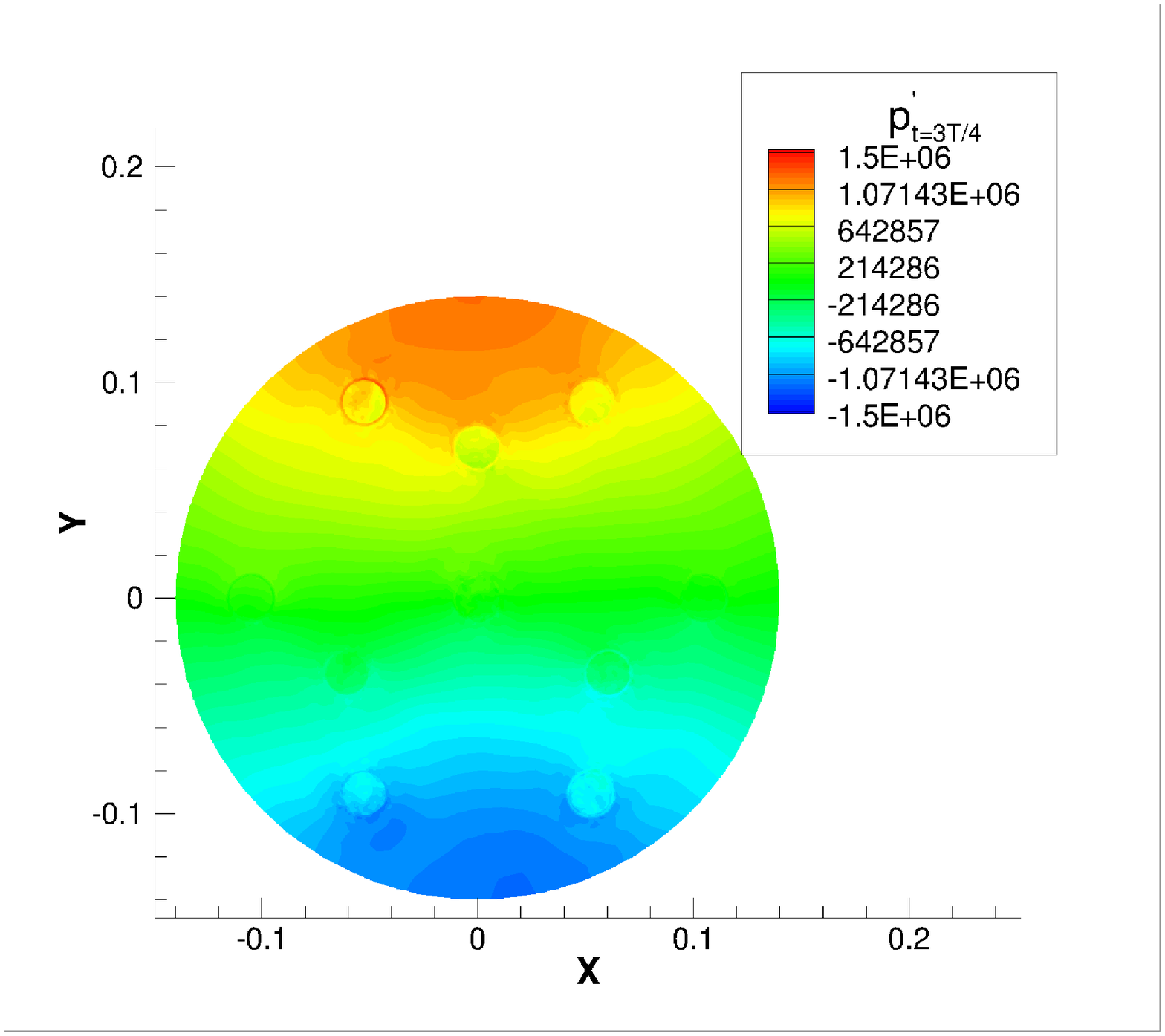}
        \label{fig:fpv_p_3qT_3200Hz_z_0cm}}
        \subfigure[$2600$ Hz mode at $t = 3T/4$ for OSK]
{\includegraphics[width=0.3\textwidth]{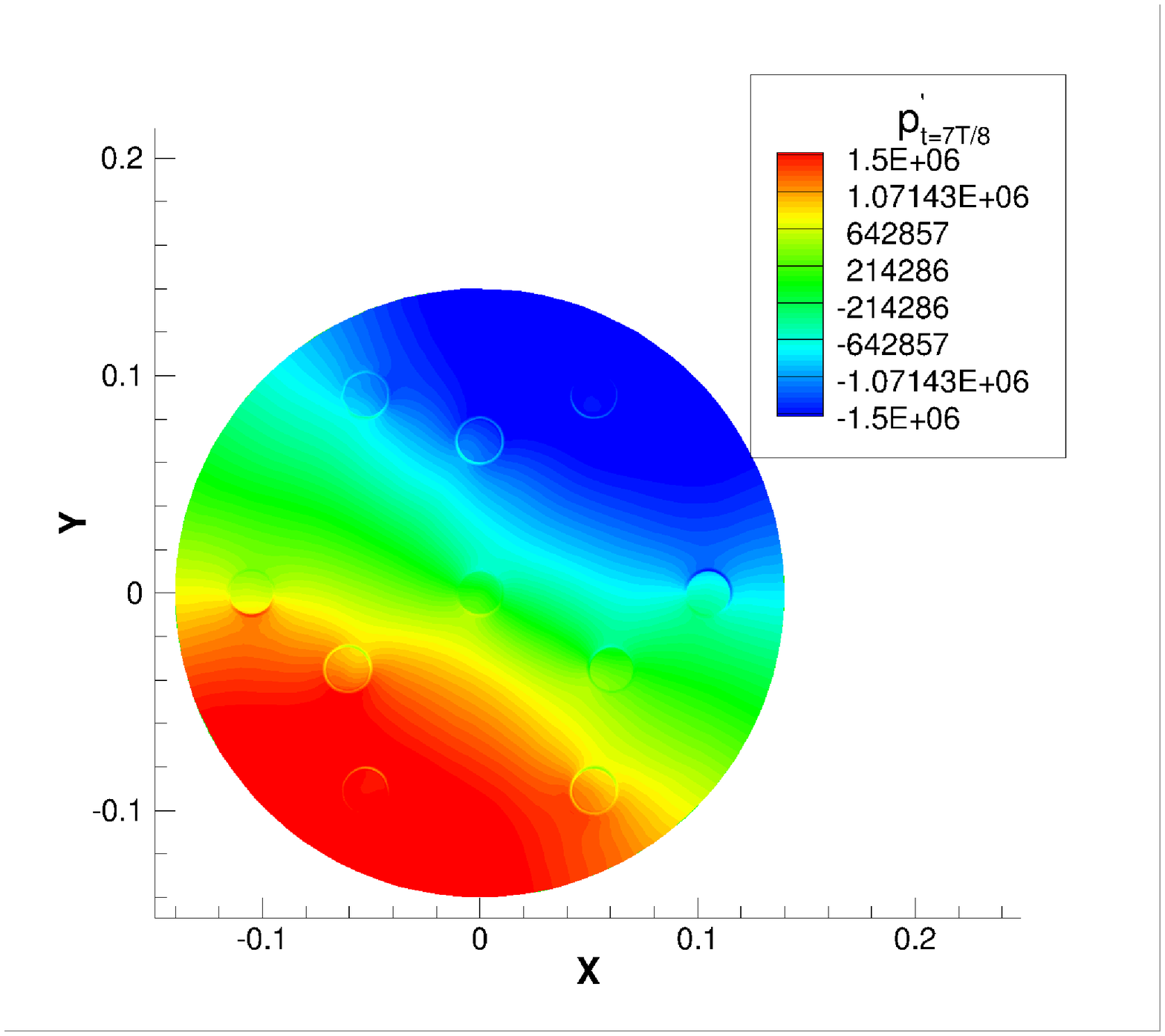}
        \label{fig:osk_p_3qT_2600Hz_z_0cm}}
        \subfigure[$3200$ Hz mode at $t = T$ for FPV]
{\includegraphics[width=0.3\textwidth]{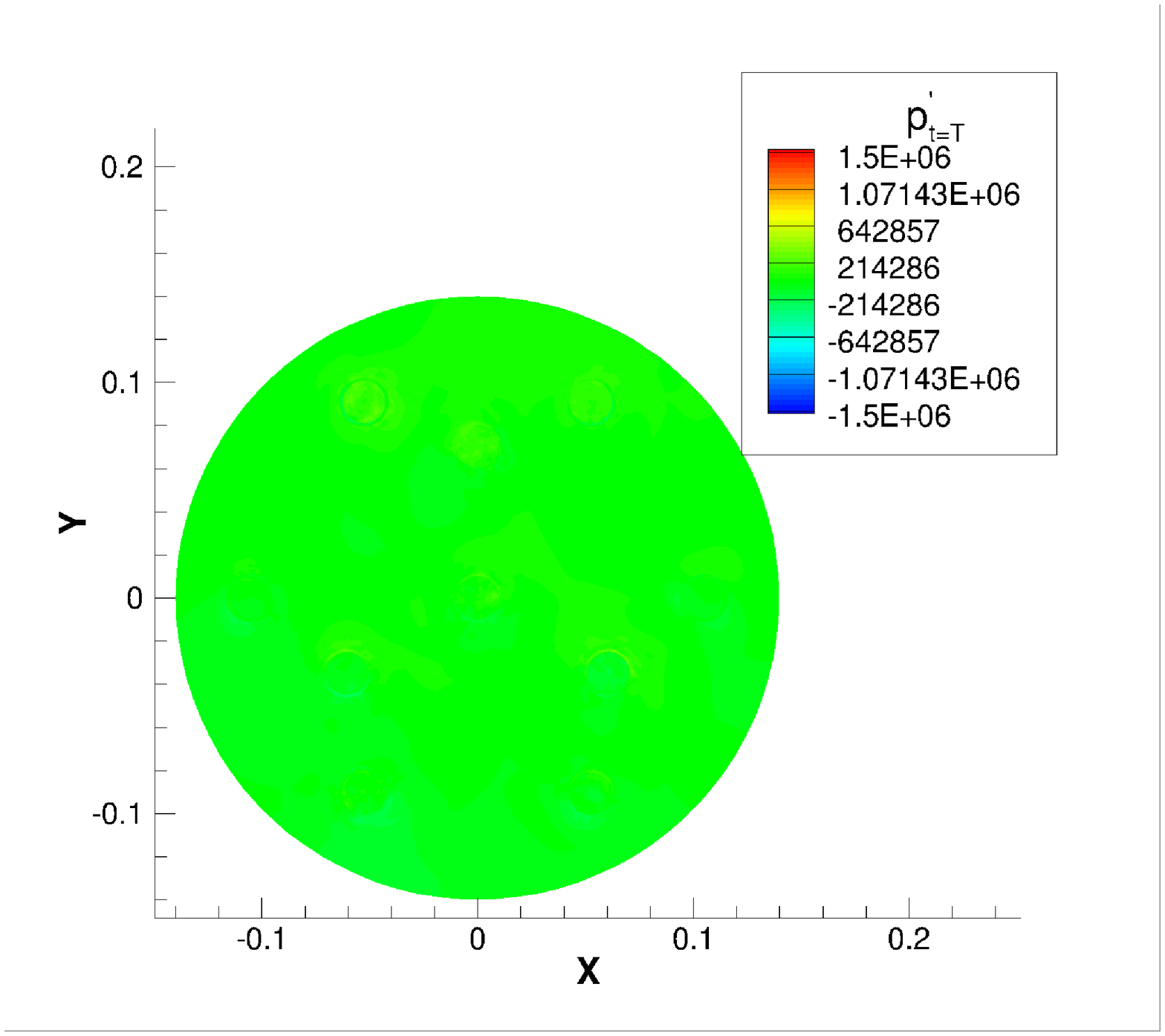}
        \label{fig:fpv_p_4qT_3200Hz_z_0cm}}
        \subfigure[$2600$ Hz mode at $t = T$ for OSK]
{\includegraphics[width=0.3\textwidth]{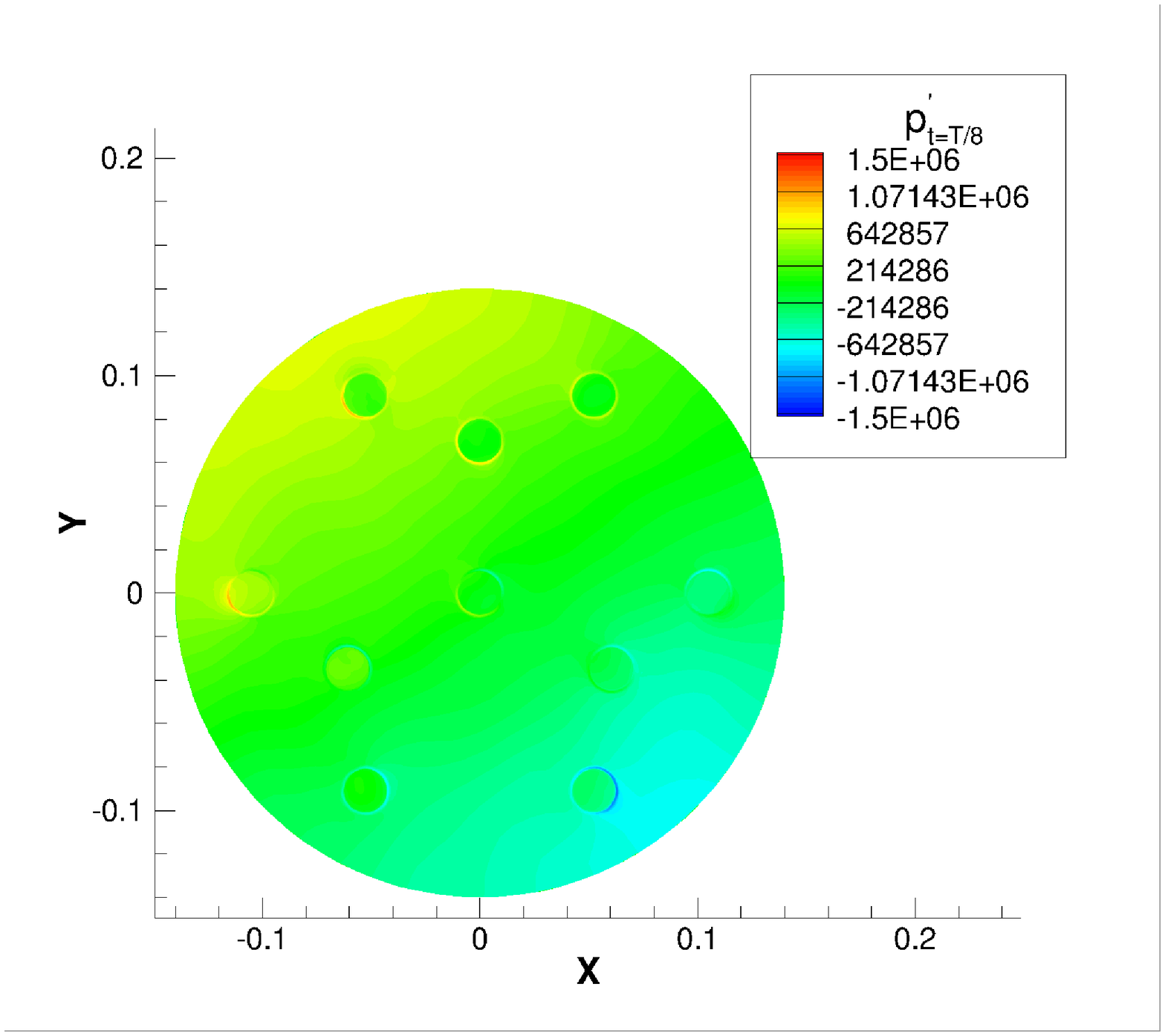}
        \label{fig:osk_p_4qT_2600Hz_z_0cm}}
    \end{subfigmatrix}
    \caption{Instantaneous contours of pressure fluctuation for
the dominant instability mode on the z = 0 cm plane}
    \label{fig:p_prime_z_0cm}
\end{figure}

The wave pattern of the dominant instability mode can be clearly illustrated through the instantaneous contours of the pressure fluctuation produced by that mode. These instantaneous contours are obtained by rebuilding the pressure time history at each spatial point using only the identified dominant Fourier mode, which is the $3200$ Hz mode for the FPV model and the $2600$ Hz mode for the OSK model. As shown in Figure~\ref{fig:p_prime_z_0cm}, on the $z = 0$ cm cross-sectional plane a pair of pressure antinodes (opposite in signs) and a pair of nodes appear at fixed azimuthal positions for both the FPV and the OSK model, indicating the characteristic of tangential standing wave for both models. However, for the FPV model, the azimuthal orientation of the pressure antinodes pair overlaps the straight line connecting probes $P2$ and $P5$, which is a symmetry axis of the injector distribution. For the OSK model, the pair of pressure nodes (rather than pressure antinodes) occur near probes $P1$ and $P4$, which also align in a direction of the injector symmetry.

\patchcmd{\subfigmatrix}{\hfill}{\hspace{0.8cm}}{}{}

\begin{figure}
    \begin{subfigmatrix}{2}
        \subfigure[$3200$ Hz mode at $t = T/4$ for FPV]
{\includegraphics[width=0.3\textwidth]{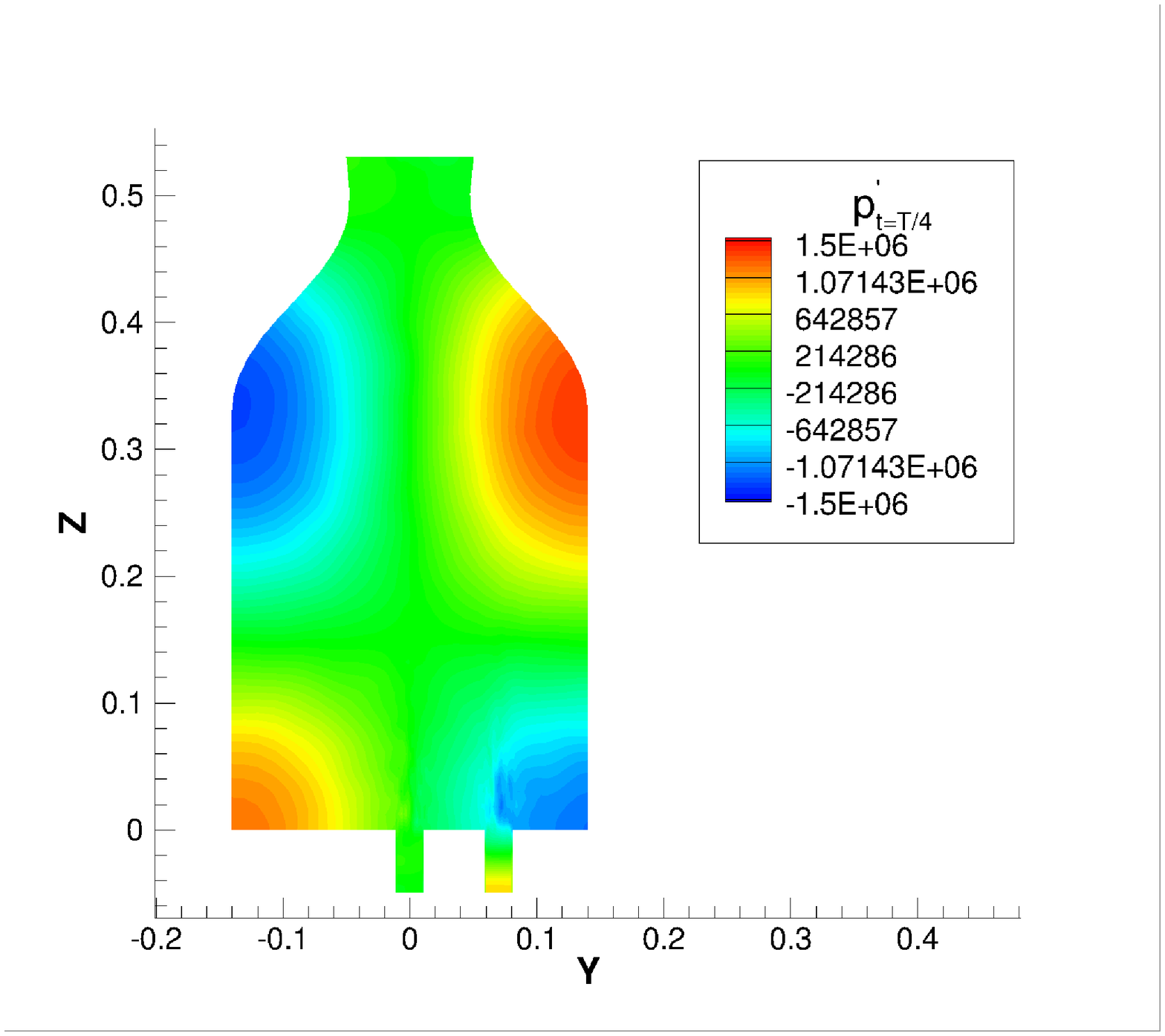}
        \label{fig:fpv_p_1qT_3200Hz_x_0cm}}
        \subfigure[$2600$ Hz mode at $t = T/4$ for OSK]
{\includegraphics[width=0.3\textwidth]{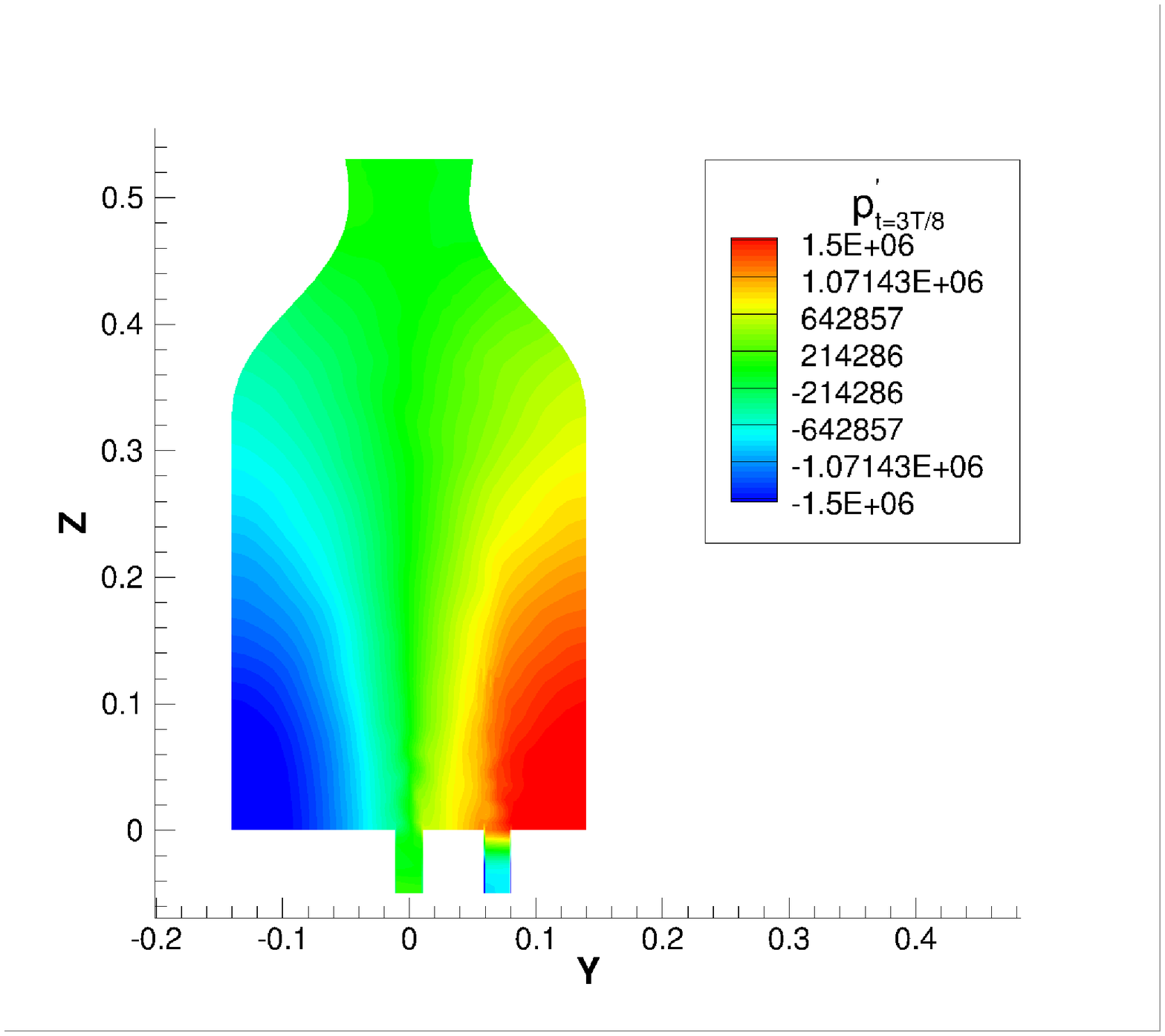}
        \label{fig:osk_p_1qT_2600Hz_x_0cm}}
        \subfigure[$3200$ Hz mode at $t = T/2$ for FPV]
{\includegraphics[width=0.3\textwidth]{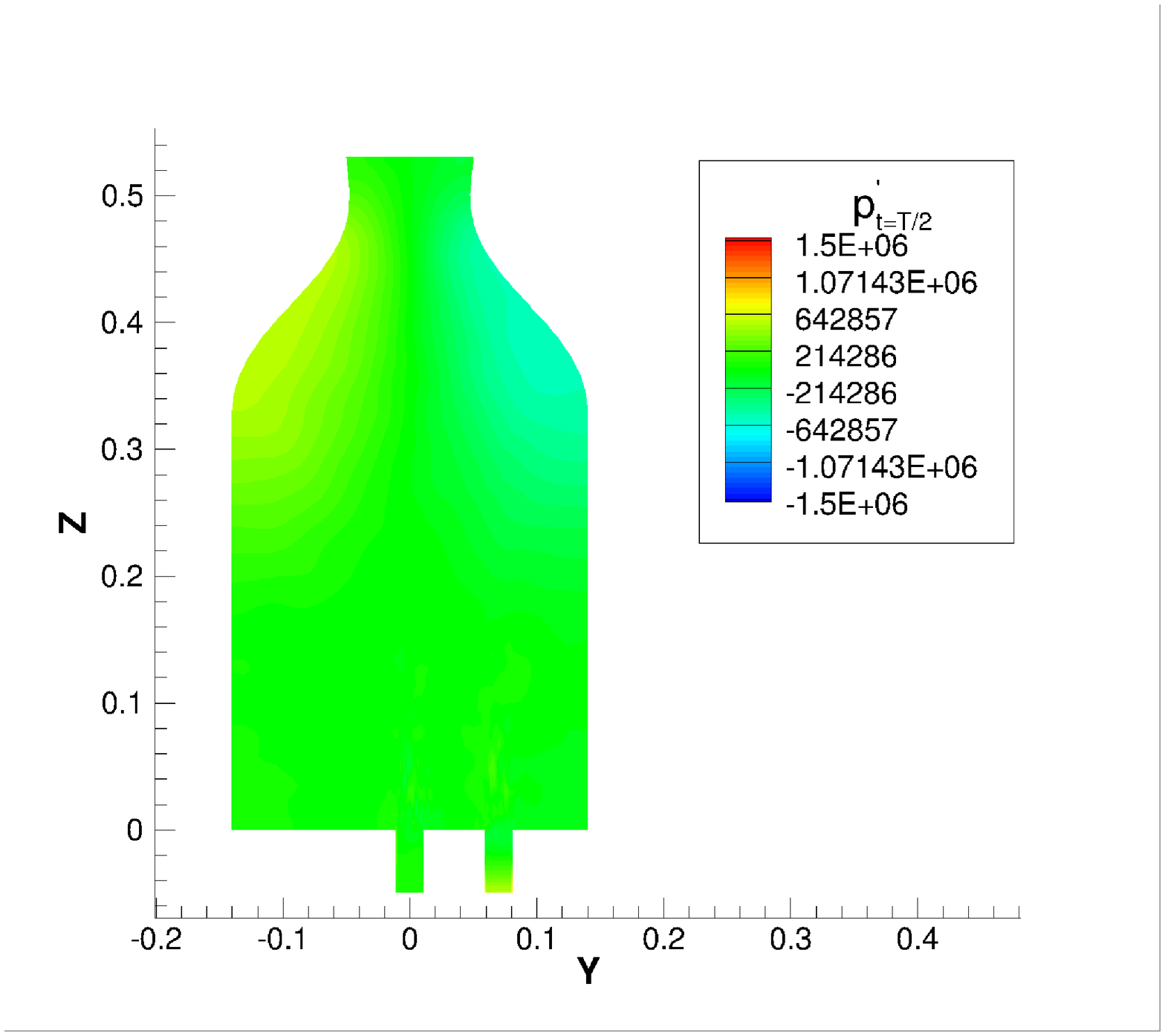}
        \label{fig:fpv_p_2qT_3200Hz_x_0cm}}
        \subfigure[$2600$ Hz mode at $t = T/2$ for OSK]
{\includegraphics[width=0.3\textwidth]{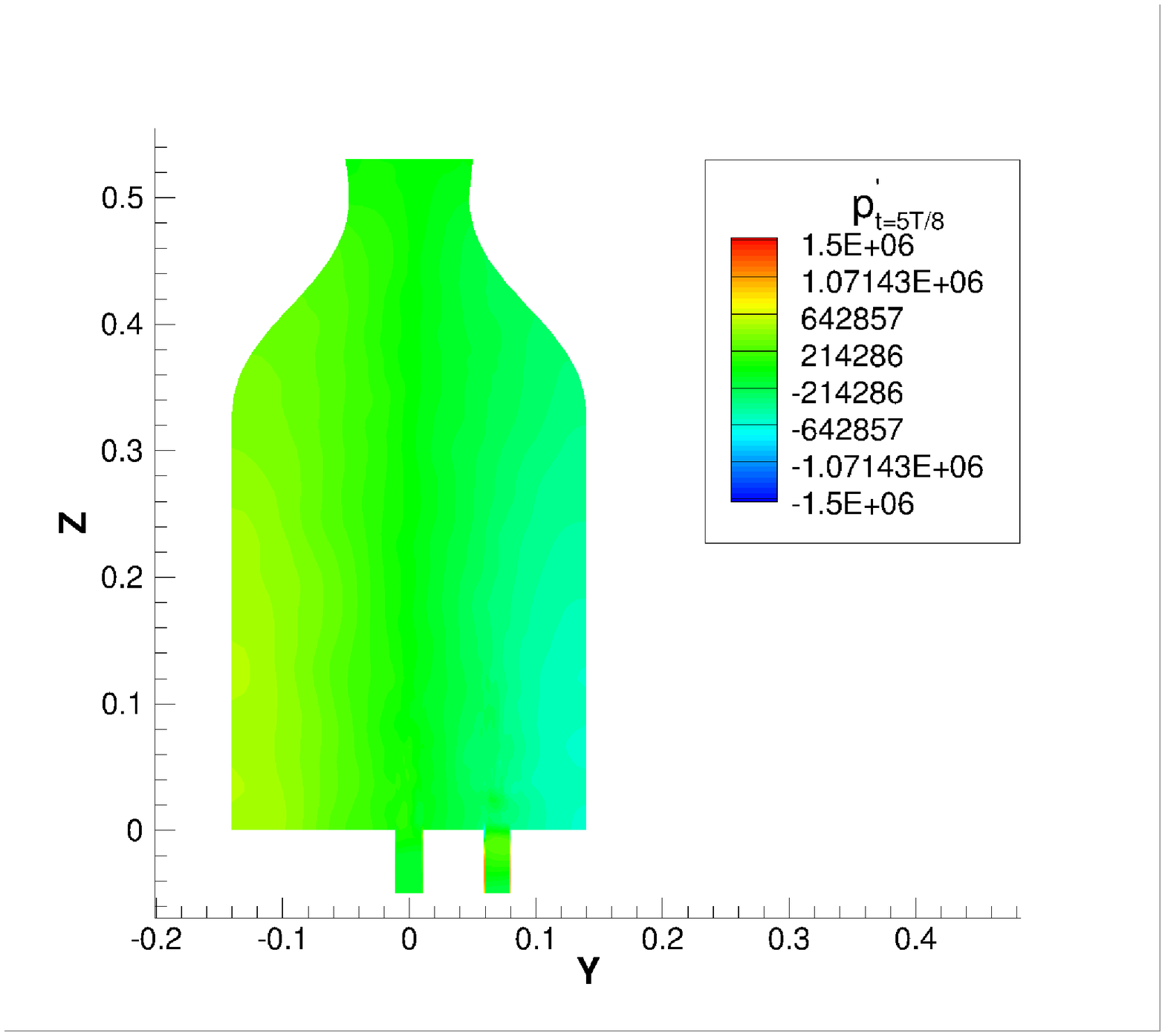}
        \label{fig:osk_p_2qT_2600Hz_x_0cm}}
        \subfigure[$3200$ Hz mode at $t = 3T/4$ for FPV]
{\includegraphics[width=0.3\textwidth]{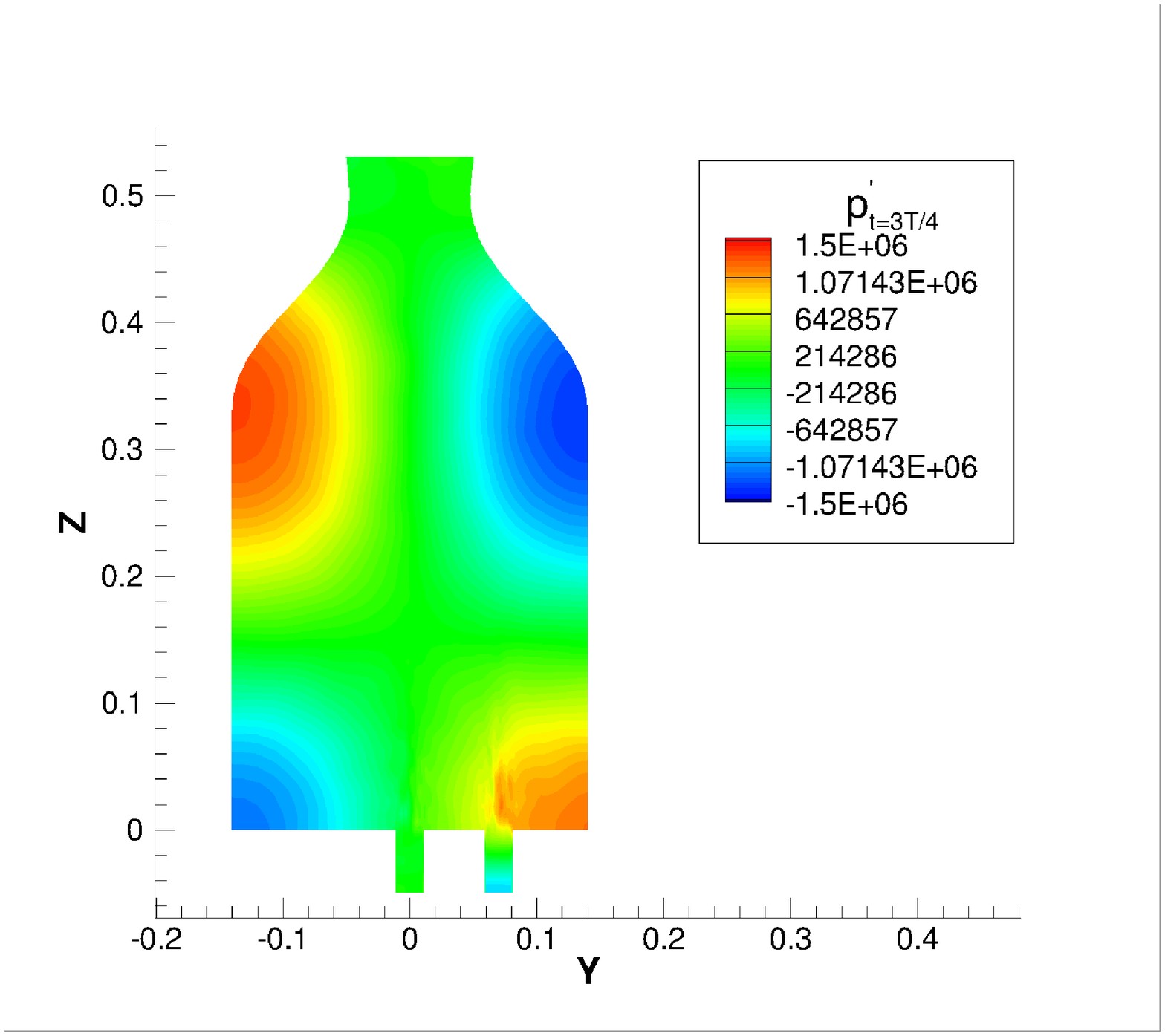}
        \label{fig:fpv_p_3qT_3200Hz_x_0cm}}
        \subfigure[$2600$ Hz mode at $t = 3T/4$ for OSK]
{\includegraphics[width=0.3\textwidth]{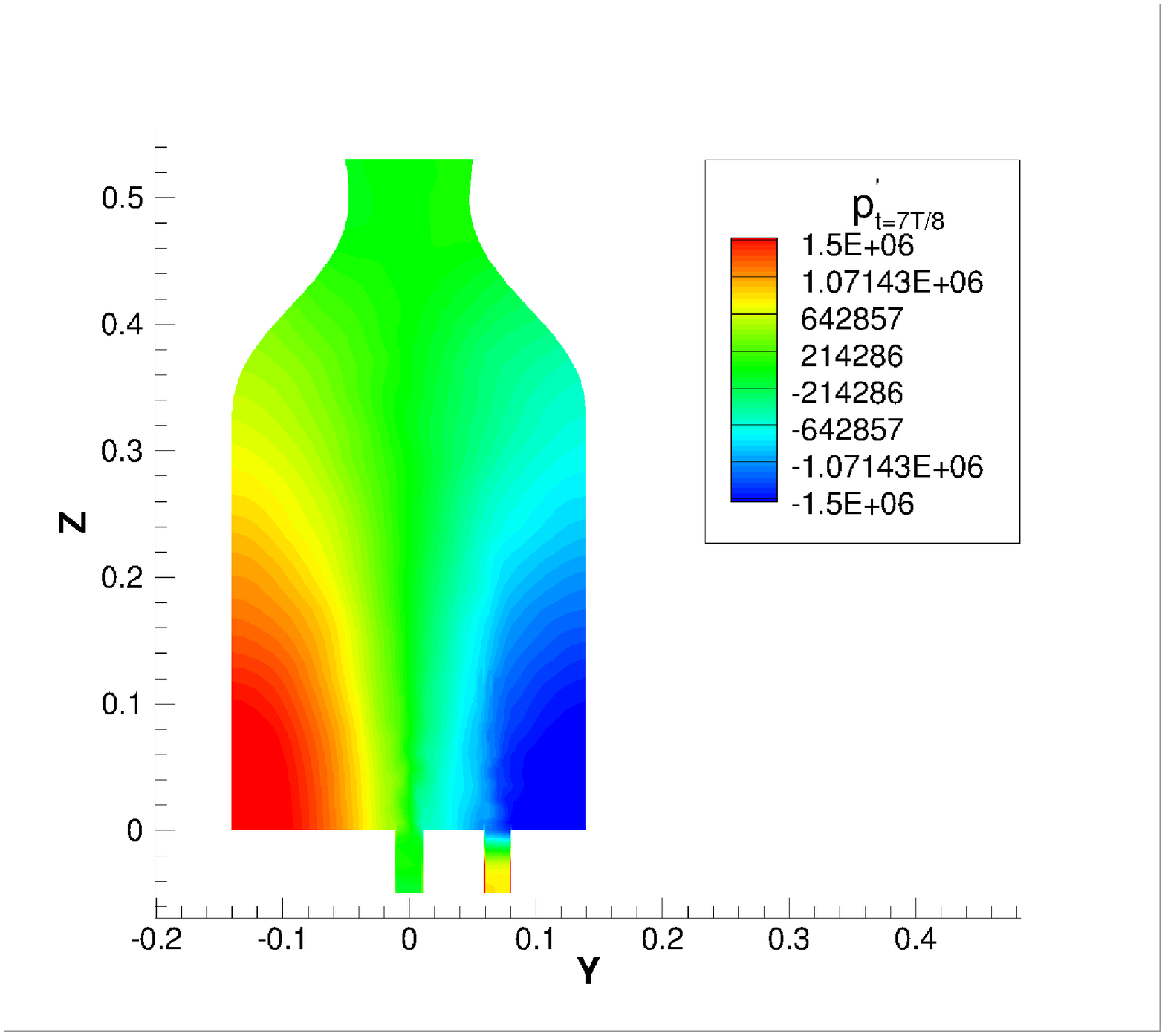}
        \label{fig:osk_p_3qT_2600Hz_x_0cm}}
        \subfigure[$3200$ Hz mode at $t = T$ for FPV]
{\includegraphics[width=0.3\textwidth]{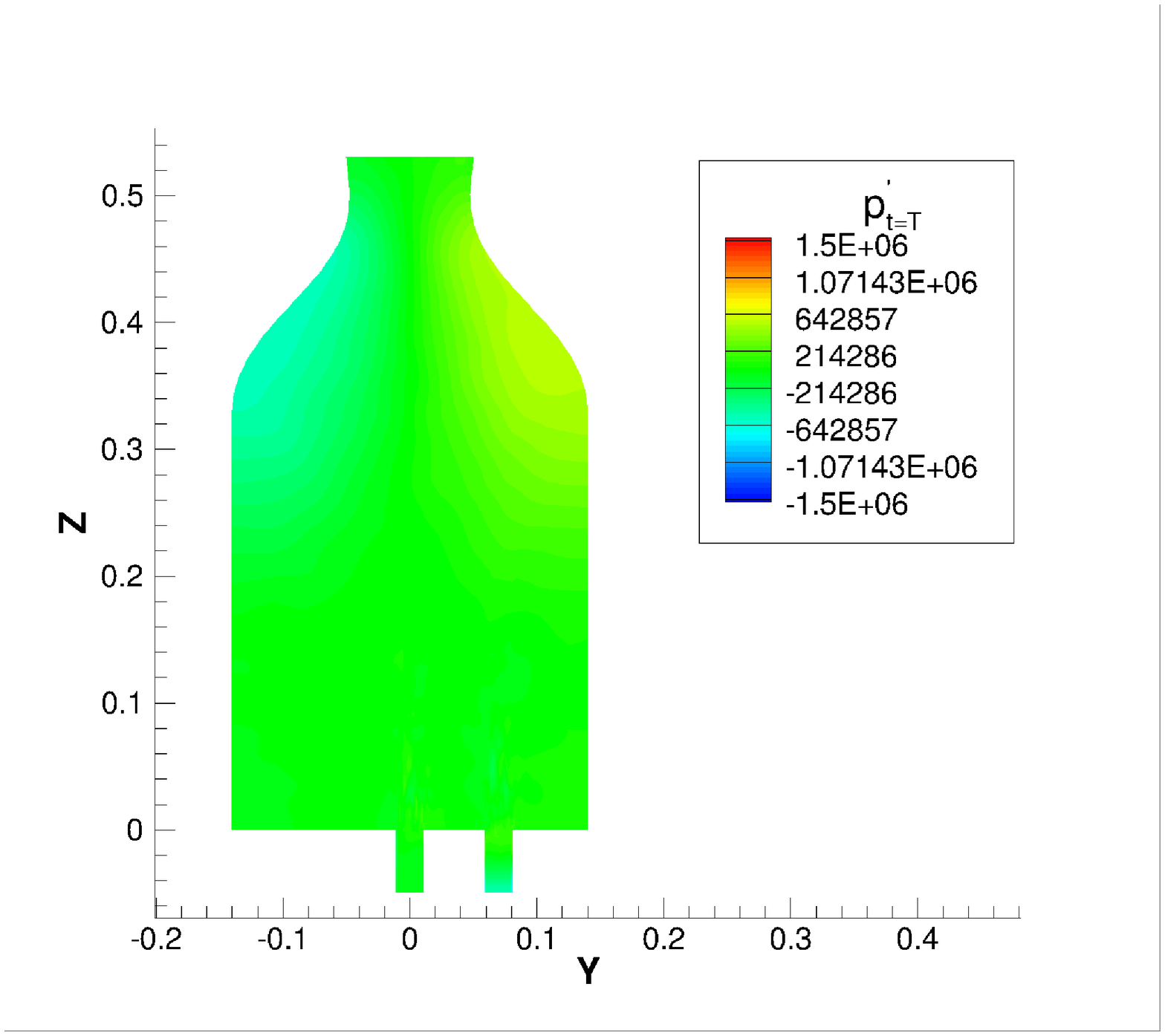}
        \label{fig:fpv_p_4qT_3200Hz_x_0cm}}
        \subfigure[$2600$ Hz mode at $t = T$ for OSK]
{\includegraphics[width=0.3\textwidth]{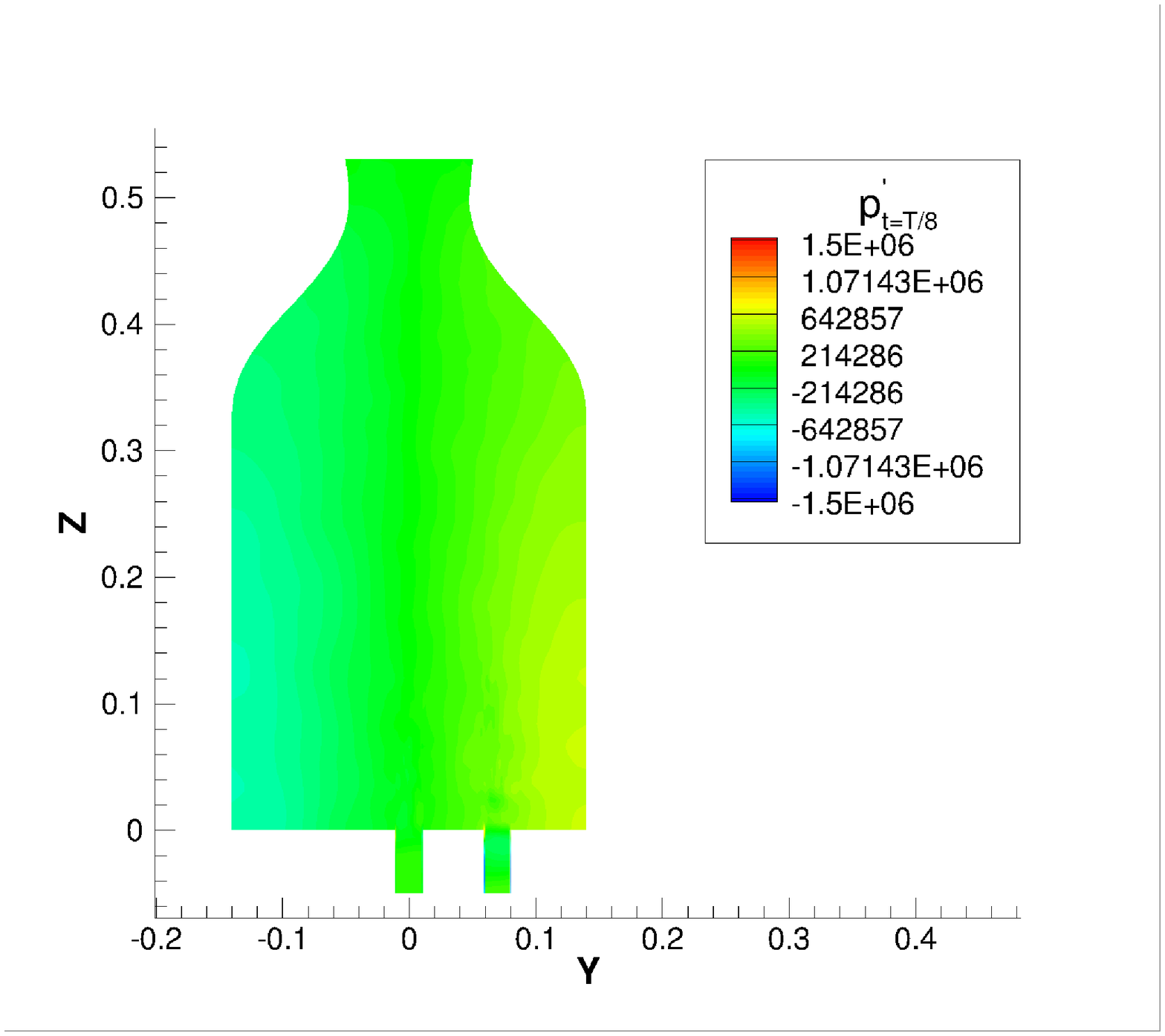}
        \label{fig:osk_p_4qT_2600Hz_x_0cm}}
    \end{subfigmatrix}
    \caption{Instantaneous contours of pressure fluctuation for
the dominant instability mode on the x = 0 cm plane}
    \label{fig:p_prime_x_0cm}
\end{figure}

For the FPV model, pressure anti-nodes formed immediately downstream of the injector plate and at the entrance of the nozzle. A pressure node appears
downstream of the injector plate by half of chamber length. These features in the axial direction of the chamber are illustrated in Figure~\ref{fig:fpv_p_1qT_3200Hz_x_0cm} 
~\ref{fig:fpv_p_2qT_3200Hz_x_0cm}~\ref{fig:fpv_p_3qT_3200Hz_x_0cm}~\ref{fig:fpv_p_4qT_3200Hz_x_0cm} and suggest a longitudinal standing wave.
Hence, the dominant instability mode predicted by the FPV model is a combined first-tangential and first-longitudinal standing wave mode. 
In contrast, the longitudinal-wave behavior is not observed for the OSK model. The pressure fluctuation
monotonically decay in the main flow direction, suggesting that the OSK model favors a pure first tangential standing wave mode.

Using the Rayleigh index, we can determine the correlation between pressure fluctuation and unsteady heat release. In 
this paper, the Rayleigh index is calculated as:

\begin{equation}
RI = \frac{1}{\tau p_{a}}\int_{t_{0}}^{t_{0}+\tau}\frac{\gamma-1}{\gamma}p^{'}\omega^{'}_{T}dt
\label{RI}
\end{equation}
where $t_{0}$ and $\tau$ are the lower limit and time interval which are determined such that the time integration is calculated over the last ten milliseconds for each model. $p^{'}$ is the local pressure
oscillation, $p_{a}$ is the local time-averaged pressure, $\gamma$ is the specific heat ratio, and $\omega^{'}_{T}$ is the local oscillation of heat release rate. 
Positive values of the Rayleigh index indicate pressure fluctuations are driven by the unsteady heat release,
while negative values suggest pressure oscillations are damped by the unsteady heat release.

\begin{figure}
    \begin{subfigmatrix}{2}
        \subfigure[$x = 0$ cm plane for FPV]
{\includegraphics[width=0.3\textwidth]{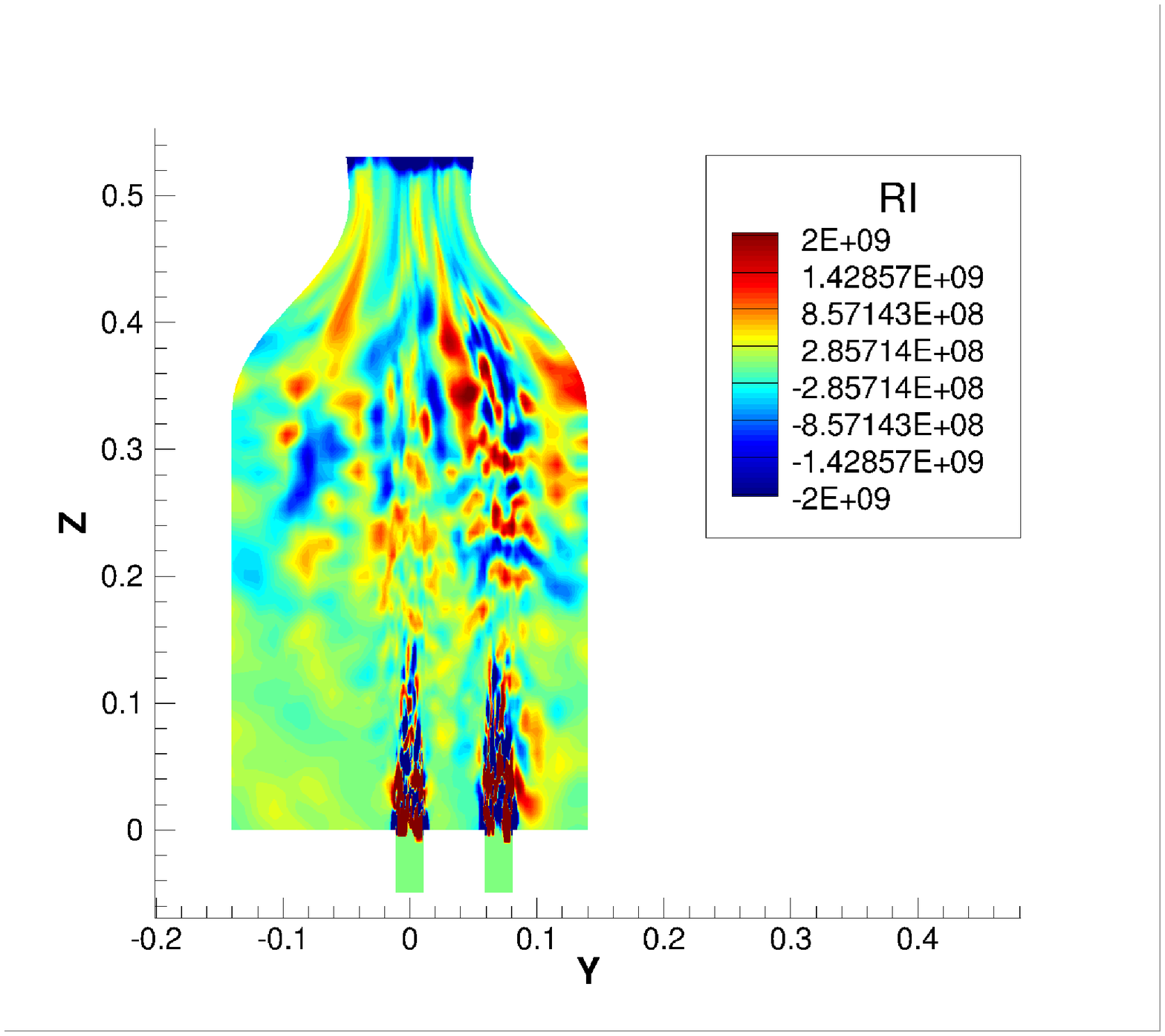}
        \label{fig:fpv_RI_x_0cm}}
        \subfigure[$x = 0$ cm plane for OSK]
{\includegraphics[width=0.3\textwidth]{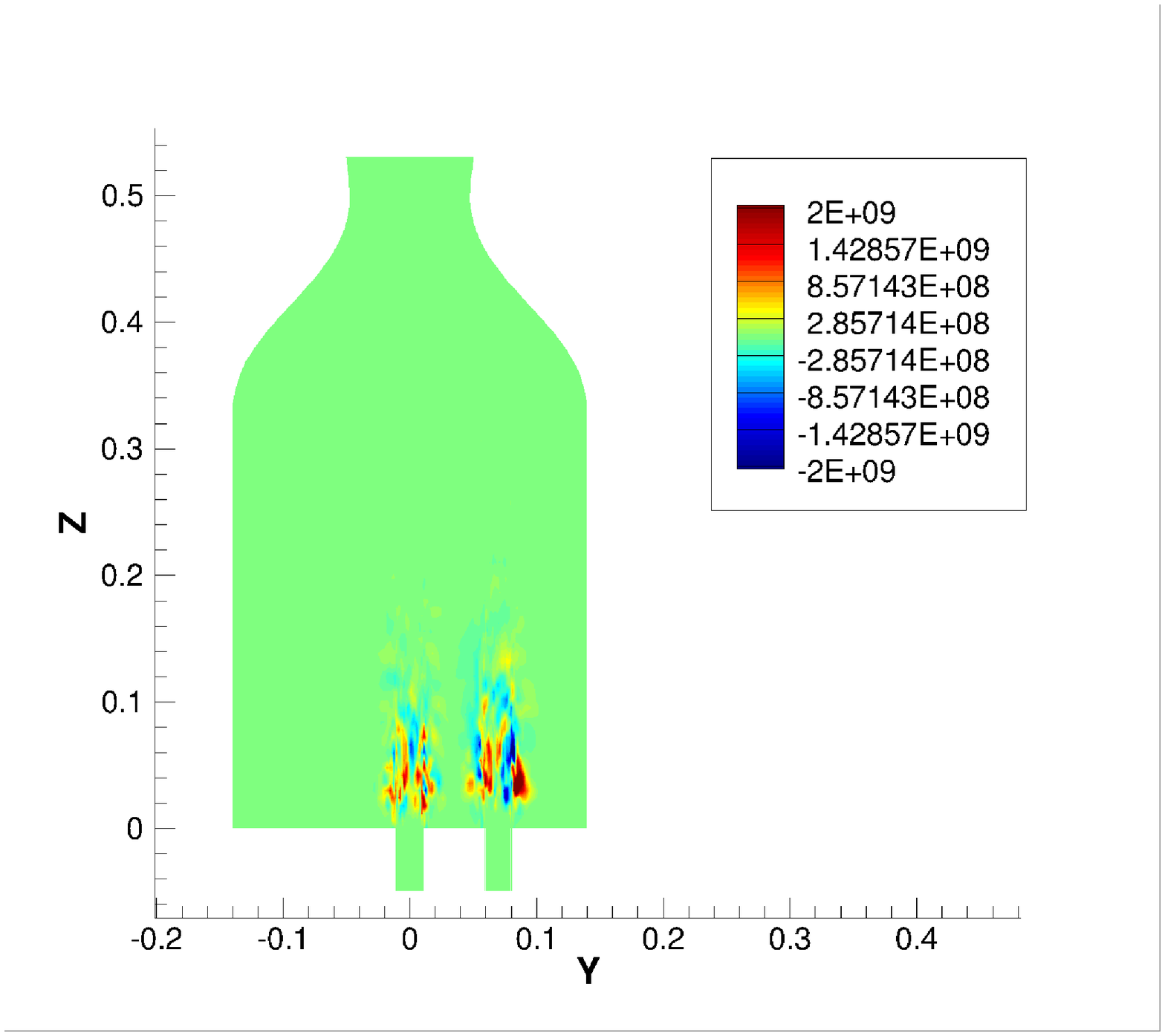}
        \label{fig:osk_RI_x_0cm}}
        \subfigure[$z = 4$ cm plane for FPV]
{\includegraphics[width=0.3\textwidth]{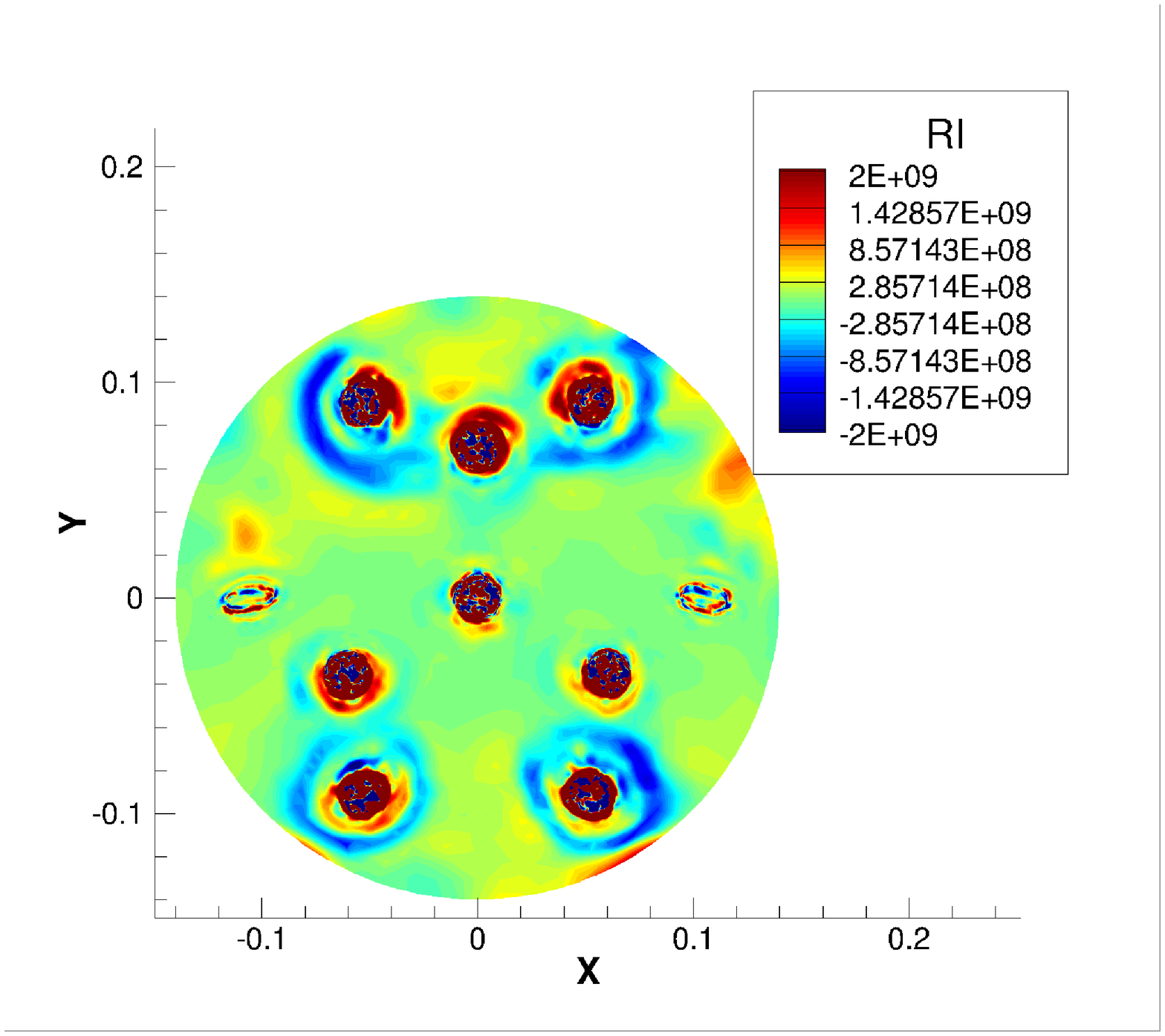}
        \label{fig:fpv_RI_z_4cm}}
        \subfigure[$z = 4$ cm plane for OSK]
{\includegraphics[width=0.3\textwidth]{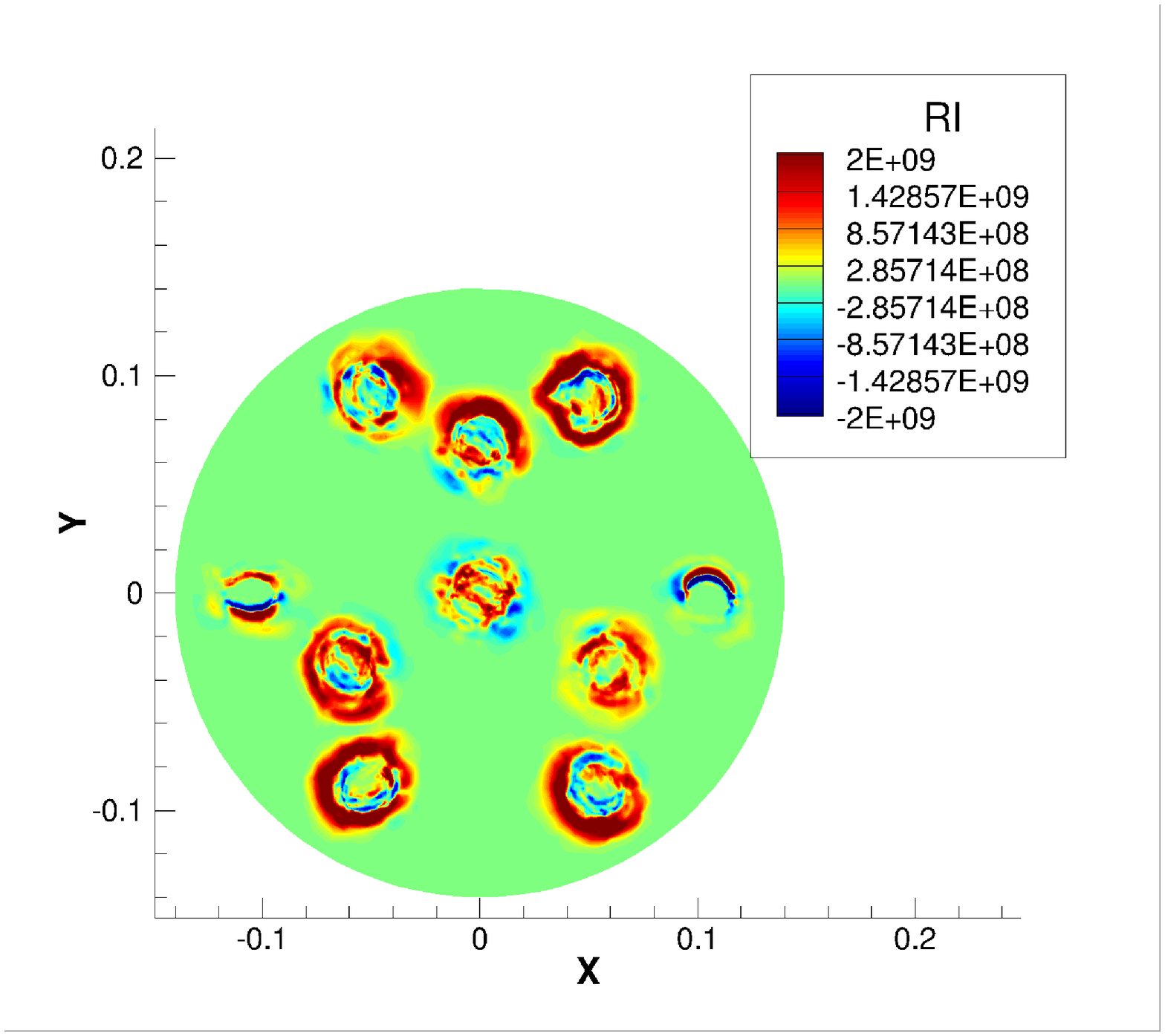}
        \label{fig:osk_RI_z_4cm}}
        \subfigure[$z = 15$ cm plane for FPV]
{\includegraphics[width=0.3\textwidth]{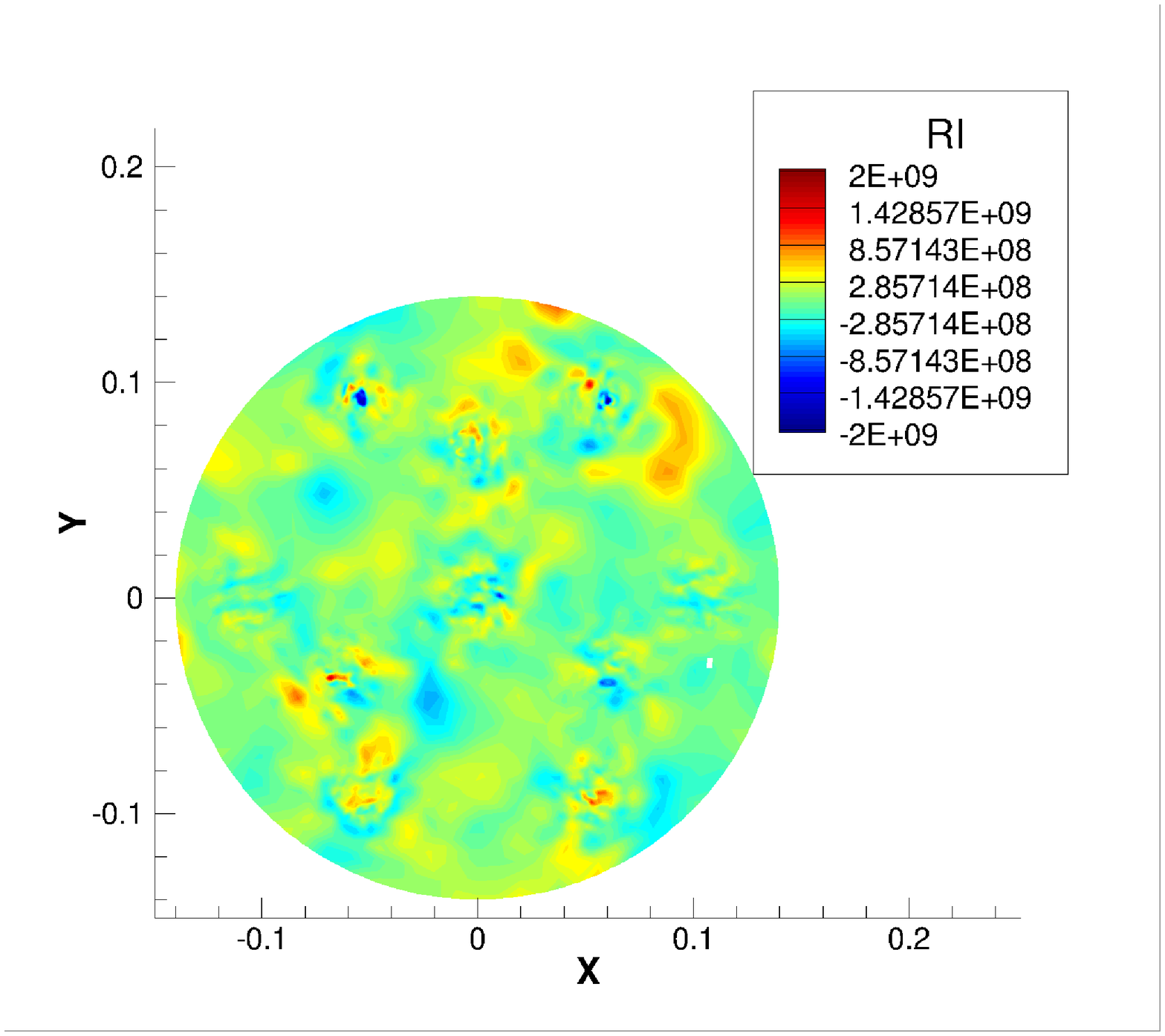}
        \label{fig:fpv_RI_z_15cm}}
        \subfigure[$z = 15$ cm plane for OSK]
{\includegraphics[width=0.3\textwidth]{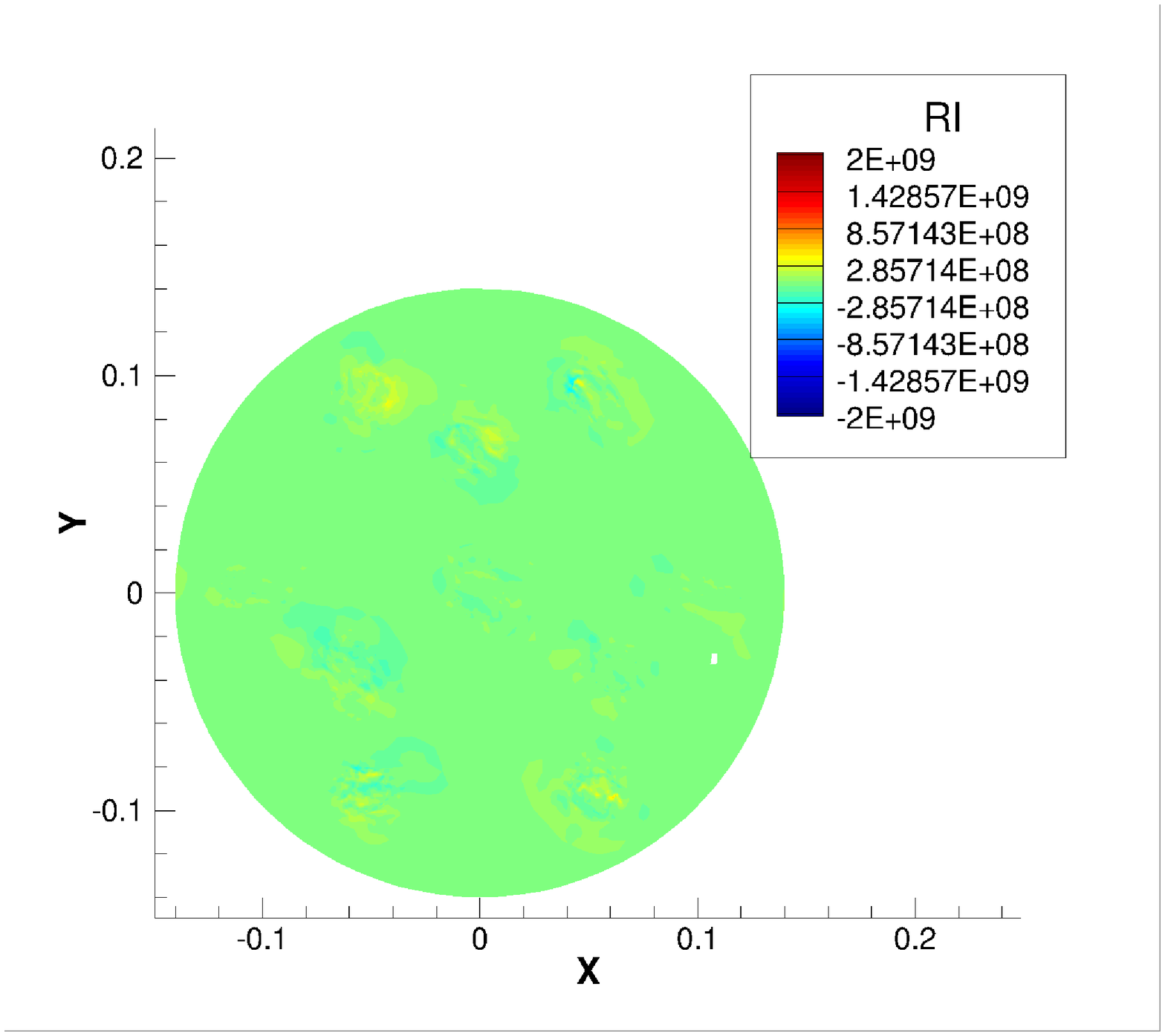}
        \label{fig:osk_RI_z_15cm}}
        \subfigure[$z = 33$ cm plane for FPV]
{\includegraphics[width=0.3\textwidth]{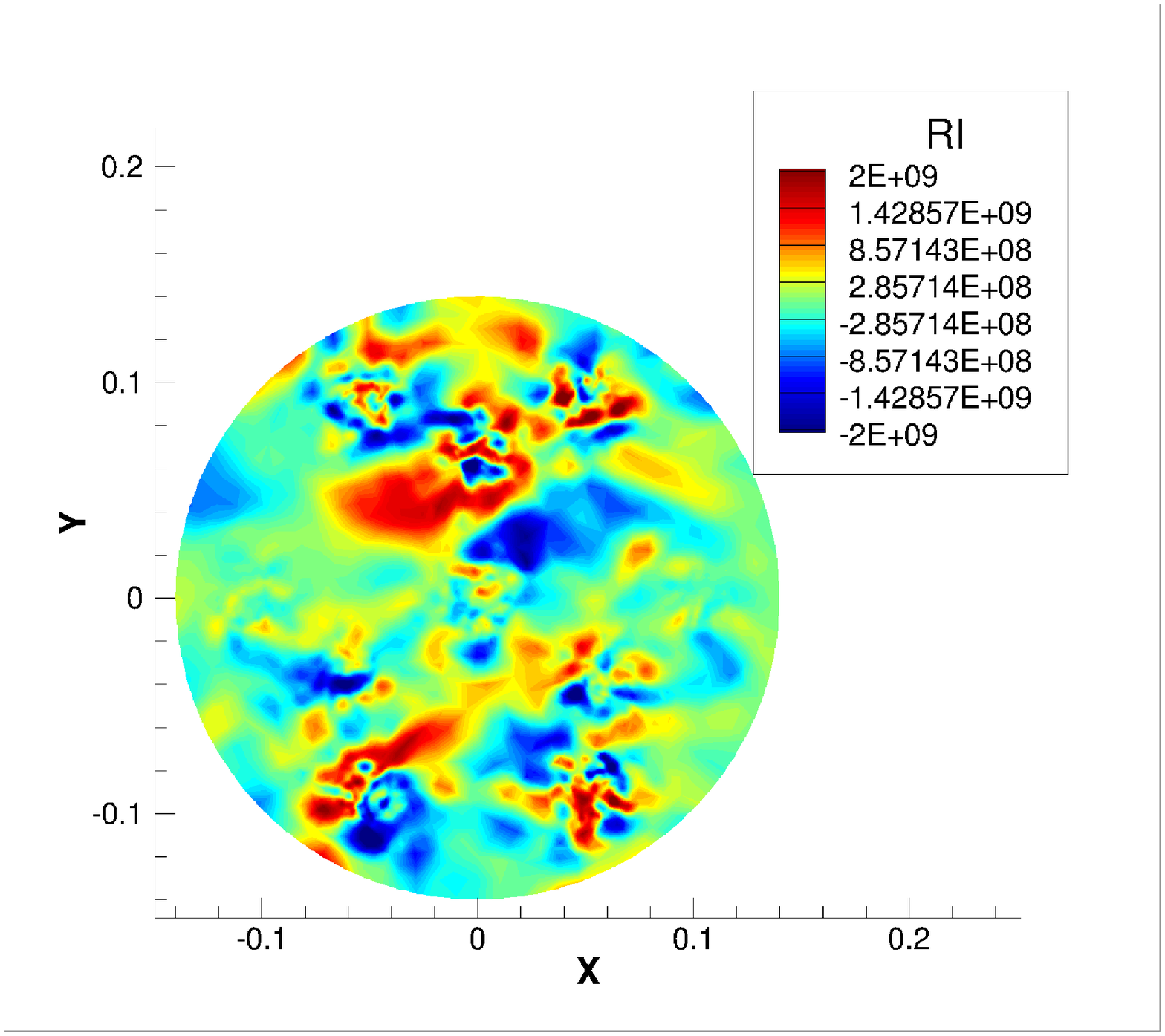}
        \label{fig:fpv_RI_z_33cm}}
        \subfigure[$z = 33$ cm plane for OSK]
{\includegraphics[width=0.3\textwidth]{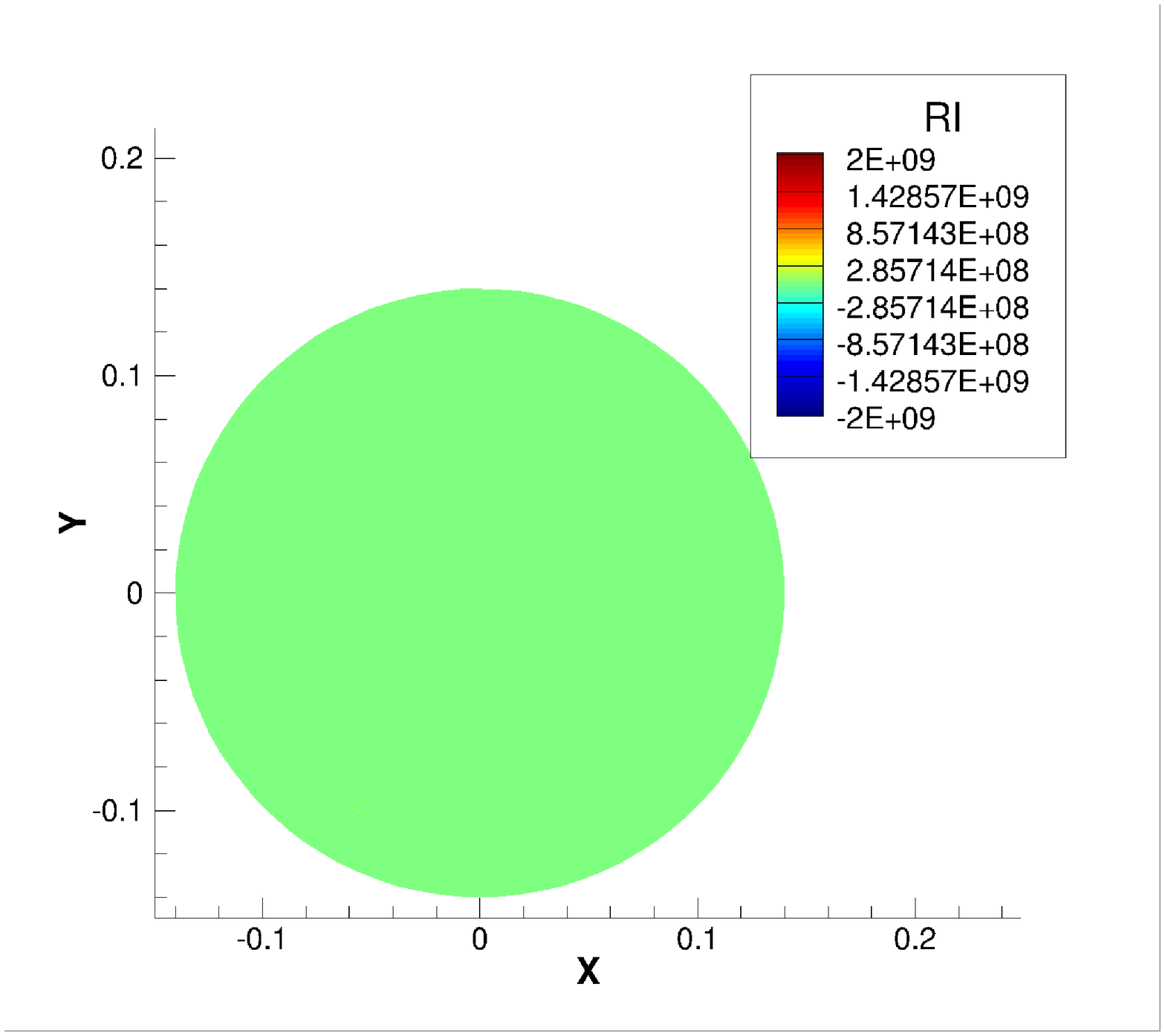}
        \label{fig:osk_RI_z_33cm}}
    \end{subfigmatrix}
    \caption{Contours of the Rayleigh index}
    \label{fig:RI}
\end{figure}

Figure~\ref{fig:RI} shows that for both the FPV and OSK models, pressure fluctuation is driven by unsteady heat release near the injector plate, and hence a tangential acoustic standing wave is formed and sustained. As shown in Figure~\ref{fig:tave_FI}, for both combustion models, the premixed flames are dominant in the immediate downstream of the injector plate where the fluctuation amplitudes of the pressure antinodes for the transverse acoustic standing wave reach maximum values (usually decay in the downstream axial direction). So, the interaction between the transverse acoustic standing wave and the premixed flames can be reasonably expected to be highly consequential. The two combustion models predict different mixing processes in the vicinity of the injector plate and hence distinct premixed flames including those stretched ones. This could be a major reason for the difference in the details of the predicted transverse wave by the two combustion models, such as the azimuthal locations of the pressure nodes/anti-nodes. At the entrance of the exhaust nozzle, the diffusion flames are dominant for both combustion models. If a longitudinal acoustic standing wave is resonant, one of its pressure antinodes often occurs at the nozzle entrance. Hence, the correlation between the longitudinal acoustic standing wave and the diffusion flames can also be expected to be noteworthy. For the FPV model, the stronger diffusion flames around the nozzle entrance indicate significant oxidation, such as reaction from $CO$ to $CO_{2}$, and hence considerable unsteady heat release, which drives the longitudinal pressure fluctuation. So, the dominant acoustic mode simulated by the FPV model is finally a combined transverse and longitudinal standing wave mode. In contrast, the extremely weak global reaction at the nozzle entrance produces diffusion flames without significant heat release for the OSK model. So, there is no driving force establishing and sustaining a longitudinal acoustic wave and the OSK model finally predicts a pure transverse mode as the dominant mode. The dominant acoustic mode may also have significant impacts on the flame dynamics in turn. For instance, the longitudinal standing wave might also help to push the premix flames towards (or even to attach) the injector lips, which is observed for the FPV model. A similar mechanism was reported by Nguyen and Sirignano \cite{tuancnf,tuanjpp}. The longitudinal acoustic standing wave could also help to enhance mixing in the immediate downstream of the injector plate, leading to accelerated burning in the region. As a result, less fuel escapes into further downstream of the combustion chamber for the FPV model as compared to the OSK model which predicts no longitudinal acoustic wave (hence zero fluctuation amplitude of the wave). This is consistent with the observation by Nguyen and Sirignano \cite{tuanaiaaj} that burning becomes more efficient as the amplitude of pressure fluctuation in the longitudinal mode increases.

\begin{figure}
    \begin{subfigmatrix}{2}
        \subfigure[$3200$ Hz mode at $t = T/4$ for FPV]
{\includegraphics[width=0.3\textwidth]{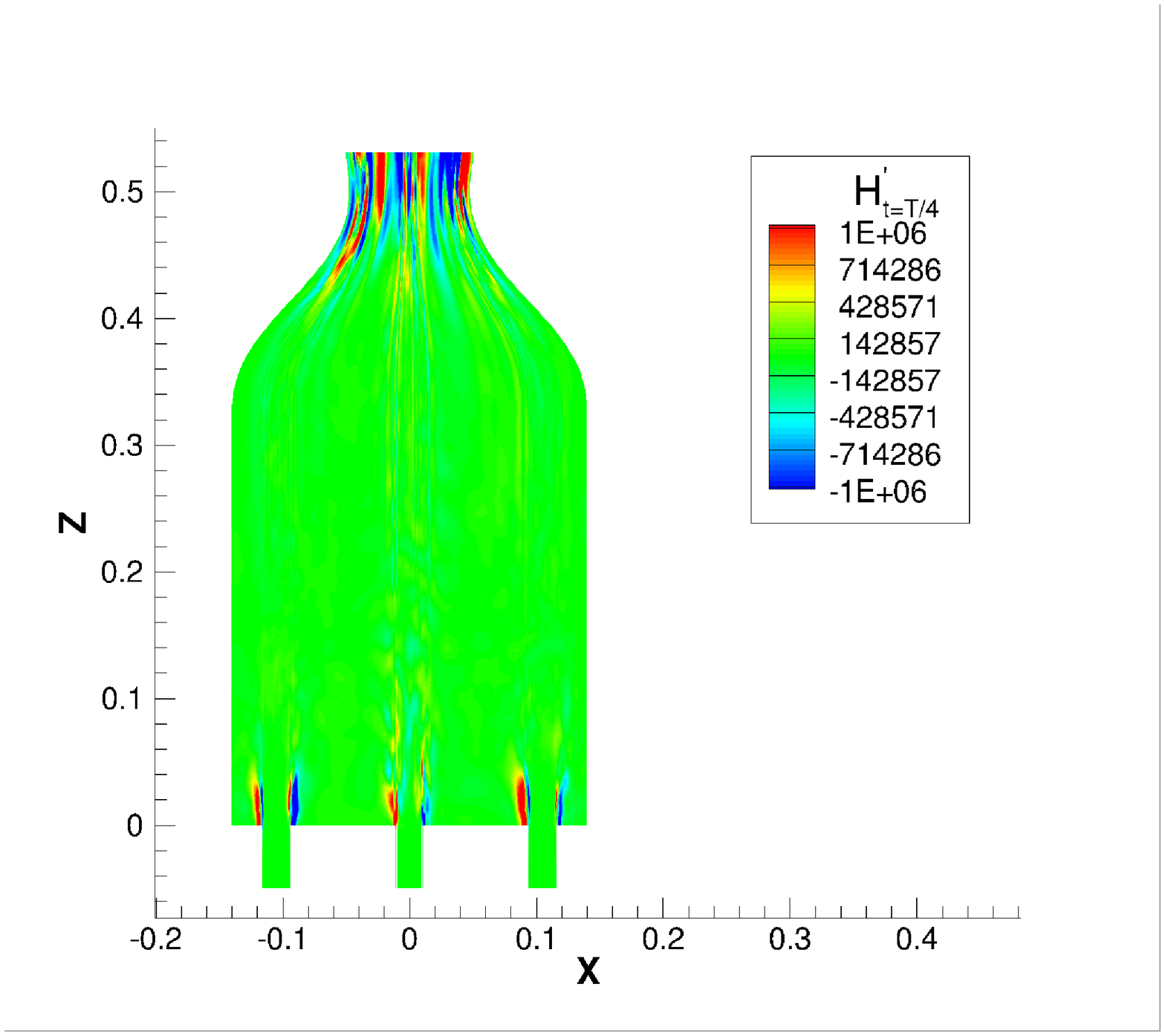}
        \label{fig:fpv_heli_1qT_3200Hz_y_0cm}}
        \subfigure[$2600$ Hz mode at $t = T/4$ for OSK]
{\includegraphics[width=0.3\textwidth]{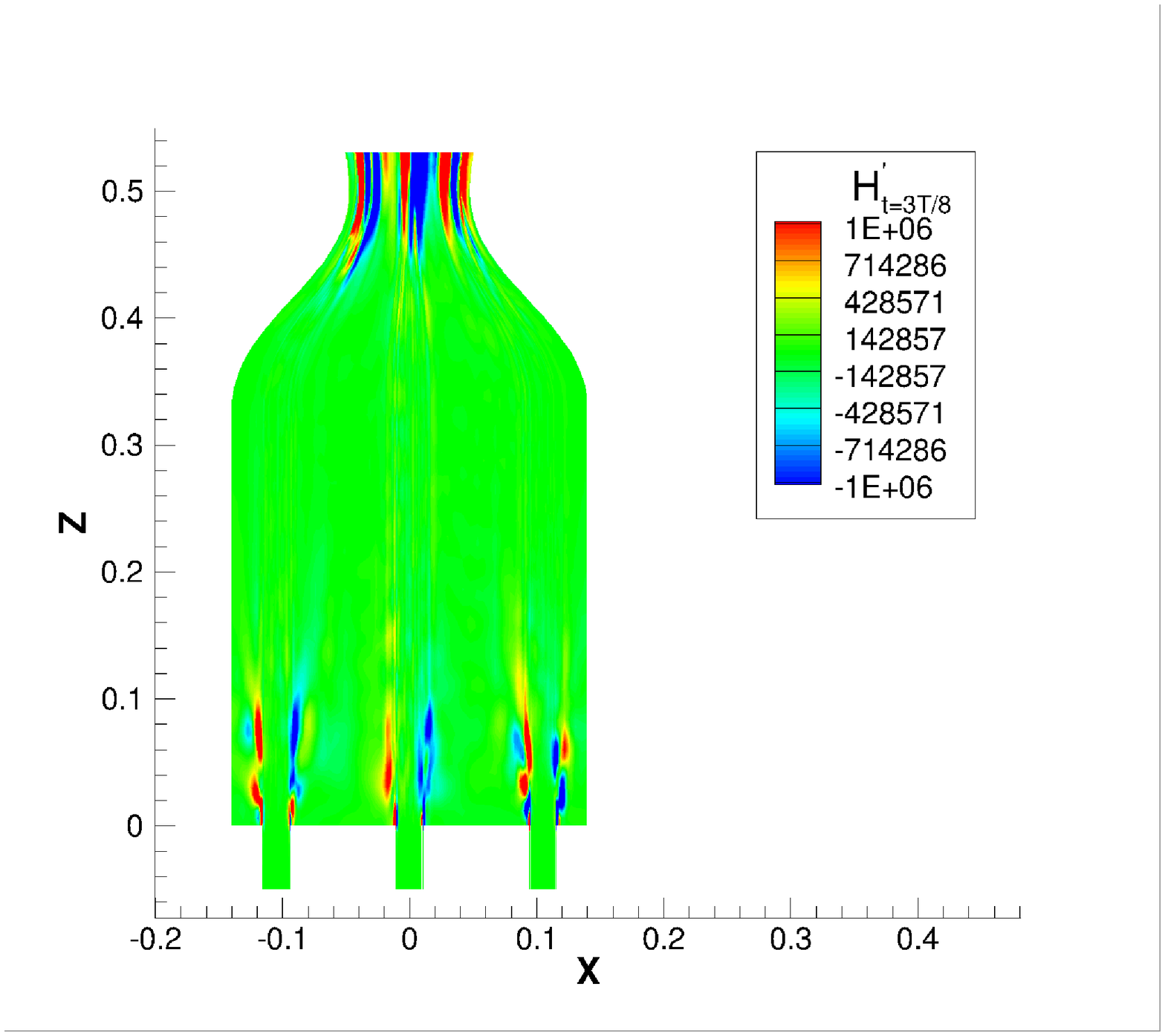}
        \label{fig:osk_heli_1qT_2600Hz_y_0cm}}
        \subfigure[$3200$ Hz mode at $t = T/2$ for FPV]
{\includegraphics[width=0.3\textwidth]{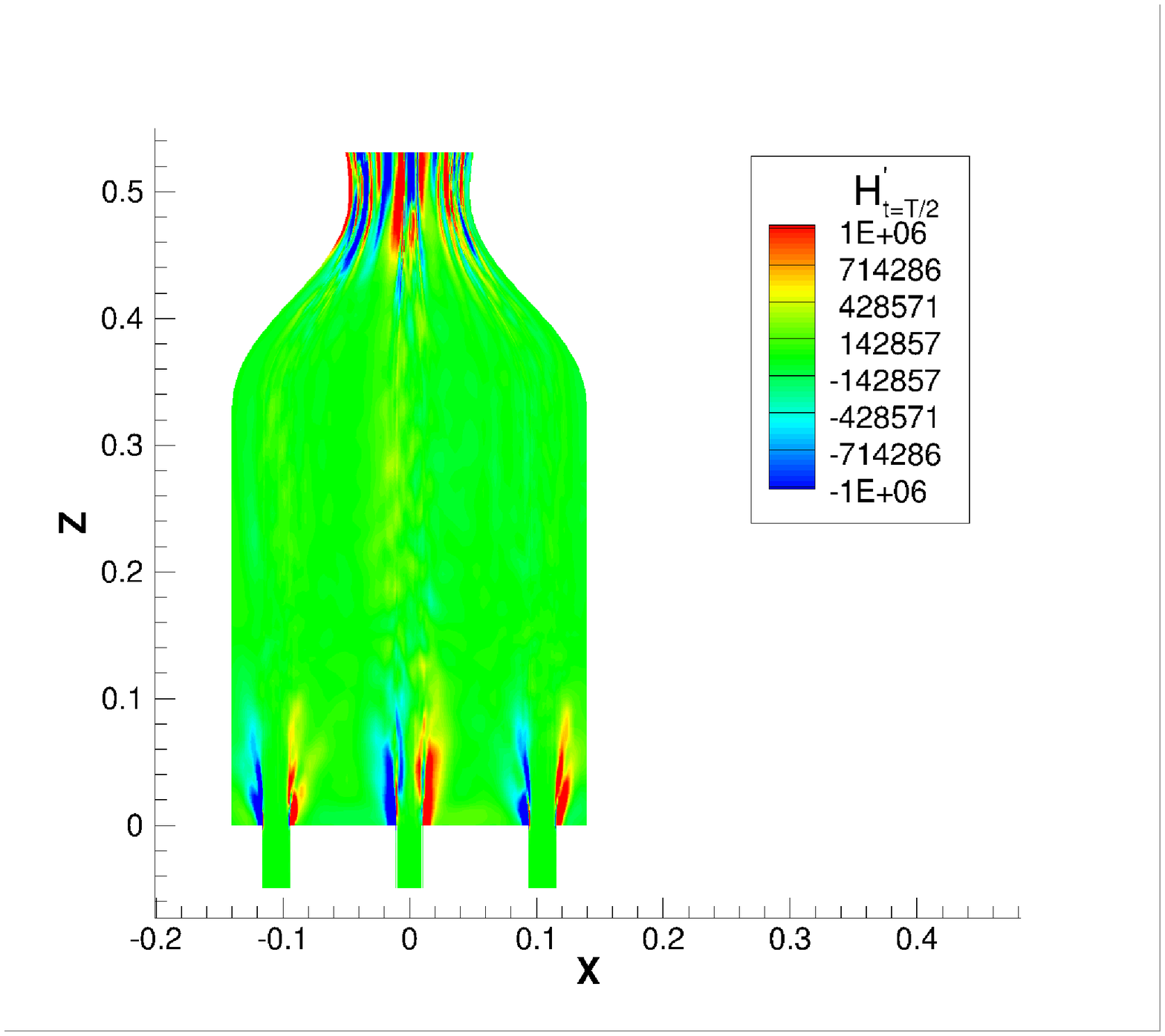}
        \label{fig:fpv_heli_2qT_3200Hz_y_0cm}}
        \subfigure[$2600$ Hz mode at $t = T/2$ for OSK]
{\includegraphics[width=0.3\textwidth]{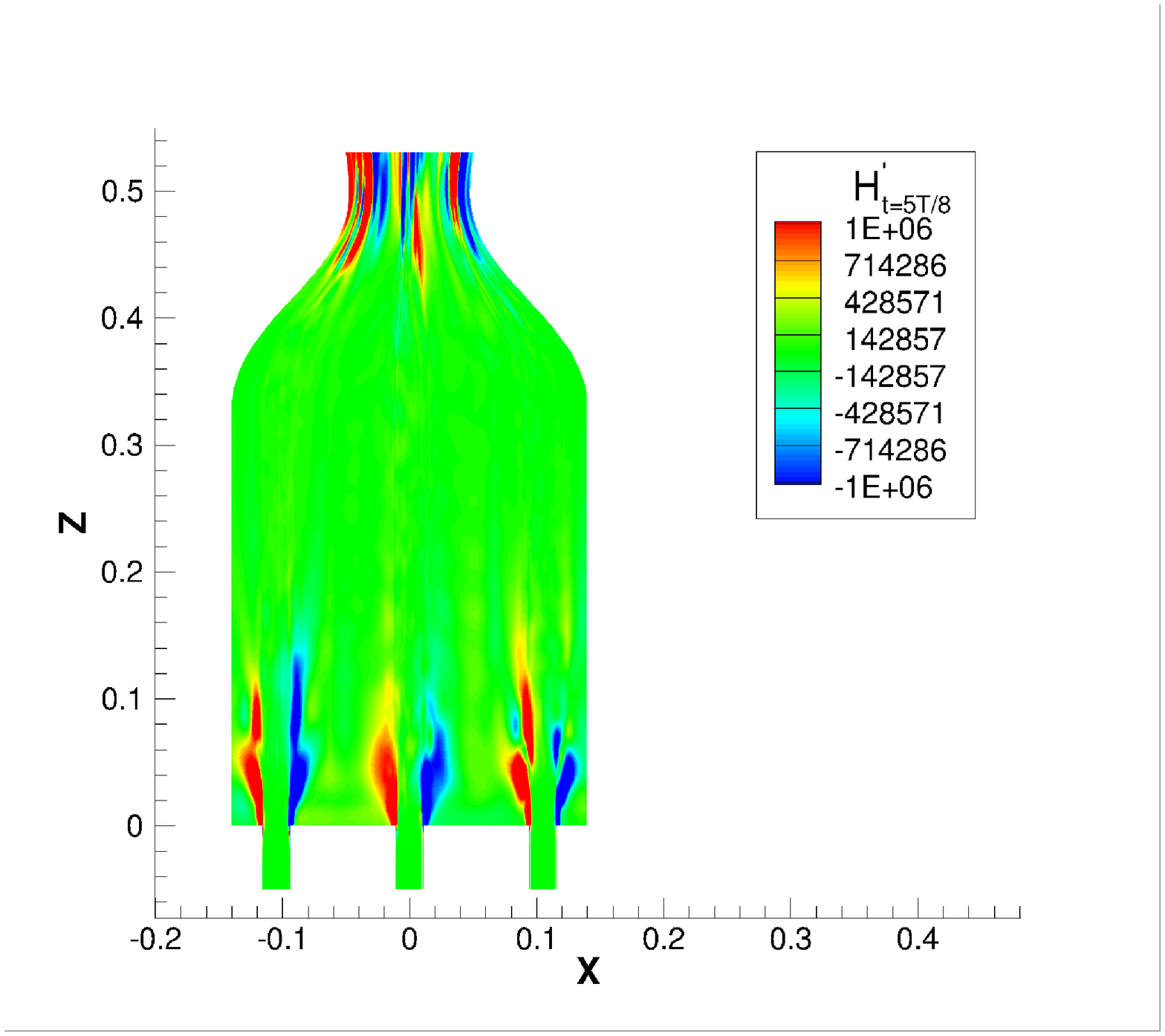}
        \label{fig:osk_heli_2qT_2600Hz_y_0cm}}
        \subfigure[$3200$ Hz mode at $t = 3T/4$ for FPV]
{\includegraphics[width=0.3\textwidth]{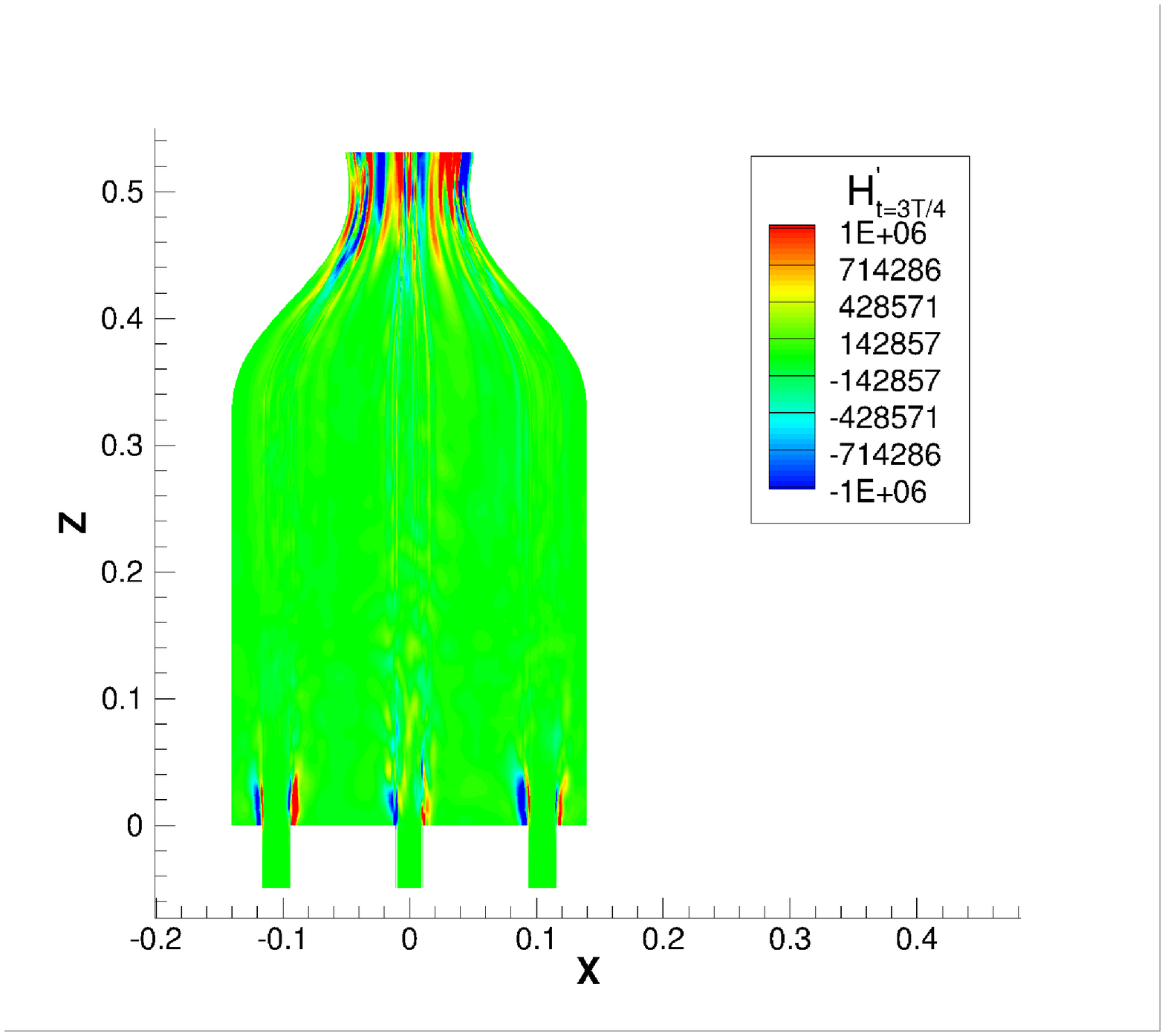}
        \label{fig:fpv_heli_3qT_3200Hz_y_0cm}}
        \subfigure[$2600$ Hz mode at $t = 3T/4$ for OSK]
{\includegraphics[width=0.3\textwidth]{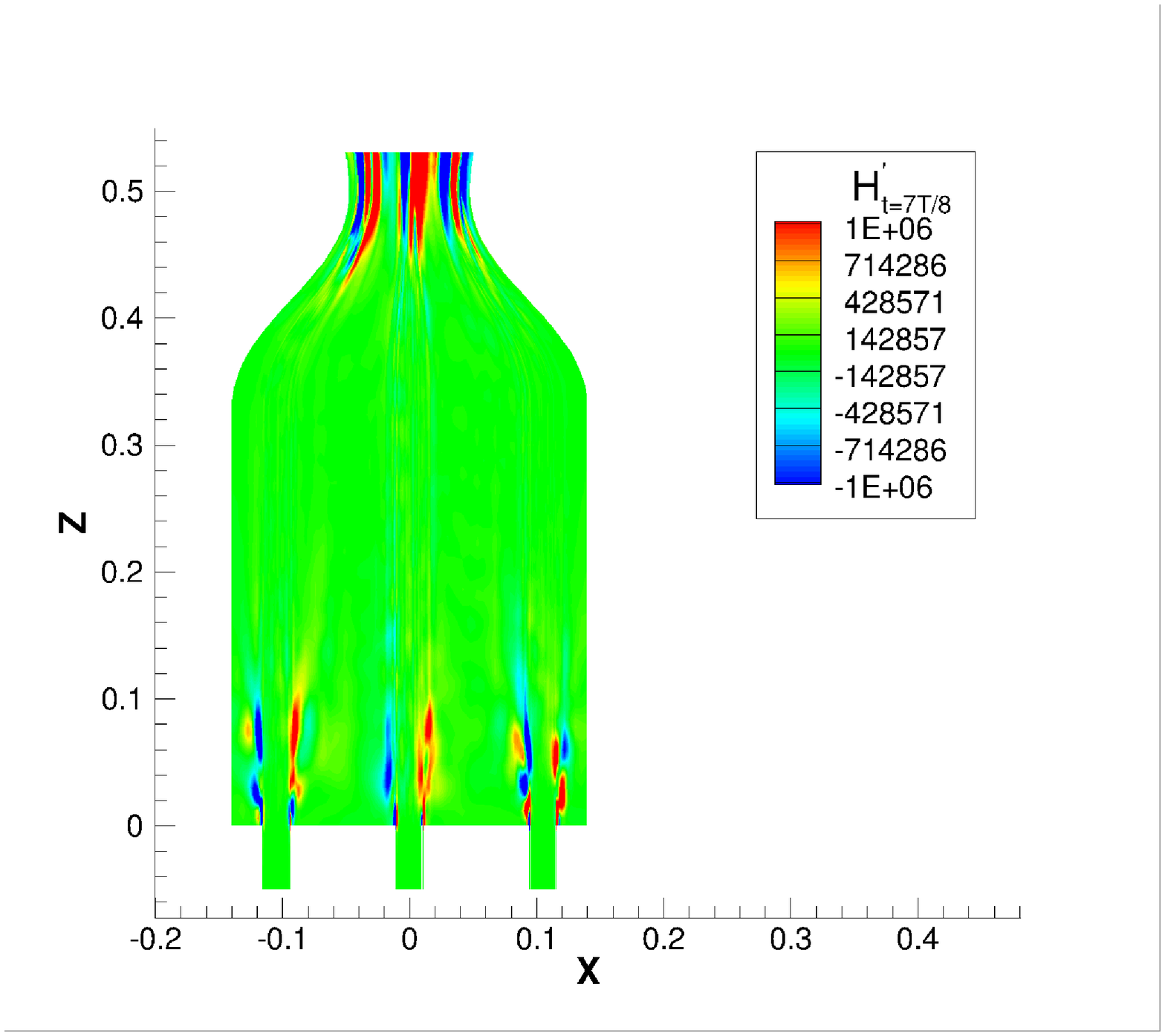}
        \label{fig:osk_heli_3qT_2600Hz_y_0cm}}
        \subfigure[$3200$ Hz mode at $t = T$ for FPV]
{\includegraphics[width=0.3\textwidth]{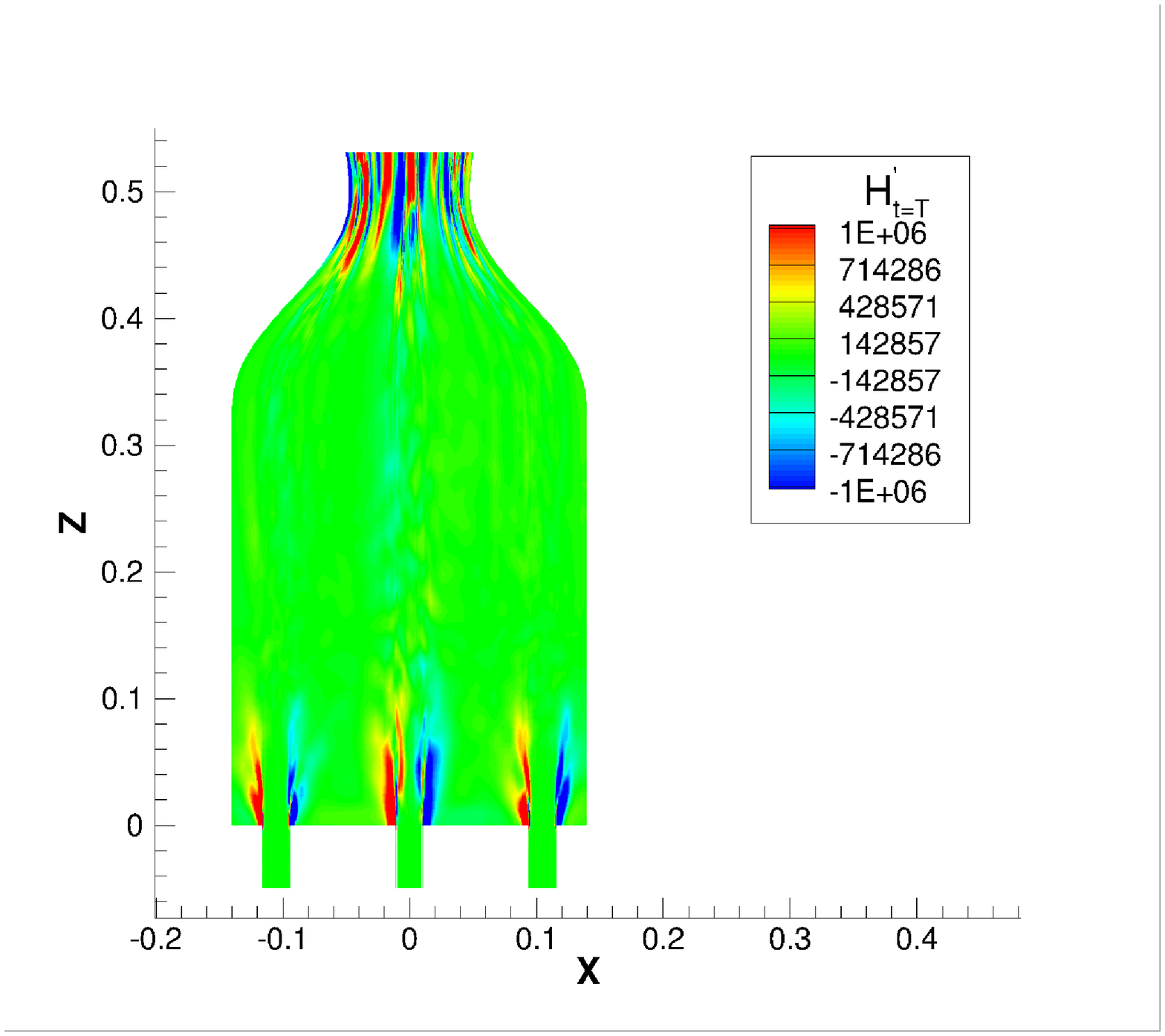}
        \label{fig:fpv_heli_4qT_3200Hz_y_0cm}}
        \subfigure[$2600$ Hz mode at $t = T$ for OSK]
{\includegraphics[width=0.3\textwidth]{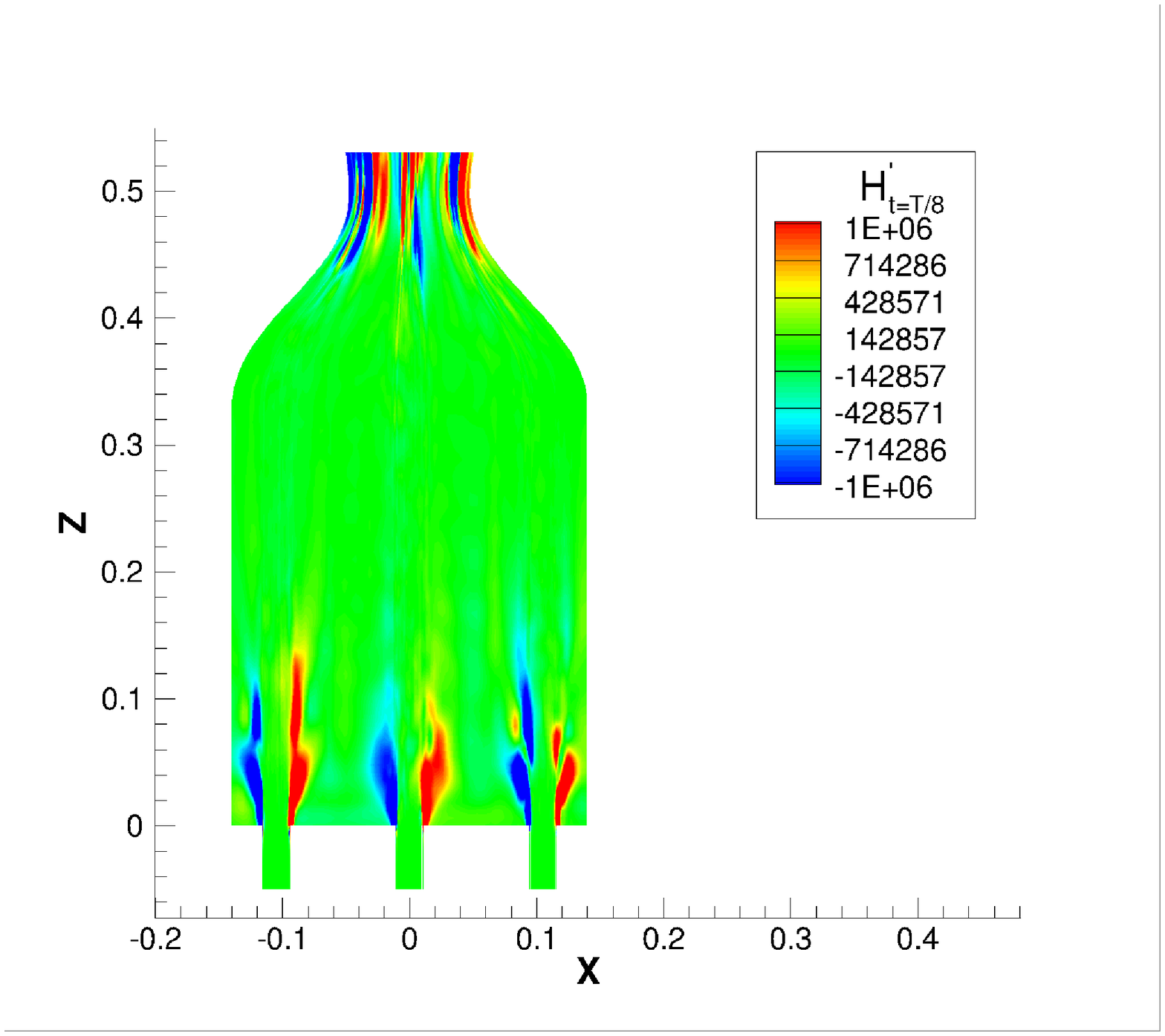}
        \label{fig:osk_heli_4qT_2600Hz_y_0cm}}
    \end{subfigmatrix}
    \caption{Instantaneous contours of helicity fluctuation for
the dominant instability mode on the y = 0 cm plane}
    \label{fig:heli_prime_y_0cm}
\end{figure}

As vorticity plays an important role in mixing and hence has impacts on combustion instability, we also study the correlation between pressure fluctuation
and the disturbance of helicity, which is the streamwise vorticity. Figure~\ref{fig:heli_prime_y_0cm} shows the instantaneous contours of helicity fluctuation for
the dominant instability mode on the y = 0 cm meridian plane. For both combustion models, helicity fluctuation is only significant 
in the vicinity of the propellant jets. So, its impact on the flame dynamics and combustion instability is only important near the injector plate. 

\begin{figure}
    \begin{subfigmatrix}{2}
        \subfigure[Helicity mode at $t = T/4$]
{\includegraphics[width=0.27\textwidth]{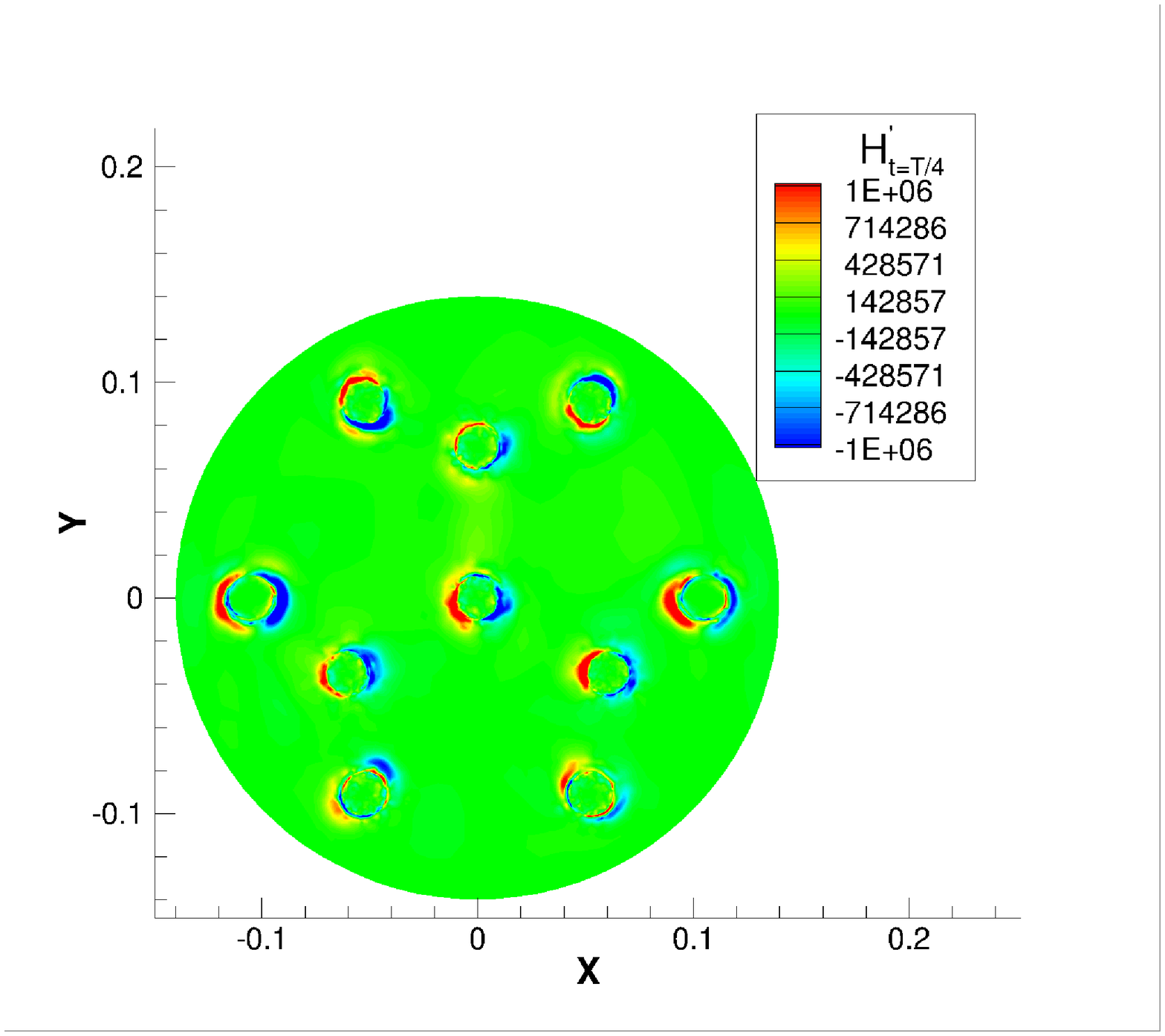}
        \label{fig:fpv_heli_1qT_3200Hz_z_1cm}}
        \subfigure[Pressure and transverse velocity modes at $t = T/4$]
{\includegraphics[width=0.27\textwidth]{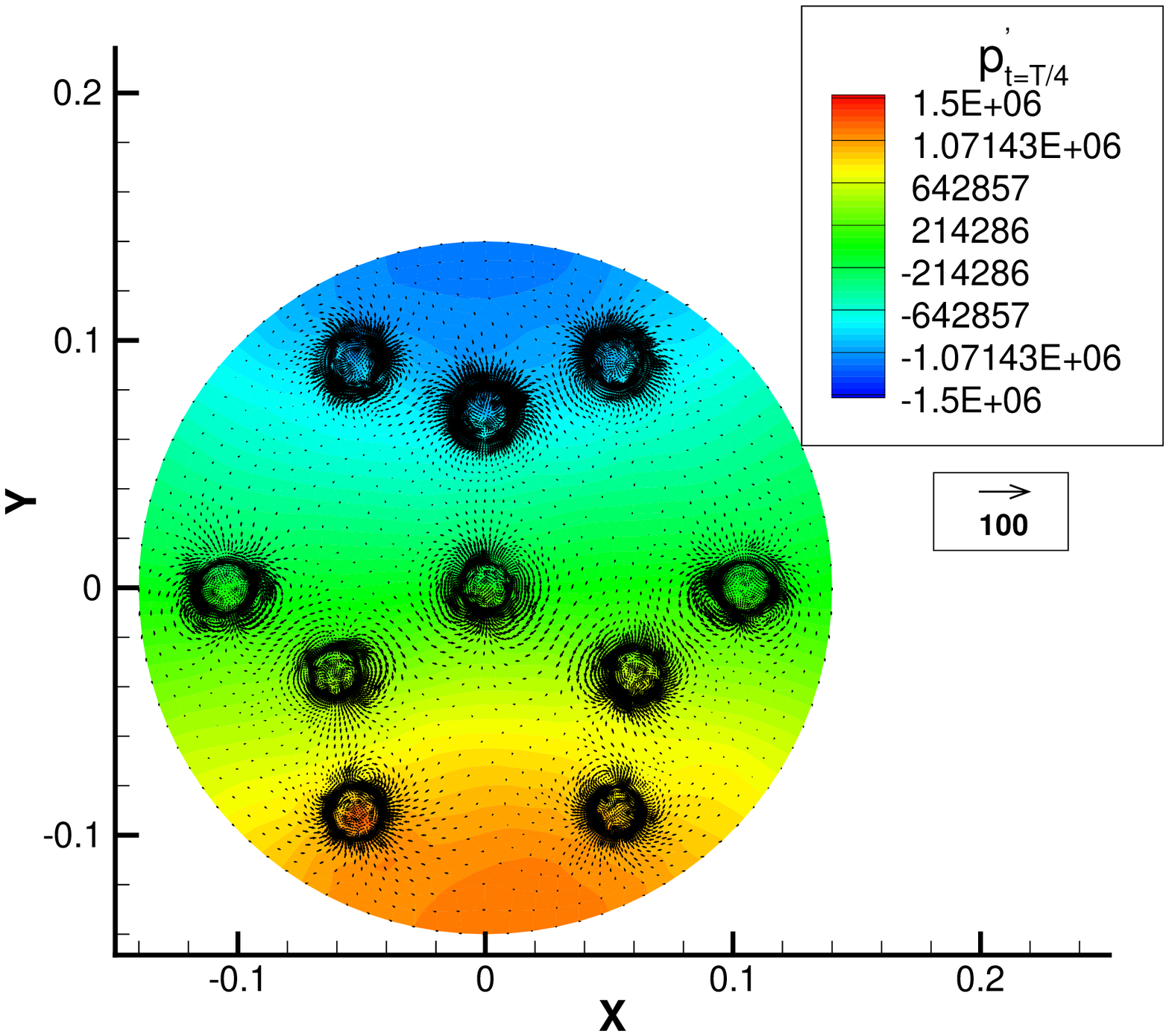}
        \label{fig:fpv_p_1qT_3200Hz_z_1cm}}
        \subfigure[Helicity mode at $t = T/2$]
{\includegraphics[width=0.27\textwidth]{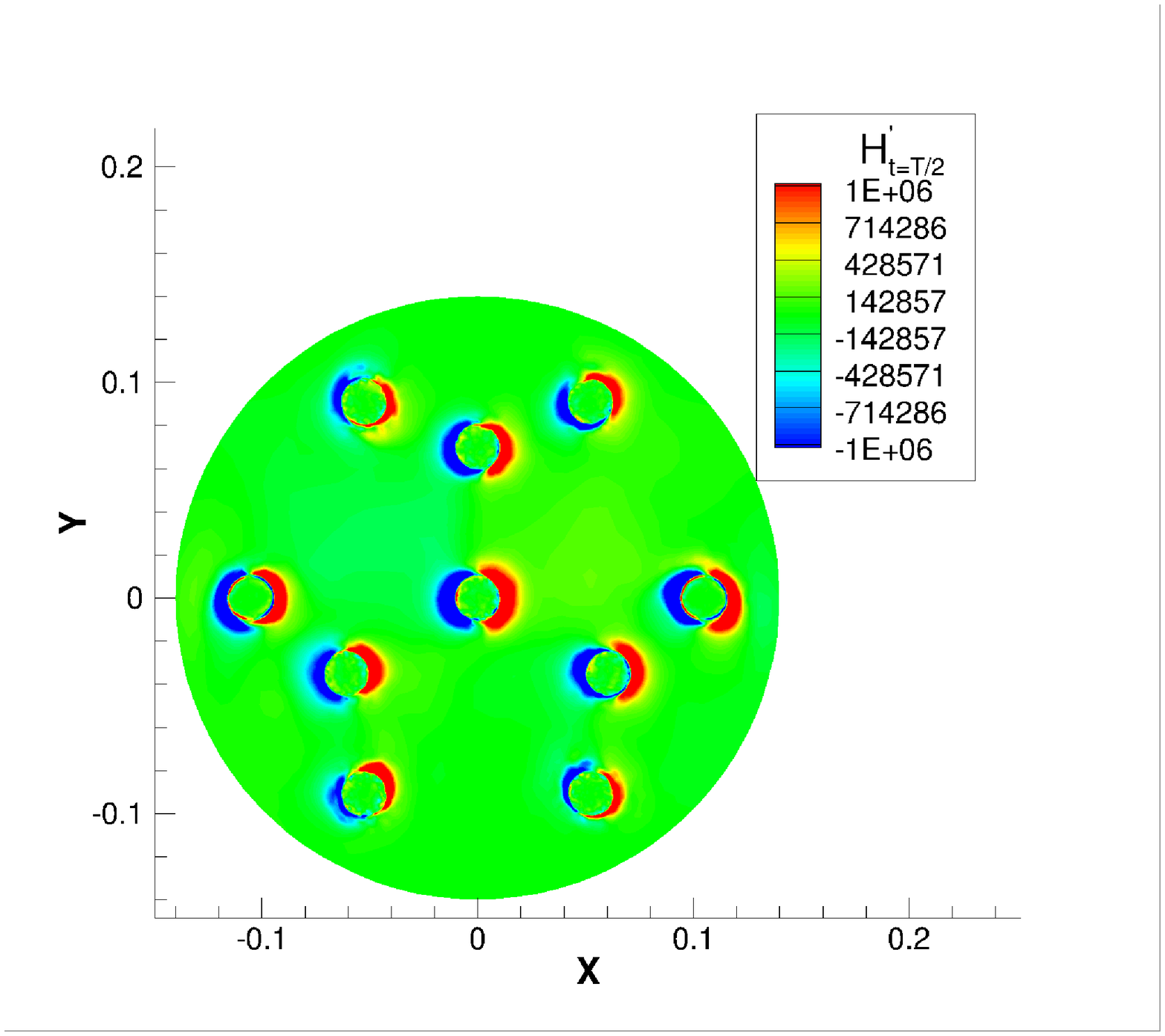}
        \label{fig:fpv_heli_2qT_3200Hz_z_1cm}}
        \subfigure[Pressure and transverse velocity modes at $t = T/2$]
{\includegraphics[width=0.27\textwidth]{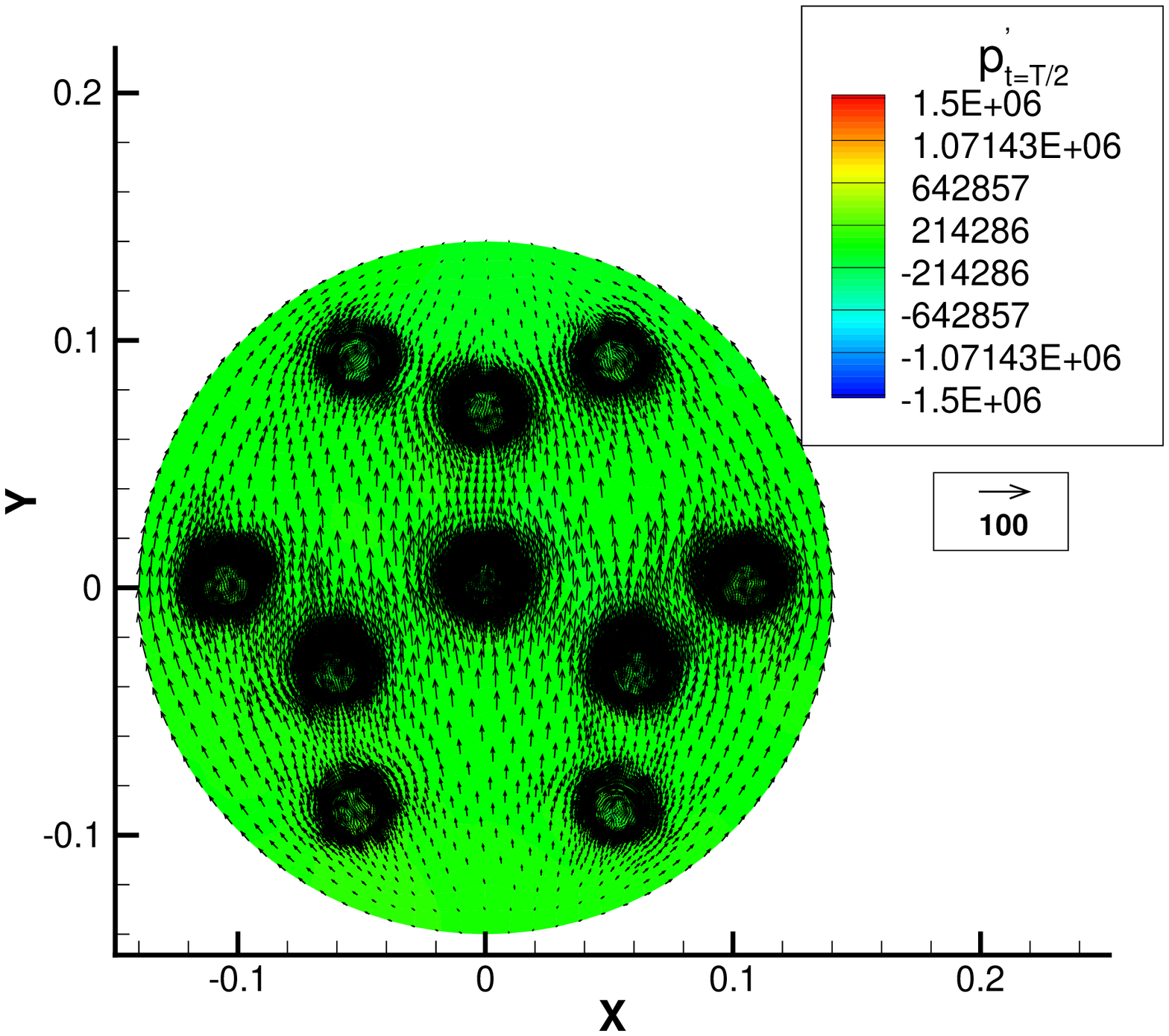}
        \label{fig:fpv_p_2qT_3200Hz_z_1cm}}
        \subfigure[Helicity mode at $t = 3T/4$]
{\includegraphics[width=0.27\textwidth]{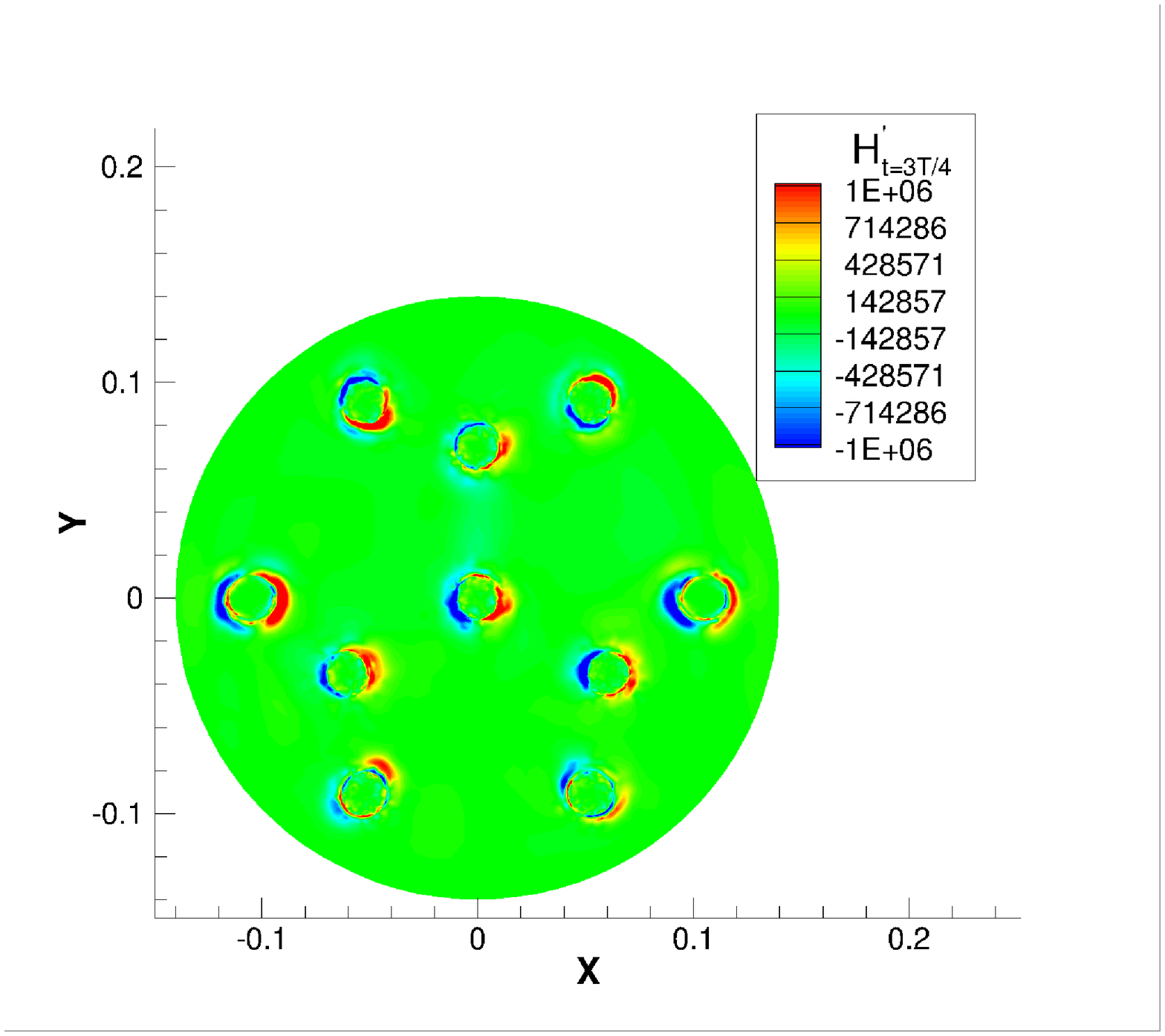}
        \label{fig:fpv_heli_3qT_3200Hz_z_1cm}}
        \subfigure[Pressure and transverse velocity modes at $t = 3T/4$]
{\includegraphics[width=0.27\textwidth]{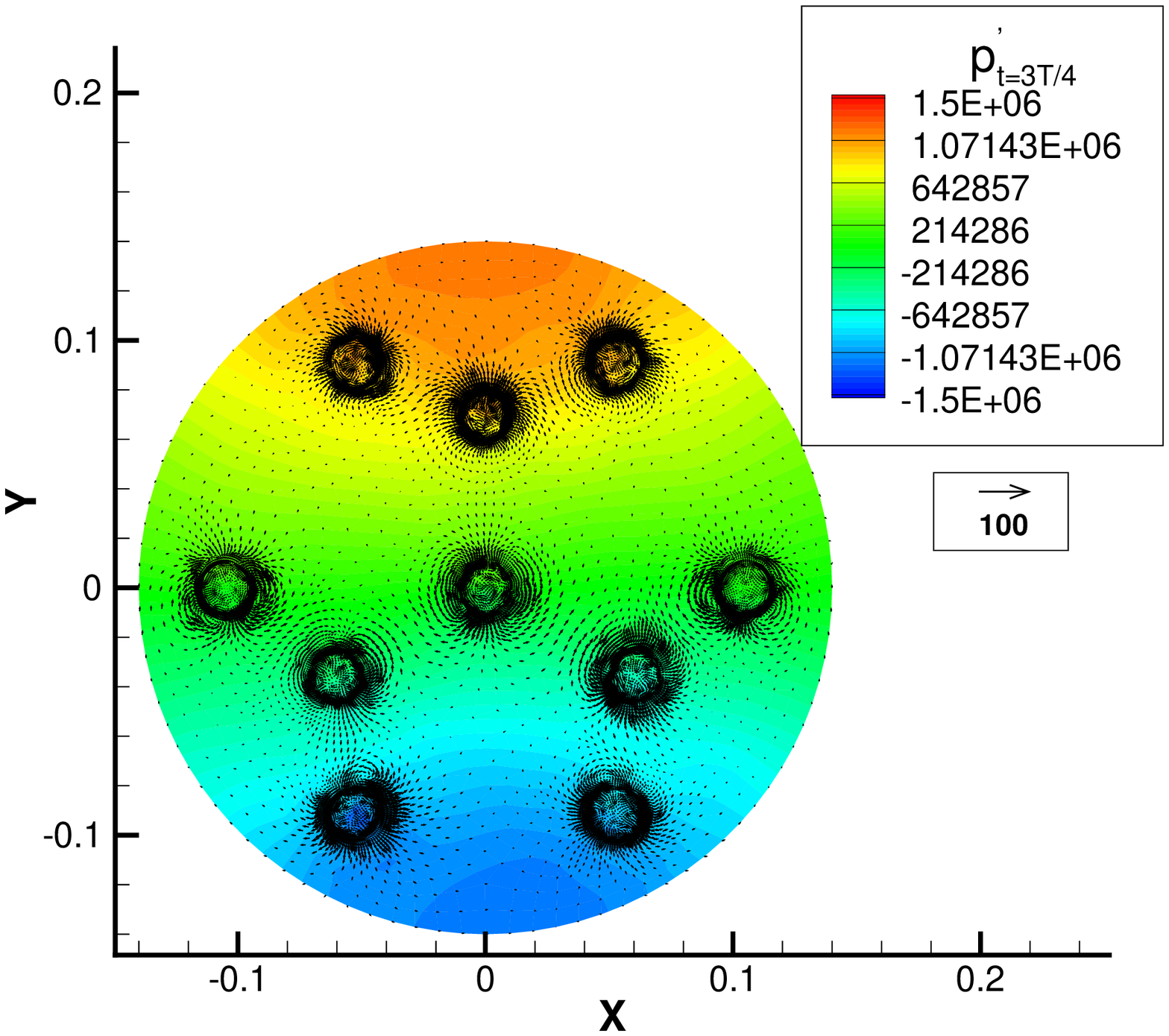}
        \label{fig:fpv_p_3qT_3200Hz_z_1cm}}
        \subfigure[Helicity mode at $t = T$]
{\includegraphics[width=0.27\textwidth]{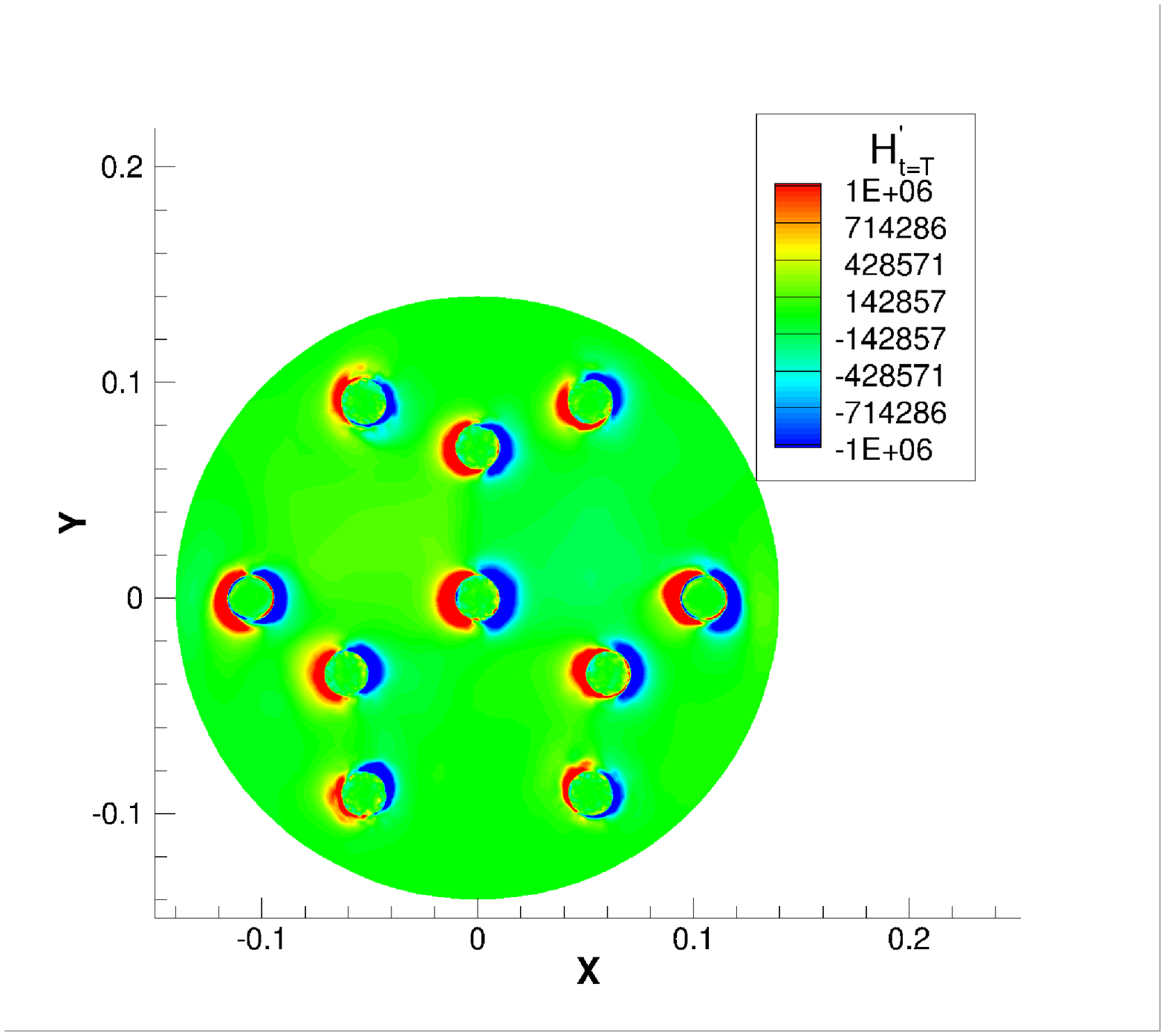}
        \label{fig:fpv_heli_4qT_3200Hz_z_1cm}}
        \subfigure[Pressure and transverse velocity modes at $t = T$]
{\includegraphics[width=0.27\textwidth]{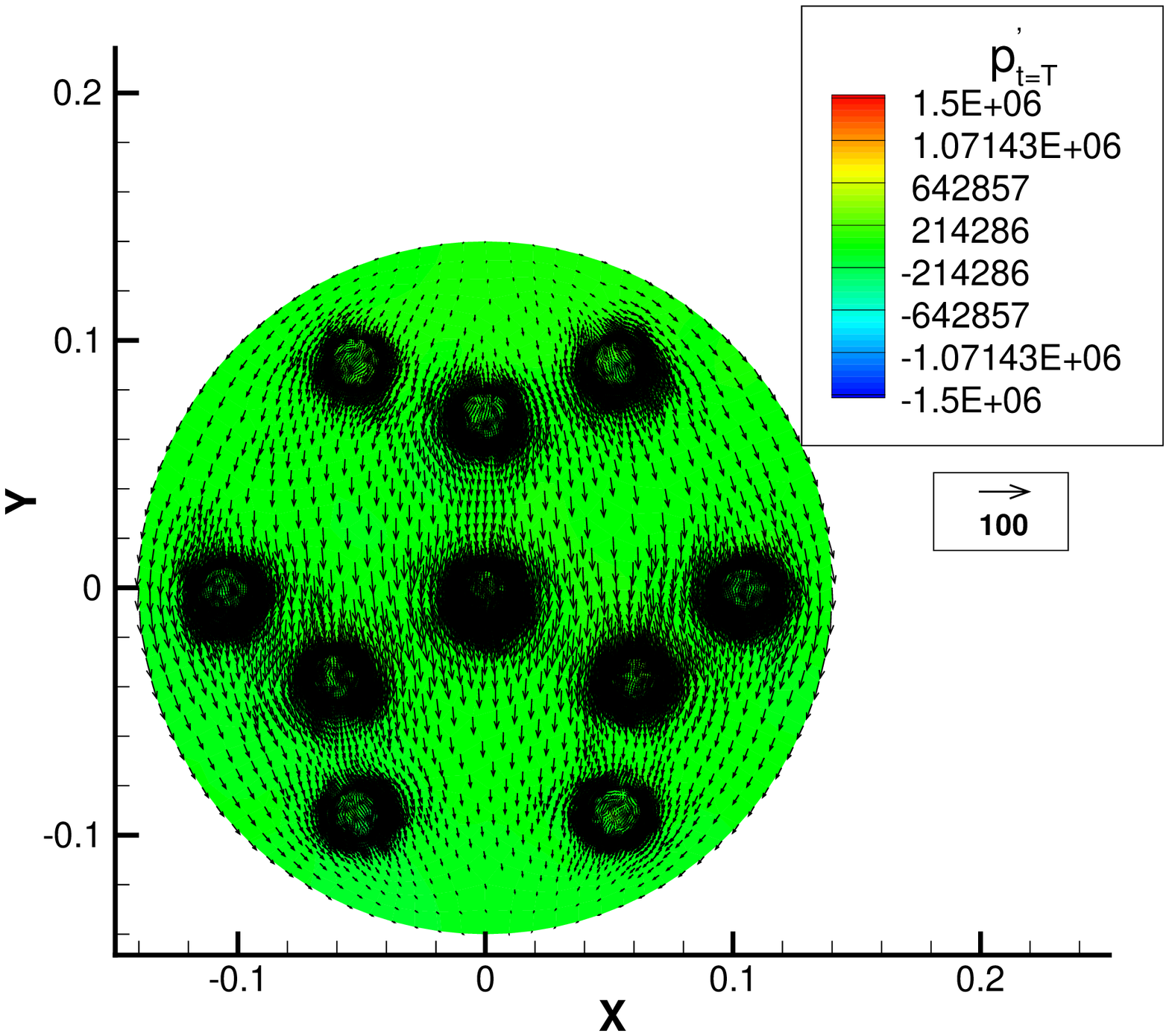}
        \label{fig:fpv_p_4qT_3200Hz_z_1cm}}
    \end{subfigmatrix}
    \caption{Correlation between pressure, transverse velocity and helicity
fluctuation (3200 Hz) for FPV on the z = 1 cm plane}
    \label{fig:corre_p_uv_heli_prime_3200Hz_z_1cm_fpv}
\end{figure}

Plotting the contours on the $z = 1$ cm cross-sectional plane, we can correlate the pressure disturbance and the helicity fluctuation. Take the computation using 
the FPV model as an example, in Figure~\ref{fig:corre_p_uv_heli_prime_3200Hz_z_1cm_fpv}, instantaneous helicity fluctuation is shown in the left column and the instantaneous
pressure disturbance is shown in the right column at the same time instants. The maximum fluctuation in helicity appears when pressure fluctuation at the pressure anti-nodes is zero, 
indicating a phase difference of about 90 degrees between the pressure disturbance and the helicity fluctuation.

The helicity fluctuation is caused by the periodic transverse flow during a cycle of the dominant instability mode. 
Figure~\ref{fig:corre_p_uv_heli_prime_3200Hz_z_1cm_fpv} also shows the instantaneous projected velocity vector on 
the $z = 1$ cm cross-sectional plane for the FPV model. 
At a $T/4$, maximum pressure fluctuations occur at the 
pressure antinodes; however, the transverse velocity can be ignored in most of the cross-sectional plane. The upward 
pressure gradient drives an upward transverse flow which achieves the maximum speed $T/4$ later. At $T/2$, the pressure gradient vanishes and after this moment pressure gradient starts to establish again but in the downward direction.
At $3T/4$, this pressure gradient reaches maximum, and the transverse flow stops temporarily. Later, the downward pressure
gradient starts to trigger a downward transverse flow. 
The fastest downward transverse flow occurs at the end of the period T and at this moment the pressure gradient no longer exists. After this moment the transverse 
flow slows down and it finally leads to an upward pressure gradient. 
This process repeat periodically and it explains how the transverse standing mode is sustained.    
The similar coupling between pressure, helicity and transverse flow is also observed for the computation using the OSK model.

\subsection{Resonance in the injectors}

\label{sec:resonance_injectors}

\patchcmd{\subfigmatrix}{\hspace{0.8cm}}{\hfill}{}{}

\begin{figure}
    \begin{subfigmatrix}{2}
        \subfigure[FPV]
{\includegraphics{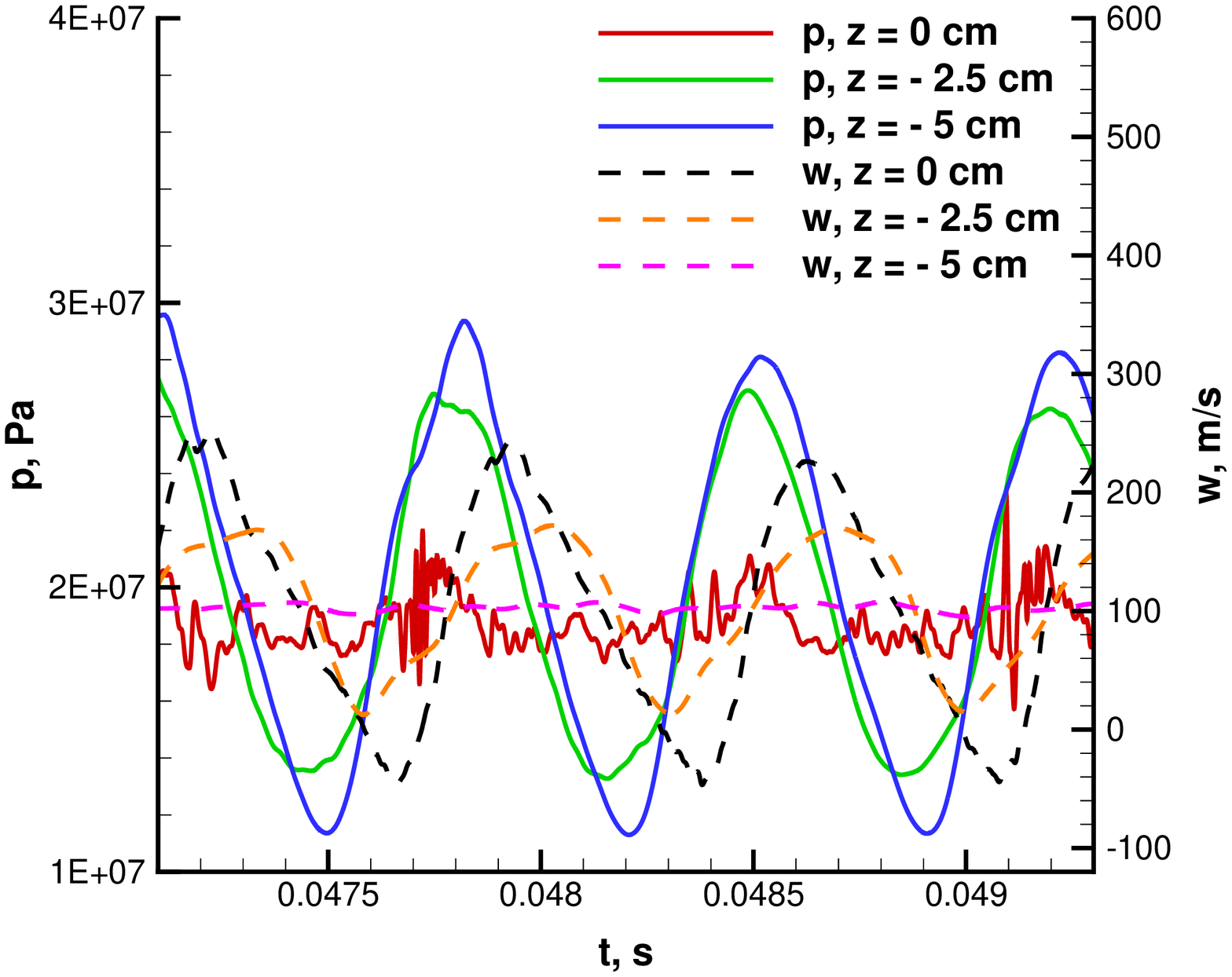}
        \label{fig:resonance_inj_p_fpv}}
        \subfigure[OSK]
{\includegraphics{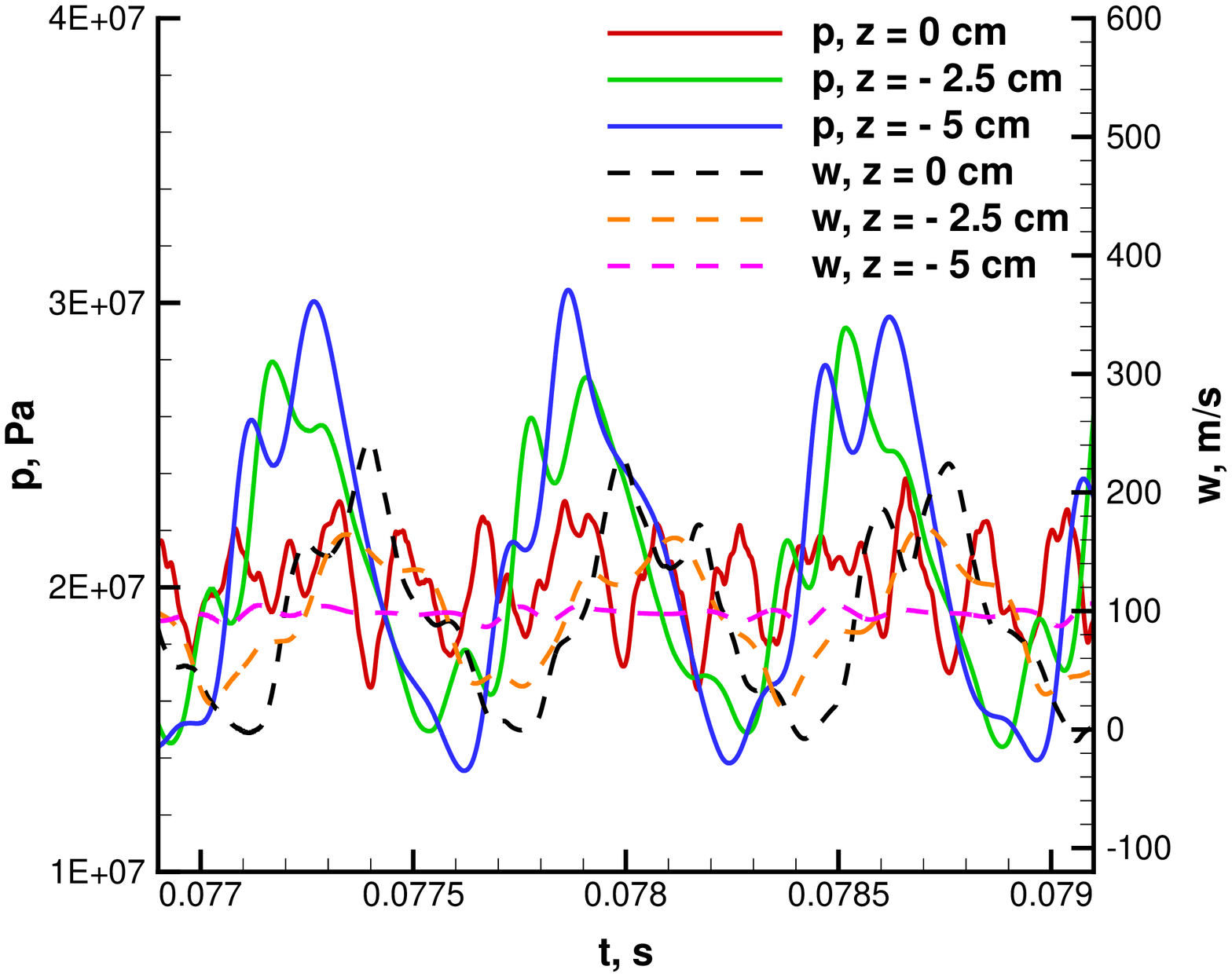}
        \label{fig:resonance_inj_p_osk}}
    \end{subfigmatrix}
    \caption{Time history of pressure and axial velocity inside the central oxidizer injector}
    \label{fig:resonance_inj_p}
\end{figure}

In this ten-injector analysis, the flow within the injector ports is simulated with a better resolution
compared to that for configurations with a larger number of injectors \cite{xiongaiaaj2020,Juntao_CST,xiongaiaaj2022}. The injector-plate face and injector
exit are at $z = 0$ cm. The coordinate $z$ inside the injector is negative. The injector port has a length of $5$
cm with $z$ varying from $- 5$ cm to $0$ cm within the port. Figure~\ref{fig:resonance_inj_p_fpv} shows the oscillatory behavior in the
central oxidizer port from an analysis using the FPV model. This resonance phenomenon is also found in other 
injectors but the oscillation magnitude can be different. The resonance inside
injectors is believed to be triggered by the acoustic instability in the combustion chamber.

A pressure antinode (velocity node) occurs at $z = - 5$ cm, the injector-port entrance, and a pressure
node (velocity antinode) occurs at $z = 0$ cm, the injector exit. Further quantitative analysis finds, at $z = - 2.5$ cm, the phase angle difference
between pressure and the axial velocity is about 90 degrees, which can be also identified in Figure~\ref{fig:resonance_inj_p_fpv}. All these features suggest that the injector
is behaving as a quarter-wave-length tube. Here, the exit of the injector behaves as the an open end while
the inlet behaves as a closed end. Although the injector inlet has mass flow, Figure~\ref{fig:resonance_inj_p_fpv} shows that the axial velocity fluctuation there is almost zero. With this key feature, we identify the similarity between the injector inlet and a closed end when discussing the behavior of the acoustic waves inside the injector port. 
For pure oxygen at $400$ K, the speed of sound
is about $379$ m/s. The theoretical frequency of this $5$ cm quarter-wave-length resonance is $1895$ Hz. The 
resonance frequency obtained in the computation is around $1400$ Hz. The discrepancy with the theoretical estimation is due to the fact
that the resonance conditions for a quarter-wave-length tube are not perfectly fulfilled at both ends of the injectors.  

Similar oscillatory behavior occurs with the OSK model as shown in Figure~\ref{fig:resonance_inj_p_osk}. Compared with the prediction of the FPV model, the resonance frequency remains almost unchanged as little reaction occurs inside the injector port with either model. However, the pressure oscillation magnitude at the injector inlet is smaller. The larger magnitude
of pressure fluctuation at an off-set frequency in the combustion chamber predicted by the OSK model might have the effect of suppressing the resonance oscillation inside the injectors.

For the injector with the current length, its resonance has no impact on the main chamber oscillation as the frequency of the former deviates dramatically from that of the latter for both models. However, a shorter injector admits higher resonant frequency according to the quarter-wave-length tube theory. For a sufficiently short injector, the acoustic oscillation in the injector might affect the dominant chamber mode since the resonant frequency in the injector might overlap with that in the chamber. On the contrary, if the current injector becomes longer, the theoretically estimated resonant frequency in the injector deviates even more from the chamber resonant frequency, and hence the resonance in the injector is more unlikely to interfere with the dominant oscillation in the chamber. This situation could be modified if overtones appear with increasing length.

\section{Conclusions}

A compressible FPV model for methane and oxygen reaction at high pressure is developed and used to study the  combustion instability for a ten-injector rocket engine. The governing equations for the FPV model are solved by a newly developed C++ code which is based on OpenFOAM 4.1. The Flamemaster code is used to generate the flamelet tables for methane/oxygen combustion at a background pressure of $200$ bar. A 12-species kinetics is adopted to represent the chemical reactions. To address the pressure effect on the reaction rate for the progress variable while reducing computational effort in looking up the tables, the reaction rate is rescaled using a power law. The OSK model is also applied in computations as a comparison to the FPV model.

In this paper, the reference stoichiometric scalar dissipation rate at the quenching limit is extremely high due to the high background pressure, leading to very small characteristic diffusion time. However, the flame structures remain qualitatively similar to those at much lower background pressures. Compared with the OSK model, the FPV model predicts lower and more realistic mean temperature in the combustion chamber since much more detailed chemical kinetics is used. Although the fuel and oxidizer are not mixed until entering the combustion chamber, combustion using both combustion models is found partially premixed.
The premixed flames dominate in the surrounding region of the propellant jets, while non-premixed flames prevails downstream the propellant jets. 
The FPV model predicts significant heat release in the downstream part of 
the combustion chamber, leading to stronger non-premixed flames in that region. The contradiction with use of FPV is noted since it is based on a non-premixed model.

Fourier analysis of the computed time histories of pressure at the near-wall probes shows that a combined first longitudinal standing wave and first tangential standing wave mode of $3200$ Hz is dominant for the FPV model while the OSK model favors a pure first tangential standing wave mode of $2600$ Hz. The significant heat release in the downstream part of the combustion chamber drives the pressure fluctuation in the region and helps to establish the longitudinal oscillation behavior for the FPV model. Helicity also fluctuates at the dominant instability frequency for both models. However, its coupling with the flame dynamics and the combustion instability is only significant in the vicinity of the propellant jets. The helicity fluctuation is in phase with the unsteady transverse flow and is 90 degrees out of phase with the pressure oscillation.

Both the FPV and OSK models predict resonance phenomenon inside the injectors. Despite slight mismatch in the resonance frequency compared with the theoretical estimation, the injectors behave similar to quarter-wave-length tubes. The resonance inside injectors is believed to be induced by the acoustic instability in the combustion chamber. The resonance inside the injectors is weaker in the OSK prediction as it is suppressed by the larger magnitude
of pressure fluctuation in the combustion chamber immediately downstream of the injector plate.

\section{Appendix: Theoretically estimated frequency of the resonant acoustic modes in a circular tube}

The linearized acoustic wave equation in cylindrical coordinates is written as
\begin{equation}
\frac{\partial^{2}\Phi}{\partial r^{2}} + \frac{1}{r}\frac{\partial \Phi}{\partial r} + \frac{1}{r^{2}}\frac{\partial^{2}\Phi}{\partial\phi^{2}}+\frac{\partial^{2}\Phi}{\partial z^{2}} = \frac{1}{c^{2}}\frac{\partial^{2}\Phi}{\partial t^{2}}
\label{awequ}
\end{equation}
where $c$ is the mean speed of sound in the chamber. The effect of the mean velocity is neglected here. Using seperation of variables, solution $\Phi$ becomes
\begin{equation}
\Phi = R(r)Z(z)\Psi(\phi)T(t)
\label{sovphi}
\end{equation}
Equation \eqref{awequ} is cast into the following form 
\begin{equation}
\left[\frac{1}{R}\frac{\partial^{2}R}{\partial r^{2}} + \frac{1}{r}\frac{1}{R}\frac{\partial R}{\partial r} + \frac{1}{\Psi r^{2}} \frac{\partial^{2}\Psi}{\partial\phi^{2}}\right] + \left[\frac{1}{Z}\frac{\partial^{2}Z}{\partial z^{2}}\right] = \frac{1}{c^{2}}\frac{1}{T}\frac{\partial^{2}T}{\partial t^{2}}
\label{newawe}
\end{equation}
For the tangential and radial direction 
\begin{equation}
\frac{1}{R}\left(\frac{\partial^{2}R}{\partial r^{2}} + \frac{1}{r}\frac{\partial R}{\partial r}\right) + \frac{1}{\Psi r^{2}} \frac{\partial^{2}\Psi}{\partial\phi^{2}} = -k^{2}_{r}
\label{trdirectione}
\end{equation}
For the axial direction 
\begin{equation}
\frac{1}{Z}\frac{\partial^{2}Z}{\partial z^{2}} = -k^{2}_{z}
\label{adirectione}
\end{equation}
In the time domain 
\begin{equation}
\frac{1}{c^{2}}\frac{1}{T}\frac{\partial^{2}T}{\partial t^{2}} = -k^{2}
\label{te}
\end{equation}
where
\begin{equation}
k^{2} = k^{2}_{z} + k^{2}_{r}
\label{kparas}
\end{equation}
Let $m$,$n$,$q$ denote the order of the mode in the tangential, radial and axial directions, respectively, then the
resonant frequency of a general acoustic mode is
\begin{equation}
f_{m,n,q} = \frac{c}{2}\sqrt{\left(\frac{\alpha_{m,n}}{\pi R_{0}}\right)^{2}+\left(\frac{q}{L_{0}}\right)^{2}}
\label{fream}
\end{equation}
where $R_{0} = 0.14 m$ is the chamber radius and $L_{0} = 0.33 m$ is the chamber length measured between the injector plate and the nozzle entrance.
$\alpha_{m,n}$ is the first roots of the derivative of the Bessel functions. 
For the first-order pure tangential mode, $m = 1$, $n = 0$, $q = 0$, $\alpha_{1,0} = 1.84$ according to Ref. \cite{foundationacoustics}, yielding 
\begin{equation}
f_{1,0,0} = \frac{c}{2} \times \frac{1.84}{\pi \times 0.14}
\label{f1T}
\end{equation}
For the first-order combined tangential and longitudinal mode, $m = 1$, $n = 0$, $q = 1$, leading to 
\begin{equation}
f_{1,0,1} = \frac{c}{2}\sqrt{\left(\frac{1.84}{\pi \times 0.14}\right)^{2}+\left(\frac{1}{0.33}\right)^{2}}
\label{f1T1L}
\end{equation}

\section{Declaration of Competing Interest}

The authors declare that they have no known competing financial interests or personal relationships that could have appeared to influence the work reported in this paper.

\section{Acknowledgments}

This research was supported by the U.S. Air Force Office of Scientific Research under grants FA9550-18-1-0392 and FA9550-22-1-0191, with Dr. Mitat Birkan as the program manager. Professor Heinz Pitsch of RWTH Aachen University is acknowledged for providing us access to the FlameMaster code. Professor Hai Wang of Stanford University is acknowledged for providing us with the FFCMy-12 chemical reaction mechanism for the methane/oxygen combustion.


\section*{References}



\begin{thebibliography}{10}

\bibliographystyle{elsarticle-num}


\bibitem{crocco_1}
L. Crocco, S. Cheng, High-frequency combustion instability in rocket motors
with concentrated combustion, J. Am. Rocket Soc. 23 (1953) 301-313.

\bibitem{crocco_2}
L. Crocco, S. Cheng, High-frequency combustion instability in rockets with
distributted combustion, Symp. (Int.) Combust. 4 (1953) 865-880.





\bibitem{crocco_3}
L. Crocco, S. Cheng, Theory of combustion instability in liquid propellant rocket
Motors, AGARDograph No. 8, Butterworths, U.K. 1956.


\bibitem{tuancnf}
T.M. Nguyen, W.A. Sirignano, The impacts of three flamelet burning regimes in nonlinear
combustion dynamics, Combust. Flame 195 (2018) 170-182. 



\bibitem{tuanaiaaj}
T.M. Nguyen, W.A. Sirignano, Spontaneous and triggered longitudinal combustion
instability in a single-injector liquid rocket combustor, AIAA J. 57 (2019) 5351-5364.


\bibitem{tuanjpp}
T.M. Nguyen, P.P. Popov, W.A. Sirignano, Longitudinal combustion instability in a rocket engine
with a single coaxial injector, J. Propul. Power 34 (2018) 354-373.


\bibitem{zeinabaiaaj}
Z. Shadram, T.M. Nguyen, A. Sideris, W.A. Sirignano, Neural network flame
closure for a turbulent combustor with unsteady pressure, AIAA J. 59 (2021) 621-635.


\bibitem{zeinabcnf}
Z. Shadram, T.M. Nguyen, A. Sideris, W. A. Sirignano, Physics-aware neural
network flame closure for combustion instability modeling in a single-injector engine, Combust. Flame 240 (2022) 111973. 








\bibitem{dlrbkdexp_a}
S. Gr\"{o}ning, D. Suslov, M. Oschwald, T. Sattelmayer, Stability behaviour of a
cylindrical rocket engine combustion chamber operated with liquid hydrogen
and liquid oxygen, $5^{th}$ European Conference for Aerospace Sciences, 2013.

\bibitem{dlrbkdexp_b}
S. Gr\"{o}ning, D. Suslov, J. Hardi, M. Oschwald, Influence of hydrogen temperature
on the acoustics of a rocket engine combustion chamber operated with LOX/H2
at representative conditions, Proceedings of Space Propulsion, 2014.

\bibitem{dlrbkdexp_c}
S. Gr\"{o}ning, J. Hardi, D. Suslov, M.Oschwald, Injector-driven combustion instabilities 
in a hydrogen/oxygen rocket combustor, J. Propul. Power  32 (2016) 560-573.


\bibitem{multicnf}
A. Urbano, L. Selle, G. Staffelbach, B. Cuenot, T. Schmitt, S. Ducruix, S. Candel, 
Exploration of combustion instability triggering using Large Eddy
Simulation of a multiple injector liquid rocket engine, Combust. Flame 169 (2016) 129-140. 

\bibitem{pci_multi_inj}
A. Urbano, Q. Douasbin, L. Selle, G. Staffelbach, B. Cuenot, T. Schmitt, S. Ducruix, S. Candel, 
Study of flame response to transverse acoustic modes
from the LES of a 42-injector rocket engine, Proc. Combust. Inst. 36 (2017) 2633-2639. 




\bibitem{energies_multi_inj}
W. Hwang, B. Sung, W. Han, K. Huh, B.J. Lee, H.S. Han, C.H. Sohn, J. Choi, 
Real-gas-flamelet-model-based numerical simulation and
combustion instability analysis of a $GH_{2}/LO_{x}$ rocket
combustor with multiple injectors, Energies, 14 (2021) 1-23. 


\bibitem{fier_multi_inj}
K. Guo, B. Xu, Y. Ren, Y. Tong, W. Nie,
Analysis of Tangential Combustion
Instability Modes in a LOX/Kerosene
Liquid Rocket Engine Based on
OpenFOAM, Front. Energy Res. 9 (2022) 1-10. 






\bibitem{xiongaiaaj2020}
J. Xiong, H. Morgan, J. Krieg, F. Liu, W.A. Sirignano,  
Nonlinear combustion instability in a multi-injector
rocket engine, AIAA J. 58 (2020) 219-235.

\bibitem{Juntao_CST}
J. Xiong, F. Liu, W.A. Sirignano, 
Combustion dynamics simulation of a 30-injector rocket
engine, Combust. Sci. Technol. 194 (2022) 1914-1942. 


\bibitem{xiongaiaaj2022}
J. Xiong, F. Liu, W.A. Sirignano, 
Combustion Simulation of an 82-injector Rocket
Engine, AIAA J. 60 (2022) 4601-4613.

\bibitem{Jensen1989}
R.J. Jensen, H.C. Dodson, and S.E. Claflin, 
LOX/Hydtocarbon Combustion lnstability
Investigation, NASA-CR-182249, 1989.








\bibitem{srinivasan}
S. Srinivasan, R. Ranjan, S. Menon, 
Flame dynamics during combustion instability in a high-pressure, shear-coaxial injector combustor, 
Flow Turbul. Combust. 94 (2015) 237-262.

\bibitem{Garby}
R. Garby, L. Selle, T. Poinsot, 
Analysis of the impact of heat losses on an unstable model
rocket-engine combustor using large-eddy simulation,
$48^{th}$ AIAA/ASME/SAE/ASEE
Joint Propulsion Conference and Exhibit, 2012.


\bibitem{sardeshmukh}
S.V. Sardeshmukh, S.D. Heister, W.E. Anderson, 
Prediction of combustion instability with
detailed chemical kinetics, 
$53^{rd}$ AIAA Aerospace Sciences Meeting, 2015.

 
\bibitem{harvazinski_aiaa_c}
M.E. Harvazinski, D.G. Talley, V. Sankaran, 
Application of detailed chemical kinetics to
combustion instability modeling, 
$54^{th}$ AIAA Aerospace Sciences Meeting, 2016.


\bibitem{harvazinski_phd_diss} 
M. E. Harvazinski, 
Modeling self-excited combustion instabilities using a combination of two
and three dimensional simulations, Ph.D. thesis, Purdue University, West Lafayette, Indiana, May, 2012.





\bibitem{pecnikaiaaj2012}
R. Pecnik, V.E. Terrapon, F. Ham, G. Iaccarino, H. Pitsch, 
Reynolds-averaged Navier-Stokes simulations of the HyShot II scramjet, AIAA J. 50 (2012) 1717-1732.

\bibitem{saghafiancnf}
A. Saghafian, V.E. Terrapon, F. Ham, G. Iaccarino, H. Pitsch, 
An efficient flamelet-based combustion
model for compressible flows, Combust. Flame 162 (2015) 652-667. 


\bibitem{piercemoinjfm}
C.D. Pierce, P. Moin, 
Progress-variable approach for large-eddy simulation of non-premixed
turbulent combustion, J. Fluid Mech. 504 (2004) 73-97. 



\bibitem{petersnotes} 
N. Peters, B. Rogg, 
Reduced kinetic mechanism for applications in combustion system,
Lecture notes in physics, Springer-Verlag, Berlin, New York, 1993.


\bibitem{pant}
T.A. Pant, C. Huang, C. Han, S.V. Sardeshmukh, W.E. Anderson, H. Wang, 
Flamelet modeling studies of a continuously variable resonance combustor, 
$54^{th}$ AIAA Aerospace Sciences Meeting, 2016.


\bibitem{nppecs}
N. Peters, 
Laminar diffusion flamelet models in non-premixed turbulent combustion,
Prog. Energy Combust. Sci. 10 (1984) 319-339.

\bibitem{nptc}
N. Peters, Turbulent combustion, Second Edition, Cambridge University Press,
Cambridge, U.K., 2000.



\bibitem{sstdes1}
F.R. Menter, M. Kuntz, R. Langtry, 
Ten Years of Industrial Experience with the SST Turbulence Model,
$4^{th}$ Internal Symposium on Turbulence, Heat and Mass Transfer (2003) 625-632.


\bibitem{Mentersstkw}
F. Menter, Two-equation eddy-viscosity turbulence models for
engineering applications, 
AIAA J. 32 (1994) 1598-1605.


\bibitem{ihme_1}
M. Ihme, C.M. Cha, H. Pitsch, Prediction of local extinction and re-ignition 
effects in non-premixed turbulent combustion using a flamelet/progress variable
approach: 1. A priori study and presumed pdf closure, Combust. Flame 155
(2008) 70-89. 

\bibitem{ihme_2}
M. Ihme, C.M. Cha, H. Pitsch, Prediction of local extinction and re-ignition 
effects in non-premixed turbulent combustion using a flamelet/progress variable
approach: 2. Application in LES of sandia flames d and e, Combust. Flame 155
(2008) 90-107.






\bibitem{flamemaster}
H. Pitsch, FlameMaster, A C++ Computer Program for 0D Combustion and 1D Laminar Flame Calculations, 
Ver. V4.0.0, https://www.itv.rwth-aachen.de/en/downloads/flamemaster/ [retrieved 7 October 2019].


\bibitem{stanfordffcmy}
G.P. Smith, Y. Tao, H. Wang, Foundational Fuel Chemistry Model Version 1.0 (FFCM-1), 
https://web.stanford.edu/group/haiwanglab/FFCM1/pages/FFCM1.html, 2016.





\bibitem{ffcmcnf}
Y. Tao, G.P. Smith, H. Wang, 
Critical kinetic uncertainties in modeling hydrogen/carbon monoxide,
methane, methanol, formaldehyde, and ethylene combustion,
Combust. Flame 195 (2018) 18-29. 


\bibitem{mullerflameletfoam}
H. Muller, F. Ferraro, F. Pfitzner, 
Implementation of a steady laminar flamelet model for
non-premixed combustion in LES and RANS simulations, 
$8^{th}$ International OpenFOAM Workshop, 2013.




\bibitem{nasa1993}
B.J. McBride, S. Gordon, M. Reno, 
Coefficients for Calculating Thermodynamic and Transport
Properties of Individual Species,
NASA-TM-4513, 1993.





\bibitem{sstdes2}
M. Strelets, Detached Eddy Simulation of Massively Separated
Flows,
$39^{th}$ AIAA Aerospace Sciences Meeting, 2001.



\bibitem{multi_regime_pitsh}
E. Knudsen, H. Pitsch, 
Capabilities and limitations of multi-regime flamelet combustion models,
Combust. Flame 159 (2012) 242-264.


\bibitem{flame_structure}
K. Seshadri, N. Peters, 
Asymptotic structure and extinction of methane-air dif-
fusion flames, Combust. Flame 73 (1988) 23-44.


\bibitem{foundationacoustics}
E. Skudrzyk, The Foundations of
Acoustics-Basic Mathematics and Basic Acoustics, First Edition, Springer-Verlag,
New York, U.S., 1971.




\end{thebibliography}
\end{document}